\documentclass[a4paper,12pt,openany]{book}
\usepackage{indentfirst}
\usepackage{pdfpages}

\usepackage[utf8]{inputenc}
\usepackage[T1]{fontenc}

\usepackage{amsmath}
\usepackage{amssymb}
\usepackage{amsthm} 
\usepackage{bbold} 

\usepackage{braket} 
\usepackage{graphicx} 
\usepackage{physics}
\usepackage{hyperref}
\usepackage[left=3cm, right=3cm, top=4 cm, bottom=4 cm]{geometry}
\usepackage{fancyhdr}
\usepackage{caption}
\usepackage{subcaption}
\usepackage[sorting=none,style=numeric-comp]{biblatex}
\usepackage{xcolor}
\usepackage{bm}

\title{Tesi completa}
\author{Gabriele Gatto}
\date{October 2024}

\def\slashchar#1{\setbox0=\hbox{$#1$}
   \dimen0=\wd0
   \setbox1=\hbox{/} \dimen1=\wd1
   \ifdim\dimen0>\dimen1
      \rlap{\hbox to \dimen0{\hfil/\hfil}}
      #1
   \else
      \rlap{\hbox to \dimen1{\hfil$#1$\hfil}}
      /
   \fi}

\def\nn{\nonumber}

\def\bei{\begin{itemize}}
\def\ei{\end{itemize}}

\def\beeq{\begin{eqnarray}} 
\def\beqa{\begin{eqnarray}}
\def\bea{\begin{eqnarray}}

\def\eea{\end{eqnarray}}
\def\eqa{\end{eqnarray}}
\def\eeeq{\end{eqnarray}}

\def\eqar{\end{array}}
\def\beqar{\begin{array}}

\def\beas{\begin{eqnarray*}}
\def\beqas{\begin{eqnarray*}}

\def\eqas{\end{eqnarray*}}
\def\eeas{\end{eqnarray*}}

\def\beq{\begin{equation}} 
\def\be{\begin{equation}}

\def\ee{\end{equation}}
\def\eq{\end{equation}}
\def\eeq{\end{equation}}

\def\beqd{\begin{displaymath}}
\def\eeqd{\end{displaymath}}
\def\eqd{\end{displaymath}}

\def\beeq{\begin{eqnarray}} \def\eeeq{\end{eqnarray}}

\newcommand{\fin}{\end{document}}

\newcommand{\QXQq}{X_{Q\bar{Q}q\bar{q}}}

\newcommand{\QXbq}{X_{b\bar{b}q\bar{q}}}
\newcommand{\QXQu}{X_{Q\bar{Q}u\bar{u}}}
\newcommand{\QXQs}{X_{Q\bar{Q}s\bar{s}}}
\newcommand{\QXcu}{X_{c\bar{c}u\bar{u}}}
\newcommand{\QXcs}{X_{c\bar{c}s\bar{s}}}
\newcommand{\QXbu}{X_{b\bar{b}u\bar{u}}}
\newcommand{\QXbs}{X_{b\bar{b}s\bar{s}}}
\newcommand{\TQQ}{T_{4Q}}
\newcommand{\TQQZpp}{T_{4Q}(0^{++})}
\newcommand{\TQQTpp}{T_{4Q}(2^{++})}
\newcommand{\TQc}{T_{4c}}
\newcommand{\TQcZpp}{T_{4c}(0^{++})}
\newcommand{\TQcTpp}{T_{4c}(2^{++})}
\newcommand{\TQb}{T_{4b}}
\newcommand{\TQbZpp}{T_{4b}(0^{++})}
\newcommand{\TQbTpp}{T_{4b}(2^{++})}
\newcommand{{\symJethad}}{\tt symJETHAD}
\newcommand{{\Jethad}}{\tt JETHAD}
\newcommand{\BCs}{B_c(^1S_0)}
\newcommand{\Bss}{B_c(^3S_1)}
\newcommand{{\HFNRevo}}{\textsc{HF-NRevo}}
\newcommand{\HENLOp}{{\rm HE}\mbox{-}{\rm NLO^+}}
\newcommand{\vqTTa}{\langle {\vec q}_T^{\;2} \rangle}
\newcommand{\Jpsi}{J/\psi}
\newcommand{\as}{\alpha_s}
\newcommand{\E}{{\cal E}}
\newcommand{\B}{{\cal B}}
\newcommand{\DY}{\Delta Y}
\newcommand{\drv}{{\rm d}}
\newcommand{\CmNLLp}{{\cal C}_m^\NLLp}
\newcommand{\CmLL}{{\cal C}_m^\LL}

\newcommand{\NLLp}{{\rm NLLA/NLO^+}}
\newcommand{\LL}{{\rm LL/LO}}
\newcommand{\MSb}{\overline{\rm MS}}
\newcommand{\CmHENLOp}{{\cal C}_m^{{\rm HE}\text{-}{\rm NLO}^+}}

\begin{document}
\noindent

\includepdf[pages=-]{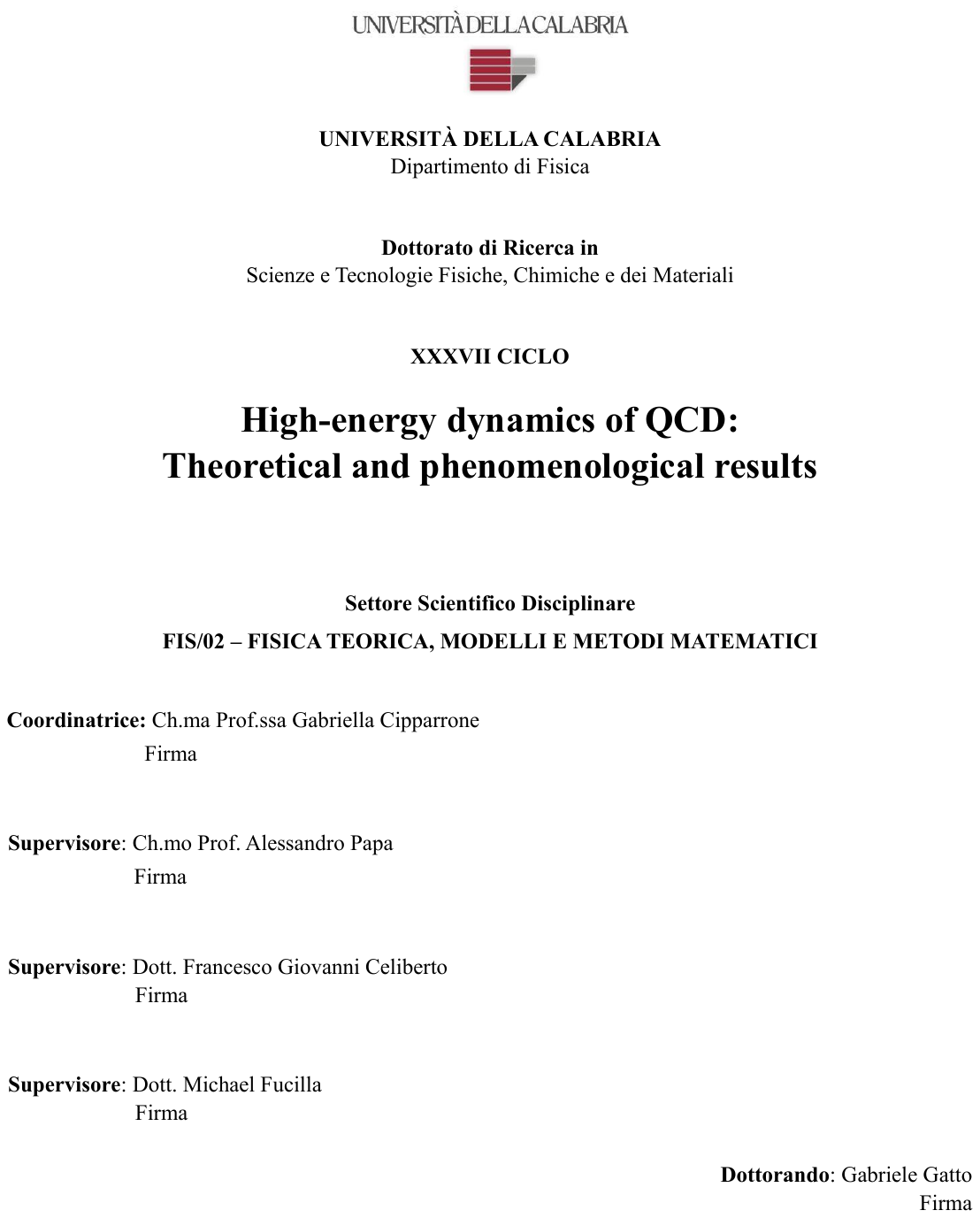}

\newpage%

\section*{The Road Not Taken}
\thispagestyle{empty} 

Two roads diverged in a yellow wood,

And sorry I could not travel both

And be one traveler, long I stood

And looked down one as far as I could

To where it bent in the undergrowth;\\

Then took the other, as just as fair,

And having perhaps the better claim,

Because it was grassy and wanted wear;

Though as for that the passing there

Had worn them really about the same,\\

And both that morning equally lay

In leaves no step had trodden black.

Oh, I kept the first for another day!

Yet knowing how way leads on to way,

I doubted if I should ever come back.\\

I shall be telling this with a sigh

Somewhere ages and ages hence:

Two roads diverged in a wood, and I—

I took the one less traveled by,

And that has made all the difference.\\

\small{\textit{Written by Robert Frost in 1916}}\\
\chapter*{Acknowledgements}
First and foremost, I would like to sincerely thank Alessandro, without whom I would never have been able to complete this wonderful journey. He has taught and shared so much with me during this path—not only new notions and concepts but also invaluable values that have helped me grow as a person. He has always supported me through my numerous challenges, and for this and many other reasons, he represents to me the very essence of what a great teacher should be. I hope that, in this world, other students will be as fortunate as I have been to meet someone like him.
Moreover, I thank Alessandro for teaching me the importance of working with dedication and passion. I hope that one day I will be able to dedicate myself to a profession that I truly love and pour into it the same passion that Alessandro manages to convey daily through physics.\\

I would like to thank Francesco, my co-supervisor, for his invaluable support throughout this journey. I have learned so much from him, particularly in the field of numerical analysis, as well as the importance of passion and dedication in one’s work. I deeply admire the work he does and the expertise he brings to his field, which has been an inspiration to me. His guidance has been a crucial part of my growth, and I am truly grateful for the contributions he has made to my academic and personal development.\\

I would like to thank Michael, who over these three years has been not only my co-supervisor but also a friend. He has been a key figure in my academic journey, showing patience in handling my messages, phone calls, silly questions, and anxieties. He has always set a good example, demonstrating that with dedication, even the most distant goals can be reached. In my pursuit of realizing my dreams, I hope to show the same determination and wisdom that have defined his inspiring path.  
Beyond all this, Michael has also been an important presence outside of physics, someone with whom I could share moments of lightness and friendship. He has been a trusted confidant, someone I have grown deeply fond of, and someone I hope to always have in my life as a friend. {Amicus fidelis protectio fortis: qui autem invenit illum invenit thesaurum}. Thanks to him, the three-month period in France was not only an opportunity for growth but also a pleasant experience filled with wonderful memories. It is no exaggeration: without Michael's help and support, I would never have been able to reach the end of this journey. I wish him nothing but the best for all the years to come. May he achieve every dream he aspires to.\\

I would like to thank Samuel and Lech for giving me the opportunity to collaborate with them during my visiting period in France at IJCLAB (Université Paris-Saclay). It was an incredible experience, not only from a professional standpoint but also on a personal level. Thanks to their kindness and understanding, I immediately felt at ease in a new environment, and they made it possible for me to carry with me not only the enchanting city of Paris but also all the memories of that time, which I will forever hold dear. I thank their entire group for the warmth with which they welcomed me.
During this period, I also had the pleasure of meeting Saad, whom I would like to thank in particular for assisting me in the project I was involved in and for teaching me so many things that will always be valuable to me.\\

I would like to thank my entire research group, and in particular Luigi, for everything he has shared with me. Through his lessons and our collaboration, he has imparted so much to me, and for that, I am deeply grateful.\\

I extend my sincere gratitude to Krzysztof Kutak and Cyrille Marquet for kindly agreeing to be the referees for this work.\\

I would like to dedicate this thesis, finally, to all the dearest people in my life.

\chapter*{Abstract}
The unprecedented center-of-mass energies achieved by modern accelerators, such as the Large Hadron Collider (LHC), and those expected from next-generation facilities like the Electron-Ion Collider (EIC) and Future Circular Collider (FCC), provide an extraordinary opportunity to probe hadronic matter under the most extreme conditions ever explored. These energy regimes enable access to the Regge-Gribov (or semi-hard) limit of quantum chromodynamics (QCD), characterized by the hierarchy $s \gg \{Q^2 \} \gg \Lambda_{{\rm{QCD}}}^2$, where $\sqrt{s}$ is the center-of-mass energy, $\{ Q \}$ represents a set of hard scales defining the process, and $\Lambda_{{\rm{QCD}}}$ is the QCD mass scale.\\

At the heart of this work lies the exploration of the Balitsky-Fadin-Kuraev-Lipatov (BFKL) formalism, a fundamental framework in high-energy QCD in the Regge limit. The discussion begins by introducing Regge theory, the foundational concept behind the analysis of high-energy scattering, emphasizing the role of Regge poles and trajectories in determining the asymptotic behavior of scattering amplitudes.
A pivotal focus is placed on the BFKL equation, which governs the evolution in the energy of scattering amplitudes with vacuum quantum numbers exchanged in the $t$-channel. The discussion elaborates on the derivation and solution of the BFKL equation, initially within the Leading Logarithmic Approximation (LLA). This solution predicts a power-law growth of the cross section with energy, a key result for understanding semi-hard processes in QCD. The exploration extends to the Next-to-Leading Logarithmic Approximation (NLLA), incorporating higher-order corrections to the gluon Regge trajectory, impact factors, and the kernel.\\

As a direct application of the theory, the computation of real corrections to the Higgs boson impact factor at next-to-leading order (NLO) is presented, explicitly incorporating the physical mass of the top quark. The analysis focuses on extending the NLO calculation in the infinite top-mass limit to physical top mass. As a critical component of high-energy resummation within the BFKL framework, the Higgs impact factor plays a pivotal role in precision studies of Higgs production processes at large rapidity separations, such as inclusive Higgs hadroproduction and Higgs production in association with jets or hadrons. These processes are of particular relevance to current LHC experiments and future colliders like the FCC.
At NLO, the Higgs impact factor was previously derived in the infinite top-mass approximation. However, extending this calculation to finite top-mass at NLO introduces additional complexities, such as two-loop amplitudes with massive propagators. This study details the computation of real corrections, providing expressions for both quark- and gluon-initiated processes. A thorough examination of infrared and rapidity divergences is undertaken, ensuring consistency with BFKL factorization and demonstrating the cancellation of divergences in the full impact factor.
The inclusion of finite-top-mass effects is essential for improving the precision of theoretical predictions, particularly in kinematic regimes where the Higgs boson transverse momentum approaches the top-quark mass. The results represent a significant step toward a complete NLO description of the Higgs impact factor, paving the way for future studies of Higgs phenomenology in high-energy collisions.\\

Also, the semi-inclusive production of exotic matter in the semi-hard regime is explored, with two primary goals identified: the investigation of exotic state formation in this kinematic domain and the use of these studies as a platform to test the BFKL dynamics at high-energy colliders. Attention is directed toward the production mechanisms of bottomonium-like states, specifically doubly bottomed tetraquarks ($X_{b\bar{b}q\bar{q}}$) and fully bottomed tetraquarks ($T_{4b}$). A hybrid framework, combining collinear factorization and high-energy BFKL dynamics (referred to as hybrid factorization), is employed to integrate perturbative and non-perturbative aspects of the hadroproduction process. Production is analyzed in a variable-flavor number scheme (VFNS), and collinear fragmentation functions (FFs) designed for tetraquark states are incorporated. Updated parametrizations, \texttt{TQHL1.1} and \texttt{TQ4Q1.1}, are provided, offering enhanced precision in predictions for tetraquark production and their associated jets at collider energies of 14 TeV and 100 TeV.\\

The study of diffractive di-hadron production is carried out in the high-energy regime, leveraging the theoretical framework of saturation physics and the Color Glass Condensate (CGC) formalism. Saturation occurs in the small Bjorken-\(x\) limit, a region where gluon densities grow exponentially with energy, resulting in a state so densely packed with gluons that further growth is suppressed due to non-linear recombination effects. The saturation scale defines the transition to this regime, depending on both the small-\(x\) gluon density and the target's mass number, with effects being more pronounced for large nuclei.
A leading-order analysis of diffractive di-hadron production is presented, employing collinear factorization and the shockwave formalism. The resulting expressions for the cross section include fragmentation functions and transverse momentum dependencies, forming the foundation for numerical computations.
This research is ongoing, with the development of numerical plots currently in progress. These plots will provide deeper insights into the dynamics of semi-inclusive diffractive processes, particularly interesting for experimental studies at facilities such as the EIC.

\chapter*{Sintesi in lingua italiana}
Le energie senza precedenti raggiunte dagli acceleratori moderni, come il Large Hadron Collider (LHC), e quelle attese dalle strutture di prossima generazione, come l'Electron-Ion Collider (EIC) e il Future Circular Collider (FCC), offrono un'opportunità straordinaria per studiare la materia adronica in condizioni estreme mai esplorate prima. Questi regimi energetici permettono di accedere al limite di Regge-Gribov (o semi-duro) della cromodinamica quantistica (QCD), caratterizzato dalla gerarchia \(s \gg \{Q^2\} \gg \Lambda_{{\rm{QCD}}}^2\), dove \(\sqrt{s}\) rappresenta l'energia nel centro di massa, \(\{Q\}\) indica un insieme di scale dure che definiscono il processo, e \(\Lambda_{{\rm{QCD}}}\) è la scala di massa della QCD.\\

Al centro di questo lavoro si trova l'esplorazione del formalismo di Balitsky-Fadin-Kuraev-Lipatov (BFKL), un quadro fondamentale nella QCD ad alta energia nel limite di Regge. La discussione inizia introducendo la teoria di Regge, il concetto di base per l'analisi degli urti ad alta energia, sottolineando il ruolo dei poli e delle traiettorie di Regge nel determinare il comportamento asintotico delle ampiezze di scattering. 
Un'attenzione particolare è dedicata all'equazione di BFKL, che regola l'evoluzione delle ampiezze di scattering con numeri quantici di vuoto scambiati nel canale \(t\). La trattazione approfondisce la derivazione e la soluzione dell'equazione di BFKL, inizialmente nell'Approssimazione dei Logaritmi Dominanti (LLA). Questa soluzione prevede una crescita a legge di potenza della sezione d'urto con l'energia, un risultato chiave per comprendere i processi semi-duri nella QCD. L'esplorazione si estende poi all'Approssimazione dei Logaritmi Sotto-dominanti (NLLA), includendo correzioni di ordine superiore alla traiettoria di Regge del gluone, ai fattori d'impatto e al kernel.\\

Come applicazione diretta della teoria, viene presentato il calcolo delle correzioni reali al fattore d'impatto del bosone di Higgs all'ordine sotto-dominante (NLO), includendo esplicitamente la massa fisica del quark top. L'analisi si concentra sull'estensione del calcolo NLO nel limite di massa infinita del top alla massa fisica del top. Come elemento cruciale della risommatoria ad alta energia nel formalismo BFKL, il fattore d'impatto di Higgs gioca un ruolo fondamentale negli studi di precisione sui processi di produzione dell'Higgs con grandi separazioni in rapidità, come la produzione inclusiva di Higgs e la produzione di Higgs associata a jet o adroni. Questi processi sono particolarmente rilevanti per gli esperimenti attuali a LHC e per i futuri collisori come FCC.\\
A NLO, il fattore d'impatto di Higgs è stato precedentemente derivato nell'approssimazione di massa infinita del top. Tuttavia, estendere questo calcolo a una massa finita del top al NLO introduce ulteriori complessità, come le ampiezze a due loop con propagatori massivi. Questo studio riguarda il calcolo delle correzioni reali, fornendo espressioni per i processi iniziati sia da quark che da gluoni. Viene condotto un esame approfondito delle divergenze infrarosse e in rapidità, garantendo la coerenza con la fattorizzazione BFKL e dimostrando la cancellazione delle divergenze nel fattore d'impatto completo.\\
L'inclusione degli effetti di massa finita del top è essenziale per migliorare la precisione delle previsioni teoriche, in particolare nei regimi cinematici in cui il momento trasverso del bosone di Higgs si avvicina alla massa del quark top. I risultati rappresentano un passo significativo verso una descrizione completa al NLO del fattore d'impatto di Higgs, aprendo la strada a futuri studi sulla fenomenologia dell'Higgs nelle collisioni ad alta energia.\\

Inoltre, viene esplorata la produzione semi-inclusiva di materia esotica nel regime semi-duro, con due obiettivi principali: indagare la formazione di stati esotici in questo dominio cinematico e utilizzare tali studi come piattaforma per testare la dinamica BFKL nei collisori ad alta energia. L'attenzione è rivolta ai meccanismi di produzione di stati simili al bottomonio, in particolare i tetraquark a doppio bottom (\(X_{b\bar{b}q\bar{q}}\)) e i tetraquark completamente bottom (\(T_{4b}\)). Un quadro ibrido, che combina la fattorizzazione collineare e la dinamica BFKL ad alta energia (denominata fattorizzazione ibrida), viene impiegato per integrare gli aspetti perturbativi e non perturbativi del processo di adroproduzione. La produzione viene analizzata in uno schema a numero di flavor variabili (VFNS), e vengono incorporati funzioni di frammentazione collineari (FF) progettate per stati di tetraquark. Sono fornite parametrizzazioni aggiornate, \texttt{TQHL1.1} e \texttt{TQ4Q1.1}, offrendo una precisione maggiore nelle previsioni per la produzione di tetraquark e i relativi jet alle energie dei collisori di 14 TeV e 100 TeV.\\

Lo studio della produzione diffrattiva di di-adroni viene condotto nel regime di alta energia, sfruttando il quadro teorico della fisica della saturazione e il formalismo del Color Glass Condensate (CGC). La saturazione si verifica nel limite di Bjorken-\(x\) piccolo, una regione in cui le densità di gluoni crescono esponenzialmente con l'energia, portando a uno stato così densamente popolato di gluoni che un'ulteriore crescita viene soppressa a causa di effetti di ricombinazione non lineari. La scala di saturazione definisce la transizione a questo regime, dipendendo sia dalla densità di gluoni a \(x\) piccolo sia dal numero di massa del bersaglio, con effetti più pronunciati per nuclei di grandi dimensioni.\\
Viene presentata un'analisi al primo ordine della produzione diffrattiva di di-adroni, impiegando la fattorizzazione collineare e il formalismo delle onde d'urto. Le espressioni risultanti per la sezione d'urto includono funzioni di frammentazione e dipendenze dal momento trasverso, formando la base per i calcoli numerici.\\
Questa ricerca è in corso, con lo sviluppo di grafici numerici attualmente in fase di completamento. Questi grafici forniranno approfondimenti più profondi sulla dinamica dei processi diffrattivi semi-inclusivi, particolarmente interessanti per studi sperimentali in strutture come l'EIC.

\newpage 

\thispagestyle{empty} 
\mbox{} 

\newpage 
\tableofcontents 
\newpage 
\thispagestyle{empty} 
\mbox{} 
\chapter*{Introduction}
\addcontentsline{toc}{chapter}{Introduction} 
\markboth{INTRODUCTION}{INTRODUCTION} 
In ancient times, Heraclitus stated that Nature loves to conceal itself. Despite lacking the knowledge and tools we possess today, he had grasped a fundamental truth: the natural world reveals itself in enigmatic forms, eluding full human understanding. Every attempt to shed light on our surroundings clashes with the inherent elusiveness of reality, which resists clear definitions and unveils itself only partially, like a riddle that defies our thirst for knowledge.

The attempts to impose order on the chaos underlying the universe and to assign it names or identities have persisted throughout history, always accompanied by the awareness that we will never attain an absolute truth. Some argue for the existence of a final purpose, a teleological principle that governs the workings of the universe; others, on the contrary, vehemently deny any rational or divine end. Among the latter was Arthur Schopenhauer, who conceived reality as the product of a blind, boundless, and unrestrainable impulse — a primordial force he called \textit{Will}. This Will is irrational and relentless, manifesting itself through the forms of time, space, and causality. Everything we perceive is merely a reflection of this untamed force, a purposeless desire that pervades all things: from inanimate objects to living beings, from the most inert matter to consciousness itself. In this perspective, the laws of physics and chemistry are nothing more than one of the many manifestations of the Will, whose culmination is revealed in human existential suffering.\\

However, while philosophy teaches us to distrust appearances and to suspect that ultimate reality may remain inaccessible, science has refined tools capable of probing the world beyond the veil of perception. Modern physics, in particular, has succeeded in translating the intuition of the invisible into a rigorous mathematical language, unveiling universal principles that govern matter on inconceivably small scales. Among these, the \textit{strong interaction} stands out as one of the four fundamental forces of nature. \textit{Quantum chromodynamics} (QCD) was developed precisely to pierce the \textit{veil of Maya} that separates us from the understanding of this enigmatic force and to shed light on its profound structural laws. Although the theory has achieved extraordinary success in describing the dynamics of strong interactions, a fully comprehensive understanding remains elusive.

Let us now examine, in broad terms, some fundamental characteristics of this discipline. QCD is based on a non-Abelian symmetry principle, represented by the $SU(3)$ group, which describes the existence of three color charges (a quantic defining property) in quarks, the ultimate constituents of matter. The formalism of the theory gradually emerged through the interplay between theoretical predictions and experimental observations. A crucial breakthrough occurred in 1973, when David Gross, Frank Wilczek, and David Politzer discovered the phenomenon of \textit{asymptotic freedom}. This principle states that the strength of the interaction between quarks and gluons decreases as energy increases, allowing the fundamental constituents of matter to appear in nearly free states at infinitesimally small distances.

Another phenomenon that challenges our logical understanding and remains one of the unresolved mysteries of QCD is the \textit{color confinement}: quarks and gluons cannot exist in isolation but always combine into color-neutral states. In other words, quarks are bound by an inescapable law — they can never be observed individually but only in combinations that give rise to hadrons, the particles that make up ordinary matter. Despite the absence of a definitive mathematical proof for this phenomenon, color confinement is strongly supported by both experimental observations and lattice QCD calculations. These findings reinforce the idea that nature imposes an impassable barrier, forcing quarks into a collective existence and preventing them from appearing individually in the phenomenal world.

The behavior of QCD varies drastically depending on the energy scale involved. In the high-energy regime, where the interaction is weak, perturbative methods can be used to describe reactions between quarks and gluons. This region falls within the domain of so-called \textit{perturbative} QCD (pQCD). Conversely, in the low-energy regime, the theory becomes strongly coupled, rendering perturbative techniques ineffective. This is where \textit{Lattice} QCD (LQCD) comes into play—a computational approach that discretizes space-time to numerically simulate the evolution of strong interactions.\\

In this context, it is crucial to mention the \textit{factorization theorems} among the most significant results of perturbative QCD. These theorems allow for the distinction and separation of high-energy interactions — calculable using perturbative methods and commonly referred to as the \textit{hard part} — from long-distance dynamics, which instead requires a non-perturbative approach. In the factorization framework, proton beams are described as ensembles of \textit{partons} — quasi-free quarks and gluons. The long-distance dynamics are encapsulated within the \textit{parton distribution functions} (PDFs), which characterize the non-perturbative structure of the proton. These distributions must either be extracted from experimental data or computed using LQCD.
These distributions are universal and, once determined in an experiment by extracting them from a given process, they can be used to predict other scattering processes. The hard parts are characterized by two fundamental energy scales: the center-of-mass energy of the entire process, \(\sqrt{s}\), and the \textit{hard scale} \(Q\), which must satisfy the following condition:  
\[
Q^2 \gg \Lambda_{\text{QCD}}^2 \, ,
\]
where \(\Lambda_{\text{QCD}}\) represents the QCD \textit{mass scale}. This parameter plays a crucial role in enabling a perturbative treatment of the theory, governing the strength of the \textit{strong coupling} \(\alpha_s(Q^2)\).\\

A delicate aspect of the perturbative approach is the handling of logarithmic corrections that arise in higher-order calculations. These terms, proportional to powers of the coupling constant \(\alpha_s\), can become dominant if the process energy varies in such a way that amplifies their contribution. In such cases, resummation techniques must be employed to sum an infinite number of terms in the perturbative series, ensuring the stability of theoretical predictions.

A key example of this issue appears in the removal of infrared (IR) divergences in QCD calculations for IR-safe observables. To address such divergences, dimensional regularization is used — a mathematical technique that makes divergent quantities well-defined by rewriting them in terms of a generic space-time dimension \(D\). A typical result of this procedure is the introduction of divergent factors of the form:
\[
    \frac{1}{\epsilon} \left( Q^2 \right)^{\epsilon} = \frac{1}{\epsilon} + \ln Q^2 + \mathcal{O}(\epsilon) \; ,
\]
where \(\epsilon\) represents the deviation from the physical space-time dimension. Although in final calculations divergences cancel out in well-defined observables, the logarithmic terms persist and must be handled carefully. In particular, in the \textit{Bjorken limit} — where \(Q^2\) grows indefinitely while the momentum fraction \(x\) remains moderately small — these logarithms can compensate for the smallness of the coupling constant \(\alpha_s\), making resummation necessary. This is achieved through the Dokshitzer-Gribov-Lipatov-Altarelli-Parisi (DGLAP)~\cite{Gribov:1972ri,Dokshitzer:1977sg,Altarelli:1977zs} evolution equations, which describe the energy dependence of parton distributions.\\

Moreover, there exists a scenario in which large logarithmic corrections emerge in a distinctive form, specifically as \(\log(s/Q^2)\). This scenario is characterized by a well-defined hierarchy among the physical scales encountered so far, given by
\[
    s \gg Q^2 \gg \Lambda_{\rm{QCD}}^2 \; .
\]
When this condition is satisfied, the system is said to be in the \textit{Regge-Gribov} region, also known as the \textit{semi-hard} regime.
To systematically address these corrections, it is essential to employ resummation techniques that account for the contributions at all orders in the strong coupling.

A well-established framework for performing this resummation is provided by the Balitsky-Fadin-Kuraev-Lipatov (BFKL) approach~\cite{Fadin:1975cb,Kuraev:1976ge,Kuraev:1977fs,Balitsky:1978ic}, which enables the resummation of large-energy logarithms at both the leading logarithmic approximation (LLA) and the next-to-leading logarithmic approximation (NLLA). Within this formalism, the cross section of hadronic processes is expressed as the convolution of two impact factors — describing the transition from each incoming particle to its respective final-state object — and a universal, process-independent Green function.

The BFKL approach has proven to be highly robust, as it successfully predicts the rapid growth of the \(\gamma^* p\) cross section at increasing energy while maintaining consistency with pre-QCD results from Regge theory. Its flexibility also makes it a powerful tool, allowing its application in a variety of different contexts.\\

In the study of hadronic production at high energies, a particularly effective approach is represented by the hybrid formalism, which combines collinear factorization with high-energy BFKL dynamics. This methodology, known as \textit{hybrid factorization}, allows for a coherent description of semi-inclusive reactions by integrating both perturbative and non-perturbative contributions of QCD. A key feature of this approach is its ability to naturally incorporate jet production and fragmentation mechanisms into hadronic states, including exotic ones, such as tetraquarks and pentaquarks, thus providing a versatile theoretical framework for describing complex processes in high-energy colliders. These exotic states differ from conventional mesons and baryons in that they contain more than two or three valence quarks, suggesting more complex internal structures and interactions.

In the context of perturbative QCD, handling heavy quarks, such as charm ($c$) and bottom ($b$), requires particular attention. Depending on the kinematic conditions, two main theoretical schemes are typically adopted:
\begin{itemize}
    \item Fixed-Flavor Number Scheme (FFNS): In this scheme, heavy quarks are always treated as massive particles, and the number of active flavors remains fixed. This description is accurate when the process scale $Q$ is close to the heavy quark mass $m_Q$, but it does not account for the resummation of logarithmic terms of the type $\ln(Q^2/m_Q^2)$, which become significant at high energies.
    \item Zero-Mass Variable-Flavor Number Scheme (VFNS): In this case, the mass of heavy quarks is neglected and treated as if it were zero. This scheme allows for the resummation of logarithmic terms $\ln(Q^2/m_Q^2)$, making it particularly effective when $Q^2 \gg m_Q^2$, but it loses accuracy at lower energies, where mass effects cannot be ignored.
\end{itemize}
Since FFNS and VFNS are valid in different kinematic regions, to bridge the gap between the two, the General-Mass Variable-Flavor Number Scheme (GM-VFNS) has been developed. This approach combines the ability of FFNS to include mass effects with the logarithmic resummation of VFNS, providing a more accurate description across the entire energy range relevant for the production of heavy tetraquarks.

One of the advancements in the application of this formalism is the construction of collinear fragmentation functions (FFs) specific to heavy tetraquarks, which enable a quantitative description of their production in current and future colliders, such as the Large Hadron Collider (LHC) and the Future Circular Collider (FCC).
Beyond the interest in discovering new exotic states, the study of heavy tetraquark production also serves as a privileged testing ground for assessing BFKL dynamics and collinear factorization in high-energy reactions. By comparing theoretical predictions with experimental data, this approach contributes not only to understanding the structure of exotic hadrons but also to refining computational techniques in perturbative QCD.\\

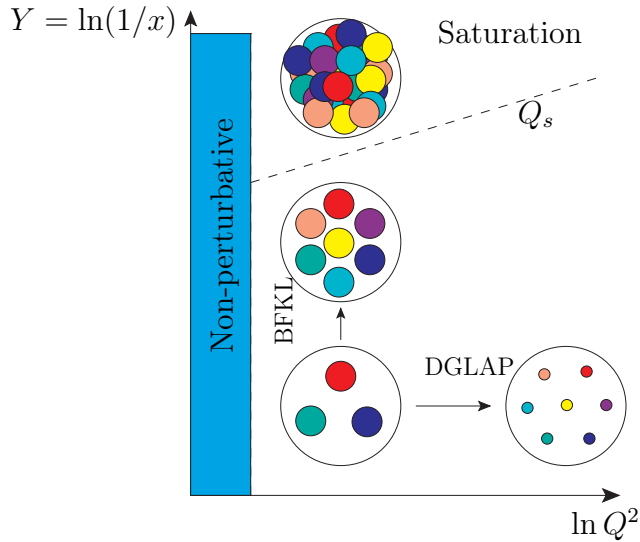
\begin{figure}
\begin{picture}(400,215)
\put(137,10){\includegraphics[width=0.4\textwidth]{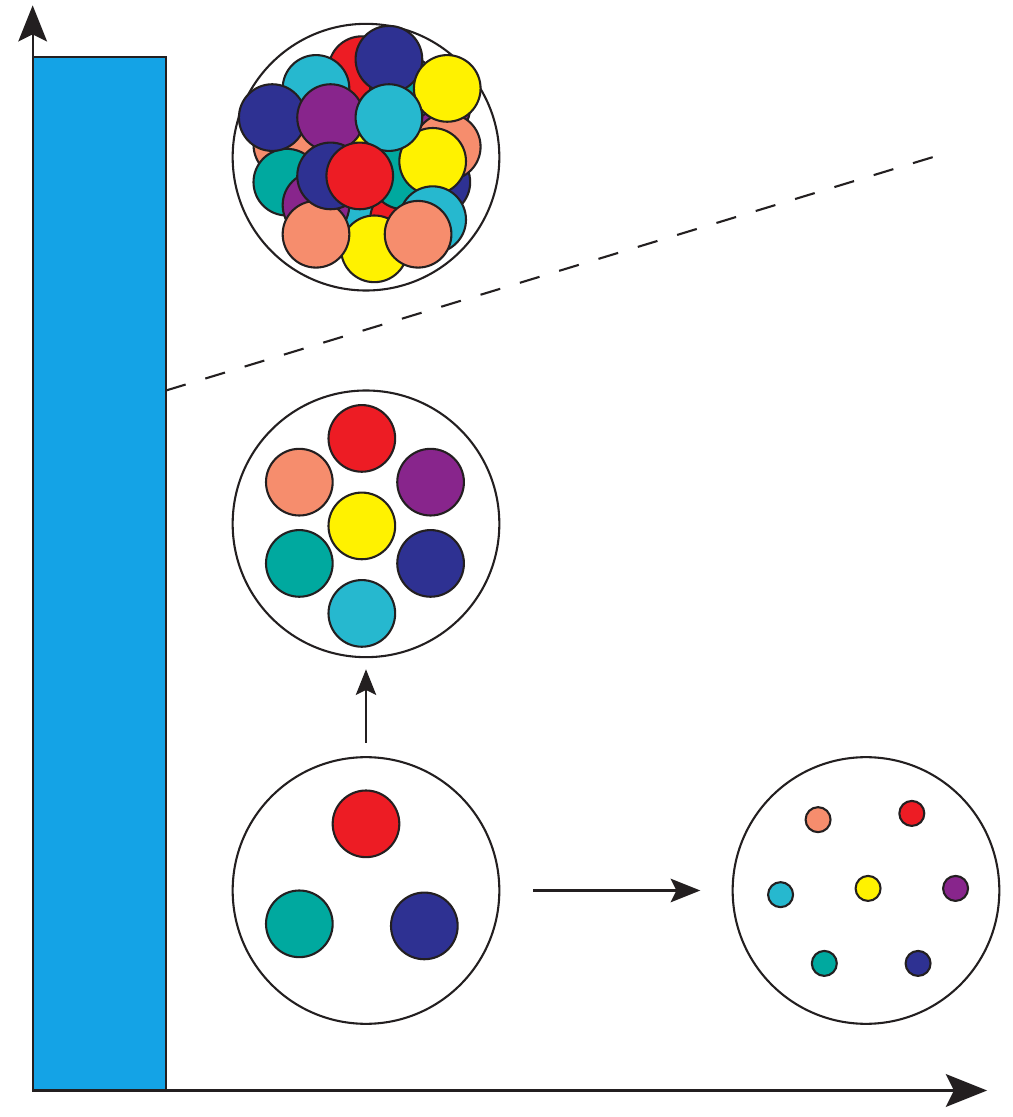}}
\put(235,185){Saturation}
\put(150,68){\rotatebox{90}{Non-perturbative}}
\put(174,71){\rotatebox{90}{\scalebox{0.8}{BFKL}}}
\put(230,60){\scalebox{0.8}{DGLAP}}
\put(265,155){$Q_s$}
\put(285,0){$\ln Q^2$}
\put(75,190){$Y = \ln (1/x)$}
\end{picture}
\caption{The parton distribution in the transverse plane as a function of $\ln (1/x)$ and $\ln Q^2$. Figure from~\cite{Fucilla:2023pma}}
\label{Int:Int:DGLAPvsBFKL}
\end{figure}

One of the main theoretical issues of the BFKL formalism is its incompatibility with the \textit{Froissart bound}, which states that total cross sections cannot grow with the energy \( s \) faster than a term proportional to \( \ln^2 s \). However, BFKL resummation leads to a power-like growth that explicitly violates this limit.
In the context of small-\( x \) logarithms, this violation has a clear physical interpretation: the gluon density increases indefinitely as \( x \) decreases, making the hadronic system appear as an increasingly dense gluon medium. If this growth were to continue unchecked, it would lead to an unphysical scenario where the density becomes infinite. Therefore, it is expected that at some point, \textit{saturation effects} must come into play to slow down this growth.
From a theoretical perspective, these effects are described by nonlinear extensions of the BFKL equation. Among them, the most general framework is the Balitsky-JIMWLK hierarchy of equations. This was derived through two distinct approaches: the \textit{Shockwave method}, developed by Balitsky~\cite{Balitsky:2001re,Balitsky:1995ub,Balitsky:1998kc,Balitsky:1998ya}, and the \textit{Color Glass Condensate} (CGC) approach, formulated by Jalilian-Marian, Iancu, McLerran, Weigert, Leonidov, and Kovner (JIMWLK)~\cite{Jalilian-Marian:1997qno,Jalilian-Marian:1997jhx,Jalilian-Marian:1997ubg,Jalilian-Marian:1998tzv,Kovner:2000pt,Weigert:2000gi,Iancu:2000hn,Iancu:2001ad,Ferreiro:2001qy}.
These approaches are essential to describe the scattering of a \textit{dilute projectile} on a \textit{dense target}, or even the interaction between two dense systems. A system becomes dense when the proton fraction \( x \) is very small, but this condition is more easily achieved in the case of nuclei. A schematic representation of the proton and nucleus structure, in terms of their constituents, is illustrated in Fig.~\ref{Int:Int:DGLAPvsBFKL}.\\
Within this theoretical framework, the present work aims to explore fundamental aspects of high-energy QCD, from some formal aspects to their phenomenological applications. More specifically, the current thesis is structured as follows:\\
\begin{itemize}
    \item The first chapter is devoted to the theoretical foundation of high-energy QCD, with a particular focus on the BFKL formalism. The discussion begins with an introduction to Regge theory, which forms the basis for the analysis of high-energy scattering processes. The role of Regge poles and trajectories in determining the asymptotic behavior of scattering amplitudes is examined in detail. The core of this chapter centers on the derivation and solution of the BFKL equation.
    \item The second chapter presents a concrete application of high-energy QCD by investigating the next-to-leading order (NLO) real corrections to the Higgs boson impact factor, explicitly incorporating the physical mass of the top quark. As a critical component in high-energy resummation within the BFKL framework, the Higgs impact factor is essential for precision studies of Higgs production at large rapidity separations.
    \item The third chapter explores the semi-inclusive production of exotic hadronic states in the semi-hard regime. This investigation serves two main objectives: studying the mechanisms underlying exotic matter formation in high-energy collisions and testing the predictions of BFKL dynamics in collider experiments. The focus is placed on the production of bottomonium-like tetraquarks. A hybrid approach, combining collinear factorization with high-energy BFKL dynamics, is employed to account for both perturbative and non-perturbative aspects of the production process. The analysis incorporates variable-flavor number schemes (VFNS) and collinear fragmentation functions (FFs) specifically parametrized for tetraquarks.
    \item The fourth and final chapter is dedicated to the study of diffractive di-hadron production in the high-energy regime, leveraging the Color Glass Condensate (CGC) framework and saturation physics. The small Bjorken-$x$ limit of QCD is explored. The saturation scale, which delineates this transition, depends on both the small-$x$ gluon density and the target’s mass number, with particularly strong effects observed in heavy nuclei. A leading-order analysis of diffractive di-hadron production is performed using collinear factorization and the shockwave formalism. This work is ongoing, with numerical computations in progress.
\end{itemize}

\newpage 
\thispagestyle{empty} 
\mbox{} 

\newpage 
\chapter{BFKL theory}

The first part of our journey concerns the approach that forms the foundation of the theory. The opening section of this chapter will introduce Regge theory, following the standard treatments found in texts~\cite{Forshaw:1997qp,Collins:1977rt}; this theory provides a broad framework for analyzing particle scattering at high center-of-mass energies. Subsequently, this chapter will focus on the BFKL approach, named after its authors Balitsky, Fadin, Kuraev, and Lipatov, as highlighted in key references~\cite{Fadin:1975cb,Kuraev:1976ge,Kuraev:1977fs,Balitsky:1978ic}. This theoretical framework is employed to analyze pQCD-scattering amplitudes in the Regge region, in which the square of the center-of-mass energy, $s$, significantly exceeds the squared momentum transfer, $|t|$. The BFKL equation, which will be presented in the chapter, systematically articulates the evolution of these amplitudes through various color states in the $t$-channel, enhancing understanding of scattering dynamics at high energies. For this section, the analysis proceeds following the Refs.~\cite{Forshaw:1997qp,Fadin:1998sh}, but, for further details and clarifications, please also refer to \cite{Fucilla:2023pma,Celiberto:2017ius}.

\section{Regge theory}
In 1959, the Italian physicist Tullio Regge discovered that, in solving the Schrödinger equation for non-relativistic potential scattering, treating angular momentum $l$ as a complex variable can be beneficial. He demonstrated that \cite{Regge:1959it}, for a broad range of potentials, the scattering amplitude $\mathcal{A}$ in the complex $l$-plane exhibits singularities exclusively as poles, which have come to be known as \textit{Regge poles}. When these poles align with integer values of $l$, they represent bound states or resonances, significantly influencing the analytic properties of the amplitudes. 
After some mathematical steps, in the specific case of simple poles in the $l$-plane and within the Regge kinematic domain $\abs{s} \gg t$, it is found that:
\begin{equation}
\label{Int:Eq:AsynAmplitude}
\mathcal{A} \left(s,t\right) \xrightarrow{s \rightarrow \infty} \frac{\eta+e^{-i\pi \alpha(t)}}{2} \beta(t) s^{\alpha(t)},
\end{equation}
where $\alpha(t)$ represents the location of the leading Regge pole in the $l$-plane; it is not constant but it is a function that depends on the momentum transfer $t$.
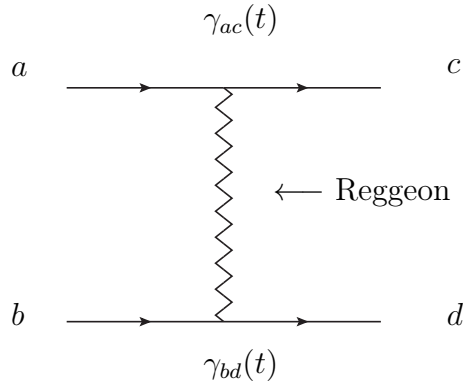
\begin{figure}
\begin{picture}(417,145)
\put(157,20){\includegraphics[width=0.3\textwidth]{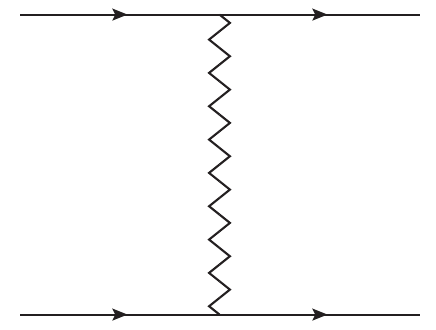}}
\put(240,70){$\longleftarrow$ Reggeon}
\put(142,117){$ a $}
\put(142,23){$ b $}
\put(305,118){$ c $}
\put(305,23){$ d $}
\put(214,135){$ \gamma_{ac} (t)$}
\put(214,5){$ \gamma_{bd} (t)$}
\end{picture}
\caption{A diagram showing the Regge exchange.}
\label{Int:Fig:Reggefigure}
\end{figure}
The amplitude in Eq.~(\ref{Int:Eq:AsynAmplitude}) can be interpreted as the exchange of an entity (that we can call \textit{Reggeon}) in the $t$-channel with an effective "angular momentum" equivalent to $\alpha(t)$, which we refer to as \textit{Regge trajectory}. A Reggeon-exchange amplitude can be viewed as a composite of amplitudes for the exchange of all possible $t$-channel particles, with quantum numbers defined by the interacting particles. The amplitude factorizes, as illustrated in Fig.~\ref{Int:Fig:Reggefigure}, into a coupling $\gamma_{ac}(t)$ of the Reggeon to particles $a$ and $c$, another coupling $\gamma_{bd}(t)$ between particles $b$ and $d$, along with a universal contribution from the Reggeon exchange. The couplings $\gamma$ depend only on $t$, meaning that the Reggeon exchange fully determines the behavior in $s$ of the total amplitude.
The index $\eta$, referred to as the \textit{signature}, plays a crucial role and can take on two distinct values, $+1$ and $-1$. The concept of signature corresponds to parity under the transformation $\cos \theta_t \leftrightarrow -\cos \theta_t$, where $\theta_t$ denotes the scattering angle in the $t$-channel.

\section{Pomeron}
A formal definition of \textit{Reggeization} can now be introduced. A particle with mass $M$ and spin $J$ is said to \textit{Reggeize} if the amplitude $\mathcal{A}$ for a process involving the exchange in the $t$-channel of the quantum numbers of that particle, behaves asymptotically in $s$ as 
\begin{equation*}
\mathcal{A} \propto s^{\alpha(t)}, 
\end{equation*}
where $\alpha(t)$ represents the trajectory and $\alpha(M^2)=J$, indicating that the particle itself lies on this trajectory.

If we apply the optical theorem starting from this amplitude in the elastic case, we can derive the cross-section of the process in which the exchange of Reggeons occurred,
\begin{equation}
\sigma_{\rm{tot}} \propto s^{(\alpha(0)-1)}.
\end{equation}  
Experimental observations show that total cross sections do not diminish asymptotically; rather, they increase gradually with $s$. If this rise is attributed to the exchange of a single Regge pole, it implies an exchange by a Reggeon with the vacuum’s quantum numbers, known as \textit{Pomeron} (where $\alpha_P(0)>1$), named after its originator, I.J. Pomeranchuk. The physical particles that could form resonances at integer values on the Pomeron trajectory have not been identified, but in QCD, hypothetical gluon bound states, known as glueballs, are considered plausible candidates. \\

It is important to reiterate that the Reggeon and Pomeron concepts introduced here are independent of perturbative methods. Moving forward, our focus will shift to the analysis of semi-hard processes in pQCD, specifically those characterized by the scale hierarchy \begin{equation}
    s \gg t \gg \Lambda_{QCD}^2 \; .
    \label{Int:Eq:Semihard}
\end{equation}
Within this framework, we will discuss gluon Reggeization in pQCD and develop the so-called hard Pomeron (or also BFKL Pomeron).

\section{Reggeization of the gluon in pQCD}
The Reggeization of the gluon within Quantum Chromodynamics unveils a remarkable conceptual depth. It signals the existence of a unique Reggeon characterized by the quantum properties of the gluon, bearing a negative signature and defined by a trajectory
\begin{equation}
\label{Int:Eq:trajectory}
\alpha(t)=1+\omega(t) \; ,
\end{equation}
which necessarily intersects at unity when $t=0$, due to the property of $\alpha(t=M^2)$ mentioned in the previous paragraph.

Consider now an elastic scattering process $ A+B \longrightarrow A'+B'$ in the limit where $s\gg \abs{t}$. In such a case, we define:
\begin{equation}
s=(p_A+p_B)^2 \; ,  \hspace{2cm}  t=q^2 \; ,  \hspace{2cm} q= p_A-p_{A'} \; .
\end{equation} 
Here, Reggeization grants the amplitude, where the gluon quantum numbers are interchanged in the $t$-channel, a distinct and beautifully factorized form:
\begin{equation}
\begin{split}
\label{Int:Eq:ReggeizedAmp}
\mathcal{A}_{AB}^{A'B'}=&\, \Gamma^i_{A'A} \left( \frac{s}{-t} \right)^{ \alpha (t) } \left[ -1 + e^{-i \pi \alpha (t)} \right] \Gamma^i_{B'B}\\ =&\, \Gamma^i_{A'A} \frac{s}{t} \left[ \left(\frac{s}{-t}\right)^{\omega(t)} + \left( \frac{-s}{-t} \right)^{\omega(t)} \right] \Gamma^i_{B'B} \; .
\end{split}
\end{equation} 
Here, $\Gamma_{P'P}^i$ (with $P=A,B$) represent the particle-particle-Reggeon (PPR), fixed and uninfluenced by $s$, while $i$ serves as the color index.
This structure, indeed, mirrors the schematic form depicted in Figure~\ref{Int:Fig:Reggefigure}: two particles couple distinctly to the Reggeon, their interactions contributing to a term that universally scales as $s^{\alpha(t)}$. 

The factorization described in Eq.~(\ref{Int:Eq:ReggeizedAmp}) captures the analytic structure of the scattering amplitude, remaining elegantly straightforward in the elastic scattering case. It holds in both the Leading Logarithmic Approximation (LLA) and the Next-to-Leading Logarithmic Approximation (NLLA) and applies even when either particle $A'$ or $B'$ is substituted by a jet. Generally, the particle-particle-Reggeon (PPR) vertex takes the form
\begin{equation}
\Gamma^C_{P^\prime P}=g_s\left\langle P^\prime|T^c|P \right\rangle\Gamma_{P^\prime P},
\label{Eq:PPR_vertex}
\end{equation}
where $g_s$ represents the QCD coupling constant, and $\left\langle P^\prime|T^c|P \right\rangle$ is the matrix element of the color-group generator in the fundamental (or adjoint) representation for quarks (or gluons).
The deviation from one of the gluon trajectory $\alpha(t)$ is
\begin{equation}
\label{Int:Eq:ReggeTraj1loop}
\omega (t) \simeq \omega^{(1)}(t) = \frac{g^2t}{(2\pi)^{(D-1)}} \frac{N}{2} \int \frac{d^{D-2}k_{\perp}}{k_{\perp}^2 \left(q-k\right)_{\perp}^2} \; .
\end{equation}
One might naturally question why the relation $\omega(t) \simeq \omega^{(1)} (t)$. This reflects that only the one-loop term, $\omega^{(1)}(t)$, has been extracted as a contribution to the full Regge trajectory. The integral in Eq.~(\ref{Int:Eq:ReggeTraj1loop}) can be computed using conventional techniques for Feynman integrals, yielding
\begin{equation}
\label{Int:Eq:ReggeTraj1loopInt}
\omega^{(1)} (t) = \frac{g^2t}{(2\pi)^{(D-1)}} \frac{N}{2} \int \frac{d^{D-2}k_{\perp}}{k_{\perp}^2 \left(q-k\right)_{\perp}^2} = -\frac{g^2N \Gamma(1-\epsilon)}{(4\pi)^{2+\epsilon}} \frac{[\Gamma(\epsilon)]^2}{\Gamma(2\epsilon)} (\vec{q}^{\; 2})^{\epsilon} \; .
\end{equation}
From Eq.~(\ref{Int:Eq:ReggeTraj1loopInt}), it is straightforward to note that for the gluon $\omega(0)=0$, leading to $\alpha(0)=1$; this result confirms that the gluon resides on the Reggeon trajectory, preserving its quantum numbers. For a more in-depth analysis, it is recommended to consult the references \cite{Fucilla:2023pma,Celiberto:2017ius}.
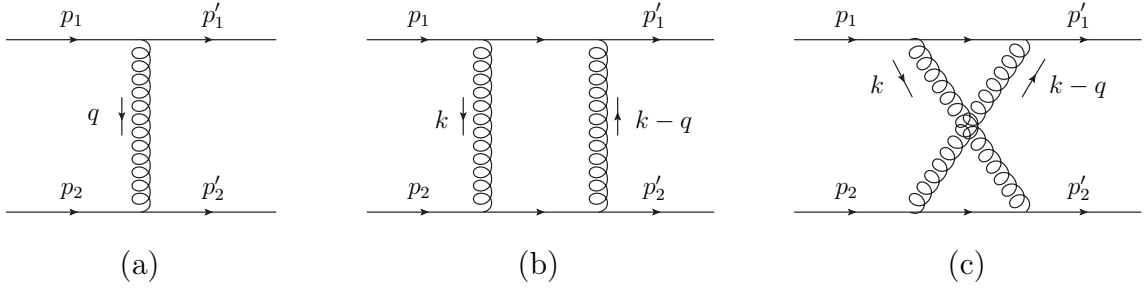
\begin{figure}
\begin{picture}(420,125)
\put(0,30){\includegraphics[scale=0.45]{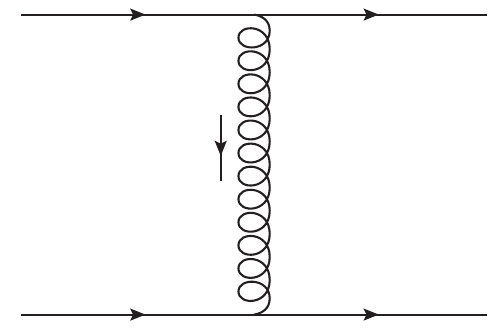}}
\put(47,10){(a)}
\put(35,68){\scalebox{0.8}{$q$}}
\put(25,40){\scalebox{0.8}{$p_2$}}
\put(25,105){\scalebox{0.8}{$p_1$}}
\put(78,40){\scalebox{0.8}{$p_2'$}}
\put(78,105){\scalebox{0.8}{$p_1'$}}
\put(135,30){\includegraphics[scale=0.45]{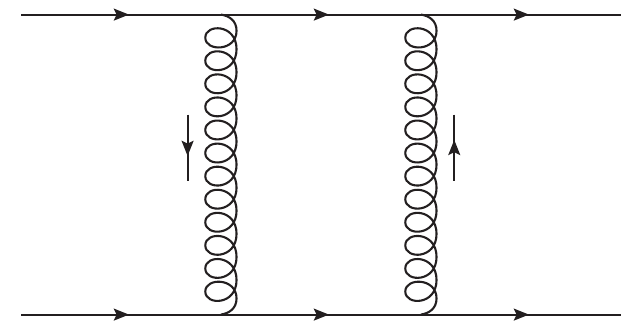}}
\put(196,10){(b)}
\put(165,65){\scalebox{0.8}{$k$}}
\put(240,65){\scalebox{0.8}{$k-q$}}
\put(155,40){\scalebox{0.8}{$p_2$}}
\put(155,105){\scalebox{0.8}{$p_1$}}
\put(243,40){\scalebox{0.8}{$p_2'$}}
\put(243,105){\scalebox{0.8}{$p_1'$}}
\put(295,30){\includegraphics[scale=0.45]{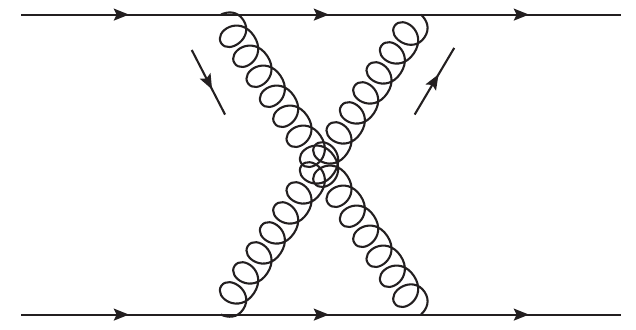}}
\put(357,10){(c)}
\put(328,78){\scalebox{0.8}{$k$}}
\put(395,78){\scalebox{0.8}{$k-q$}}
\put(315,40){\scalebox{0.8}{$p_2$}}
\put(315,105){\scalebox{0.8}{$p_1$}}
\put(402,40){\scalebox{0.8}{$p_2'$}}
\put(402,105){\scalebox{0.8}{$p_1'$}}
\end{picture}
  \caption{In (a) the leading order diagram contributing to the quark-quark scattering; (b) and (c) are the two next-leading order diagrams contributing to the Regge trajectory in the LLA.}
  \label{Int:Fig:LOqqqq}
\end{figure}

\section{The Pomeron of pQCD}
\begin{figure}
\begin{picture}(400,240)
\put(110,0){\includegraphics[width=0.375\textwidth]{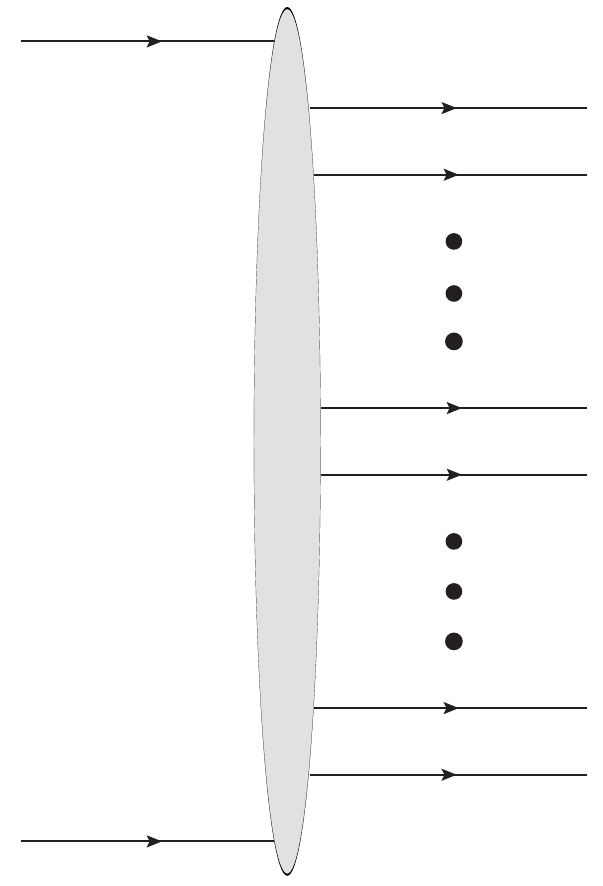}}
\put(85,10){$p_2$}
\put(85,220){$p_1$}
\put(282,205){$k_0$}
\put(282,185){$k_1$}
\put(282,125){$k_{i-1}$}
\put(282,105){$k_i$}
\put(312,115){$\bigg \} \; \; s_i = (k_{i-1} + k_i)^2$}
\put(282,45){$k_{n}$}
\put(282,25){$k_{n+1}$}
\end{picture}
\caption{Production of $n+2$ particles in the multi-Regge kinematics}
\label{Int:Fig:MRKn+2}
\end{figure}
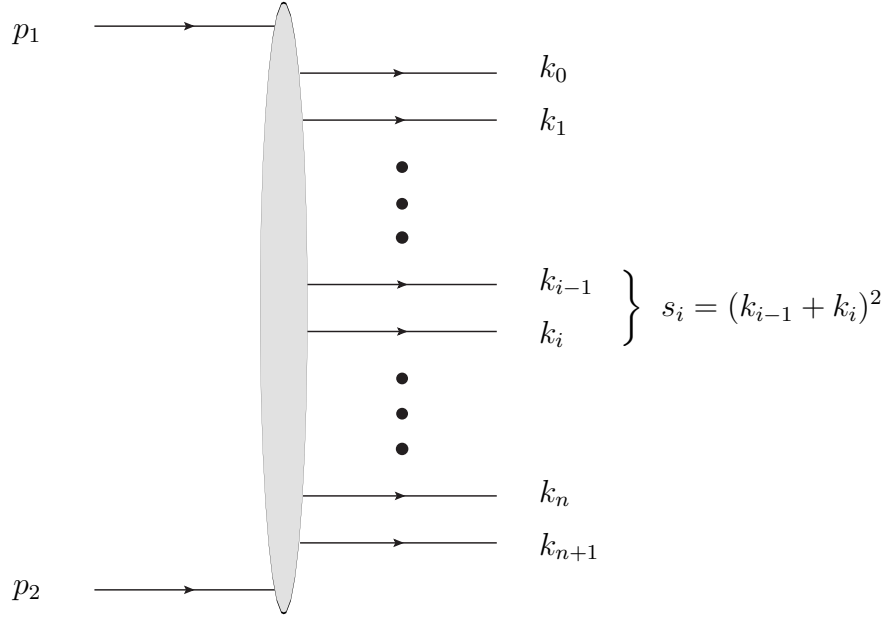

It is now time to sail into deeper waters. The imaginary part of the scattering amplitude \( \mathcal{A}_{12}^{1'2'} \) (for the process \( 1 + 2 \rightarrow 1' + 2' \)) in the \( s \)-channel can be expressed, using the Cutkosky rules. These state that, as an immediate consequence of the unitarity of the scattering matrix, the imaginary part of the generic scattering amplitude, \( \mathcal{A}_{ab} \), for scattering from an \( \text{in} \)-state \(\ket{a}\) to an \( \text{out} \)-state \(\ket{b}\), is obtained by considering the scattering amplitudes of the incoming and outgoing states into all possible "intermediate" states:
\begin{equation}
\label{Int:Eq:Cutkosky}
2 \Im m \mathcal{A}_{ab}=(2\pi)^{4} \delta^{4} \left( \sum_{a} p_a -\sum_{b} p_b \right) \sum_{c} \mathcal{A}_{ac} \mathcal{A}^{\dagger}_{cb} \; ,
\end{equation}  
where $p_a,p_b$ are the 4-momenta of the particles in the states $\ket{a},\ket{b}$, respectively. Using these rules for the process \( 1 + 2 \rightarrow 1' + 2' \), we obtain:
\begin{equation}
\label{Unitary}
\Im_s \mathcal{A}_{12}^{1'2'} = \frac{1}{2} \sum_{n=0}^{\infty} \sum_{\{f\}} \int \mathcal{A}_{12}^{\tilde{1} \tilde{2} + n} \left(\mathcal{A}_{1'2'}^{\tilde{1} \tilde{2}+n}\right)^* d\Phi_{n+2} \; .
\end{equation}
Here, \( \mathcal{A}_{12}^{\tilde{1} \tilde{2} + n} \) represents the amplitude for the production of \( n+2 \) particles with momenta \( k_i \) (where \( i = 0, 1, \ldots, n, n+1 \)), and we set \( p_{\tilde{1}} = k_0 \) and \( p_{\tilde{2}} = k_{n+1} \). This result was obtained by generalizing Eq.~(\ref{Int:Eq:ReggeizedAmp}) to the case of an inelastic \(2 \rightarrow 2 + n\) amplitude in the Regge limit discussed in the previous section; in this context, we refer to multi-Regge kinematics (for further details and deeper insights, please refer to \cite{Fucilla:2023pma}). The term \( d\Phi_{n+2} \) denotes the phase space element for the intermediate particles, and \( \sum_{{f}} \) signifies the summation over all discrete quantum numbers of the intermediate particles (as illustrated in Figure~\ref{Int:Fig:MRKn+2}).

Now, it proves useful the \textit{Sudakov} decomposition to express a momentum $k$ in terms of
\begin{equation}
    k^{\mu} = \beta p_1^{\mu} + \alpha p_2^{\mu} + k_{\perp}^{\mu} \; ,
\end{equation}
where $k_{\perp}=(0, \vec{k}, 0)$ is a four-vector, completely transverse with respect to the plane identified by the two light-cone vectors $p_1$ and $p_2$.

Let us imagine an elastic process, $A+B \rightarrow A+B$, in which we allow ourselves to simplify further by considering the phenomenon of forward scattering, where the momentum transfer $q$ vanishes. Through the \textit{optical theorem}, we can establish a deep and significant connection between the imaginary part (in the $s$-channel of the related amplitude and the total cross section of the process $A+B \rightarrow X$:
\begin{equation}
\label{Int:Eq:HighEneOpThe}
\sigma_{AB}(s) = \frac{\Im_s \mathcal{A}_{AB}^{AB}}{s} \; .
\end{equation} 
Through this theorem, it is possible to rewrite this cross section as follows (see \cite{Fucilla:2023pma}):
\begin{equation}
\begin{split}
\label{Int:Eq:TotCross}
\sigma_{AB}(s) =& \int_{\delta-i\infty}^{\delta+i\infty} \frac{d\omega}{2 \pi i} \frac{1}{(2 \pi)^{D-2}} \\ \times&\,\int d^{D-2} q_{A\perp} d^{D-2} q_{B\perp} \left( \frac{s}{s_0} \right)^{\omega} \frac{\Phi_A(\vec{q}_A)}{(\vec{q}_A^{\;2})^2}G_{\omega}(\vec{q}_A, \vec{q}_B) \frac{\Phi_B(-\vec{q}_B)}{(\vec{q}_B^{\;2})^2} ,
\end{split}
\end{equation}  
where $s_0$ is an arbitrary energy scale introduced in the context of the inverse Mellin transform; $G_{\omega}$ is the \textit{Green function} for the scattering of two Reggeized gluons and, not depending on the specific process considered, is universal; $\Phi_A(\vec{q}_A)$ and $\Phi_B(\vec{q}_B)$ are the so-called \textit{impact factors}, specific since they depend on the particles involved in the process. The Green function governs the behavior of the amplitude with respect to $s$, while the impact factors are independent of this variable. Their general definition reads
\begin{equation}
\label{Int:Eq:ImpactFactSingGen}
    \Phi_{A} (\vec{q}_A) = \langle cc|\hat{\mathcal{P}}_0|0 \rangle \sum_{\{ f \}} \int \frac{d s_{PR}}{2 \pi} d \rho_f \Gamma_{\{ f \} A}^c \left( \Gamma_{\{ f \} A}^{c'}  \right)^* \; ,
\end{equation}
where $ \langle cc|\hat{\mathcal{P}}0|0 \rangle$ denotes a color singlet projector, $\Gamma_{\{f\} A}^c$ represents an effective vertex that generalizes the one in (\ref{Eq:PPR_vertex}), the summation runs over all possible intermediate states $\{f \}$, and the integration is carried out over the phase space of the intermediate state $d \rho_f$ and the particle-Reggeon invariant mass $s_{PR}$. \\

\begin{figure}
\begin{picture}(400,212) 
\put(130,12){\includegraphics[width=0.4\textwidth]{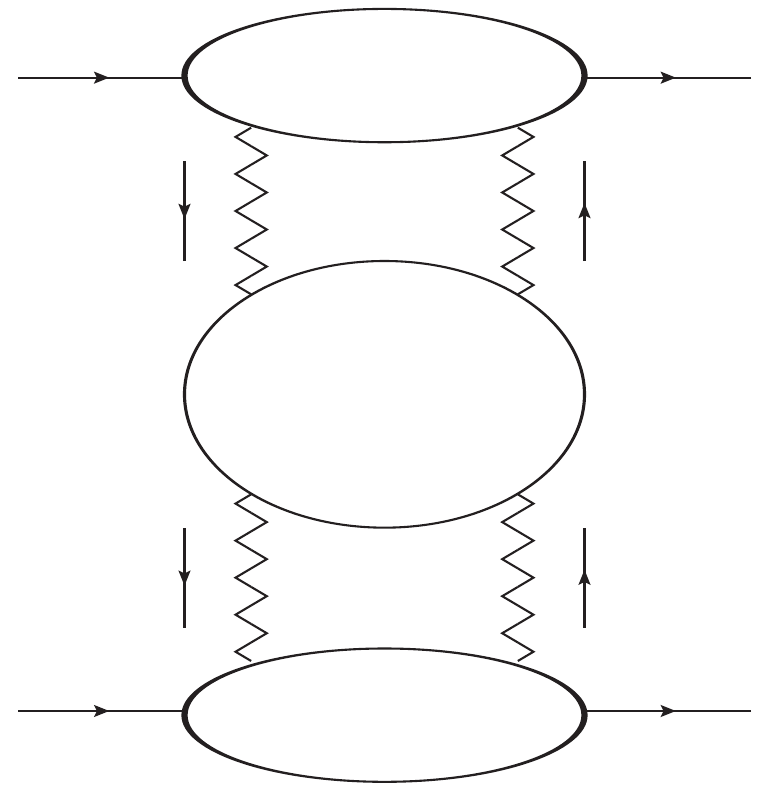}} 
\put(208,168){\scalebox{1.2}{$\Phi_{AA}$}} 
\put(210,102){\scalebox{1.2}{$G_{\omega}$}} 
\put(208,24){\scalebox{1.2}{$\Phi_{BB}$}} 
\put(112,29){$p_B$}
\put(112,170){$p_A$} 
\put(306,29){$p_B$} 
\put(306,170){$p_A$} 
\put(157,62){$q_B$}  
\put(157,148){$q_A$} 
\put(276,62){$q_B$} 
\put(276,148){$q_A$}
\end{picture}
\caption{Schematic representation of the factorized amplitude.}
\label{convolution}
\end{figure}
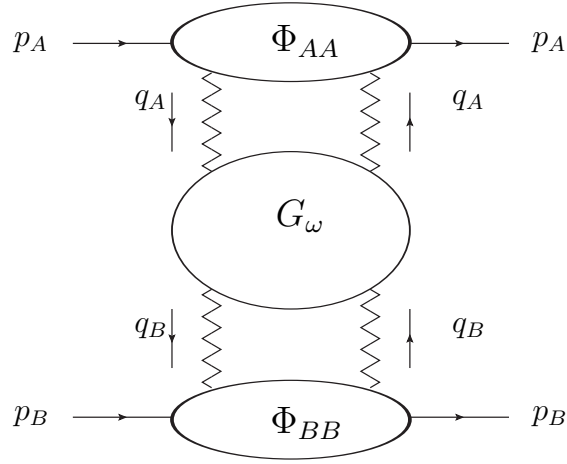

Therefore, in this context, the amplitude for a high-energy scattering process can be interpreted as a convolution between the Green function and the two impact factors related to the colliding particles (see Fig.~\ref{convolution}).
\section{The BFKL equation}
The Green function that appears in \eqref{Int:Eq:TotCross} satisfies the following integral equation:
\begin{equation}
\begin{split}
\label{BFKLequation}
\omega G_{\omega}^{(R)}(\vec{q}_1,\vec{q}_2;\vec{q})&= \vec{q}_1^{\; 2}(\vec{q}_1-\vec{q})^{2} \delta^{(D-2)}(\vec{q}_1-\vec{q}_2)\\ &+ \int \frac{d^{D-2}q'_{1 \perp}}{\vec{q}_1^{\; ' 2}(\vec{q}_1^{\; '}-\vec{q})^2} \mathcal{K}^{(R)}(\vec{q}_1, \vec{q}_1^{\; '} ; \vec{q}) G_{\omega}^{(R)}(\vec{q}_1^{\; '} , \vec{q}_2; \vec{q}),
\end{split}
\end{equation}
where the function $\mathcal{K}^{(R)}(\vec{q}_1, \vec{q}_1^{; '}; \vec{q})$, known as the kernel of the integral equation, is given by
\begin{equation}
\label{kernel}
\begin{split}
\mathcal{K}^{(R)}(\vec{q}_1, \vec{q}_2; \vec{q}) = \left[ \omega(q_{1\perp}^2) + \omega((q_1-q)_{\perp}^2) \right] \vec{q}_1^{\;2} \left( \vec{q}_1-\vec{q} \right)^{2} \delta^{(D-2)} (\vec{q}_1-\vec{q}_2) + \mathcal{K}_r^{(R)}(\vec{q}_1, \vec{q}_2; \vec{q}).
\end{split}
\end{equation}
In Eq.(\ref{kernel}), two distinct terms can be identified; the first term is referred to as ``virtual'' and is expressed in terms of the gluon’s Regge trajectory, while the second term pertains to the production of real particles.
The equation (\ref{BFKLequation}), when considering $R=0$ (singlet) and $t=0$, is known as the \textit{BFKL equation}. In its most general form, it is termed the \textit{generalized BFKL equation}. This equation is iterative in nature.

The kernel $\mathcal{K}^{(R)}$ and the Green function $G_{\omega}^{(R)}$ can be regarded as operators acting within the space of transverse momentum defined by
\begin{equation*} \hat{\vec{q}} \ket{\vec{q}_i} = \vec{q}_i \ket{\vec{q}_i}, \end{equation*}
\begin{equation} \label{Int
} \Braket{\vec{q}_1  |\vec{q}_2}= \delta^{(2)} \left( \vec{q}_1-\vec{q}_2 \right), \quad \quad \Braket{A|B} = \int d^2 \vec{k}  A(\vec{k} ) B(\vec{k} ). \end{equation}
In this formulation, the equation (\ref{BFKLequation}) can be represented in operator form as:
\begin{equation} \omega \hat{G}_{\omega}^{(R)} = 1 + \hat{ \mathcal{K} }^{(R)} \hat{G}_{\omega}^{(R)}  , \label{Int:Eq
} \end{equation}
which leads to
\begin{equation}
    \hat{G}_{\omega}^{(R)} = \frac{1}{\omega-\hat{ \mathcal{K} }^{(R)} }\; .
\end{equation}
\subsection{Solution at $t=0$}
The $s$-behavior of the total cross section is established by solving the BFKL equation (\ref{BFKLequation}) and determining the leading singularity $\omega_0$. Subsequently, by performing the anti-Mellin transformation, one can obtain the asymptotic behavior of cross sections. To begin with, Eqs.(\ref{BFKLequation}, \ref{kernel}) must be considered in the scenario of $R=0$ (singlet) and $t=0$. Redefining
\begin{equation}
\frac{G_{\omega}^{(0)} (\vec{q}_1,\vec{q}_2;\vec{0})}{\vec{q}_1^{\;2} \vec{q}_2^{\; 2}} \equiv G_{\omega}\left(\vec{q}_1,\vec{q}_2\right),
\end{equation}    
\begin{equation}
\frac{\mathcal{K}^{(0)}(\vec{q}_1,\vec{q}_2,\vec{0})}{\vec{q}_1^{\; 2} \vec{q}_2^{\; 2}} \equiv \mathcal{K} (\vec{q}_1,\vec{q}_2),
\end{equation}
and thus (see Ref.\cite{Fucilla:2023pma})
\begin{equation}
\frac{\mathcal{K}_{r}^{(0)}(\vec{q}_1,\vec{q}_2,\vec{0})}{\vec{q}_1^{\; 2} \vec{q}_2^{\; 2}} \equiv \mathcal{K}_{r} (\vec{q}_1,\vec{q}_2) = \frac{g^2N}{(2\pi)^{D-1}} \frac{2}{(\vec{q}_1-\vec{q}_2)^2}  \; ,
\label{Int:Eq:RealBornKert0}
\end{equation}
it is straightforward to conclude that the BFKL equation assumes the following form:
\begin{equation}
\omega G_{\omega}\left(\vec{q}_1,\vec{q}_2\right)=\delta^{(D-2)}\left(\vec{q}_1-\vec{q}_2\right)+ \int d^{D-2} k \;  \mathcal{K}(\vec{q}_1, \vec{k}) G_{\omega} (\vec{k}, \vec{q}_2)  ,
\end{equation}
which can be rewritten as
\begin{equation}
\omega G_{\omega}\left(\vec{q}_1,\vec{q}_2\right)=\delta^{(2)}\left(\vec{q}_1-\vec{q}_2\right)+\mathcal{K} \bullet G_{\omega} \left( \vec{q}_1, \vec{q}_2 \right) \; ,
\end{equation}
where we have defined the convolution operator as
\begin{equation}
    A \bullet B \; (q_1, q_2) = \int d^{D-2} k \; A (q_1,k) B(k,q_2) \; .
\end{equation}
The expression of the kernel function reads
\begin{equation}
    \mathcal{K} (\vec{q}_1, \vec{q}_2) = \frac{\mathcal{K}^{(0)} (\vec{q}_1, \vec{q}_2)}{\vec{q}_1^{\; 2} \vec{q}_2^{\; 2}} = 2 \omega (- \vec{q}_1^{\; 2}) \delta^{D-2} (\vec{q}_1 - \vec{q}_2) +  \frac{g^2N}{(2\pi)^{D-1}} \frac{2}{(\vec{q}_1-\vec{q}_2)^2} \; .
\end{equation}
To solve the eigenvalue problem of the kernel means finding also the solution for the Green function. In fact, solved
\begin{equation}
\label{eigenvalueproblem}
\mathcal{K} \bullet \phi_i(\vec{q}) =  \lambda_i \phi_i(\vec{q})\;,
\end{equation}
it is possible to reconstruct the Green function through the spectral representation: 
\begin{equation}
G_{\omega} \left( \vec{q}_1 , \vec{q}_2 \right)= \sum_i \frac{\phi_i(\vec{q}_1)\phi_i^*(\vec{q}_2)}{\omega-\lambda_i} \; .
\end{equation}
Note that the kernel is a scale-invariant operator, which means that, for any $\lambda\in{\rm I\!R}$, it holds
\begin{equation}
     \mathcal{K} ( \lambda \vec{q}_1, \lambda \vec{q}_2) = \lambda^{-2} \mathcal{K} (\vec{q}_1, \vec{q}_2) \; , \hspace{0.5 cm} \vec{q}_1, \vec{q}_2 \in \mathcal{V} \; ,
\end{equation}
where $\mathcal{V}$ is the vector space in which the kernel operator acts. In addiction, the kernel is rotationally invariant\footnote{Please note that the Jacobian of a rotation is one.}, \textit{i.e.}
\begin{equation}
    \mathcal{K} (R \vec{q}_1, R \vec{q}_2) = \mathcal{K} (\vec{q}_1, \vec{q}_2) \; , \hspace{0.5 cm} \vec{q}_1, \vec{q}_2 \in \mathcal{V} \; ,
\end{equation}
with $R$ a generic rotation operator in $\mathcal{V}$. As a consequence of this feature, the set of eigenfunctions has the following form:
\begin{equation}
    \phi_{\gamma}^{n} (\vec{q}) = f_{\gamma} (|\vec{q} \;|) Y_n (\phi) \; .
\end{equation}
Here, $\phi$ on the right-hand side denotes the set of angular variables and $Y_n$ is a spherical harmonic, the exponential function $e^{in\phi}$. Due to scale invariance, the function $f_{\gamma} (|\vec{q} \;|)$ must be a power of $|\vec{q} \;|$. We will consider:
\begin{equation}
    f_{\gamma} (|\vec{q} \;|) = (\vec{q}^{\; 2})^{\gamma} \; .
\end{equation}
The exponent $\gamma$ is obtained by imposing the normalization condition of the set,
\begin{equation}
\int d^2 \vec{q} \; \phi_{\nu}^n (\vec{q}) \phi_{\nu'}^{n'} (\vec{q}) = \delta (\nu-\nu') \delta (n-n')\;;
\end{equation}
it is $\gamma=i\nu-1/2$.
Therefore, the set of eigenfunctions of the kernel that solves the problem is
\begin{equation}
\label{Int:Eq:LOBFKLEigenfuntions}
\left \{ \phi_{\nu}^n (\vec{q} \;)= \frac{1}{\pi \sqrt{2}} (\vec{q}^{\;2} )^{-\frac{1}{2}+i\nu} e^{in \theta} \; : \; \nu \in {\rm I\!R} \; , \; n \in \mathbb{Z} \right \}.
\end{equation}
By applying the operator to these eigenfunctions, we can find the corresponding eigenvalues. The result is as follows:
\begin{equation}
    \mathcal{K} \bullet \phi_{\nu}^n = \bar{\alpha}_s \chi(n, \nu)\phi_{\nu}^n \equiv \omega_n (\nu)\phi_{\nu}^n \; ,
\end{equation}
where
\begin{equation}
\label{Int:Eq:AlphaStrongBar}
    \bar{\alpha}_s = \frac{\alpha_s N}{\pi} 
\end{equation}
and
\begin{equation*}
    \chi(n, \nu) =  2 \psi (1) -\psi ( 1/2 + n/2 + i \nu)-\psi ( 1/2 + n/2 - i \nu)
\end{equation*}
\begin{equation}
\label{Int:Eq:CharacteristicFunction}
 = 2 (-\gamma_E - \Re [\psi((n+1)/2 + i\nu)]).  
\end{equation}
The function $\chi(n, \nu)$ is a significant result of L. Lipatov — it is hence referred to as the \textit{Lipatov characteristic function} — and defines the behavior predicted by the BFKL approach in the leading-logarithmic approximation.

At this point, as previously mentioned, we can derive the spectral representation for the Green function, which will therefore be written as
\begin{equation}
\label{sol}
G_{\omega} \left( \vec{q}_1,\vec{q}_2 \right) = \sum_n^{\infty} \int_{-\infty}^{\infty} d\nu \left( \frac{q_1^2}{q_2^2} \right)^{i \nu} \frac{e^{in(\theta_1-\theta_2)}}{2 \pi^2 q_1 q_2} \frac{1}{\omega-\bar{\alpha}_s \chi(n,\nu)}\; .
\end{equation}
Since here we are interested in the leading $\ln s$ behavior, which corresponds to the singularity with the largest real part in the $\omega$-plane, we can make some simplifications. Firstly, the Lipatov characteristic function is decreasing with respect to $n$; consequently, we can restrict ourselves to the case where $n=0$ and disregard the other values in the sum. Moreover, the obtained $\chi(0,\nu)$ decreases as increases $\abs{\nu}$; this allows us to expand the function into a power series in the variable $\nu$ and to retain only the first two terms, thus obtaining
\begin{equation}
\chi(0, \nu)= 4\ln2 - 14 \zeta (3) \nu^2 +... \; .
\end{equation}
In this approximation, the expression \eqref{sol} simply becomes
\begin{equation}
\label{approxsol}
G_{\omega} \left( \vec{q}_1, \vec{q}_2 \right) \approx \frac{1}{\pi q_1 q_2} \int_{-\infty}^{\infty} \frac{d\nu}{2 \pi} \left( \frac{q_1^2}{q_2^2} \right)^{i \nu} \frac{1}{(\omega - \omega_0 + a^2 \nu^2)} ,
\end{equation}
having defined
\begin{equation}
\omega_0 = 4\bar{\alpha}_s \ln2 
\end{equation}
and 
\begin{equation}
a^2 = 14 \bar{\alpha}_s \zeta(3).
\end{equation} 
Eq.~(\ref{approxsol}) can be inverted by performing the anti-Mellin transform,
\begin{equation}
\tilde{G}_{s}(\vec{q}_1, \vec{q}_2) = \frac{1}{2 \pi i} \oint_{C} d\omega \left(\frac{s}{s_0}\right)^{\omega} \frac{1}{\pi q_1 q_2} \int_{-\infty}^{+\infty} \frac{d\nu}{2 \pi} \left( \frac{q_1^2}{q_2^2} \right)^{i \nu} \frac{1}{\omega-\omega_0+a^2 \nu^2}\;,
\end{equation}
where the contour $C$ is to the right of the singularity of the integrand function in the $\omega$-plane. Utilizing the residue theorem, one obtains
\begin{equation}
\begin{split}
\tilde{G}_{s}(\vec{q}_1, \vec{q}_2)& = \frac{1}{2 \pi^2 q_1 q_2} \int_{-\infty}^{+\infty} d\nu \left( \frac{s}{s_0} \right)^{\omega_0 - a^2\nu^2} \left( \frac{q_1^2}{q_2^2} \right)^{i \nu} \\ & = \frac{1}{2 \pi^2 q_1 q_2} \int_{-\infty}^{+\infty} d\nu e^{\left( \omega_0 - a^2 \nu^2 \right) \ln\left( \frac{s}{s_0} \right)+i \nu \ln \left( \frac{q_1^2}{q_2^2} \right)} \; .
\end{split}
\end{equation}
An exponential factor, independent of $\nu$, can be factored out of the integral, while in the remaining term, one can complete the square to obtain
\begin{equation}
\begin{split}
\tilde{G}_{s}(\vec{q}_1, \vec{q}_2) = & \frac{1}{2 \pi^2 q_1 q_2} \left( \frac{s}{s_0} \right)^{\omega_0} \int_{-\infty}^{+\infty} d\nu \\ & \times \exp \left(-a^2 \ln\left( \frac{s}{s_0} \right) \nu^2 + i \ln \left(\frac{q_1}{q_2}\right)\nu + \frac{\ln^2\left(\frac{q_1^2}{q_2^2}\right)}{4a^2\ln\left(\frac{s}{\vec{q}_2^{\; 2}}\right)} - \frac{\ln^2\left(\frac{q_1^2}{q_2^2}\right)}{4a^2\ln\left(\frac{s}{\vec{q}_2^{\; 2}}\right)} \right) \\ & = \frac{1}{2 \pi^2 q_1 q_2} \left( \frac{s}{s_0} \right)^{\omega_0} \int_{-\infty}^{+\infty} d\nu \\ & \times \exp \left( - \frac{\ln^2\left(\frac{q_1^2}{q_2^2}\right)}{4a^2\ln\left(\frac{s}{\vec{q}_2^{\; 2}}\right)} \right) \exp \left( - \left(a \sqrt{\ln\left( \frac{s}{s_0} \right)} \nu - \frac{i \ln \left(\frac{q_1^2}{q_2^2} \right)}{2a \sqrt{\ln \left(\frac{s}{s_0}\right)}} \right)^2 \right) \; .
\end{split}
\end{equation}
Applying the substitution
\begin{equation}
z = \nu - i \frac{\ln \left( \frac{q_1^2}{q_2^2} \right)}{2a^2 \ln \left( \frac{s}{s_0} \right)} \; ,
\end{equation} 
and remembering the solution of the Gaussian integral, 
\begin{equation}
\int_{-\infty}^{\infty} dz e^{-Az^2} = \sqrt{\frac{\pi}{A}} \; ,
\end{equation}
the result is
\begin{equation}
\tilde{G}_{s}(\vec{q}_1, \vec{q}_2) \approx \frac{1}{\sqrt{\vec{q}_1^{\; 2} \vec{q}_2^{\; 2}}} \left( \frac{s}{s_0}\right)^{\omega_0} \frac{1}{\sqrt{\pi \ln(s/s_0)}} \frac{1}{2 \pi a} \exp \left(-\frac{\ln^2(\vec{q}_1^{\; 2}/\vec{q}_2^{\; 2})}{4a^2 \ln(s/s_0)} \right).
\end{equation}
Using the power series expansion of the exponential, one has
\begin{equation}
\begin{split}
\tilde{G}_{s}(\vec{q}_1, \vec{q}_2) \approx & \frac{1}{\sqrt{\vec{q}_1^{\; 2} \vec{q}_2^{\; 2}}} \left( \frac{s}{s_0}\right)^{\omega_0} \frac{1}{\sqrt{\pi \ln(s/s_0)}} \\ & \frac{1}{2 \pi a} \left[1 + \left(-\frac{\ln^2(\vec{q}_1^{\; 2}/\vec{q}_2^{\; 2})}{4a^2 \ln(s/s_0)} \right) + \frac{1}{2} \left(-\frac{\ln^2(\vec{q}_1^{\; 2}/\vec{q}_2^{\; 2})}{4a^2 \ln(s/s_0)} \right)^2 + ... \right].
\end{split}
\end{equation}
With $\alpha_s$ fixed and increasing $s$, all terms of the expansion are suppressed with respect to one, and the leading contribution is represented by
\begin{equation}
\tilde{G}_{s}(\vec{q}_1, \vec{q}_2) \approx \frac{1}{\sqrt{\vec{q}_1^{\; 2} \vec{q}_2^{\; 2}}} \left( \frac{s}{s_0}\right)^{\omega_0} \frac{1}{\sqrt{\pi \ln(s/s_0)}} \frac{1}{2 \pi a}\;.
\end{equation}
This important result indicates that the total cross section, calculated in LLA, $\sigma_{tot}^{LLA}$ increases at large center-of-mass energies as
\begin{equation}
\label{Int:Eq:AsyPomerSing}
\sigma_{tot}^{LLA} \sim \frac{s^{\omega_0}}{\sqrt{\ln s}}\;,
\end{equation}
where $\omega_0 = (g^2 N \ln 2)/\pi^2$, with $N$ representing the number of colors in QCD, denotes the LLA position of the rightmost singularity in the complex momentum plane of the $t$-channel partial wave with vacuum quantum numbers.
\subsection{The BFKL equation in the NLLA}
In the preceding sections, we developed the BFKL approach in the LLA. Let us continue our journey with the most natural next step, that is the discussion about the BFKL equation in the NLLA. While the underlying approach remains largely similar to what has been outlined above, the technical work required to finalize the calculation program is considerable. Therefore, in the subsequent sections, we will restrict our discussion to a brief introductory overview, with full details available in \cite{Fadin:1998sh}.
In the NLLA, Eq.~(\ref{Int:Eq:ReggeizedAmp}) has been confirmed through the first three orders of perturbation theory and assumed valid across all orders. It was only afterward that this result was established to hold universally across perturbative orders. Proving Reggeization in the NLLA remains a complex challenge; for detailed discussions, refer to \cite{Ioffe:2010zz} and references therein. \\
Here, we focus on extracting the Regge trajectory under this approximation. To this end, let us revisit quark-quark scattering and expand Eq.~(\ref{Int:Eq:ReggeizedAmp}) up to $\alpha_s^3$:
\begin{equation*}
    \Gamma^a_{1'1} \frac{s}{t} \left[ \left(\frac{s}{-t}\right)^{\omega(t)} + \left( \frac{-s}{-t} \right)^{\omega(t)} \right] \Gamma^a_{2'2} \simeq  \bigg \{ \Gamma^{a(0)}_{1'1} \frac{2s}{t} \Gamma^{a(0)}_{2'2} \bigg \}_{\color{red} _{LLA}} 
\end{equation*}
\begin{equation*}
\begin{split}
  +& \bigg \{ \Gamma^{a(0)}_{1'1} \frac{s}{t} \left[ \omega^{(1)}(t) \ln \left(\frac{s}{-t}\right) + \omega^{(1)}(t) \ln \left( \frac{-s}{-t} \right) \right] \Gamma^{a(0)}_{2'2} \bigg \}_{\color{red} _{LLA}} \hspace{-0.4 cm}  \\+&\bigg \{ \Gamma^{a(1)}_{1'1} \frac{2s}{t} \Gamma^{a(0)}_{2'2} + \Gamma^{a(0)}_{1'1} \frac{2s}{t} \Gamma^{a(1)}_{2'2} \bigg \}_{\color{blue} _{NLLA}}
  \end{split}
\end{equation*}
\begin{equation*}
    + \bigg \{ \Gamma^{a(0)}_{1'1} \frac{s}{t} \left[ \frac{(\omega^{(1)}(t))^2}{2} \ln^2 \left(\frac{s}{-t}\right) + \frac{(\omega^{(1)}(t))^2}{2} \ln^2 \left( \frac{-s}{-t} \right) \right] \Gamma^{a(0)}_{2'2} \bigg \}_{\color{red} _{LLA}}
\end{equation*}
\begin{equation*}
\begin{split}
   +\bigg \{ \Gamma^{a(1)}_{1'1} \frac{s}{t} \omega^{(1)}(t) \left[ \ln \left(\frac{s}{-t}\right) + \ln \left( \frac{-s}{-t} \right) \right] \Gamma^{a(0)}_{2'2} \\+ \Gamma^{a(0)}_{1'1} \frac{s}{t} \omega^{(1)}(t) \left[  \ln \left(\frac{s}{-t}\right) +  \ln \left( \frac{-s}{-t} \right) \right] \Gamma^{a(1)}_{2'2}
   \end{split}
\end{equation*}
\begin{equation}
   + \Gamma^{a(0)}_{1'1} \frac{s}{t} \left[ \omega^{(2)}(t) \ln \left(\frac{s}{-t}\right) + \omega^{(2)}(t) \ln \left( \frac{-s}{-t} \right) \right] \Gamma^{a(0)}_{2'2} \bigg \}_{\color{blue} _{NLLA}} \hspace{-0.4 cm} + \hspace{0.1 cm} \bigg \{ \Gamma^{a(2)}_{1'1} \frac{2s}{t} \Gamma^{a(0)}_{2'2} + ... \bigg \}_{\color{purple} _{NNLLA}} \; . 
\label{Int:Eq:ReggeExpAlpha3}
\end{equation}
In Eq.~(\ref{Int:Eq:ReggeExpAlpha3}), we reintroduce the superscript $(i)$ to specify $i$-loop corrections for each term, alongside a subscript identifying the respective approximation level. From analyzing Eq.~(\ref{Int:Eq:ReggeExpAlpha3}), we find that
\begin{itemize}
    \item The first line shows the sole term of order $\alpha_s$, directly proportional to the Born vertices and contributing within the LLA.
    \item The second line has three terms of order $\alpha_s^2$: the first, proportional to the $1$-loop Regge trajectory, includes an energy logarithm balancing the additional $\alpha_s$ factor, thus contributing within LLA. The other two terms relate to the $1$-loop corrections to each effective vertex, presenting an extra $\alpha_s$ power without a compensating energy logarithm; these contribute within NLLA. 
    \item The remaining lines list terms of order $\alpha_s^3$. The initial term, proportional to the squared $1$-loop Regge trajectory, includes two energy logarithms balancing the $\alpha_s^2$ factor, thus contributing within LLA. The next two terms combine the $1$-loop Regge trajectory, $1$-loop vertex corrections, and an energy logarithm compensating an $\alpha_s$ factor, placing them within NLLA. The fourth term involves a $2$-loop correction to the Regge trajectory with a large energy logarithm, contributing within NLLA. Lastly, terms involving products of $1$-loop vertex corrections or $2$-loop corrections for each effective vertex contribute within NNLLA.
\end{itemize}
\begin{figure}
\begin{picture}(400,280)
\put(30,227){ $\bullet$ \; $\omega^{(1)} (t) \; \longrightarrow \; \omega^{(2)} (t)$}
\put(230,200){\includegraphics[width=0.05\textwidth]{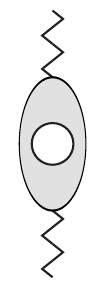}}
\put(255,247){1-loop}
\put(375,247){2-loop}
\put(30,137){ $\bullet$ $ \Gamma_{P'P}^{c (0)} \; \longrightarrow \; \Gamma_{P'P}^{c (1)} $}
\put(300,227){$\longrightarrow$}
\put(350,200){\includegraphics[width=0.05\textwidth]{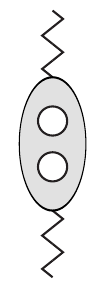}}
\put(210,120){\includegraphics[width=0.15\textwidth]{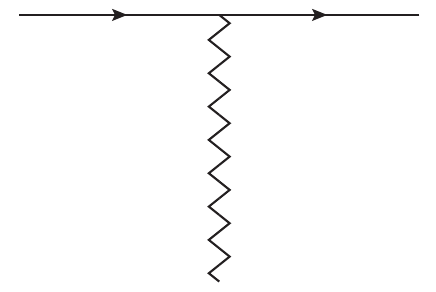}}
\put(260,148){Born}
\put(300,143){$\longrightarrow$}
\put(330,120){\includegraphics[width=0.15\textwidth]{images/PPRVertex.pdf}}
\put(380,148){1-loop}
\put(30,57){ $\bullet$ $ \gamma_{c_i c_{i+1}}^{G_i (0)} \; \longrightarrow \; \gamma_{c_i c_{i+1}}^{G_i (1)}  $}
\put(235,30){\includegraphics[width=0.1\textwidth]{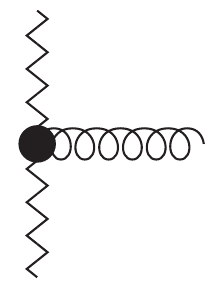}}
\put(255,77){Born}
\put(300,57){$\longrightarrow$}
\put(355,30){\includegraphics[width=0.1\textwidth]{images/LipatovVertex.pdf}}
\put(375,77){1-loop}
\end{picture}
\caption{Next-to-leading corrections to Regge trajectory, PPR vertices, and central gluon production vertex in the multi-Regge kinematics.}
\label{MRK}
\end{figure}
This procedure now becomes evident. We need to return to our $1$-loop calculation of quark-quark scattering and include corrections previously omitted in LLA, which do not produce large energy logarithms. Among these are contributions from box and cross diagrams (Fig.~\ref{Int:Fig:LOqqqq}), which originate from previously ignored phase space regions, and vertex corrections, which, in Feynman gauge, do not result in large energy logarithms. \\
Comparing these results to the $\alpha_s$ correction in Eq.~(\ref{Int:Eq:ReggeExpAlpha3}), we can identify effective vertices up to $1$-loop precision. Generally, they are expressed as follows:
\begin{equation}
\label{PPRNLLA}
\Gamma_{P'P} = \delta_{\lambda_P, \lambda_{P'}} \Gamma_{P'P}^{(+)} + \delta_{\lambda_P, -\lambda_{P'}} \Gamma_{P'P}^{(-)} \; ,
\end{equation}  
where in the first term of Eq.~(\ref{PPRNLLA}), helicity is conserved, while a second term allows for non-conservation of helicity~\cite{Fadin:1992zt,Fadin:1993rc,Fadin:1993qb,Fadin:1994fj,Fadin:1995km}. \\
Once vertex corrections are obtained, only $\omega^{(2)} (t)$ remains to be determined. To achieve this, we compute the $2$-loop correction for the quark-quark scattering amplitude in the high-energy limit, isolating contributions that lead to energy logarithms and comparing them with Eq.~(\ref{Int:Eq:ReggeExpAlpha3}). The result can be found in \cite{Fadin:1998sh}, initially derived in \cite{Fadin:1995km,Fadin:1994uz,Kotsky:1996xm,Fadin:1995xg,Fadin:1996tb}.

\subsubsection{Multi-Regge kinematics}
In the MRK and NLLA frameworks, as well as in the LLA, only amplitudes with gluon quantum numbers in the momentum transfer channels \( q_i \) contribute. The inclusion of amplitudes on the right-hand side of Eq.~(\ref{Unitary}) with quantum numbers in \( t_i \)-channels that differ from those of gluons results in the loss of at least two significant logarithms. Consequently, these terms can be neglected in the NLLA. As previously noted, the central aspect of calculating the amplitudes involved in the unitarity relation (\ref{Unitary}) is the Reggeization of gluons. Within MRK, the real parts of the contributing amplitudes (the only parts relevant in NLLA, as LLA amplitudes are purely real) are expressed in the same manner as in LLA.

For this kinematic configuration, the two-loop contribution \(\omega^{(2)}(t)\) to the gluon Regge trajectory \(\omega(t)\) and the corrections to the real parts of the PPR vertices, previously discussed, are required. Additionally, it is necessary to determine the one-loop corrections to the Lipatov vertex~\cite{Fadin:1993rc,Fadin:2000yp,Fadin:2023roz}. The workflow for computing these corrections is illustrated in Fig.~\ref{MRK}. 

\subsubsection{Quasi-multi-Regge kinematics}
One of the most notable aspects of the NLLA is the emergence of a new type of kinematics. Unlike in the LLA, where MRK is the sole kinematics contributing to the unitarity relation (\ref{Unitary}), this is no longer the case in the NLLA. The relaxation of constraints allows for the "loss" of a single large logarithm compared to the LLA. As a result, the strict ordering of gluons in rapidity space is no longer mandatory. Specifically, any single pair of produced particles can exhibit a fixed invariant mass (independent of \( s \)), meaning the rapidities of the particles in this pair can be of a similar magnitude. This configuration is referred to as quasi-multi-Regge kinematics (QMRK).
\begin{figure}
\begin{picture}(400,260)
\put(30,217){ $\bullet$ $ \Gamma_{P'P}^{c (0)} \; \longrightarrow \; \Gamma_{ \{f \} P}^{c (0)} $}
\put(210,200){\includegraphics[width=0.15\textwidth]{images/PPRVertex.pdf}}
\put(30,137){ $\bullet$ $ \gamma_{c_i c_{i+1}}^{G_i (0)} \; \longrightarrow \; \gamma_{c_i c_{i+1}}^{G G (0)}  $}
\put(300,227){$\longrightarrow$}
\put(330,200){\includegraphics[width=0.15\textwidth]{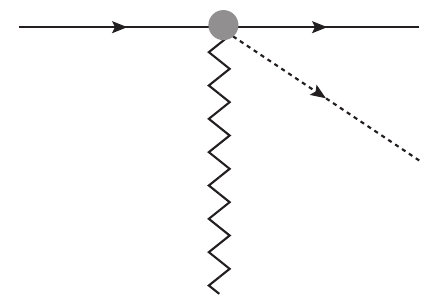}}
\put(235,120){\includegraphics[width=0.1\textwidth]{images/LipatovVertex.pdf}}
\put(300,143){$\longrightarrow$}
\put(355,120){\includegraphics[width=0.1\textwidth]{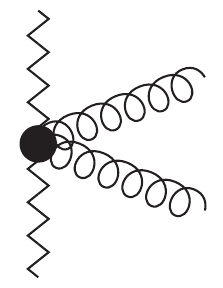}}
\put(30,57){ $\bullet$ $ \gamma_{c_i c_{i+1}}^{G_i (0)} \; \longrightarrow \; \gamma_{c_i c_{i+1}}^{Q Q (0)}  $}
\put(235,30){\includegraphics[width=0.1\textwidth]{images/LipatovVertex.pdf}}
\put(300,57){$\longrightarrow$}
\put(355,30){\includegraphics[width=0.1\textwidth]{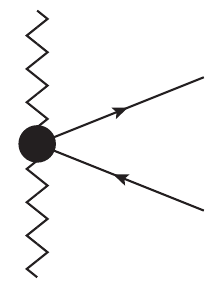}}
\end{picture}
\caption{Next-to-leading corrections to the vertices PPR and RRG in the quasi-multi-Regge kinematics.}
\label{QMRK}
\end{figure}
This kinematics can be addressed by incorporating, alongside the production of a single gluon, the production of more complex states in Reggeon-Reggeon (RR) and Reggeon-particle (RP) interactions, which were omitted in the LLA. Specifically, the following must be considered:  
\begin{itemize}  
    \item Account for the production of gluon-gluon (GG)~\cite{Fadin:1989kf,Fadin:1996nw,Fadin:1996zv} and quark-antiquark ($\bar{Q}Q$)~\cite{Catani:1990xk,Catani:1990eg,Camici:1996st,Camici:1997ta,Fadin:1997hr} states in Reggeon-Reggeon collisions.  
    \item Include the production of states with a higher number of particles in Reggeon-particle collisions, occurring in the fragmentation region of one of the initial particles.  
\end{itemize}  
The workflow for computing these corrections is illustrated in Fig.~\ref{QMRK}.

\subsubsection{BFKL kernel and impact factors in the NLLA}
As previously noted, the factorization of the cross section (\ref{Int:Eq:TotCross}) can still be expressed in the same form within the NLLA, albeit with modified impact factors and Green functions. This is a crucial point, as it demonstrates that even in the NLLA, the amplitudes retain the same factorized structure as depicted in Fig.~\ref{convolution}.
\begin{figure}
\begin{picture}(400,110)
\put(30,0){\includegraphics[scale=0.5]{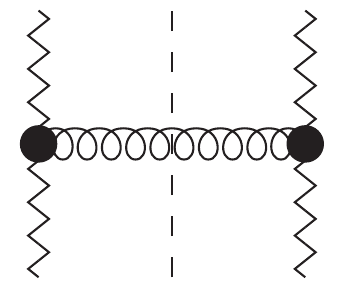}}
\put(55,85){$ \mathcal{K}_{RRG}^{(1)} $}
\put(115,0){\includegraphics[scale=0.5]{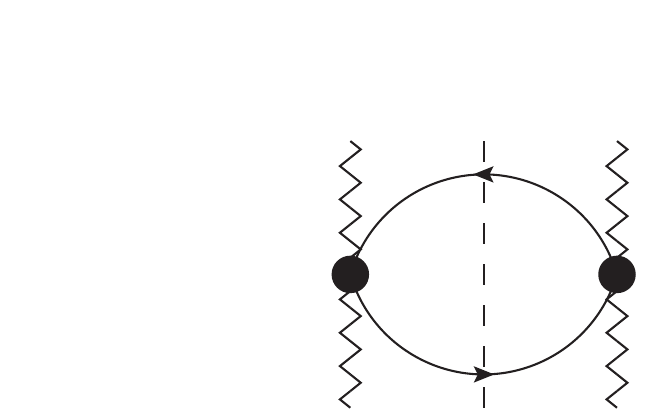}}
\put(212,85){$ \mathcal{K}_{RRQQ}^{(0)} $}
\put(270,0){\includegraphics[scale=0.5]{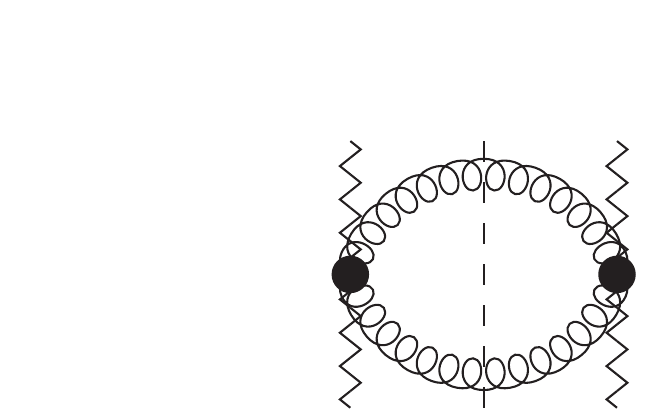}}
\put(367,85){$ \mathcal{K}_{RRGG}^{(0)} $}
\end{picture}
\caption{Schematic representation of the real part of the NLO BFKL kernel. In the first contribution, one Lipatov vertex is to be taken at one-loop.}
\label{Int:Fig:RealKernelcorrectionsNLO}
\end{figure}
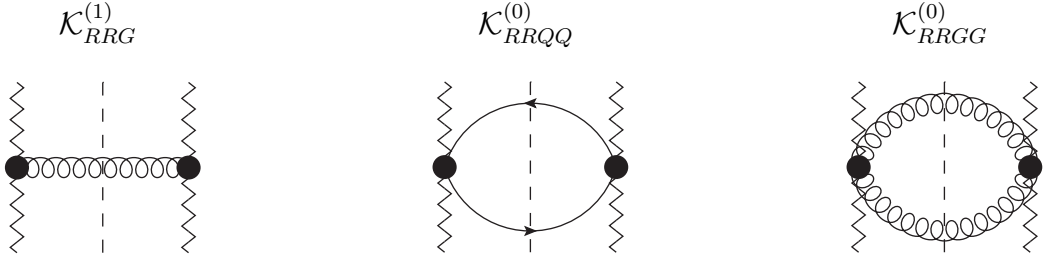
In defining the impact factors (\ref{Int:Eq:ImpactFactSingGen}), it is necessary to account for radiative corrections to the PPR vertices as well as the contributions from "excited" states in the fragmentation region. The Green function equation (\ref{BFKLequation}) remains unaltered, along with the kernel representation (\ref{kernel}). However, the gluon trajectory must be considered at the two-loop approximation,
\begin{equation}
\omega(t)= \omega^{(1)}(t)+ \omega^{(2)}(t)\;.
\end{equation}
,Additionally, the part associated with real particle production must include not only the contribution from single-gluon production in RR collisions (evaluated at the one-loop level) but also contributions from two-gluon and quark-antiquark productions. The single-gluon contribution must be computed with one-loop precision, while the two-gluon and quark-antiquark contributions should be considered in the Born approximation. For the case of forward scattering, the real production component is represented as shown in Fig.~\ref{Int:Fig:RealKernelcorrectionsNLO}:
\begin{equation}
\mathcal{K}_r \left( \vec{q}_1, \vec{q}_2 \right) = \mathcal{K}_{RRG}^{(1)} \left( \vec{q}_1, \vec{q}_2 \right) + \mathcal{K}_{RRGG}^{(0)} \left( \vec{q}_1, \vec{q}_2 \right) + \mathcal{K}_{RRQ\bar{Q}}^{(0)} \left( \vec{q}_1, \vec{q}_2 \right)\;.
\end{equation}
It is crucial to highlight that the introduction of new kinematics brings technical challenges. When calculating the two-gluon production contribution to the kernel and the contribution to the impact factor from gluon production in the fragmentation region, divergences arise in the integrals over the invariant masses of the produced particles at their upper limits\footnote{These will be referred to as \textit{rapidity divergences} in the next chapter.}. These divergences occur due to the lack of a natural boundary separating MRK and QMRK. To handle these contributions rigorously and assign them a well-defined meaning, an artificial boundary, denoted as \( s_{\Lambda} \), is introduced~\cite{Fadin:1998fv}. Naturally, the dependence on this artificial parameter vanishes in the final results.

As a result of the above considerations, the next-to-leading order definition of the impact factor, in the singlet case, is given by
\begin{equation*}
\Phi_{AA}(\vec q_1; s_0) = \left( \frac{s_0}
{\vec q_1^{\:2}} \right)^{\omega( - \vec q_1^{\:2})}
\sum_{\{f\}}\int\theta(s_{\Lambda} -
s_{AR})\frac{ds_{AR}}{2\pi}\ d\rho_f \ \Gamma_{\{f\}A}^c
\left( \Gamma_{\{f\}A}^{c^{\prime}} \right)^* 
\langle cc^{\prime} | \hat{\cal P}_0 | 0 \rangle
\end{equation*}
\begin{equation}
-\frac{1}{2}\int d^{D-2}q_2\ \frac{\vec q_1^{\:2}}{\vec q_2^{\:2}}
\: \Phi_{AA}^{(0)}(\vec q_2)
\: {\cal K}^{(0)}_r (\vec q_2, \vec q_1)\:\ln\left(\frac{s_{\Lambda}^2}
{s_0(\vec q_2 - \vec q_1)^2} \right)~.
\label{Int:Eq:ImpactproNext}
\end{equation}
The first term in Eq.~(\ref{Int:Eq:ImpactproNext}) closely resembles the corresponding term in Eq.~(\ref{Int:Eq:ImpactFactSingGen}), with a significant difference: the inclusion of a Heaviside theta function. This function creates a clear separation between MRK and QMRK, effectively regularizing the rapidity divergence (as \( s_{AR} \) approaches infinity). In this framework, a contribution involving the emission of an additional gluon is assigned to the "QMRK part" of the impact factor only if \( s_{AR} < s_{\Lambda} \), meaning that the particle (or particle system) present at leading order and the additional emitted gluon are not strongly separated in rapidity.
The second term, on the other hand, arises from an MRK contribution to the amplitude in the NLLA~\cite{Fadin:1989kf}. In alignment with the definition of QMRK kinematics using the cutoff \( s_{\Lambda} \), the integration over invariant masses \( s_i \) in MRK kinematics is restricted by the condition \( s_i > s_{\Lambda} \).  
In the LLA, contributions from the lower bound of integration over invariant masses are negligible within this approximation. However, in the NLLA, terms originating from the lower limit of integration must be taken into account. Specifically, for the invariant mass \( s_{AR} \), the resulting term corresponds precisely to the second line in Eq.~(\ref{Int:Eq:ImpactproNext}). 
The \( s_{\Lambda} \) parameter is to be understood as tending to infinity. While both expressions in the first and second lines of Eq.~(\ref{Int:Eq:ImpactproNext}) are individually divergent, their combination eliminates any dependence on \( s_{\Lambda} \), yielding a finite result. Moreover, as can be intuitively inferred, the kernel contribution from the emission of two gluons also includes a term analogous to the second term in Eq.~(\ref{Int:Eq:ImpactproNext}). For a detailed calculation of the separation between MRK and QMRK, refer to \cite{Fadin:1998sh}.\\

This paragraph concludes the discussion of the BFKL formalism within the NLLA, which represents the level of accuracy adopted for both the calculations and the phenomenological analyses presented in this thesis.

\chapter{The real corrections to the Higgs Impact Factor at Next-to-Leading Order with a physical top mass}
\chaptermark{Real corrections to the Higgs Impact Factor}

In this chapter, we focus on calculating the impact factor for the production of a forward Higgs boson at next-to-leading-order accuracy. This calculation serves as a key component for various studies centered on Higgs phenomenology (see, e.g.,~\cite{Bonvini:2018ixe}). Specifically, it can be applied to investigate the inclusive hadroproduction of a forward Higgs boson. Additionally, it plays a crucial role in describing Higgs production in association with jets~\cite{DelDuca:1993ga,Celiberto:2020tmb,Andersen:2022zte,Andersen:2023kuj} or hadrons~\cite{Celiberto:2022zdg} at large rapidity separations. As such, it provides a robust theoretical foundation for analyzing Higgs production in kinematic regimes accessible at the current LHC and the future FCC~\cite{FCC:2018byv,FCC:2018evy,FCC:2018vvp,FCC:2018bvk}.\\

The leading-order impact factor for Higgs production was computed in~\cite{DelDuca:1993ga} (see also~\cite{Celiberto:2020tmb}). Extending this calculation to next-to-leading order presents significant challenges, as it involves two-loop amplitudes with massive propagators. The NLO impact factor was derived using the Lipatov effective action formalism in~\cite{Hentschinski:2020tbi,Nefedov:2019mrg} and through the formalism outlined in the first chapter of~\cite{Celiberto:2022fgx}. Both calculations are consistent with each other but are conducted in the infinite top-mass approximation, where the coupling between gluons and the Higgs boson is described by a dimension-5 non-renormalizable operator.\\

Relaxing the infinite-top-mass approximation and computing the NLO Higgs impact factor at the physical top-quark mass is a necessary step forward for at least two reasons. The first is straightforward: the need for improved precision. Top-mass effects are expected to become significant when the Higgs boson's transverse momentum approaches the top mass (see, e.g.,~\cite{Maltoni:2018dar}). Incorporating heavy-quark finite mass corrections into high-energy resummation provides a valuable tool for exploring higher center-of-mass energy regimes, such as those anticipated at the FCC.\\

A key element in expressing high-energy distributions in the BFKL factorized form is the Mellin representation of impact factors. The first Mellin projection of the LO forward-Higgs impact factor onto BFKL kernel eigenfunctions was introduced in Ref.~\cite{Celiberto:2020tmb}. This work also provided BFKL predictions for azimuthal-angle correlations and transverse-momentum distributions in Higgs-plus-jet production, including NLLA contributions via the {\tt JETHAD} interface~\cite{Celiberto:2020wpk,Celiberto:2022rfj,Celiberto:2023fzz,Celiberto:2024mrq,Celiberto:2024swu}, combining the Green function, LO impact factors, and NLO universal terms inferred through RG analysis.
The study highlighted two major findings. First, azimuthal correlations showed strong stability against next-to-leading corrections and scale variations, suggesting that the large transverse masses in Higgs emissions act as natural stabilizers for BFKL. This contrasts with previous studies on light jets and hadrons~\cite{Celiberto:2015yba,Celiberto:2020wpk}, where large higher-order corrections caused significant instabilities. Similar stabilization trends were later confirmed for heavy-flavor emissions~\cite{Celiberto:2021dzy,Celiberto:2021fdp,Celiberto:2022dyf,Celiberto:2022rfj}.
Second, differential cross sections in Higgs transverse momentum proved to be effective in distinguishing BFKL predictions from fixed-order results, with the difference growing at higher momenta.
Preliminary studies on Higgs transverse-momentum in the inclusive Higgs-plus-jet channel revealed stability under corrections, but a significant discrepancy between NLO fixed-order and high-energy NLLA resummation led to the development of a matching method to resolve double counting.\\

The second reason is more subtle and pertains to formal issues~\cite{Celiberto:2022fgx,Fucilla:2022whr} that emerged during the NLO Higgs impact factor calculation in the infinite-top-mass approximation, which were addressed in a recent study~\cite{Fucilla:2024cpf}. A full calculation is essential to validate the consistency of the results obtained using the infinite-top-mass approximation, particularly in relation to gluon Reggeization. At the one-loop level, this consistency was shown to hold in a highly non-trivial manner due to the involvement of the effective non-renormalizable Higgs-gluon coupling~\cite{Fucilla:2024cpf}. \\

In this chapter, we take the initial step in this direction by presenting the calculation of the real corrections to the impact factor, explicitly incorporating the dependence on the top-quark mass. 

\section{General definition of BFKL impact factors}
In this section, we revisit key concepts from the previous chapter, as they are essential for the discussion in the current one. To provide a comprehensive and thorough examination, we will proceed by following the calculations and analyses conducted in the work presented in ~\cite{Celiberto:2024bbv}. The starting point of our analysis is the NLO impact factor of the Higgs, as presented in Ref.~\cite{Fadin:1998fv}:
$$
\Phi_{AA}(\vec q_1; s_0) = \left( \frac{s_0}
{\vec q_1^{\:2}} \right)^{\omega( - \vec q_1^{\:2})}
\sum_{\{f\}}\int\theta(s_{\Lambda} -
s_{AR})\frac{ds_{AR}}{2\pi}\ d\rho_f \ \Gamma_{\{f\}A}^c
\left( \Gamma_{\{f\}A}^{c^{\prime}} \right)^* 
\langle cc^{\prime} | \hat{\cal P}_0 | 0 \rangle
$$
\begin{equation}
-\frac{1}{2}\int d^{D-2}q_2\ \frac{\vec q_1^{\:2}}{\vec q_2^{\:2}}
\: \Phi_{AA}^{(0)}(\vec q_2)
\: {\cal K}^{(0)}_r (\vec q_2, \vec q_1)\:\ln\left(\frac{s_{\Lambda}^2}
{s_0(\vec q_2 - \vec q_1)^2} \right)~,
\label{ImpactUnpro}
\end{equation}
where \( \omega(t) \) denotes the Reggeized gluon trajectory, which appears in this expression at LO, given by
\begin{equation}
    \omega^{(1)}(t) = \frac{g^2 t}{(2 \pi)^{D-1}} \frac{N}{2} \int \frac{d^{D-2}k_{\perp}}{k_{\perp}^2 (q-k)_{\perp}^{2}} = - \frac{g^2 N \Gamma(1+\epsilon)(\vec{q}^{\; 2})^{-\epsilon}}{(4 \pi)^{2-\epsilon}} \frac{\Gamma^2(-\epsilon)}{\Gamma(-2\epsilon)} \; ,
    \label{ReggeTraj}
\end{equation}
with $t=q^2=-\vec q^{\:2}$ and $N$ the number of colors.
\begin{figure}
\begin{center}
\includegraphics[scale=0.50]{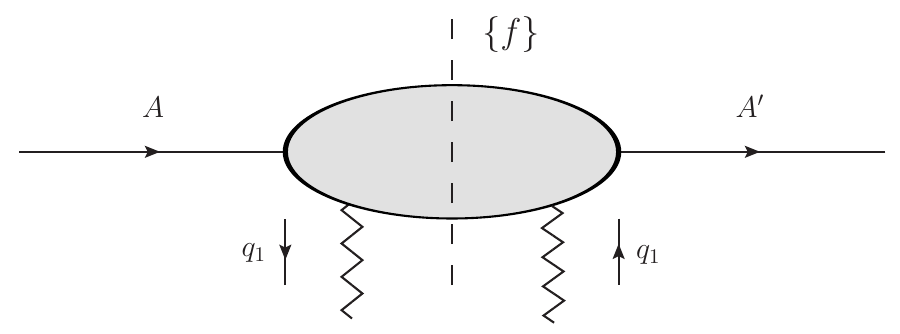}
\end{center}
\caption{Schematic description of an impact factor.}
\label{ImpactFactorRepres}
\end{figure}
$\Gamma_{\{f\}A}^c$ is the effective vertex for production of the
system ${\{f\}}$ (see Fig.~\ref{ImpactFactorRepres}) in the collision of the particle $A$ off the Reggeized gluon with color index $c$ and momentum $-q_1$, with
\begin{equation}
q_1 = \alpha k_2 + {q_1}_{\perp}~,\ \ \ \ \ \ \ \alpha 
\approx -\left( s_{AR} - m_A^2 + \vec q_1^{\:2} \right)/s \ll 1~, 
\end{equation}
and \( s_{AR} \) represents the squared invariant mass of the particle-Reggeon system. In the fragmentation region of particle \( A \), where all transverse momenta as well as the invariant mass \( \sqrt{s_{AR}} \) do not increase with \( s \), we find \( q_1^2 = - \vec{q}_1^{\:2} \). The factor
\begin{equation}
    \langle cc^{\prime} | \hat{\cal P}_{0} | 0 \rangle = \frac{\delta^{c c'}}{\sqrt{N^2-1}}
\end{equation}
is the projector on the singlet color state representation. The summation in Eq.~(\ref{ImpactUnpro}) is performed over all systems \( \{f\} \) that can be produced in the NLLA, with integration carried out over the phase space volume of the produced system. For an \( n \)-particle system (if identical particles are present, the appropriate symmetry factors should also be included), this phase space volume is given by:
\begin{equation}
d\rho_f = (2\pi)^D\delta^{(D)}\biggl(p_A - q_1 - \sum_{m=1}^nk_m\biggr)
\prod_{m=1}^n\frac{d^{D-1}k_m}{2E_m(2\pi)^{D-1}}~,
\label{GenPhasSpa}
\end{equation}
as well as over the particle-Reggeon invariant mass. The average over initial-state color and spin degrees of freedom is implicitly assumed.
The parameter \( s_{\Lambda} \), which limits the integration region over the invariant mass in the first term on the right-hand side of Eq.~(\ref{ImpactUnpro}), is introduced to separate the contributions of multi-Regge and quasi-multi-Regge kinematics (MRK and QMRK). This parameter is to be considered as tending to infinity. The dependence of the impact factors on \( s_{\Lambda} \) vanishes due to the cancellation between the first and second terms on the right-hand side of Eq.~(\ref{ImpactUnpro}).The second term, commonly referred to as the "BFKL counter-term", includes \(\Phi_{AA}^{(0)}\), the Born-level contribution to the impact factor, which is independent of \( s_0 \). It also contains \({\cal K}^{(0)}_r\), the part of the BFKL kernel in the Born approximation associated with real particle production, given by
\begin{equation}
{\cal K}^{(0)}_r(\vec q_1, \vec q_2) =
\frac{2 g^2 N}{(2\pi)^{D-1}} \frac{1}{(\vec q_1 - \vec q_2)^2}\;.
\label{BornKer}
\end{equation}

\section{Amplitude calculation}
In this part, we re-calculate the amplitudes necessary for constructing the impact factor, following the original calculations presented in Ref.\cite{DelDuca:2001fn}, which we use for comparison. The amplitudes are generated with {\tt FormCalc} \cite{Hahn:1998yk}, and then the results are organized as done in Ref.~\cite{DelDuca:2001fn}.

\subsection{The off-shell $ggH$ amplitude}
\label{ssec:ggH_vertex}

The off-shell two gluons-Higgs vertex can be expanded on the tensor basis 
\beqa
 \left[ g^a(q_1^\mu) g^b(q_2^\nu) H (q_3) \right] &\equiv&  i \delta^{ab} T^{\mu\nu}(q_1, q_2) \nn \\ &=& i \delta^{ab} \left\{  F_T(q_1, q_2)  t_T^{\mu\nu}(q_1, q_2)  + F_L(q_1, q_2)  t_L^{\mu\nu}(q_1, q_2)  \right\}  \nn \;,
\eqa
with
\beqa
t_T^{\mu\nu}(q_1, q_2) &=& q_1 \cdot q_2 \, g^{\mu\nu} - q_2^\mu q_1^\nu \;,\nn \\
t_L^{\mu\nu}(q_1, q_2) &=& q_1^2 \, q_2^2  g^{\mu\nu}  +  q_1 \cdot q_2 \, q_1^\mu q_2^\nu  -  q_2^2 \, q_1^\mu q_1^\nu -  q_1^2  \, q_2^\mu q_2^\nu \;,
\label{Eq:FundamentalTriangularAmo}
\eqa
which can be obtained as solutions of the two Ward-identities $q_{1 \mu} T^{\mu\nu} = q_{2 \nu} T^{\mu\nu} = 0$.
The validity of the two QED-like Ward identities, regardless of whether the gluons are on-shell, arises because Higgs production through two-gluon fusion is essentially Abelian, given the absence of a triple-gluon vertex. The coefficients $F_T$ and $F_L$ are free from both IR and UV divergences. Additionally, they remain finite and non-zero when either $q_1^2 = 0$ or $q_2^2 = 0$. If one of the gluons is both on-shell and transverse, the longitudinal component vanishes,
\beqa
\epsilon_\mu(q_1) t_L^{\mu\nu}(q_1, q_2) |_{q_1^2 = 0} = \epsilon_\nu(q_2) t_L^{\mu\nu}(q_1, q_2) |_{q_2^2 = 0} = 0 \, .
\eqa 
The full expressions for the two coefficients are
\beqa
F_L(q_1,q_2) &=& \frac{\alpha_s m_t^2}{\pi v} \bigg\{  \frac{1}{2 \det \mathcal G} \left\{ \left[ 2 - \frac{3 q_1^2 \, q_2 \cdot p_H}{\det \mathcal G} \right] ( B_0(q_1^2) -  B_0(m_H^2))  \right. \nn \\
&+& \left[ 2 - \frac{3 q_2^2 \, q_1 \cdot p_H}{\det \mathcal G} \right] ( B_0(q_2^2) -  B_0(m_H^2)  )   \nn \\
 &-& \left. \left[ 4 m_t^2 + q_1^2 + q_2^2 + m_H^2 - 3 m_H^2 \frac{q_1^2 q_2^2 }{\det \mathcal G} \right]  C_0(q_1^2, q_2^2) - 2   \right\} \bigg\} \,, 
\label{Eq:FLtri}
\eqa
\beqa
F_T(q_1,q_2) &=& \frac{\alpha_s m_t^2}{\pi v} \bigg\{ \frac{1}{2 \det \mathcal G} \left\{  m_H^2 \left[  B_0(q_1^2) +  B_0(q_2^2)  - 2  B_0(m_H^2) - 2 q_1 \cdot q_2  C_0(q_1^2, q_2^2)   \right] \right. \nn \\
&+& \left. (q_1^2  - q_2^2) ( B_0(q_1^2) -  B_0(q_2^2)) \right\}  \bigg\}    - q_1 \cdot q_2  F_L(q_1^2,q_2^2) \;.
\label{Eq:FTtri}
\eqa
In the given expressions, we have defined \( p_H = q_1 + q_2 \) and denote the Gram determinant as \( \det \mathcal{G} = q_1^2 q_2^2 - (q_1 \cdot q_2)^2 \). The terms \( B_0 \) and \( C_0 \) represent the bubble and triangle master integrals, respectively. Furthermore, we introduce \( v^2 = 1 / (G_F \sqrt{2}) \), where \( G_F \) is the Fermi constant. The above formulas employ a more concise notation for clarity and brevity:             
\beqa
B_0(p^2) \equiv B_0(p^2, m_t^2, m_t^2) \,, \;\;\;  C_0(p_1^2, p_2^2) \equiv C_0(p_1^2, p_2^2, (p_1 + p_2)^2, m_H^2 ,m_t^2, m_t^2) \;,
\eqa
with
\beqa
B_0(p^2, m_1^2, m_2^2) &=& \frac{\mu^{4-D}}{i \pi^{D/2} r_\Gamma} \int d^D q \frac{1}{[q^2 - m_1^2] [(q+p)^2 - m_2^2]} \;,
\eqa
\begin{gather}
    C_0(p_1^2, p_2^2, (p_1+p_2)^2,m_1^2, m_2^2, m_3^2) = \frac{\mu^{4-D}}{i \pi^{D/2} r_\Gamma} \nn \\ \times \int d^D q \frac{1}{[q^2 - m_1^2] [(q+p_1)^2 - m_2^2] [(q+p_1+p_2)^2 - m_3^2]} \;, 
\end{gather}
and
\beqa
r_\Gamma = \frac{\Gamma^2(1-\epsilon) \Gamma(1 + \epsilon)}{\Gamma(1-2\epsilon)}, \qquad D = 4 - 2 \epsilon \; .
\eqa
Bubble integrals inherently possess ultraviolet (UV) singularities; however, these singularities are immediately canceled out in the specific combinations employed. As a result, the coefficients \( F_T \) and \( F_L \) remain free of infrared singularities. Additionally, the obtained result is consistent with Ref.~\cite{DelDuca:2001fn}.

\noindent In the limit of a large top-quark mass, \( F_T \) contributes dominantly, whereas \( F_L \) becomes suppressed by a factor of \( 1/m_t^2 \). Through explicit calculations, we derive that
\beqa
F_T(q_1, q_2) &=&  \frac{\alpha_s}{\pi v}   \left[ -\frac{1}{3} - \frac{7 m_H^2 + 11 q_1^2 + 11 q_2^2}{360 m_t^2} \right]+  \mathcal O (m_t^{-4}) \;,
\nn \\
F_L(q_1, q_2) &=&  \frac{\alpha_s}{180 \pi  v m_t^2}  + \mathcal O (m_t^{-4}) \,.
\eqa
The result aligns with the vertex derived from the effective Lagrangian
\beqa
\mathcal L_{ggH} = \frac{\alpha_s}{12 \pi} F^{a \, \mu\nu}F^a_{\mu\nu} \frac{H}{v} \,.
\eea

\subsection{The off-shell $gggH$ amplitude}
\label{ssec:gggH_vertex}
The six diagrams contributing to the four-point \( gggH \) amplitude are grouped into pairs due to charge conjugation symmetry, which ultimately determines the color structure of the box by selecting only the antisymmetric structure functions \( f^{abc} \). Regarding the tensor structure, we use the following general decomposition:
\begin{equation}
 \left[ g^a(q_1^\alpha) g^b(q_2^\beta) g^c(q_3^\gamma) H (k_4) \right] \equiv g f^{abc} B^{\alpha\beta\gamma}(q_1, q_2, q_3) = 
g f^{abc} \left[ g^{\alpha \beta} q_{i}^{\gamma} +  q_{i}^{\alpha} q_{j}^{\beta} q_{k}^{\gamma} + {\rm{perm}}. \right] \; .
\end{equation}
Since, in our case, the three gluons are transverse\footnote{Note that for the \( t \)-channel gluon, its effective (\textit{non-sense}) polarization remains transverse to its momentum.}, this general and intricate form undergoes significant simplification:
\beqa
B^{\alpha\beta\gamma}(q_1, q_2, q_3) &=& B_a(q_1,q_2,q_3) \, g^{\alpha \beta} q_1^\gamma  +   B_a(q_2,q_3,q_1) \,    g^{\beta\gamma} q_2^\alpha   + B_a(q_3,q_1,q_2) \,    g^{\alpha\gamma} q_3^\beta     \nn \\
&-&    B_a(q_2,q_1,q_3) \,   g^{\alpha\beta} q_2^\gamma  -   B_a(q_1,q_3,q_2) \,    g^{\alpha\gamma} q_1^\beta         -   B_a(q_3,q_2,q_1) \,    g^{\beta\gamma} q_3^\alpha     \nn \\
&+& B_b(q_1,q_2,q_3) \, q_3^{\alpha} q_3^{\beta} q_1^{\gamma}  + B_b(q_2,q_3,q_1) \,   q_1^{\beta} q_1^{\gamma} q_2^{\alpha}   +   B_b(q_3,q_1,q_2) \,   q_2^{\alpha} q_2^{\gamma} q_3^{\beta} \nn \\
&-& B_b(q_2,q_1,q_3) \,  q_3^{\alpha} q_3^{\beta} q_2^{\gamma}  -  B_b(q_1,q_3,q_2) \, q_2^{\alpha} q_2^{\gamma} q_1^{\beta}    -  B_b(q_3,q_2,q_1) \, q_1^{\beta} q_1^{\gamma} q_3^{\alpha}         \nn \\
&+& B_c(q_1,q_2,q_3)  q_1^{\gamma}  q_2^{\alpha}  q_3^{\beta}    -  B_c(q_2, q_1, q_3)   q_1^{\beta} q_2^{\gamma}  q_3^{\alpha}  \; , \label{Eq:General_box_amplitude}
\eqa
We have additionally utilized the symmetry of the amplitude under the exchange of any two pairs of gluons. In the aforementioned expression, all momenta are considered incoming.
The three independent coefficients in this expression are explicitly provided by
\beqa
B_a(q_1,q_2,q_3)  &=& \frac{\alpha_s m_t^2}{\pi v} \bigg\{ - \frac{1}{2} q_2 \cdot q_3 \left[ D_0(q_1,q_2,q_3) + D_0(q_2,q_3,q_1) + D_0(q_3,q_1,q_2) \right]  \nn \\
&+& q_1 \cdot q_2 \left[  D_{13}(q_2,q_3,q_1) + D_{12}(q_3,q_1,q_2) - D_{13}(q_3, q_2,q_1) \right] \nn \\
&-& 4 \left[ D_{313}(q_2,q_3,q_1)  +  D_{312}(q_3,q_1,q_2)  - D_{313}(q_3,q_2,q_1)   \right] \nn \\&-& C_0(q_1, q_2+q_3) \bigg\} \;, \nn \\
B_b(q_1,q_2,q_3)  &=& - \frac{\alpha_s m_t^2}{\pi v} \bigg\{  
D_{13}(q_1,q_2,q_3) + D_{12}(q_2,q_3,q_1) - D_{13}(q_2,q_1,q_3) \nn \\
&+& 4 \left[  D_{37}(q_1,q_2,q_3) + D_{23}(q_1,q_2,q_3)  + D_{38}(q_2,q_3,q_1)   \right. \nn \\
&+& \left. D_{26}(q_2,q_3,q_1) -  D_{39}(q_2,q_1,q_3)  -  D_{23}(q_2,q_1,q_3)     \right]
\bigg\} \;,\nn\\
B_c(q_1,q_2,q_3)  &=& - \frac{\alpha_s m_t^2}{\pi v} \bigg\{
  - \frac{1}{2} \left[ D_0(q_1,q_2,q_3) +  D_0(q_2,q_3,q_1)  + D_0(q_3,q_1,q_2) \right] \nn \\ 
  &+&   4   \left[  D_{26}(q_1, q_2, q_3) + D_{26}(q_2,q_3,q_1) + D_{26}(q_3,q_1,q_2)  \right. \nn \\
  &+& \left.  D_{310}(q_1, q_2, q_3) + D_{310}(q_2,q_3,q_1)  + D_{310}(q_3,q_1,q_2) \right]
 \bigg\} \;,
\eqa
Here, we have used the notation from~\cite{Passarino:1978jh,DelDuca:2001fn}\footnote{With a different sign convention for \( C_0, D_{312} \), and \( D_{313} \).}. We carried out a numerical comparison with the {\tt VBFNLO} code, achieving complete agreement~\cite{Baglio:2011juf,Baglio:2024gyp}. Additionally, we verified analytically that, in the on-shell limit, our result matches the helicity amplitudes calculated in Refs.~\cite{Baur:1989cm,Rozowsky:1997dm}. 

\noindent Alternatively, the results can be reformulated using the following mapping:
\beqa
\label{eq:D_PV}
D_{12}(q_1, q_2, q_3) &\equiv& D_{2}(\textrm{arg}) + D_{3}(\textrm{arg}) \;,\nn \\
D_{13}(q_1, q_2, q_3) &\equiv& D_{3}(\textrm{arg}) \;,\nn \\
D_{23}(q_1, q_2, q_3) &\equiv& D_{33}(\textrm{arg}) \;,\nn \\
D_{26}(q_1, q_2, q_3) &\equiv& D_{23}(\textrm{arg}) + D_{33}(\textrm{arg}) \;,\nn \\
D_{37}(q_1, q_2, q_3) &\equiv& D_{133}(\textrm{arg}) + D_{233}(\textrm{arg})  + D_{333}(\textrm{arg}) \;, \nn \\
D_{38}(q_1, q_2, q_3) &\equiv& D_{223}(\textrm{arg}) + 2 D_{233}(\textrm{arg})  + D_{333}(\textrm{arg}) \;, \nn \\
D_{39}(q_1, q_2, q_3) &\equiv& D_{233}(\textrm{arg}) + D_{333}(\textrm{arg}) \;, \nn \\
D_{310}(q_1, q_2, q_3) &\equiv& D_{123}(\textrm{arg}) + D_{133}(\textrm{arg}) + D_{223}(\textrm{arg}) + 2 D_{233}(\textrm{arg}) + D_{333}(\textrm{arg}) \;, \nn \\
D_{312}(q_1, q_2, q_3) &\equiv& D_{002}(\textrm{arg}) + D_{003}(\textrm{arg}) \;,\nn \\
D_{313}(q_1, q_2, q_3) &\equiv& D_{003}(\textrm{arg}) \;,
\eqa
with $\textrm{arg} = q_1^2, q_2^2, q_3^2, q_4^2, (q_1+q_2)^2, (q_2+q_3)^2, m_t^2,m_t^2,m_t^2,m_t^2$.

\noindent The \( D \) coefficients on the right-hand sides of Eq.~(\ref{eq:D_PV}) are derived from the expansions of the four-point tensor integrals:
\beqa
D_{\mu_1 \ldots \mu_n} = \frac{\mu^{4-D}}{i \pi^{D/2} r_\Gamma} \int d^D \ell \frac{\ell_{\mu_1} \ldots \ell_{\mu_n}}{\mathcal D(0, m_0) \mathcal D(p_1, m_1) \mathcal D(p_2, m_2) \mathcal D(p_3, m_3)} \;.
\label{Eq:DTensDefinition}
\eqa
Here, the denominators are given by \( \mathcal{D}(p, m) = (\ell + p)^2 - m^2 \). Specifically, up to rank-3, the expansion includes:
\beqa
D_\mu &=& \sum_{i = 1}^3 p_{i \mu} D_i \,, \nn \\
D_{\mu\nu} &=& g_{\mu\nu} D_{00} + \sum_{i,j = 1}^3 p_{i, \mu}p_{j, \nu} D_{ij} \,, \nn \\
D_{\mu\nu\rho} &=& \sum_{i=1}^3 \left( g_{\mu\nu} p_{i,\rho} + g_{\nu\rho} p_{i,\mu} + g_{\mu\rho} p_{i,\nu}  \right) D_{00i} + \sum_{i,j,l =1}^3 p_{i, \mu}p_{j, \nu} p_{l, \rho} D_{ijl} \;,
\eqa
with $p_1 = q_1$, $p_2 = q_1 + q_2$, $p_3 = q_1 + q_2 + q_3$ and $p_4 = q_1 + q_2 + q_3 + q_4$. The primary distinction between the notation used here and that in \cite{DelDuca:2001fn} lies in the momenta utilized in the tensor expansion. In this context, we used the momenta appearing in the denominators, denoted as \( p_i \), whereas \cite{DelDuca:2001fn} employed the external momenta \( q_i \).

\noindent In the large-top-mass limit, the three coefficients are expressed as:
\beqa
B_a(q_1,q_2,q_3) &=& \frac{\alpha_s}{\pi v} \bigg[ \frac{1}{3} +  \frac{1}{360 m_t^2}  
( 11 q_1^2 + 40 q_2^2 + 27 q_3^2 - 4 q_4^2 + 11 (q_1 + q_3)^2 \nn  \\
&-& 14 (q_2 + q_3)^2 )
 \bigg]   + \mathcal O (m_t^{-4}) \;, \nn \\
B_b(q_1,q_2,q_3) &=& \frac{11 \alpha_s }{180 \pi v m_t^2}  + \mathcal O (m_t^{-4}) \;,\nn \\
B_c(q_1,q_2,q_3) &=& \frac{ \alpha_s }{5 \pi v m_t^2}  + \mathcal O (m_t^{-4}) \;,
\eqa
where \( B_a \) represents the leading contribution.

\section{Leading-Order Calculation}
\label{sec:kinematicsLO}

\subsection{Kinematics}
\label{ssec:kinematics}
To begin, it is often advantageous to decompose any four-vector using the Sudakov basis, defined by two light-like vectors \( k_1 \) and \( k_2 \) with the property \( k_1 \cdot k_2 = s/2 \). In this basis, we conventionally employ the Sudakov decomposition as follows:
\bea
&& p_H = z_H \, k_1 + \frac{m_H^2 + \vec{p}_H^{\; 2}}{z_H s} k_2 + p_{H_\perp} \,, \qquad  p_{H_\perp}^2 = - \vec{p}_H^{\; 2}\,, \qquad  p_H^2 = m_H^2 \;, \nn \\
&& q = - \alpha k_2 + q_\perp \,, \qquad q^{\; 2} = q_\perp^2 = - \vec{q}^{\; 2} \;, \nn \\
&& p_p = z_p \, k_1 + \frac{{\vec{p}_p}^{\;2}}{z_p s} k_2 + p_{p_\perp} \,, \qquad  p_{p_\perp}^{2} = - \vec{p}_p^{\; 2} \,, \qquad  p_p^2 = 0 \; .
\eea
Here, \( p_H \) denotes the Higgs momentum, \( q \) represents the \( t \)-channel Reggeon momentum, and \( p_p \) is the momentum of the additional particle produced at NLO (with \( p = q \) for the quark and \( p = g \) for the gluon). The relevant scalar products among these momenta can be described using the variables \( z_H, z_q, \vec{q}, \vec{p}_H, \vec{p}_q \). For real corrections, the invariant mass of the particle-Reggeon system can be conveniently expressed as
\begin{equation}
   s_{AR} = (k_1+q)^2 = (p_p + p_H)^2 = \frac{z_p (z_H + z_p) m_H^2 + (z_p \vec{p}_H   -    z_H   \vec{p}_p)^2}{z_H z_p} \; .
\end{equation}
Consequently, the integration measure of the NLO impact factor can be expressed as follows:
\begin{equation}
\frac{d s_{pR}}{2 \pi} d \rho_{ \{H q \}} = \delta(1 - z_p - z_H) \ \delta^{(2)}(\vec{p}_p + \vec{p}_H - \vec{q}) \ \frac{d z_p d z_H}{z_p z_H}\  \frac{d^{D-2}p_p \, d^{D-2} p_H}{2 (2\pi)^{D-1}} \;.  
\label{Eq:PhasePhaseInt}
\end{equation}
To derive an impact factor that is differential in the Higgs kinematic variables, we will only integrate over the longitudinal fraction and the transverse momentum of the additional produced particle, utilizing the Dirac delta functions present in Eq.~(\ref{Eq:PhasePhaseInt}). Consequently, when calculating effective vertices, we can directly apply the constraints
\begin{equation}
    z_p = 1- z_H \; , \hspace{3 cm} \vec{p}_p = \vec{q} - \vec{p}_H \; .
\end{equation}

\subsection{Computation}
\label{ssec:LO computation}

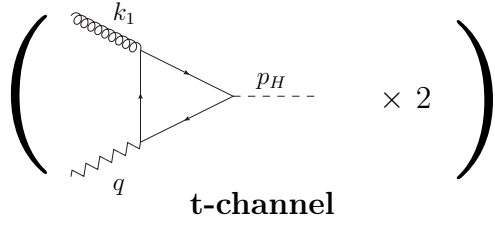
\begin{figure}
\begin{picture}(430,72) 
\put(120,25){ \scalebox{5.0}{( }} 
\put(150,7){\includegraphics[scale=0.27]{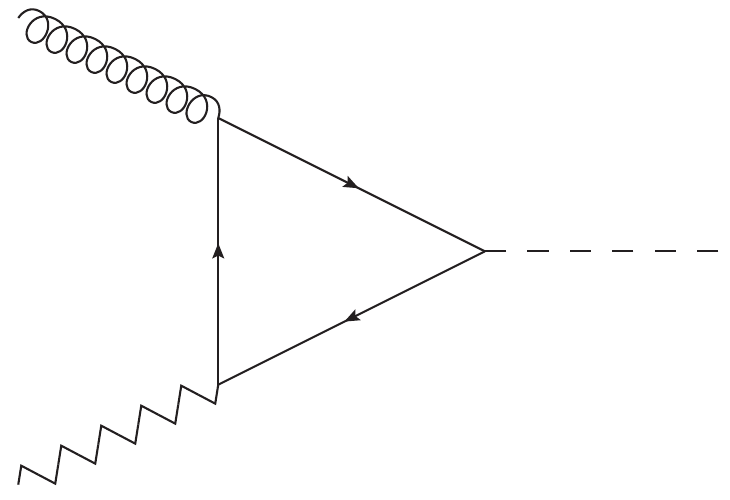}} 
\put(168,68){\scalebox{0.8}{$k_1$}} 
\put(168,4){\scalebox{0.8}{$q$}} 
\put(222,44){\scalebox{0.8}{$p_H$}} 
\put(193,-5){ \textbf{t-channel} } 
\put(264,35){ $\times \hspace{0.15 cm} 2$} 
\put(270,23){ \scalebox{5.0}{ ) }} 
\end{picture}
\caption{The two triangular-like diagrams contribute to the Higgs impact factor at leading order (LO). The “\( \times 2 \)” indicates the diagram in which the direction of the fermion lines is reversed. The momenta of the initial-state particles (the collinear parton and the Reggeized gluon) are treated as incoming, while the momenta of the final-state particles (in this case, the Higgs) are considered outgoing.}
\label{Fig:2DiagramsTriLO}
\end{figure}
The leading-order impact factor for Higgs production in a gluon-Reggeon collision is given by
\begin{equation}
\Phi_{gg}^{ \{ H \} }(\vec q) = \frac{\langle cc^{\prime} | \hat{\cal P}_0 | 0 \rangle }{2(1-\epsilon) (N^2-1)}
\sum_{a, \lambda } \int \frac{ds_{AR}}{2\pi}\ d\rho_f \ \Gamma_{\{ H \}g}^{ac} (\vec q \; )
\left( \Gamma_{\{ H \}g}^{ac^{\prime}} (\vec q \; ) \right)^* \; .
\label{Eq:LoImpactD4-2E}
\end{equation}
Here, we average over the color and polarizations of the incoming gluon in \( D = 4 - 2\epsilon \) dimensions. The effective interaction vertex used in Eq.~(\ref{Eq:LoImpactD4-2E}) is given by
\begin{equation}
  \Gamma_{\{ H \}g}^{ac} (\vec q \; ) =  (-i) i \delta^{ac} T^{\mu \nu} (k_1,q) \varepsilon_{\perp , \mu }(k_1) \left( -\frac{k_{2, \nu }}{s} \right) = \frac{F_T ( k_1 , q ) (q_{\perp} \cdot \varepsilon_{\perp} (k_1) ) \delta^{ac}}{2} \; ,
  \label{EffectivegRH}
\end{equation}
where \( T^{\mu \nu} \) represents the tensor defined in Eq.~(\ref{Eq:FundamentalTriangularAmo}), and \( F_T \) corresponds to the function in~(\ref{Eq:FTtri}). For the physical gluon, we adopt the gauge condition \( \varepsilon (k_1) \cdot k_2 = 0 \), a choice that will consistently apply to all physical gluons in the NLO calculation. Notably, in general, \( F_T \) can be expressed as a function of \( k_1^2, q^2 = -\vec{q}^{\; 2} \), and \( (k_1 + q) = p_H^2 = m_H^2 \), {\it i.e.}\footnote{The dependence on \( m_t^2 \) is implicit.}
\begin{equation}
    F_T (k_1 , q) \equiv  F_T \left(k_1^2, q^{2}, (k_1+q)^2 \right) = F_T \left(0, -\vec{q}^{\; 2}, m_{H}^2  \right)  \; ,
\end{equation}
utilizing the on-shell condition for the collinear gluon, \( k_1^2 = 0 \). We find that, in the infinite-top-mass limit, the effective vertex in Eq.~(\ref{EffectivegRH}) simplifies to
\begin{equation}
  \Gamma_{\{ H \}g}^{ac} (\vec q \; ) = - \frac{g_H (q_{\perp} \cdot \varepsilon_{\perp} (k_1) ) \delta^{ac}}{2} \; ,
  \label{EffectivegRHMtInfinity}
\end{equation}
consistent with Ref.~\cite{Celiberto:2022fgx}.
The gluon-initiated impact factor, differential in the Higgs kinematic variables, is given by
\begin{equation}
  \frac{ d \Phi_{gg}^{ \{ H \} (0) }(\vec q) }{d z_H d^{2} \vec{p}_H } = \frac{| F_T \left( 0, -\vec{q}^{\; 2}, m_{H}^2  \right) |^2 \vec{q}^{\; 2} }{8 (1-\epsilon) \sqrt{N^2-1}}  \delta (1-z_H) \delta^{(2)} \left( \vec{q} -\vec{p}_H \right) \; .
  \label{Eq:LoImpactD4-2EPart}
\end{equation}
By convolving with the gluon parton distribution function, we can define the proton-initiated impact factor as
\begin{equation}
\begin{split}
  \frac{ d \Phi_{PP}^{ \{ H \} (0) }(\vec q \; ) }{d x_H d^{2} \vec{p}_H } &= \int_{x_H}^1 \hspace{-0.2 cm} \frac{d z_H}{z_H} f_g \hspace{-0.1 cm}  \left( \hspace{-0.05 cm}  \frac{x_H}{z_H} \hspace{-0.05 cm}  \right)  \frac{ d \Phi_{gg}^{ \{ H \} (0) }(\vec q \; ) }{d z_H d^{2} \vec{p}_H } \\&= \frac{ |F_T \left( 0, -\vec{q}^{\; 2}, m_{H}^2  \right)|^2 \vec{q}^{\; 2} f_g (x_H) }{8 (1-\epsilon) \sqrt{N^2-1}} \delta^{(2)} \left( \vec{q} -\vec{p}_H \right),
  \label{Eq:LoImpactD4-2EHadro}
  \end{split}
\end{equation} 
which, in the infinite-top-mass limit, reproduces the result previously obtained in Ref.~\cite{Celiberto:2022fgx}.
\section{Next-to-Leading-Order computation}
\label{sec:NLOcomputation}

\subsection{Quark-initiated contribution}
\label{ssec:NLOquark}

\begin{figure}
\begin{picture}(430,70) 
\put(120,25){ \scalebox{5.0}{( }} 
\put(150,10){\includegraphics[scale=0.36]{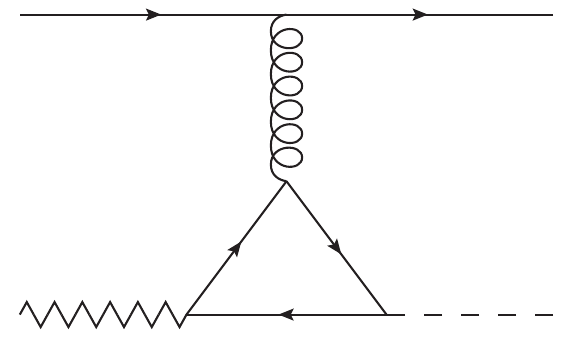}} 
\put(165,75){\scalebox{0.8}{$k_1$}} 
\put(224,75){\scalebox{0.8}{$p_q$}} 
\put(165,5){\scalebox{0.8}{$q$}} 
\put(224,5){\scalebox{0.8}{$p_H$}} 
\put(193,-6){ \textbf{t-channel} } 
\put(264,37){ $\times \hspace{0.15 cm} 2$} 
\put(270,25){ \scalebox{5.0}{ ) }} 
\end{picture}
\caption{The two triangular-like diagrams contribute to the quark-initiated part of the Higgs impact factor at NLO. Here, the notation “\( \times 2 \)” denotes the diagram where the direction of the fermion lines is reversed. The momenta of the initial-state particles (the collinear parton and the Reggeized gluon) are treated as incoming, while the momenta of the final-state particles (the Higgs and the outgoing quark) are considered outgoing.}
\label{Fig:2DiagramsTriQ}
\end{figure}
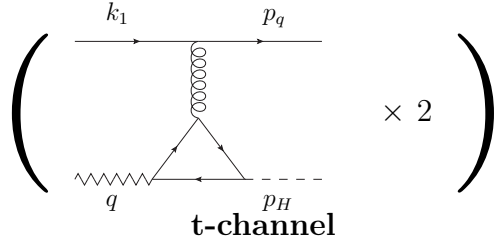
For the quark-initiated process, the impact factor is expressed as
\bea
d \Phi_{qq}^{\{H q\}} = \langle cc' | \hat{\mathcal P}_0 | 0 \rangle \frac{1}{2 N} \sum_{i,j, \lambda, \lambda'} \int \frac{ds_{qR}}{2\pi} d \rho_{\{Hq\}} \Gamma^{c(0)}_{\{Hq\}q}(q) \left( \Gamma^{c'(0)}_{\{Hq\}q}(q) \right)^* \;,
\eea
Here, the averaging (or summing) over the spin states and color configurations of the incoming (or outgoing) quark has already been performed. The vertex \( \Gamma^{c(0)}_{\{Hq\}q} \) is obtained from the diagrams shown in Fig.~\ref{Fig:2DiagramsTriQ}, using the results presented in Sec.~\ref{ssec:ggH_vertex}. The final expression is given by
\bea
\Gamma^{c(0)}_{\{Hq\}q}(q) \! &=& \! - g t^c_{ij}  \frac{ 1-z_H}{(\vec{p}_H - \vec{q})^2} \bar u(p_q) \bigg\{  F_T((p_H - q)^2, q^2, m_H^2)\nn \\ &\times& \bigg[   (p_H - q) \cdot q  \frac{\hat{k}_2}{s}  -  (p_H - q)\cdot \frac{k_2}{s} \hat{q} \bigg]      \nn \\
&-&  F_L((p_H - q)^2, q^2, m_H^2)  \frac{(\vec{p}_H - \vec{q})^2}{ 1-z_H} q^2 \frac{\hat{k}_2}{s}   \bigg\} u(k_1) \;,
\eea
(here \( \hat{a} \equiv \gamma^\mu a_\mu \)), from which the differential impact factor can be readily evaluated,
\bea
&& \frac{d \Phi_{qq}^{\{Hq\}} (z_H, \vec{p}_H, \vec{q} ) }{d z_H d^2 \vec{p}_H} = 
 \frac{g^2  \sqrt{N^2-1}}{16 z_H N (2 \pi)^{D-1} (\vec{q}-\vec{p}_H )^4}   \nn \\
&& \times \bigg\{  |F_T((p_H - q)^2, q^2, m_H^2)|^2   [4 (1-z_H) [ ( \vec{q}  -  \vec{p}_H )\cdot \vec{q} \; ]^2 + z_H^2 \vec{q}^{\; 2}  ( \vec{q}  -  \vec{p}_H )^2  ]    \nn \\
&&+ 4 \Re\{ F_T((p_H - q)^2, q^2, m_H^2)^* F_L((p_H - q)^2, q^2, m_H^2)\}(2 - z_H) ( \vec{q}  -  \vec{p}_H )^2 \vec{q}^{\; 2}   \,  \nn \\ &&\times ( \vec{q}  -  \vec{p}_H ) \cdot \vec{q}  
+ 4 |F_L((p_H - q)^2, q^2, m_H^2)|^2  ( \vec{q} - \vec{p}_H )^4 \vec{q}^{\; 4}  \bigg\}  \,.
\eea 
As required by gauge invariance, both the vertex and the impact factor vanish when \( \vec{q} \to \vec{0} \). A quick comparison with the computation performed in the infinite-top-mass limit~\cite{Celiberto:2022fgx} reveals the emergence of a genuinely new contribution, represented by the longitudinal form factor \( F_L \), when retaining the full top-mass dependence. At leading order in the heavy-top expansion, we recover the result from~\cite{Celiberto:2022fgx}. The impact factor exhibits a collinear singularity when \( \vec{p}_q = \vec{q} - \vec{p}_H \rightarrow \vec{0} \), which is entirely contained within the first term of the curly brackets. Therefore, as anticipated, the infrared structure of the impact factor remains unaffected by the specific details of the coupling between the real gluon and the Reggeon that produces the Higgs, making it fully analogous to the structure obtained using the infinite-top-mass approximation~\cite{Celiberto:2022fgx}. 

\subsection{Gluon-initiated contribution}
\label{ssec:NLOgluon}

\subsubsection{$gR \rightarrow gH$ \textit{via} a triangular-like contribution}
\label{sssec:triangle}
\begin{figure}
\vspace{6mm} 
\begin{picture}(430,135)
\put(30,100){\includegraphics[scale=0.36]{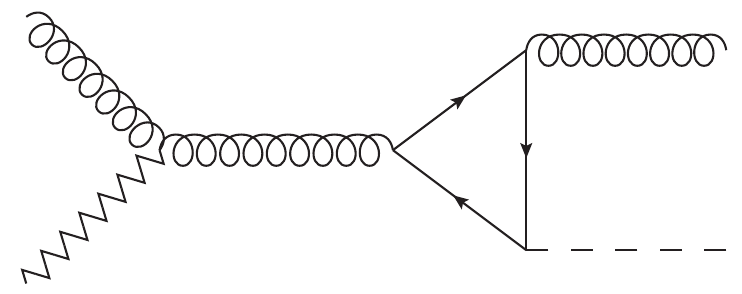}}
\put(5,112){ \scalebox{5.0}{( }}
\put(164,122){ $\times \hspace{0.15 cm} 2$}
\put(50,140){\scalebox{0.8}{$k_1$}}
\put(50,110){\scalebox{0.8}{$q$}}
\put(137,150){\scalebox{0.8}{$p_g$}}
\put(137,100){\scalebox{0.8}{$p_H$}}
\put(160,112){ \scalebox{5.0}{ ) }}
\put(203,125){ \scalebox{1.5}{ + }}
\put(230,112){ \scalebox{5.0}{( }}
\put(275,160){\scalebox{0.8}{$k_1$}}
\put(347,160){\scalebox{0.8}{$p_H$}}
\put(263,95){\includegraphics[scale=0.36]{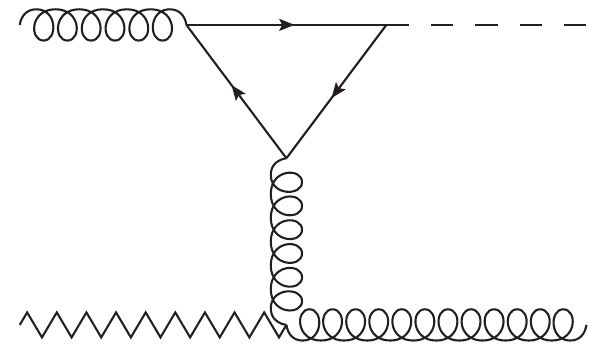}}
\put(275,90){\scalebox{0.8}{$q$}}
\put(347,90){\scalebox{0.8}{$p_g$}}
\put(370,122){ $\times \hspace{0.15 cm} 2$}
\put(375,112){ \scalebox{5.0}{ ) }}
\put(80,77){ \textbf{s-channel} }
\put(298,77){ \textbf{u-channel} }
\put(120,20){ \scalebox{5.0}{( }}
\put(150,5){\includegraphics[scale=0.36]{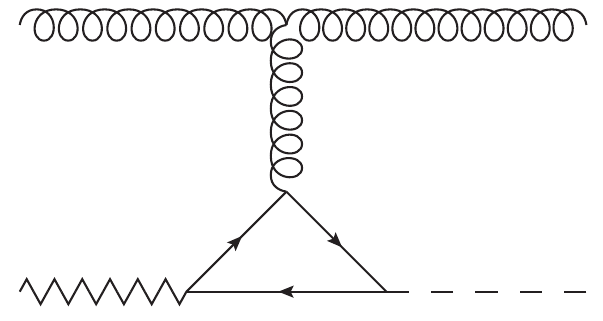}}
\put(165,66){\scalebox{0.8}{$k_1$}}
\put(224,66){\scalebox{0.8}{$p_g$}}
\put(165,0){\scalebox{0.8}{$q$}}
\put(224,0){\scalebox{0.8}{$p_H$}}
\put(193,-15){ \textbf{t-channel} }
\put(264,32){ $\times \hspace{0.15 cm} 2$}
\put(270,20){ \scalebox{5.0}{ ) }}
\put(85,30){ \scalebox{1.5}{ + }}
\end{picture}
\vspace{1mm} 
\caption{The six triangular-like diagrams that contribute to the gluon-initiated part of the Higgs impact factor at next-to-leading order. The notation "$\times 2$" represents the diagram where the direction of the fermionic lines is reversed. The momenta of the initial state particles (the collinear parton and the Reggeized gluon) are treated as incoming, whereas the momenta of the final state particles (the Higgs boson and the outgoing quark) are considered as outgoing.}
\label{Fig:6DiagramsTri}
\end{figure}
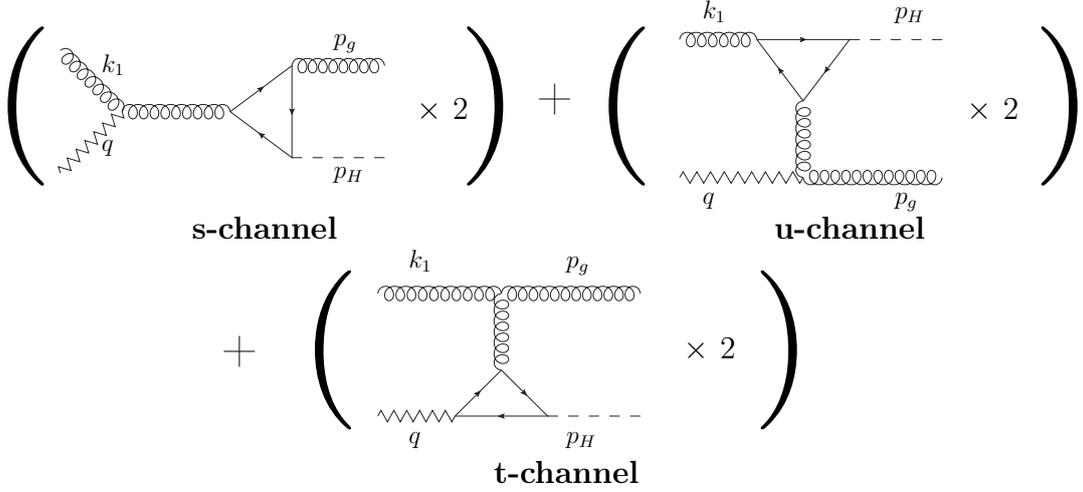

In this section, we explore the role of the \( ggH \) vertex in the amplitudes for the process \( gR \to gH \). The Feynman diagrams shown in Fig.~\ref{Fig:6DiagramsTri} are categorized into \( s \)-, \( t \)-, and \( u \)-channel diagrams. In the \( s \)- and \( u \)-channel diagrams, one of the two gluon lines emerging from the triangle loop (labeled as \( p_g \) and \( k_1 \) respectively) is on-shell and transverse. Consequently, the longitudinal form factor is irrelevant, as discussed in Sec.~\ref{ssec:ggH_vertex}. In this case, the tensor structure of the \( ggH \) vertex mirrors that in the heavy-top limit, enabling a straightforward comparison with the results presented in \cite{Celiberto:2022fgx}. On the other hand, for the \( t \)-channel diagrams, both gluon lines linked to the quark loops are off-shell (even though the \textit{effective} polarization vector of the Reggeized gluon satisfies the transversality condition \( \frac{k_2}{s} \cdot q = 0 \)), and the \( ggH \) vertex contributes with
\bea
F_L(p^2, q^2, (p+q)^2) \left[ p^2 \, q^2  g^{\rho\nu}    -  q^2 \, p^\rho p^\nu    \right] \frac{k_{2,\nu}}{s} \;,
\eea
where $p = k_1 - p_g = p_H - q$. The Feynman diagrams depicted in Fig.~\ref{Fig:6DiagramsTri} yield
\bea
\Gamma_\triangle = \Gamma_s + \Gamma_t + \Gamma_u \;,
\eea
with 
\begin{gather}
\Gamma_s = i \frac{g f^{abc} F_T((p_g + p_H)^2, 0, m_H^2) }{2 [ m_H^2 (1-z_H) +  (\vec{p}_H - z_H \vec{q})^2 ]} \left[ 
(m_H^2 (1- z_H)^2 + (\vec{p}_H - z_H \vec{q})^2 )   g^{\mu\nu}   \right. \nn \\
+ \left. 2 z_H (1 - z_H)^2  p_H^{\mu} p_H^{\nu}
- 2 z_H^2 (1 - z_H)  p_g^{\mu} p_H^{\nu}
 \right] \epsilon_{\mu}(k_1)\epsilon^*_{\nu}(p_g) \;, 
\end{gather}
 \begin{gather}
 \Gamma_u = -i \frac{g f^{abc} F_T(0, (p_H - k_1)^2, m_H^2) }{2 [  m_H^2 (z_H - 1) -   \vec{p}_H^{\; 2} ]} \nn \\ \times \left[
  (m_H^2 + \vec{p}_H^{\; 2}) (1 - z_H) g^{\mu\nu} - 2 z_H p_H^{\mu} (p_H^{\nu} - z_H k_1^{\nu})
 \right] \epsilon_{\mu}(k_1)\epsilon^*_{\nu}(p_g) \; ,
 \end{gather} 
 \begin{gather}
  \Gamma_t = i  \frac{g f^{abc} F_T((k_1- p_g)^2, q^2, m_H^2) }{(\vec{q} - \vec{p}_H)^2} (1 \hspace{-0.05 cm} - \hspace{-0.05 cm} z_H) \hspace{-0.05 cm} \left[
  (\vec{q} - \vec{p}_H) \hspace{-0.05 cm}  \cdot \hspace{-0.05 cm}  \vec{q} g^{\mu\nu} \hspace{-0.1 cm} + \hspace{-0.1 cm}  z_H (p_H^{\mu} k_1^{\nu} + p_g^{\mu} p_H^{\nu})
  \right] \nn \\
  \times\,\epsilon_{\mu}(k_1)\epsilon^*_{\nu}(p_g)- i \frac{g f^{abc} F_L((k_1- p_g)^2, q^2, m_H^2)}{2} (z_H - 2) \vec{q}^{\; 2}  \epsilon(k_1) \cdot \epsilon^*(p_g) \;,
\end{gather}
which are consistent with those in~\cite{Celiberto:2022fgx} in the infinite-top-mass limit where \( F_T(p^2, q^2) = - \frac{\alpha_s}{3 \pi v} \) and \( F_L = 0 \). By representing the amplitudes in terms of the transverse degrees of freedom, we ultimately obtain
\begin{gather}
\Gamma_s = i \frac{g f^{abc} F_T ( (p_g + p_H)^2, 0, m_H^2) }{2 [ m_H^2 (1-z_H) +  (\vec{p}_H - z_H \vec{q})^2 ]} \left[ 
(m_H^2 (1- z_H)^2 + (\vec{p}_H - z_H \vec{q})^2 )   g^{\mu\nu}   \right. \nonumber \\
+ \left. 2 z_H (p_{\perp H} - z_H q_\perp)^{\mu} (p_{\perp H} - z_H q_\perp)^{\nu}
 \right] \epsilon_{\perp \mu}(k_1)\epsilon^*_{\perp \nu}(p_g) \; , 
\label{Eq:Gammas}
\end{gather}
\begin{gather}
 \Gamma_u = -i \frac{g f^{abc} F_T (0, (p_H - k_1)^2, m_H^2) }{2 [  m_H^2 (z_H - 1) -   \vec{p}_H^{\; 2} ]} \nn \\ \times \left[
  (m_H^2 + \vec{p}_H^{\; 2}) (1 - z_H) g^{\mu\nu} - 2 z_H p_{\perp H}^{\mu} p_{\perp H}^{\nu} 
 \right] \epsilon_{\perp \mu}(k_1) \epsilon^*_{\perp \nu}(p_g) \; , 
 \label{Eq:Gammau}
\end{gather}
\begin{gather}
   \Gamma_t = i  \frac{g f^{abc} F_T ((k_1- p_g)^2, q^2, m_H^2)  }{(\vec{q} - \vec{p}_H)^2}  \left[
  (1-z_H) (\vec{q} - \vec{p}_H) \cdot \vec{q} g^{\mu\nu} 
  \hspace{-0.1 cm} +\hspace{-0.1 cm}  z_H  q_{\perp}^{\mu}  p_{\perp H}^{\nu} 
  \hspace{-0.05 cm} \right. \nonumber \\ 
  - \hspace{-0.1 cm}  z_H ( 1 - z_H) p_{\perp H}^{\mu} q_{\perp}^{\nu}- \left. z_H^2 q_{\perp}^{\mu} q_{\perp}^{\nu}
  \right] \epsilon_{\perp \mu}(k_1)\epsilon^*_{\perp \nu}(p_g) \nn\\
  - i \,\frac{g f^{abc} F_L ((k_1- p_g)^2, q^2, m_H^2)  }{2} (z_H - 2) \vec{q}^{\; 2}  \epsilon_{\perp}(k_1) \cdot \epsilon_{\perp}^*(p_g)   \;.
  \label{Eq:Gammat}
\end{gather}
To derive the infinite-top-mass result, it suffices to set
\begin{equation}
 F_T ( (p_g + p_H)^2, 0, m_H^2)  = F_T  (0, (p_H - k_1)^2, m_H^2) = \nn
\end{equation}
\begin{equation}
F_T((k_1- p_g)^2, q^2, m_H^2)  = - \frac{\alpha_s}{3 \pi v}  \equiv - g_H 
\end{equation}
and
\begin{equation}
    F_L ((k_1- p_g)^2, q^2, m_H^2) = 0 \; 
\end{equation}
in eqs.~(\ref{Eq:Gammas}), (\ref{Eq:Gammau}) and (\ref{Eq:Gammat}).
The result is consistent with Ref.~\cite{Celiberto:2022fgx}.

\subsubsection{$gR \rightarrow gH$ \textit{via} a box-like contribution}
\label{sssec:box}

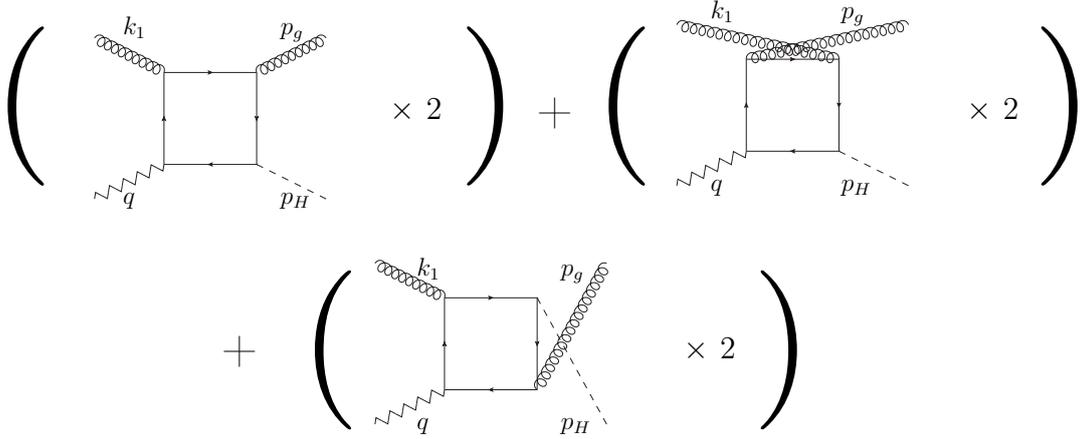
\begin{figure}
\vspace{6mm} 
\begin{picture}(430,150)
\put(45,105){\includegraphics[scale=0.27]{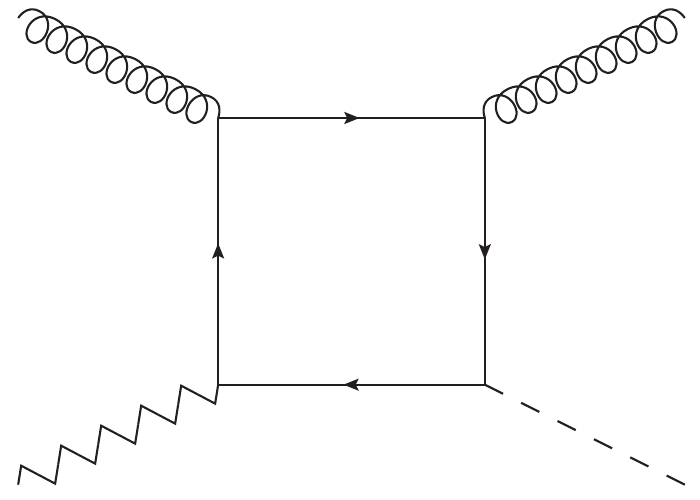}}
\put(58,169){\scalebox{0.8}{$k_1$}}
\put(58,105){\scalebox{0.8}{$q$}}
\put(117,169){\scalebox{0.8}{$p_g$}}
\put(117,105){\scalebox{0.8}{$p_H$}}
\put(5,127){ \scalebox{5.0}{( }}
\put(154,137){ $\times \hspace{0.15 cm} 2$}
\put(160,127){ \scalebox{5.0}{ ) }}
\put(203,135){ \scalebox{1.5}{ + }}
\put(230,127){ \scalebox{5.0}{( }}
\put(263,110){\includegraphics[scale=0.27]{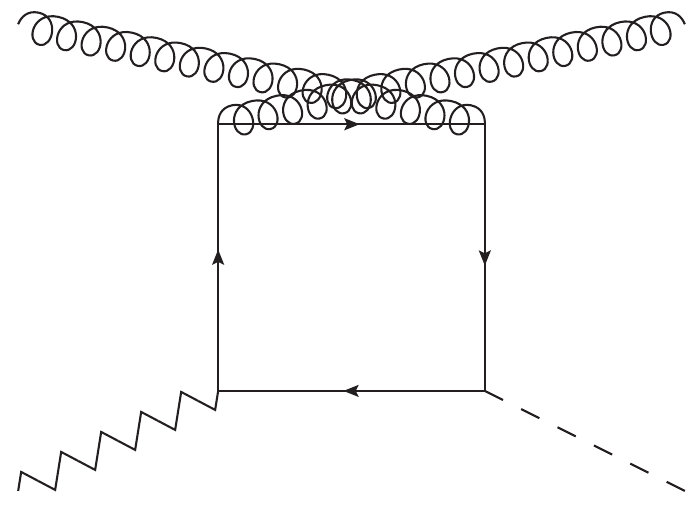}}
\put(278,175){\scalebox{0.8}{$k_1$}}
\put(278,110){\scalebox{0.8}{$q$}}
\put(327,175){\scalebox{0.8}{$p_g$}}
\put(327,110){\scalebox{0.8}{$p_H$}}
\put(370,137){ $\times \hspace{0.15 cm} 2$}
\put(375,127){ \scalebox{5.0}{ ) }}
\put(120,35){ \scalebox{5.0}{( }}
\put(150,20){\includegraphics[scale=0.27]{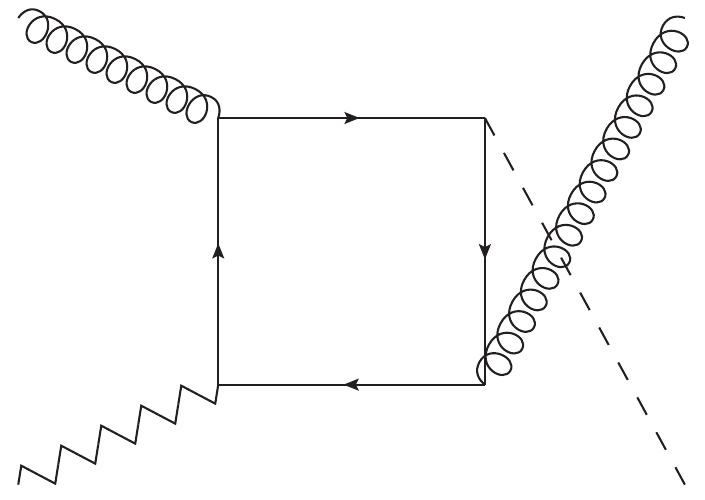}}
\put(168,78){\scalebox{0.8}{$k_1$}}
\put(168,20){\scalebox{0.8}{$q$}}
\put(222,78){\scalebox{0.8}{$p_g$}}
\put(222,20){\scalebox{0.8}{$p_H$}}
\put(264,47){ $\times \hspace{0.15 cm} 2$}
\put(270,35){ \scalebox{5.0}{ ) }}
\put(85,45){ \scalebox{1.5}{ + }}
\end{picture} 
\caption{The six box-like diagrams contribute to the gluon-initiated part of the Higgs impact factor at NLO. The notation “\( \times 2 \)” again indicates the diagram where the direction of the fermion lines is reversed. The momenta of the initial-state particles (the collinear parton and the Reggeized gluon) are treated as incoming, while the momenta of the final-state particles (the Higgs and the outgoing quark) are considered outgoing.}
\label{Fig:6DiagramsBox}
\end{figure}
By applying Eq.~(\ref{Eq:General_box_amplitude}) and expressing the result in terms of transverse polarization vectors, we obtain
\beqa
\Gamma_\Box &=& -i  \left[ g^a(k_1^\mu) g^c(q^\rho) g^b(-p_g^\nu) H (k_4) \right] \nn \\
&=& -i g f^{abc}   \frac{1}{2} \bigg\{ -\left[ B_a(k_1, -p_g, q) + (1 - z_H) B_a(-p_g, k_1, q)  \right] g^{\mu\nu}     \nn \\
 &-& [B_b(k_1, -p_g, q)  + (1 - z_H) B_b(-p_g, k_1, q)]  q_\perp^{\mu} q_\perp^{\nu}  \nn \\
&-&   (1 - z_H) B_b(q, k_1, -p_g) r_\perp^{\mu} q_\perp^{\nu}  - \frac{1}{1 - z_H} B_b(q, -p_g, k_1) q_\perp^{\mu} r_\perp^{\nu} \nn \\
&-&  \left[ B_b(k_1, q, -p_g)  + \frac{1}{1 - z_H}  B_b(-p_g, q, k_1) \right] r_\perp^{\mu} r_\perp^{\nu} 
    \nn \\
&+&   B_c(q, k_1, -p_g)  r_\perp^{\mu} q_\perp^\nu   +  B_c(k_1, q, -p_g) q_\perp^\mu    r_\perp^\nu       \bigg\} \epsilon_{\perp \mu}(k_1) \epsilon^*_{\perp \nu}(p_g) \;,
\label{Eq:Box_Amplit}
\eqa
where \( r_\perp^\mu = q_\perp^\mu - p_{\perp H}^\mu \). In the large-top-mass limit, \( \Gamma_{\Box} \) simplifies to
\beqa
 \Gamma_\Box \big |_{m_t \rightarrow \infty} = i g g_H f^{abc}  \frac{2 - z_H}{2}  \epsilon_{\perp}(k_1) \cdot \epsilon^*_{\perp}(p_g) \;,
\eqa
which precisely matches the results obtained in \cite{Celiberto:2022fgx}.

\subsubsection{Infrared structure and verification of gauge invariance}
\label{sssec:infraredgauge}
As anticipated, the real contribution to the impact factor exhibits infrared phase-space singularities. The soft singularities will cancel out in the combination of real and virtual corrections, while any remaining divergence of a purely collinear nature will be eliminated during the renormalization of the gluon PDF. The pertinent infrared singularities can be categorized as
\begin{itemize}
    \item[\textbullet] \textbf{Collinear singularity}: occurs when \( \vec{p}_g = \vec{q} - \vec{p}_H \rightarrow \vec{0} \) while \( z_g = 1 - z_H \) remains fixed;
    \item[\textbullet] \textbf{Soft singularity}: arises after parameterizing \( \vec{p}_g = \vec{q} - \vec{p}_H = (1 - z_H) \vec{u} \), with \( z_g = 1 - z_H \rightarrow 0 \).
\end{itemize}
We exclude from this list the rapidity divergence characteristic of the BFKL approach:
\begin{itemize}
    \item[\textbullet] \textbf{Rapidity singularity}: occurs when \( z_H \rightarrow 1 \) with \( (\vec{q} - \vec{p}_H) \) held fixed,
\end{itemize}
which is addressed by the BFKL counter-term and discussed in detail in Section~\linebreak\ref{sssec:Rapiditydivergence}. It is important to emphasize that this limit is the only case where the Higgs-gluon invariant mass diverges, corresponding to a large separation in rapidity.\\

The infrared divergences categorized above are entirely contained within the interference term arising from the multiplication of the \( F_T \)-part of the \( \Gamma_t \)-term with its complex conjugate. This contribution, in particular, is proportional to the square of
\begin{equation}
    F_T( k_1- p_g , q) =  F_T((k_1- p_g)^2 , -\vec{q}^{\; 2} , m_H^2) \xrightarrow[\text{collinear}]{\text{soft and/or}} F_T(0, -\vec{p}_H^{\; 2} , m_H^2)  \; .
\end{equation}
This structure coincides with the analogous form present in the leading-order impact factor. Once this structure is factorized, it becomes evident that the infrared-singular part of the contribution is simply a generalization of the result obtained in the infinite-top-mass limit (see the last line of Eq.~(3.17) in~\cite{Celiberto:2022fgx}), where \( g_H^2 \) is replaced by \( |F_T(0, -\vec{p}_H^{\; 2}, m_H^2)|^2 \). It is not surprising that the soft and collinear divergences originate from the \( t \)-channel diagram. The expected collinear divergence is, in fact, associated with the initial-state radiation of the collinear gluon. Even in the soft case, this is the only diagram where the additional gluon is emitted from an external on-shell gluon line. In all other diagrams, the soft limit is "protected" either by the top mass or by the virtuality of the \( t \)-channel Reggeon.\\

Given the complexity of the final result, it is beneficial to explicitly extract its infrared singularities. The complete expression for the infrared limit of the impact factor is
\begin{gather}
\frac{d \Phi_{gg}^{ \{ Hg \} } \left( z_H, \vec{p}_H, \vec{q}; s_0 \right)}{d z_H d^2 \vec{p}_H}  \xrightarrow{\vec{q} \; \sim \; \vec{p}_H} \frac{g^2 \left| F_T \left(0, -\vec{p}_H^{\; 2}, m_H^2 \right) \right|^2 N}{4 (1-\epsilon) \sqrt{N^2-1} (2 \pi)^{D-1}} \frac{\vec{p}_H^{\; 2}}{(\vec{q}- \vec{p}_H)^{2}} \nonumber \\ \times \,\frac{z_H}{1 - z_H} \theta (s_{\Lambda}-s_{gR}) + \frac{g^2 \left| F_T \left(0, -\vec{p}_H^{\; 2}, m_H^2 \right) \right|^2 N}{4 (1-\epsilon) \sqrt{N^2-1} (2 \pi)^{D-1}} \frac{1}{(\vec{q} - \vec{p}_H)^2} \nn\\ \times \,\left[ z_H (1-z_H) \vec{p}_H^{\; 2} + 2 (1-\epsilon) \frac{1-z_H}{z_H} \frac{(\vec{p}_H \cdot (\vec{q}- \vec{p}_H))^2}{(\vec{q} - \vec{p}_H)^2} \right] . 
\label{Eq:IFinTheCollLim}
\end{gather}
The first term contains a soft divergence, while the second and third terms exhibit collinear divergences. It is essential to highlight that the first divergence is purely of a soft nature and is not a rapidity divergence. Specifically, in a cut-off regularization scheme for singularities related to the longitudinal fraction \( z_g \), terms that scale inversely with \( z_g \) correspond to both rapidity and soft divergences. However, in the current scenario, at high rapidity with \( z_H \rightarrow 1 \) while keeping \( \vec{q} - \vec{p}_H \) fixed, the \( t \)-channel diagrams do not contribute to the rapidity divergence of the impact factor, due to a \( (1 - z_H) \)-suppressing factor introduced by the structure \( F_T \) (see Section \ref{sssec:Rapiditydivergence}). For added clarity, we emphasize that, while in the collinear/soft limit
\begin{gather}
    F_T((k_1- p_g)^2 , -\vec{q}^{\; 2} , m_H^2) = F_T \left( - \frac{(\vec{q} - \vec{p}_H)^2}{1-z_H} , -\vec{q}^{\; 2} , m_H^2 \right) \xrightarrow[\text{collinear}]{\text{soft and/or}} F_T(0, -\vec{p}_H^{\; 2} , m_H^2) \; ,
\end{gather}
in the rapidity limit
\begin{gather}
    F_T((k_1- p_g)^2 , -\vec{q}^{\; 2} , m_H^2) \nonumber  \\ = F_T \left( - \frac{(\vec{q} - \vec{p}_H)^2}{1-z_H} , -\vec{q}^{\; 2} , m_H^2 \right)  \xrightarrow{\text{rapidity}} F_T( \infty, -\vec{p}_H^{\; 2} , m_H^2) \sim (1-z_H) \; .
\end{gather}
Therefore, the divergence in the first term of Eq.~(\ref{ImpactUnpro}) is soft. When considering the infrared approximation of the impact factor, the dependence on the parameter \( s_{\Lambda} \) is eliminated through its combination with the BFKL counter-term (second term of Eq.~(\ref{ImpactUnpro})), which, in the same approximation, also exhibits a soft divergence. As previously discussed, our goal is to add and subtract a term that renders the impact factor completely finite, thereby removing both rapidity and infrared divergences. By anticipating the result from Section \ref{sssec:Rapiditydivergence} concerning the rapidity divergence, we can construct a subtraction term,
\begin{gather}
\frac{d \Phi_{gg}^{ \{ Hg \} } \left( z_H, \vec{p}_H, \vec{q}; s_0 \right)}{d z_H d^2 \vec{p}_H} \bigg |_{\rm div.} \hspace{-0.5 cm} = \frac{g^2 \left| F_T \left(0, -\vec{p}_H^{\; 2}, m_H^2 \right)  \right|^2 N}{4 (1-\epsilon) \sqrt{N^2-1} (2 \pi)^{D-1}} \frac{\vec{q}^{\; 2}}{(\vec{q}- \vec{p}_H)^{2}} \nonumber \\ \times\,\frac{z_H}{1 - z_H} \theta \left(s_{\Lambda} - \frac{(\vec{q}-\vec{p}_H)^2}{1-z_H} \right) + \frac{g^2 \left| F_T \left(0, -\vec{p}_H^{\; 2}, m_H^2 \right)  \right|^2 N}{4 (1-\epsilon) \sqrt{N^2-1} (2 \pi)^{D-1}} \frac{1}{(\vec{q} - \vec{p}_H)^2} \nn \\ \times\,\left[ z_H (1-z_H) \vec{q}^{\; 2} + 2 (1-\epsilon) \frac{1-z_H}{z_H} \frac{(\vec{q} \cdot (\vec{q}- \vec{p}_H))^2}{(\vec{q} - \vec{p}_H)^2} \right] \; ,
\label{Eq:IFCombinedSub}
\end{gather}
capable of entirely eliminating both the rapidity and infrared divergences in the impact factor. It is important to emphasize that, even though the terms in eqs.~(\ref{Eq:IFinTheCollLim}) and~(\ref{Eq:IFCombinedSub}) are equivalent in the collinear limit \( q \rightarrow p_H \)\footnote{The \(\theta\)-function is required only in the region where \( z_H \rightarrow 1 \).}, the rapidity divergence in~(\ref{Eq:IFCombinedSub}) does not originate from the diagrams of \( \Gamma_t \).\\

A stringent validation of the result is the demonstration of gauge invariance. Specifically, the definition of the impact factor used here is explicitly gauge invariant, meaning that for \( \vec{q} = \vec{0} \), the entire amplitude must vanish. Consequently, in this limit, the triangular contribution, \( \Gamma_{\triangle} \), should precisely cancel out the box contribution, \( \Gamma_{\Box} \). To explicitly demonstrate the gauge invariance of the calculation, specifically the cancellation of various contributions as \( \vec{q} \to \vec{0} \), we utilize the large-top-mass expansion. To next-to-leading order, the triangular amplitude is given by
\begin{equation}
\begin{split}
& \Gamma_{\triangle}|_{\vec{q} = 0} =  -i  g f^{abc} \frac{\alpha_s}{3 \pi v}   \frac{2- z_H}{2}  \epsilon_{\perp}(k_1) \cdot \epsilon^*_{\perp}(p_g) + i  g f^{abc} \left( -\frac{\alpha_s }{360 \pi v m_t^2}  \right) \nn \\
& \times \frac{2-z_H}{2(1 - z_H)} \left[  ( 11 \vec{p}_H^{\; 2} + 7 m_H^2 (1-z_H))   g^{\mu\nu}  + 22  p_{\perp H}^{\mu} p_{\perp H}^{\nu} \right] \epsilon_{\perp \mu}(k_1) \epsilon^*_{\perp \nu}(p_g) + \mathcal O(m_t^{-4}) \nn \; ,
\end{split}
\end{equation}
while the box amplitude gives
\begin{equation}
\begin{split}
& \Gamma_{\Box}|_{\vec{q} = 0} = i  g f^{abc} \frac{\alpha_s}{3 \pi v}   \frac{2- z_H}{2}  \epsilon_{\perp}(k_1) \cdot \epsilon^*_{\perp}(p_g) -i g f^{abc}  \left( -\frac{\alpha_s }{360 \pi v m_t^2}  \right) \nn \\ & \times \frac{2 - z_H}{2(1 - z_H)}  \left[  (11 \vec{p}_H^{\; 2} + 7 m_H^2 (1 - z_H))  g^{\mu\nu} + 22 p_{H \perp}^{\mu} p_{H \perp}^{\nu} \right]  \epsilon_{\perp \mu}(k_1) \epsilon^*_{\perp \nu}(p_g) + \mathcal O(m_t^{-4}) \, .
\end{split}    
\end{equation}
The sum of these two terms cancels exactly, order by order, as expected. Utilizing {\tt Package X}, we have confirmed this cancellation up to terms of order \( m_t^{-4} \) (next-to-next-to-leading order).

\subsubsection{The gluon-initiated impact factor}
\label{sssec:gluonIF}
The total amplitude of the \( gR \to gH \) process can be represented in a concise manner as
\[
\Gamma^{abc(0)}_{\{Hg\}g}(q) = i g f^{abc}  \left[ C_{00} g_{\mu\nu} + C_{11} p_{\perp H}^\mu p_{\perp H}^\nu + C_{12} p_{\perp H}^\mu q_{\perp}^\nu \right.
\]
\beq
\left.
+ C_{21} q_{\perp}^\mu p_{\perp H}^\nu + C_{22} q_{\perp }^\mu q_{\perp}^\nu \right]  \epsilon_{\perp \mu}(k_1) \epsilon^*_{\perp \nu}(p_g) \;,
\eeq
where the coefficients $C_{ij}$ are
\begin{gather}
C_{00} =   \frac{m_H^2 (1-z_H)^2 +  (\vec{p}_H - z_H \vec{q})^2}{2[ m_H^2 (1-z_H) +  (\vec{p}_H - z_H \vec{q})^2 ]} F_T((p_g + p_H)^2, 0, m_H^2) - \frac{(1 - z_H) (m_H^2 + \vec{p}_H^{\,2} )}{2 [ m_H^2 (z_H - 1) -   \vec{p}_H^{\,2}] }\nn \\
\times\, F_T(0, (p_H - k_1)^2, m_H^2) 
+ \! (1-z_H) \frac{(\vec{q} - \vec{p}_H) \cdot \vec{q} }{(\vec{q} - \vec{p}_H)^2} F_T((k_1- p_g)^2, q^2, m_H^2)  \nn \\
+ \frac{ (2-z_H) \vec{q}^{\,2} }{2}   F_L((k_1- p_g)^2, q^2, m_H^2) + \frac{1}{2} \left[ B_a(k_1, -p_g, q) + (1 - z_H) B_a(-p_g, k_1, q)  \right]  \; , 
\end{gather}
\begin{gather}
    C_{11} = z_H \frac{F_T((p_g + p_H)^2, 0, m_H^2)}{ m_H^2 (1-z_H) +  (\vec{p}_H - z_H \vec{q})^2 }  + z_H \frac{F_T(0, (p_H - k_1)^2, m_H^2)}{m_H^2 (z_H - 1) -   \vec{p}_H^2} \nn \\
+ \frac{1}{2} \left[ B_b(k_1, q, -p_g)  + \frac{1}{1 - z_H}  B_b(-p_g, q, k_1) \right], 
\end{gather}
\begin{gather}
  C_{12} = - z_H^2 \frac{F_T((p_g + p_H)^2, 0, m_H^2)}{ m_H^2 (1-z_H) +  (\vec{p}_H - z_H \vec{q})^2 }  - \frac{z_H (1 - z_H)}{(\vec{q} - \vec{p}_H)^2} F_T((k_1- p_g)^2, q^2, m_H^2)  \nn \\
- \frac{1}{2}   \left[    B_b(k_1, q, -p_g) +  \frac{1}{ 1 - z_H}   B_b(-p_g, q, k_1) + (1 - z_H) B_b(q, k_1, -p_g)     -  B_c(q, k_1, -p_g) \right],   
\end{gather}
\begin{gather}
    C_{21} = - z_H^2 \frac{F_T((p_g + p_H)^2, 0, m_H^2)}{ m_H^2 (1-z_H) +  (\vec{p}_H - z_H \vec{q})^2 }  + \frac{z_H }{(\vec{q} - \vec{p}_H)^2} F_T((k_1- p_g)^2, q^2, m_H^2) \nn \\
- \frac{1}{2} \left[ B_b(k_1, q, -p_g)  +  \frac{1}{1 - z_H}B_b(-p_g, q, k_1)  +  \frac{1}{1 - z_H} B_b(q, -p_g, k_1)    -    B_c(k_1, q, -p_g)   \right],
\end{gather}
\bea
C_{22} &=&  z_H^3 \frac{F_T((p_g + p_H)^2, 0, m_H^2)}{ m_H^2 (1-z_H) +  (\vec{p}_H - z_H \vec{q})^2 }  -  \frac{z_H^2 }{(\vec{q} - \vec{p}_H)^2} F_T((k_1- p_g)^2, q^2, m_H^2)  \nn \\
&+&  \frac{1}{2}  \bigg[  B_b(k_1, -p_g, q) + B_b(k_1, q, -p_g) + (1 - z_H) B_b(-p_g, k_1, q) \nn \\ &+& (1 - z_H) B_b(q, k_1, -p_g)
+  \frac{1}{1 - z_H}  (B_b(-p_g, q, k_1) + B_b(q, -p_g, k_1)) \nn\\ &-& 
   B_c(k_1, q, -p_g) - B_c(q, k_1, -p_g)   \bigg] \,.
\eea
The summation over the transverse polarizations of the gluons is carried out using
\bea
\sum_\lambda \epsilon^{\mu}_{\perp, \lambda}(k)\epsilon^{\nu *}_{\perp, \lambda}(k) = - g^{\mu\nu}_{\perp \perp} = - g^{\mu\nu} + \frac{k_1^{\mu} k_2^{\nu}+ k_2^{\mu} k_1^{\nu}}{k_1 \cdot k_2} \;,
\eea
where \( g^{\mu\nu}_{\perp \perp} \) denotes the metric tensor in the transverse space. We also perform the convolution with the gluon PDF to obtain the proton-initiated impact factor:
\begin{gather}
\frac{d \Phi_{PP}^{\{H g\}} (x_H, \vec{p}_H, \vec{q}) }{  d z_H d^2 p_H} \!=   \frac{g^2 N  }{ 4 (2\pi)^{D-1} (1-\epsilon) \sqrt{N^2-1} } \int_{x_H}^1 \frac{d z_H}{z_H^2} f_g \left( \frac{x_H}{z_H} \right) \frac{1}{(1-z_H)}\nn\\ \times\, \bigg \{ 2 (1-\epsilon) |C_{00}|^2   + \vec{p}_H^{\; 4} |C_{11}|^2 +  \vec{q}^{\; 4} |C_{22}|^2 + 2 (\vec{p}_H \cdot \vec{q})^2   \Re\{ C_{12}^* C_{21} + C_{11}^* C_{22} \} 
 \nonumber \\ + \vec{p}_H^{\; 2} \vec{q}^{\; 2} (|C_{12}|^2 + |C_{21}|^2)+ 2 \vec{p}_H^{\; 2} \, \vec{p}_H \cdot \vec{q} \, \Re \{C_{11}^* (C_{12} + C_{21}) \} \nonumber  \\ + 2\vec{q}^{\; 2} \, \vec{p}_H \cdot \vec{q} \, \Re \{C_{22}^* (C_{12} + C_{21}) \} -  2 \vec{p}_H^{\; 2} \, \Re \{ C_{00}^* C_{11} \}
 - 2 \vec{q}^{\; 2} \, \Re \{ C_{00}^* C_{22} \} \nn\\ - 2 \vec{p}_H \cdot \vec{q} \, \Re \{C_{00}^* (C_{12} + C_{21}) \}  \bigg\}\; \theta \left( s_{\Lambda} - \frac{(1-z_H) m_H^2 + \vec{\Delta}^2}{z_H (1-z_H)} \right) \,.
\label{eq:gRgH_square}
\end{gather}
The contribution from a real gluon emission also exhibits a divergence as \( z_H \to 1 \) (\( z_g \to 0 \)) for any value of the outgoing gluon transverse momenta \( \vec{q} - \vec{p}_H \). This rapidity divergence is regulated by the parameter \( s_{\Lambda} \). In the final result, it should be canceled by the BFKL counter-term included in the definition of the NLO impact factor. However, unlike the infrared case, managing this (high-energy) singularity is considerably more complex and does not proceed in the same manner as in the infinite-top-mass limit scenario. 

\subsubsection{Rapidity divergence}
\label{sssec:Rapiditydivergence}
In Ref.~\cite{DelDuca:2003ba}, the high-energy factorization for Higgs-plus-two-jet production was analyzed in two distinct scenarios: (a) with the Higgs boson centrally positioned in rapidity between the two jets and located far from either one, and (b) where the Higgs boson is near in rapidity to one \textit{identified} jet, with both being far from the other jet. The latter scenario enables the extraction of the impact factor for Higgs production in association with a jet. The key difference between this case and our present calculation is that, since we are more inclusive regarding the final state, we can explore the region where \( z_H \rightarrow 1 \) (\( z_g \rightarrow 0 \)), which is affected by the rapidity divergence caused by the gluon-Higgs invariant mass diverging. This singularity is a natural consequence of the separation of contributions associated with MRK and QMRK kinematics~\cite{Fadin:1998fv}, and in our definition of impact factors, Eq.~(\ref{ImpactUnpro}), it cancels out in the combination of the first (QMRK) and second (MRK) terms. However, to explicitly demonstrate this cancellation, it is necessary to study the limit in which the Higgs and additional gluon are highly separated in rapidity. Revisiting the example of Higgs-plus-two-jet production, this corresponds to the kinematic configuration (c) where the Higgs boson is emitted forward and well-separated from the centrally emitted jet, which is itself strongly separated in rapidity from the backward jet. In such a case, the amplitude should assume a Regge form with the central gluon emission described by the Lipatov vertex.

To extract the rapidity divergence from our result and verify that it aligns with that predicted by BFKL factorization, thus being canceled by the BFKL counter-term (i.e., the final term in Eq.~(\ref{ImpactUnpro})), we need to expand the contributions in eqs.~(\ref{Eq:Gammas}), (\ref{Eq:Gammau}), (\ref{Eq:Gammat}), (\ref{Eq:Box_Amplit}) around \( z_H = 1 \). Comparing the infinite and finite top-mass cases reveals interesting differences. In the former~\cite{Celiberto:2022fgx}, we have \( F_T = -B_a = -g_H \), while \( F_L = B_b = B_c = 0 \). Due to the constant nature of these terms and the phase-space factor in Eq.~(\ref{Eq:PhasePhaseInt}), all diagrams contribute in the \( z_H \rightarrow 1 \) limit. However, thanks to substantial simplification, the impact factor in this limit is solely given by the contribution from the \( \Gamma_t \) term, indicating that contributions from all other diagrams cancel out in the \( z_H \rightarrow 1 \) limit.

In the finite-top-mass scenario, the structures \( F_T, F_L, B_a, B_b \), and \( B_c \) are non-trivial functions of \( z_H \), and their asymptotic behavior as \( z_H \rightarrow 1 \) must be carefully analyzed to determine the rapidity limit of the impact factor. We begin by considering the triangular functions \( F_T((p_g + p_H)^2, 0, m_H^2), F_T(0, (p_H - k_1)^2, m_H^2), F_T((k_1 - p_g)^2, 0, m_H^2) \) and \( F_L((k_1 - p_g)^2, 0, m_H^2) \). The asymptotic behavior of the "external masses" entering these functions is
\begin{gather}
(p_g+p_H)^2 \xrightarrow{z_H \sim 1} \frac{(\vec{q} - \vec{p}_H)^2}{1-z_H} \; , \hspace{1 cm} (k_1-p_g)^2  \xrightarrow{z_H \sim 1} - \frac{(\vec{q} - \vec{p}_H)^2}{1-z_H} \nonumber \\ 
(p_H-k_1)^2 \xrightarrow{z_H \sim 1} - \vec{p}_H^{\; 2} \; , \hspace{1 cm}  p_g^2 = k_1^2 = 0 \; , \hspace{1 cm} p_H^2 = m_H^2 \; , \hspace{1 cm} q^2 = -\vec{q}^{\; 2}  \;.
\end{gather}
From the definitions of \( F_L \) and \( F_T \) (eqs.~(\ref{Eq:FLtri}) and~(\ref{Eq:FTtri}), respectively), it is apparent that they scale inversely with their external masses. Consequently, the diagrams for \( \Gamma_t \) and \( \Gamma_s \) experience a linear suppression in \( (1 - z_H) \), ensuring that they do not contribute in the high-rapidity limit. Thus, the entire contribution from the triangular diagrams in the \( z_H \to 1 \) limit arises from \( \Gamma_u \), which does not involve any suppressing scale. The phenomenon where diagrams that remain finite in the high-energy limit become divergent when the infinite-top-mass approximation is initially taken demonstrates the non-commutativity of the limits \( s \rightarrow \infty \) and \( m_t \rightarrow \infty \).\\

In the case of box-type contributions, analytically extracting the \( z_H \to 1 \) limit is highly intricate, prompting us to use the symbolic assistance of {\tt Package X}. The LoopRefineSeries function of {\tt Package X}, combined with the standard Series function of {\tt Mathematica}, enables the expansion of all structures contributing to the impact factor, except for the box-type scalar integrals
\begin{gather}
  D_0 (0, 0, m_H^2, q^2, -s, s, m_t^2, m_t^2, m_t^2, m_t^2) \; , \nonumber \\
  D_0 (0, 0, m_H^2, q^2, -s, u, m_t^2, m_t^2, m_t^2, m_t^2) \; , \hspace{1 cm} D_0 (0, 0, m_H^2, q^2, s, u, m_t^2, m_t^2, m_t^2, m_t^2) \; ,
\label{EQ:variousBox}
\end{gather}
with
\begin{gather}
   D_0(p_1^2, p_2^2, p_3^2, p_4^2, (p_1+p_2)^2, (p_2+p_3)^2, m_0^2, m_1^2, m_2^2, m_3^2) = \frac{\mu^{4-D}}{i \pi^{D/2} r_\Gamma} \nn \\ \times \int d^D q \frac{1}{[q^2 - m_0^2] [(q+p_1)^2 - m_1^2] [(q+p_1+p_2)^2 - m_2^2] [(q+p_1+p_2+p_3)^2 - m_3^2]}  
\end{gather}
and
\begin{equation}
s = (k_1 + q)^2 \xrightarrow{z_H \sim 1} \frac{(\vec{q} - \vec{p}_H)^2}{1-z_H} \; , \hspace{1 cm} u = (k_1 - p_H)^2 \xrightarrow{z_H \sim 1} - \vec{p}_H^{\; 2} \; .
\label{sAndu}
\end{equation}
For the last two terms in (\ref{EQ:variousBox}), it suffices to note that they scale inversely with \( s \) (or linearly in \( (1 - z_H) \)) and thus do not contribute. Conversely, the first term must be evaluated in the \( s \rightarrow \infty \) limit and yields
\begin{gather}
  D_0 (0, 0, m_H^2, q^2, -s, s, m_t^2, m_t^2, m_t^2, m_t^2) = - \frac{1}{s^2} \left[ \left( \ln \left( \frac{s}{m_t^2} \right) - i \pi \right)^2 + \ln^2 \left( \frac{s}{m_t^2} \right) \right. \nonumber \\
 \left. - \ln^2 \left( - \frac{1 + \sqrt{ 1 + 4 m_t^2 / \vec{q_{}}^{\; 2} } }{1 - \sqrt{ 1 + 4 m_t^2 / \vec{q_{}}^{\; 2}  }} \right) -  \ln^2 \left( - \frac{ 1 + \sqrt{   1 - 4 m_t^2 / m_H^{2}  } }{ 1 - \sqrt{ 1 - 4 m_t^2 / m_H^{2} }} \right) \right] \; .
 \label{BoxIntegralAsymptotic}
\end{gather} 
From eqs.~(\ref{sAndu}) and (\ref{BoxIntegralAsymptotic}), it is immediately evident that, besides appearing as a power, \( s \sim 1/(1 - z_H) \) also appears within the argument of logarithms. In the box contributions, this manifests in the dominant terms of the \( z_H \rightarrow 1 \) expansion, resulting in rapidity divergences of the form
\begin{equation}
\frac{1}{1-z_H} \; , \hspace{1 cm} \frac{1}{1-z_H} \ln (1-z_H) \; , \hspace{1 cm} \frac{1}{1-z_H} \ln^2 (1-z_H) \; .
\label{Rapidity divergences}
\end{equation}
The last two types of terms are inconsistent with those predicted by BFKL factorization and, in fact, cancel individually for all ten coefficients in Eq.~(\ref{Eq:Box_Amplit}). Utilizing the results outlined above and leveraging {\tt Package X}, we can finally compute the high-rapidity limit of the impact factor, which is given by
\begin{equation*}
   \frac{d \Phi_{gg}^{ \{H g \}} (z_H, \vec{p}_H, \vec{q} ; s_0)}{d z_H d^2 \vec{p}_H} \bigg|_{z_H \rightarrow 1} 
\end{equation*}
\begin{equation*}
   = \frac{\langle cc'|\hat{\mathcal{P}}|0 \rangle }{2(1-\epsilon)(N^2-1)} 
    \left[ \sum_{\{ f \}} \int \frac{d s_{PR} d\rho_f}{2 \pi} \Gamma^c_{P \{ f \}} \left( \Gamma^c_{P \{ f \}} \right)^{*} \theta \left( s_{\Lambda} - s_{PR} \right) \right]_{z_H \rightarrow 1}
\end{equation*}
\begin{equation}
    = \frac{g^2 |F_T ( 0, -\vec{p}_H^{\; 2}, m_{H}^2  )|^2 N}{4 (1-\epsilon) \sqrt{N^2-1} (2 \pi)^{D-1} } \frac{\vec{q}^{\; 2}}{(\vec{q}- \vec{p}_H)^2} \frac{1}{(1-z_H)} \theta \left( s_{\Lambda} - \frac{(\vec{q}-\vec{p}_H)^2}{(1-z_H)} \right) \; .
\label{Eq:HighRapidityLimitIF}
\end{equation}
It is important to emphasize that, while the expression in Eq.~(\ref{Eq:HighRapidityLimitIF}) may appear to be a straightforward extension of the infinite-top-mass limit case~\cite{Celiberto:2022fgx}, the mechanisms leading to this result are fundamentally different. We can now express Eq.~(\ref{Eq:IFCombinedSub}) as
\begin{gather}
\frac{d \Phi_{gg}^{ \{ Hg \} } \left( z_H, \vec{p}_H, \vec{q}; s_0 \right)}{d z_H d^2 \vec{p}_H} \bigg |_{\rm div.} \hspace{-0.5 cm} = z_H \frac{d \Phi_{gg}^{ \{H g \}} (z_H, \vec{p}_H, \vec{q} ; s_0)}{d z_H d^2 \vec{p}_H} \bigg|_{z_H \rightarrow 1}  \nonumber \\ + \frac{g^2 \left| F_T ( 0, -\vec{p}_H^{\; 2}, m_{H}^2  ) \right|^2 N}{4 (1-\epsilon) \sqrt{N^2-1} (2 \pi)^{D-1}} \left[ z_H (1-z_H) \vec{q}^{\; 2} + 2 (1-\epsilon) \frac{1-z_H}{z_H} \frac{(\vec{q} \cdot (\vec{q}- \vec{p}_H))^2}{(\vec{q} - \vec{p}_H)^2} \right] \; .
\end{gather}
After the convolution with the gluon PDF, the subtraction term becomes
\begin{gather}
 \int_{x_H}^1 \frac{d z_H}{z_H} f_g \left( \frac{x_H}{z_H} \right) \frac{d \Phi_{gg}^{ \{ Hg \} } \left( z_H, \vec{p}_H, \vec{q}; s_0 \right)}{d z_H d^2 \vec{p}_H} \bigg |_{\rm div.} \hspace{-0.3 cm} =\nonumber \\ \int_{x_H}^1 d z_H f_g \left( \frac{x_H}{z_H} \right)  \frac{d \Phi_{gg}^{ \{ H g \}}(z_H, \vec{p}_H, \vec{q};s_0)}{d z_H d^2 \vec{p}_H} \Bigg |_{z_H=1} \nonumber \\ 
 + \int_{x_H}^1 \frac{d z_H}{z_H} f_g \left( \frac{x_H}{z_H} \right) \frac{g^2 \left| F_T ( 0, -\vec{p}_H^{\; 2}, m_{H}^2  ) \right|^2 N}{4 (1-\epsilon) \sqrt{N^2-1} (2 \pi)^{D-1}} \nonumber \\ \times \left[ z_H (1-z_H) \vec{q}^{\; 2} + 2 (1-\epsilon) \frac{1-z_H}{z_H} \frac{(\vec{q} \cdot (\vec{q}- \vec{p}_H))^2}{(\vec{q} - \vec{p}_H)^2} \right] \nonumber \\
 \equiv \int_{x_H}^1 d z_H f_g \left( \frac{x_H}{z_H} \right)  \frac{d \Phi_{gg}^{ \{ H g \}}(z_H, \vec{p}_H, \vec{q};s_0)}{d z_H d^2 \vec{p}_H} \Bigg |_{z_H=1} + \frac{d \Phi^{\{Hg\}{\rm no \; plus}}_{PP}(x_H, \vec{p}_H,\vec q;s_0)}{d x_H d^2 p_H} \;.
\end{gather}
Next, we combine the BFKL counter-term with the gluon-initiated contribution in a manner that immediately reveals the cancellation of the rapidity divergence and isolates the IR-singular sector. To begin, we rewrite~(\ref{eq:gRgH_square}) in an equivalent form by adding and subtracting the following three terms:
\begin{gather*}
\int_{x_H}^1 \frac{d z_H}{z_H} f_g \left( \frac{x_H}{z_H} \right) \frac{d \Phi_{gg}^{ \{ Hg \} } \left( z_H, \vec{p}_H, \vec{q}; s_0 \right)}{d z_H d^2 \vec{p}_H} \bigg |_{\rm div.} \; , \\ \int_{x_H}^1 d z_H f_g (x_H)  \frac{d \Phi_{gg}^{ \{H g \}} (z_H, \vec{p}_H, \vec{q} ; s_0)}{d z_H d^2 \vec{p}_H} \bigg|_{z_H \rightarrow 1} 
\end{gather*}
and
\begin{gather*}
\int^{x_H}_0 d z_H f_g (x_H)  \frac{d \Phi_{gg}^{ \{H g \}} (z_H, \vec{p}_H, \vec{q} ; s_0)}{d z_H d^2 \vec{p}_H} \bigg|_{z_H \rightarrow 1} \; .
\end{gather*}
Then, we get
\[
\frac{d \Phi^{\{Hg\}}_{PP}(x_H, \vec{p}_H,\vec q;s_0)}{d x_H d^2 p_H}
= \frac{d \tilde{\Phi}^{\{Hg\}}_{PP}(x_H, \vec{p}_H,n,\vec q;s_0)}{d x_H d^2 p_H}
\]
\[
+ \frac{d \Phi^{\{Hg\} (1-x_H)}_{PP}(x_H, \vec{p}_H,\vec q;s_0)}{d x_H d^2 p_H}
+ \frac{d \Phi^{\{Hg\}{{\rm real} \; P_{gg}}}_{PP}(x_H, \vec{p}_H,\vec q;s_0)}{d x_H d^2 p_H}
\]
\begin{equation}
+ \int_{x_H}^1 d z_H f_g (x_H)  \frac{d \Phi_{gg}^{ \{H g \}} (z_H, \vec{p}_H, \vec{q} ; s_0)}{d z_H d^2 \vec{p}_H} \bigg|_{z_H \rightarrow 1} \; ,
\label{subtractions}
\end{equation}
where
\begin{equation*}
    \frac{d \tilde{\Phi}^{\{Hg\}}_{PP}(x_H, \vec{p}_H,\vec q;s_0)}{d x_H d^2 p_H}
    =\int_{x_H}^1 \frac{d z_H}{z_H} f_g \left( \frac{x_H}{z_H} \right) \frac{d \Phi_{gg}^{ \{ H g \}}(z_H, \vec{p}_H, \vec{q};s_0)}{d z_H d^2 \vec{p}_H} 
\end{equation*}
\begin{equation}
    - \int_{x_H}^1 d z_H f_g \left( \frac{x_H}{z_H} \right)  \frac{d \Phi_{gg}^{ \{ H g \}}(z_H, \vec{p}_H, \vec{q};s_0)}{d z_H d^2 \vec{p}_H} \Bigg |_{z_H=1} -  \frac{d \Phi^{\{Hg\}{\rm no \; plus}}_{PP}(x_H, \vec{p}_H,\vec q;s_0)}{d x_H d^2 p_H} \; ,
\label{Eq:TermDiff}
\end{equation}
\begin{equation}
\frac{d \Phi^{\{Hg\} (1-x_H)}_{PP}(x_H, \vec{p}_H,\vec q;s_0)}{d x_H d^2 p_H}
=\int_{0}^{x_H} d z_H f_g (x_H) \frac{d \Phi_{gg}^{ \{ H g \}}(z_H, \vec{p}_H, \vec{q};s_0)}{d z_H d^2 \vec{p}_H} \Bigg |_{z_H=1} 
\label{Eq:Term(1-zH)}
\end{equation}
and
\begin{equation*}
\frac{d \Phi^{\{Hg\}{\rm real} \; P_{gg}}_{PP}(x_H, \vec{p}_H,\vec q;s_0)}{d x_H d^2 p_H}
= - \int_{0}^{x_H} d z_H f_g (x_H) \frac{d \Phi_{gg}^{ \{ H g \}}(z_H, \vec{p}_H, \vec{q};s_0)}{d z_H d^2 \vec{p}_H} \Bigg |_{z_H=1}
\end{equation*}
\begin{equation}
\begin{split}
    +& \int_{x_H}^1 d z_H \left( f_g \left( \frac{x_H}{z_H} \right) - f_g (x_H) \right) \frac{d \Phi_{gg}^{ \{ H g \}}(z_H, \vec{p}_H, \vec{q};s_0)}{d z_H d^2 \vec{p}_H} \Bigg |_{z_H=1} \\ +& \frac{d \Phi^{\{Hg\}{\rm no \; plus}}_{PP}(x_H, \vec{p}_H,\vec q;s_0)}{d x_H d^2 p_H} .
\label{Eq:TermPlus}
\end{split}
\end{equation}
The terms \( d\tilde{\Phi} \), \( d\Phi^{\{Hg\}(1-x_H)} \), and \( d\Phi^{\{Hg\} {\rm real} \; P_{gg}} \) are free from divergences as \( z_H \to 1 \). Consequently, in their expressions, the limit \( s_{\Lambda} \rightarrow \infty \) can be safely taken, meaning that \( \theta(s_{\Lambda} - s_{PR}) \) can be set to one. Furthermore, \( d\tilde{\Phi} \) is also infrared-safe and can be expressed in a more explicit form as
\begin{equation*}
    \frac{d \tilde{\Phi}^{\{Hg\}}_{PP}(x_H, \vec{p}_H,\vec q;s_0)}{d x_H d^2 p_H}
    =\int_{x_H}^1 \frac{d z_H}{z_H} f_g \left( \frac{x_H}{z_H} \right) \left[ \frac{d \Phi_{gg}^{ \{ H g \}}(z_H, \vec{p}_H, \vec{q};s_0)}{d z_H d^2 \vec{p}_H} - \frac{g^2  N}{4 (1-\epsilon) \sqrt{N^2-1}} \right.
\end{equation*}
\begin{equation}
    \left. \times \frac{\left| F_T ( 0, -\vec{p}_H^{\; 2}, m_{H}^2  ) \right|^2}{{ (2 \pi)^{D-1} (\vec{q}-\vec{p}_H)^2}} \left( \frac{z_H}{1-z_H} \vec{q}^{\; 2} + z_H (1-z_H) \vec{q}^{\; 2} + 2 (1-\epsilon) \frac{1-z_H}{z_H} \frac{(\vec{q} \cdot (\vec{q}- \vec{p}_H))^2}{(\vec{q} - \vec{p}_H)^2} \right) \right] .
\end{equation}
We now demonstrate the explicit cancellation of rapidity divergences. The final term in Eq.~(\ref{subtractions}) can be readily evaluated using Eq.~(\ref{Eq:HighRapidityLimitIF}), yielding
\begin{equation*}
    \int_{x_H}^1 d z_H f_g (x_H)  \frac{d \Phi_{gg}^{ \{H g \}} (z_H, \vec{p}_H, \vec{q} ; s_0)}{d z_H d^2 \vec{p}_H} \bigg|_{z_H \rightarrow 1}
\end{equation*}
\begin{equation*}
=    \frac{g^2 |F_T ( 0, -\vec{p}_H^{\; 2}, m_{H}^2  )|^2 N}{4 (1-\epsilon) \sqrt{N^2-1} (2 \pi)^{D-1} } \frac{\vec{q}^{\; 2}}{(\vec{q}- \vec{p}_H)^2} \int_{x_H}^1 d z_H \frac{1}{(1-z_H)} f_g (x_H) \theta \left( s_{\Lambda} - \frac{(\vec{q}-\vec{p}_H)^2}{(1-z_H)} \right) 
\end{equation*}
\begin{equation}
    = \frac{g^2 |F_T ( 0, -\vec{p}_H^{\; 2}, m_{H}^2  )|^2 N}{4 (1-\epsilon) \sqrt{N^2-1} (2 \pi)^{D-1} } \frac{\vec{q}^{\; 2}}{(\vec{q}- \vec{p}_H)^2} f_g (x_H) \left[ \ln (1-x_H) - \frac{1}{2} \ln \left( \frac{\left[(\vec{q}-\vec{p}_H)^2 \right]^2}{s_{\Lambda}^2} \right) \right] \; .
\label{Phi_g_rap}
\end{equation}
Next, let us examine the BFKL counter-term, which is represented by the last term in Eq.~(\ref{ImpactUnpro}),
\begin{equation}
\begin{split}
   \frac{d \Phi^{\rm{BFKL \ c.t.}}_{gg}(z_H, \vec{p}_H, \vec q;s_0)}{d z_H d^2 p_H} &= - \frac{1}{2} \int d^{D-2} q' \frac{ \vec{q}^{\; 2}}{\vec{q}^{\; '2}}  \frac{d \Phi_{gg}^{ \{H \} (0)} (\vec{q} \; ' )}{d z_H d^2 p_H} \\ &\times \mathcal{K}^{(0)}_r (\vec{q} \; ', \vec{q} \; ) \ln \left( \frac{s_{\Lambda}^2}{(\vec{q} \; ' - \vec{q} \; )^2 s_0} \right) \; .
\end{split}
\end{equation}
Using Eq.~(\ref{Eq:LoImpactD4-2EPart}) and Eq.~(\ref{BornKer}), we find
\begin{equation}
\begin{split}
   \frac{d \Phi^{\rm{BFKL \ c.t.}}_{gg}(z_H, \vec{p}_H, \vec q;s_0)}{d z_H d^2 p_H} &\!=\! \frac{-g^2 |F_T ( 0, -\vec{p}_H^{\; 2}, m_{H}^2  )|^2 N}{8 (2 \pi)^{D-1} (1-\epsilon) \sqrt{N^2-1}} \frac{\vec{q}^{\; 2}}{(\vec{q}-\vec{p}_H)^2}\\ &\times \ln \left( \frac{s_{\Lambda}^2}{(\vec{q} - \vec{p}_H )^2 s_0}  \right) \!\delta (1-z_H)
\end{split}
\end{equation}
and, after convolution with the gluon PDF, we get
\[
 \frac{d \Phi^{{\rm{BFKL\ c.t.}}}_{PP}(x_H,\vec p_H,\vec q;s_0)}{d x_H d^2 p_H} = \int_{x_H}^1 \frac{d z_H}{z_H} f_g \left( \frac{x_H}{z_H} \right) \frac{d \Phi_{gg}^{{\rm{BFKL\ c.t.}}} (z_H, \vec{p}_H, \vec{q})}{d z_H d^2 \vec{p}_H} 
 \]
 \begin{equation}
= - \frac{g^2 |F_T ( 0, -\vec{p}_H^{\; 2}, m_{H}^2  )|^2 N}{8 (2 \pi)^{D-1} \sqrt{N^2-1}} \frac{\vec{q}^{\; 2}}{(\vec{q}-\vec{p}_H)^2} \frac{f_g (x_H)}{(1-\epsilon)} \ln \left( \frac{s_{\Lambda}^2}{(\vec{q} - \vec{p}_H )^2 s_0}  \right) \; .
 \label{BFKLct}
\end{equation}
When we combine the final term of Eq.~(\ref{subtractions}), as given in~(\ref{Phi_g_rap}), with the BFKL counter-term presented in~(\ref{BFKLct}), we obtain:
\begin{equation}
\begin{split}
\frac{d \Phi^{{\rm{BFKL}}}_{PP}(x_H, \vec{p}_H, \vec q;s_0)}{d x_H d^2 p_H} &\equiv \frac{g^2 |F_T ( 0, -\vec{p}_H^{\; 2}, m_{H}^2  )|^2 N}{4 (2 \pi)^{D-1} (1-\epsilon) \sqrt{N^2-1}} \frac{\vec{q}^{\; 2}}{(\vec{q}-\vec{p}_H)^2}\\ &\times f_g (x_H) \ln \left( \frac{(1-x_H) \sqrt{s_0}}{|\vec{q} - \vec{p}_H|}  \right) \,.
 \label{CountFin}
\end{split}
\end{equation}
It is worth noting that this term remains finite with respect to the high-energy divergence. The BFKL term in Eq.~(\ref{CountFin}) includes a soft singularity, which is expected to cancel out when combined with the virtual corrections, as well as a collinear singularity proportional to \( \ln (1 - x_H) \) that cancels a corresponding term in Eq.~(\ref{Eq:Term(1-zH)}). The residual collinear singularity, associated with the initial-state radiation, is eliminated by employing the proper NLO definition of the gluon PDF. Specifically, the term in Eq.~(\ref{Eq:TermPlus}) yields the real part of the \( P_{gg} (z_H) \) DGLAP splitting function~\cite{Celiberto:2022fgx}.

\section{Conclusion and outlook}

We computed the real corrections to the next-to-leading-order Higgs impact factor, arising from the emission of an additional parton in the fragmentation region of Higgs production. Our work advances beyond the infinite-mass approximation~\cite{Nefedov:2019mrg, Hentschinski:2020tbi, Celiberto:2022fgx} by incorporating the finite top-quark mass into the Higgs impact factor calculation. We verified gauge invariance, the absence of rapidity divergences, and indications of correct infrared behavior. 

From a theoretical standpoint, the next and final step in this research line will involve computing the virtual corrections, an aspect to be addressed in an upcoming publication. On the phenomenological side, once complete, the impact factor will enable the extension of previous studies on Higgs production~\cite{DelDuca:1993ga,Celiberto:2020tmb,Andersen:2022zte,Andersen:2023kuj,Celiberto:2022zdg} by including next-to-leading-order corrections with full top-mass dependence.

\chapter{Tetraquark-plus-jet in a Variable-Flavor Number Scheme}
\chaptermark{Tetraquark-plus-jet in a VFNS}
In this chapter, we explore the semi-inclusive production of exotic matter in the semi-hard regime, focusing on two main objectives: first, to investigate the formation of exotic states in this kinematic domain, and second, to utilize these studies as a testing ground for the BFKL dynamics at high-energy colliders. We analyze the production mechanisms of doubly bottomed tetraquarks ($X_{b\bar{b}q\bar{q}}$) and fully bottomed tetraquarks ($T_{4b}$), collectively referred to as \textit{bottomonium-like} states.

More specifically, we focus on a class of partially inclusive processes featuring a forward-plus-backward two-particle final-state configuration. To this aim, we employ a hybrid framework that combines collinear factorization with high-energy BFKL dynamics, a methodology referred to as \textit{hybrid factorization}. This approach enables a comprehensive treatment of the relevant dynamics, integrating both perturbative and non-perturbative aspects of the hadroproduction process~\cite{Colferai:2010wu,Celiberto:2020wpk,Celiberto:2020tmb,Bolognino:2021mrc,Celiberto:2022rfj,Celiberto:2022dyf} (see also~\cite{vanHameren:2022mtk,Golec-Biernat:2018kem,Bonvini:2018ixe,Silvetti:2022hyc,Silvetti:2023suu,Rinaudo:2024hdb} for similar approaches to single-particle detections). \\

All considered reactions involve the production of heavy-flavored hadrons, whose theoretical description depends on the kinematical conditions. In our studies, we will always consider the hadroproduction of heavy flavor through the collinear fragmentation of a light quark. This mechanism is dominant at high transverse momentum of the produced particle. Studies of semi-hard reactions at the NLO accuracy in the same spirit have been conducted for other heavy-flavored hadrons in refs.~\cite{Celiberto:2021dzy,Celiberto:2021fdp,Celiberto:2022dyf,Celiberto:2022rfj,Celiberto:2023fzz,Celiberto:2024mrq,Celiberto:2024swu}. Complementarily, at low transverse momentum, the most suitable description is the short-distance production, which involves the production of heavy quarks at the level of the hard coefficient functions\footnote{In the case of BFKL, at the level of impact factors~\cite{Boussarie:2017oae,Bolognino:2019yls,Bolognino:2021mrc}.}. To perform our large-transverse-momentum analysis, we present a set of collinear fragmentation functions (FFs) tailored for these tetraquark states. We introduce~\cite{Celiberto:2024beg} two updated parametrizations, \texttt{TQHL1.1} and \texttt{TQ4Q1.1}, which replace and enhance the earlier \texttt{1.0} versions~\cite{Celiberto:2023rzw,Celiberto:2024mrq,Celiberto:2024mab}. The \texttt{TQHL1.1} functions focus on the fragmentation of doubly heavy tetraquarks, incorporating an advanced formulation of the Suzuki model~\cite{Suzuki:1977km,Suzuki:1985up,Amiri:1986zv,Nejad:2021mmp} for heavy-quark contributions. On the other hand, the \texttt{TQ4Q1.1} functions describe the fragmentation of fully heavy tetraquarks, integrating initial-scale inputs derived from potential nonrelativistic QCD for both gluon and heavy-quark channels. These new sets of functions allow for precise predictions of tetraquark production and their associated jets at collider energies of 14~TeV and 100~TeV\footnote{The {\tt TQHL1.1} and {\tt TQ4Q1.1} collinear FF families can be publicly accessed from the following url: \url{https://github.com/FGCeliberto/Collinear_FFs/}.}.\\

In the first section of this chapter, we review different flavor number schemes to describe heavy quarks in perturbative QCD.
We also briefly give highlights on heavy-flavor fragmentation, from singly heavy-flavored hadrons to quarkonia and exotic matter. 
The subsequent section is dedicated to detailing the construction of the fragmentation functions that will be employed in our analysis. In the third section, we detail the formulas describing the tetraquarks-plus-jet production in the hybrid factorization. Finally, in the last section, we showcase our results, discussing their implications in the context of exotic matter production at high-energy colliders.\\

All Feynman diagrams and plots presented in this sections are taken from~\cite{Celiberto:2024beg}.

\section{Tetraquark production in a VFNS}
Exotic hadrons, such as tetraquarks and pentaquarks, are fundamental to understanding multiquark systems beyond traditional mesons and baryons. These states, characterized by more complex internal structures with respect to ordinary hadrons, offer insight into the dynamics of strong interactions and color confinement. High-energy colliders like the LHC, and future facilities such as the EIC~\cite{AbdulKhalek:2021gbh,AbdulKhalek:2022hcn,Hentschinski:2022xnd,Amoroso:2022eow,Abir:2023fpo,Allaire:2023fgp} and the FCC~\cite{FCC:2018byv,FCC:2018evy,FCC:2018vvp,FCC:2018bvk}, will provide opportunities to study these particles' internal quark-gluon configurations and production mechanisms. Recent advances in QCD factorization and all-order perturbative techniques enable precise cross section calculations, crucial for comparing theory with experimental data and understanding exotic matter dynamics.

Exotic hadrons can be divided into states involving gluons and those containing multiple quarks, like tetraquarks and pentaquarks. Recent observations, such as $X(3872)$~\cite{Belle:2003nnu} and the doubly charmed $T_{cc}^+$~\cite{LHCb:2021vvq,LHCb:2021auc}, have provided valuable insights into their structures. Various theoretical models, including compact diquark configurations, meson molecules, and hadroquarkonium states, aim to describe the internal composition and interactions of these particles. Heavy tetraquarks, particularly $\QXQq$ and $\TQQ$, provide a unique perspective on QCD dynamics. However, understanding their production mechanisms remains challenging, with limited experimental confirmation, especially in the bottom-quark sector.\\

The $X(3872)$, despite having conventional quantum numbers, exhibits isospin-violating decays, indicating a more complex structure than traditional quarkonium. Alternative models supporting the tetraquark hypothesis include:
\begin{itemize}
    \item \textbf{Compact diquarks}: A tightly bound diquark-antidiquark pair (see Refs.~\cite{Maiani:2004vq,Mutuk:2021hmi,Wang:2013vex,Grinstein:2024rcu} and references therein).
    \item \textbf{Meson molecules}: A loosely bound state of two mesons (see Refs.~\cite{Tornqvist:1993ng,Braaten:2003he,Guo:2013sya,Esposito:2023mxw} and references therein).
    \item \textbf{Hadroquarkonium}: A quarkonium core surrounded by an orbiting light meson (see Refs.~\cite{Dubynskiy:2008mq,Voloshin:2013dpa,Ferretti:2020ewe} and references therein).
\end{itemize}

In this work, we focus on their production at high transverse momentum ($p_T$), adopting a collinear-fragmentation-based approach. Fragmentation is particularly suited for the study of heavy particles, such as tetraquarks, where a high-$p_T$ regime justifies the use of a \textbf{variable flavor number scheme} (VFNS). Indeed, the treatment of heavy quarks, such as $c$ and $b$, in perturbative QCD requires special attention, particularly in schemes involving parton distribution functions and fragmentation functions. Depending on kinematic conditions, two main frameworks are typically employed:
\begin{itemize}
    \item In the \textbf{Fixed-Flavor Number Scheme (FFNS)}, heavy quarks are always treated as massive particles (see, \emph{e.g.}, Ref.~\cite{Alekhin:2009ni} for more details). Their mass effects are fully included, and the number of active flavors remains fixed. This approach is accurate when the scale $Q$ is close to the heavy quark mass $m_Q$ (i.e. $Q\lesssim m_Q$), but it does not include the resummation of large logarithmic terms like $\ln(Q^2/m_Q^2)$, which can arise at high $Q^2$.
    \item The \textbf{Zero-Mass Variable-Flavor Number Scheme (VFNS)} ignores the heavy quark mass effects entirely~\cite{Mele:1990cw,Cacciari:1993mq,Buza:1996wv}. In this scheme, it is considered a standard fragmentation function, except that the initial scale of evolution (typically denoted as $\mu_0$) is conventionally taken at the scale of the charm quark mass. Consequently, this leads to the resummation of large logarithms of the type $\ln(Q^2/m_Q^2)$. However, this approximation is valid only when $Q^2 \gg m_Q^2$, where power corrections related to the quark mass become negligible.
\end{itemize}
The FFNS and VFNS approaches complement each other and are valid in different kinematic regions. While VFNS is appropriate for large values of $Q^2$, the FFNS is more reliable when the energy scale is close to the quark mass. To bridge the gap between these two schemes, the \textbf{General-Mass Variable-Flavor Number Scheme (GM-VFNS)} is introduced (see Refs.~\cite{Kramer:2000hn,Forte:2010ta,Blumlein:2018jfm,Aivazis:1993pi,Thorne:1997ga} and references therein). This framework combines the accuracy of FFNS in including mass effects with the logarithmic resummation capabilities of VFNS, providing a consistent description across all relevant energy scales. 


\section{Fragmentation functions}
\label{sec:frag_funcs}

To enable our VFNS analysis, we introduce a set of fragmentation functions specifically designed to model the formation of these exotic particles. These functions, presented in~\cite{Celiberto:2024beg}, will be discussed in detail in the present Section. The fragmentation functions consist of two families:
\begin{enumerate}
    \item The \texttt{TQHL1.1} set, which describes the fragmentation of doubly heavy tetraquarks such as $\QXcu$, $\QXcs$, $\QXbu$, and $\QXbs$. These functions are built on an enhanced version of the Suzuki model~\cite{Suzuki:1977km,Suzuki:1985up,Amiri:1986zv,Nejad:2021mmp}, which provides initial energy-scale inputs tailored to the heavy-quark channel.
    \item The \texttt{TQ4Q1.1} set, which models the fragmentation of fully heavy tetraquarks, including $\TQcZpp$, $\TQbZpp$, and their radial excitations, $\TQcTpp$ and $\TQbTpp$. This family leverages nonrelativistic QCD (NRQCD) approaches~\cite{Caswell:1985ui,Thacker:1990bm,Bodwin:1994jh,Cho:1995vh,Cho:1995ce,Leibovich:1996pa,Bodwin:2005hm} to incorporate inputs for both gluon and heavy-quark channels~\cite{Feng:2020riv,Bai:2024ezn}.
\end{enumerate}
These updated \texttt{1.1} sets~\cite{Celiberto:2024beg} supersede their \texttt{1.0} predecessors~\cite{Celiberto:2023rzw,Celiberto:2024mrq,Celiberto:2024mab} by refining normalization procedures and parameter definitions. Given the limited experimental data available for tetraquark fragmentation, these functions provide valuable guidance for studies at both current and next-generation colliders.\\

Our fragmentation functions evolve in energy following the DGLAP equations~\cite{Gribov:1972ri,Gribov:1972rt,Lipatov:1974qm,Altarelli:1977zs,Dokshitzer:1977sg}, ensuring consistency across varying energy scales. The analysis spans a wide range of center-of-mass energies, from the 14~TeV regime accessible at the LHC to the 100~TeV nominal energy anticipated at the FCC. This comprehensive energy range enables robust predictions and facilitates a deeper understanding of exotic matter production mechanisms.\\

All symbolic computations were carried out using the {\symJethad}~\cite{Celiberto:2020wpk,Celiberto:2022rfj,Celiberto:2023fzz,Celiberto:2024mrq,Celiberto:2024swu} plugin for \textsc{Mathematica}, designed to manipulate analytic expressions for hadronic structure and high-energy QCD.\\

For heavy-light hadrons, like $D$ or $B$ mesons or $\Lambda_Q$ baryons, the initial-scale fragmentation input can be envisioned as a two-step process~\cite{Cacciari:1996wr,Cacciari:1993mq,Kniehl:2005mk,Helenius:2018uul,Helenius:2023wkn}. 
In the first step, a parton $i$, produced in a hard scattering event with large transverse momentum $|\vec \kappa| \gg m_Q$, fragments into a heavy quark $Q$ with mass $m_Q$: charm or anticharm for $D$ mesons and $\Lambda_c$ baryons, bottom or antibottom for $B$ mesons and $\Lambda_b$ baryons. 
Given that $\alpha_s(m_Q) < 1$, we can compute this step perturbatively at an initial reference scale of ${\cal O}(m_Q)$. 
Since its time scale is shorter than that of hadronization, this part is often referred to as the short-distance coefficient (SDC) of the $(i \to Q)$ fragmentation process. 
The first next-to-leading-order (NLO) calculation of SDCs for singly heavy hadrons can be found in Ref.~\cite{Mele:1990yq,Mele:1990cw}.
Related studies at next-to-NLO were performed in Refs.~\cite{Rijken:1996vr,Mitov:2006wy,Blumlein:2006rr,Melnikov:2004bm,Mitov:2004du,Biello:2024zti}.\\

The formation of quarkonia and charmed $B$ mesons involves intricate dynamics governed by the NRQCD framework~\cite{Caswell:1985ui,Thacker:1990bm,Bodwin:1994jh,Cho:1995vh,Cho:1995ce,Leibovich:1996pa,Bodwin:2005hm}. 
A proper description of quarkonium state, whose lowest Fock state is composed of a heavy quark $Q$ and its antiquark $\bar{Q}$, requires the separation of a short-distance perturbative process, described by SDCs, from the nonperturbative hadronization, governed by long-distance matrix elements (LDMEs). NRQCD enables this factorization by treating heavy quarks as nonrelativistic degrees of freedom and accounting for both color-singlet~\cite{Berger:1980ni,Baier:1981uk} and color-octet contributions~\cite{Bodwin:1992ye}.
The latter are needed to observe a complete cancellation of singularities arising in NLO calculations of $P$-wave-quarkonium hard factors~\cite{Barbieri:1976fp,Bodwin:1992ye}.\\

Charmed $B$ mesons, such as $\BCs$ and $\Bss$, extend this complexity with their lowest Fock state containing both a charm and a bottom quark.
Charmed $B$ mesons, owing to the presence of two heavy quarks, can be regarded as generalized quarkonium states.
In contrast to charmonia and bottomonia, these mesons cannot annihilate into gluons, rendering them exceptionally stable with remarkably narrow decay widths~\cite{Alonso:2016oyd,Aebischer:2021eio,Aebischer:2021ilm}.
With top quarks being too short-lived to hadronize, charmed $B$ mesons represent the ultimate frontier in the study of meson spectroscopy~\cite{Ortega:2020uvc}.\\

NRQCD permits to investigate quarkonium formation mechanisms across both low and moderate-to-high transverse momentum ($p_T$) ranges. 
At low $p_T$, the primary mechanism involves the \emph{short-distance} creation of a $(Q\bar{Q})$ pair during the hard scattering process, followed by its hadronization into physical quarkonium at larger distances. As $p_T$ increases, the fragmentation of a \emph{single parton} into the observed hadron, accompanied by inclusive radiation, becomes competitive with the short-distance mechanism and eventually predominates.\\

The short-distance production can be interpreted as FFNS two-parton fragmentation process, embodying genuine higher-power corrections (see Refs.~\cite{Fleming:2012wy,Kang:2014tta,Echevarria:2019ynx,Boer:2023zit,Celiberto:2024mex,Celiberto:2024bxu,Celiberto:2024rxa} for more details). In contrast, single-parton fragmentation aligns with a VFNS fragmentation approach, with its energy evolution governed by the DGLAP equations.\\

In the early 1990s, leading-order (LO) calculations were performed to determine the initial-scale inputs for gluon and heavy-quark channels into $S$-wave vector quarkonia within color-singlet configurations~\cite{Braaten:1993rw,Braaten:1993mp}. More recently, next-to-leading order (NLO) analyses have been conducted~\cite{Zheng:2019gnb,Zheng:2021sdo}. Building upon these inputs, a pioneering set of new VFNS, DGLAP-evolving FFs for vector quarkonia, named {\tt ZCW19$^+$}, was developed~\cite{Celiberto:2022dyf,Celiberto:2023fzz}. This work was soon extended to include $\BCs$ and $\Bss$ states in the {\tt ZCFW22} set~\cite{Celiberto:2022keu,Celiberto:2024omj}.\\

A key result was achieved in Ref.~\cite{Celiberto:2024omj}.
In that work, it was pointed out that the behavior of high-energy resummed rapidity and transverse-momentum distributions of these charmed $B$ mesons, whose fragmentation production was described via {\tt ZCFW22} determinations, align with the LHCb Collaboration's observation that the production rate of $\BCs$ mesons relative to singly bottomed $B$ mesons remains below 0.1\%~\cite{LHCb:2014iah,LHCb:2016qpe}.
This served as a simultaneous benchmark for both the hybrid factorization and the NRQCD fragmentation applied to charmed $B$ mesons.\\

The integration of NRQCD with collinear factorization, as designed within the novel \emph{heavy-flavor nonrelativistic evolution} ({\HFNRevo}) method~\cite{Celiberto:2024mex,Celiberto:2024bxu,Celiberto:2024rxa}, enables threshold-consistent DGLAP evolution and enhances our understanding of fragmentation mechanisms. 
These advancements lay the groundwork for extending this framework to exotic states, such as doubly charmed or fully heavy tetraquarks, offering deeper insights into the dynamics of heavy-flavored hadron production.\\

{\HFNRevo} builds upon three core ingredients: \emph{interpretation}, \emph{evolution}, and \emph{uncertainties}.
As mentioned before, the first ingredient facilitates the interpretation of the short-distance mechanism, which dominates at low transverse momentum, as a FFNS two-parton fragmentation process that extends beyond the leading-power approximation.
The second component involves the DGLAP evolution of {\HFNRevo} FFs, which unfolds in two stages.
First, an \emph{expanded} and semi-analytic \emph{decoupled} evolution ({\tt EDevo}) accurately accounts for the evolution thresholds across all parton channels.
Then, a standard \emph{all-order} evolution ({\tt AOevo}) is numerically applied.
Finally, the third pillar addresses the quantification of Missing Higher-Order Uncertainties (MHOUs) stemming from scale variations associated with evolution thresholds.

\subsection{Doubly Heavy Tetraquarks}
\label{ssec:Doubly Heavy Tetraquarks}

Our modeling of the initial-scale inputs for the fragmentation of heavy quarks into color-singlet $S$-wave $\QXQq$ tetraquarks is based on a method inspired by the Suzuki framework~\cite{Suzuki:1977km,Suzuki:1985up,Amiri:1986zv}, as already mentioned. This approach, rooted in spin physics, incorporates transverse-momentum dependence while retrieving the collinear limit by neglecting the relative motion of the constituent quarks within the bound state. 

Initial inputs for the $[c \to \QXcu]$ and $[b \to \QXbs]$ fragmentation channels were first developed and encoded in the {\tt TQHL1.0} set, with subsequent extensions covering $[c \to \QXcs]$ and $[b \to \QXbu]$ channels. These fragmentation functions serve as a foundation for describing the hadronization of doubly heavy tetraquarks.
\begin{figure*}[!t]
\centering
\includegraphics[width=0.6\textwidth]{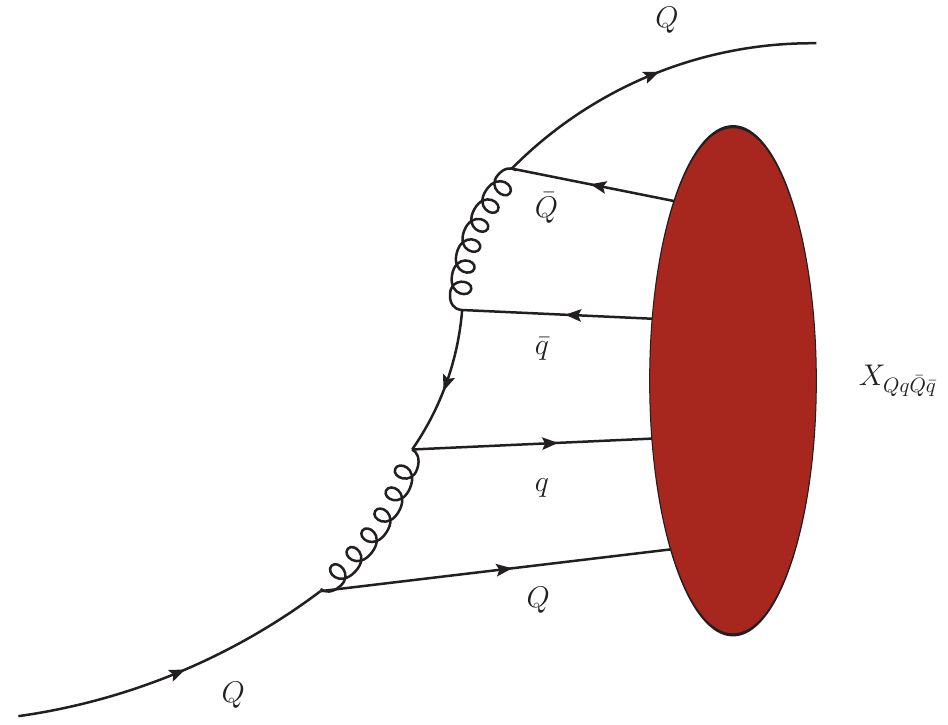}
\caption{A leading-order representative diagram illustrating the collinear fragmentation of a constituent heavy quark into a color-singlet $S$-wave doubly heavy tetraquark. The firebrick-colored blob denotes the nonperturbative hadronization component of the associated fragmentation function.}
\label{fig:XQq_FF_diagrams}
\end{figure*}
The methodology parallels the factorization structure of nonrelativistic QCD (NRQCD). In this framework, the $(Q\bar{Q})$ pair is initially produced perturbatively, with the hadronization process governed by long-distance matrix elements (LDMEs). Similarly, for tetraquarks, the $(Q\bar{Q}q\bar{q})$ system is generated perturbatively through the splitting of a heavy quark above threshold. The production amplitude is then combined with a bound-state wave function, which encapsulates the non-perturbative dynamics of tetraquark hadronization, following the guidelines established by Suzuki.\\

Assuming full symmetry between the $Q$ and $\bar{Q}$ fragmentation channels, the initial-scale fragmentation function $[Q \to \QXQq]$ in the {\tt TQHL1.1} parametrization depends on several components, which we define below. The normalization factor ${\cal N}_{X}^{(Q)}$ is given by (Ref.~\cite{Nejad:2021mmp}):
\begin{equation}
 \label{XQq_FF_initial-scale_Q_N}
 {\cal N}_{X}^{(Q)} \, = \, \left\{ 128 \, \pi^2 \, f_{\cal B} \, C_F \big[ \alpha_s(2\mu_X + m_Q) \big]^2 \right\}^2
 \,,
\end{equation}
where $\mu_X = m_Q + m_q$, $f_{\cal B} = 0.25$~GeV is the hadron decay constant~\cite{ParticleDataGroup:2020ssz}, $C_F = (N_c^2-1)/(2N_c)$ is the Casimir constant for gluon emission from a quark, $m_Q$ is the heavy quark mass ($m_Q = m_c = 1.5$~GeV or $m_Q = m_b = 4.9$~GeV), and $m_q$ is the light quark mass ($m_q = m_{u,d,s}$).

The function $\Xi^{(Q,q)}(z)$, which incorporates the relevant kinematic factors, is defined as~\cite{Nejad:2021mmp}:
\begin{equation}
 \label{XQq_FF_initial-scale_Q_Xi}
 \Xi^{(Q,q)}(z) \,=\,
 \frac{{\cal F}_X^{(Q,q)}(z)}
 {\mu_X^2 \, m_Q^5 \, m_q}
 \;,
\end{equation}
where ${\cal F}_X^{(Q,q)}(z)$ represents the production dynamics and is expressed as:
\begin{equation}
\begin{split}
 \label{XQq_FF_initial-scale_Q_F}
 {\cal F}_X^{(Q,q)}(z) \,=\, & \left\{ [ (2-z)\,m_Q + 2\,m_q ]^2 + z^2 \vqTTa \right\} \\[0.10cm]
 &\times \left\{ [ (2-z)\,m_Q + (1-z)\,m_q ]^2 + z^2 \vqTTa \right\}^{2} \\[0.15cm]
 &\times \left\{ 4 (1-z) \mu_X^2 + z^2 (m_Q^2 + \vqTTa) \right\}
 \,.
\end{split}
\end{equation}

Using these definitions, the explicit form of the $[Q \to \QXQq]$ fragmentation function at the initial scale $\mu_{F,0}$ is expressed as~\cite{Nejad:2021mmp}:
\begin{equation}
 \label{XQq_FF_initial-scale_Q}
 D^{\QXQq}_Q(z,\mu_{F,0}) \,=\,
 {\cal N}_{X}^{(Q)} \,
 \frac{(1 - z)^5}{\left[\Xi^{(Q,q)}(z)\right]^2}\sum\limits_{l=0}^3 \, z^{2(l+2)} \rho_l^{(Q,q)}(z) \left\{ \frac{\vqTTa}{m_Q^2} \right\}^{\!l}
 \,.
\end{equation}
This expression incorporates the combined dynamics of the perturbative and non-perturbative processes, fully describing the fragmentation of a heavy quark into a doubly heavy tetraquark.
The coefficients $\rho_l^{(Q,q)}(z)$ in Eq.~\eqref{XQq_FF_initial-scale_Q} are expressed as follows~\cite{Nejad:2021mmp}:
\begin{equation}
\begin{split}
 \label{XQq_FF_initial-scale_Q_rho_0}
 \rho_0^{(Q,q)}(z) &\,=\,
 32\, {\cal R}_{q/Q}^6\, (2 - 3 z + z^3) 
 \,-\, 64\, {\cal R}_{q/Q}^5\, (6 - 10 z + 3 z^2 + z^3) \\ 
 &\,+\, 8\, {\cal R}_{q/Q}^4\, (-120 + 236 z - 158 z^2
 \,+\, 42 z^3 - 6 z^4 + z^5) \\ 
 &\,+\, 16\, {\cal R}_{q/Q}^3\, (-80 + 192 z - 196 z^2 
 \,+\, 98 z^3 - 24 z^4 + 3 z^5) \\ 
 &\,-\, 2\, {\cal R}_{q/Q}^2\, (480 - 1424 z + 1872 z^2  
 \,-\, 1232 z^3 + 422 z^4 - 71 z^5 + 6 z^6) \\
 &\,-\, 4\, {\cal R}_{q/Q}\, (96 - 352 z + 544 z^2  
 \,-\, 424 z^3 + 182 z^4 - 40 z^5 + 3 z^6) \\
 &\,-\, (4 - 2 z + z^2) (4 - 8 z + 3 z^2)^2 \;,
\end{split}
\end{equation}
\begin{equation}
\begin{split}
 \label{XQq_FF_initial-scale_Q_rho_1}
 \rho_1^{(Q,q)}(z) \,=\, 
 &-8\, {\cal R}_{q/Q}^4\, (-6 + 6 z - 6 z^2 + z^3)  
 \,-\, 
 16\, {\cal R}_{q/Q}^3\, (-12 + 14 z - 12 z^2 + 3 z^3) \\
 &\,+\, 
 4\, {\cal R}_{q/Q}^2\, (72 - 112 z + 102 z^2 - 37 z^3 + 6 z^4)  \\
 &\,+\, 
 8\, {\cal R}_{q/Q}\, (24 - 52 z + 54 z^2 - 22 z^3 + 3 z^4) \\ 
 &\,+\, 48 - 144 z + 168 z^2 - 84 z^3 + 19 z^4 \;,
\end{split}
\end{equation}
\begin{equation}
\begin{split}
 \label{XQq_FF_initial-scale_Q_rho_2}
 \rho_2^{(Q,q)}(z) &\,=\, 
 6\, {\cal R}_{q/Q}^2\, (2 - z + 2 z^2)  
 \,+\, 4\, {\cal R}_{q/Q}\, (6 - 4 z + 3 z^2)\,+\, 12 - 18 z + 11 z^2 \;,
\end{split}
\end{equation}
and
\begin{equation}
\begin{split}
 \label{XQq_FF_initial-scale_Q_rho_3}
 \rho_3^{(Q,q)}(z) \,=\, 1 \;,
\end{split}
\end{equation}
with ${\cal R}_{q/Q} \equiv m_q/m_Q$\,. The value of the parameter $\vqTTa$ is determined through a numerical scan, based on its correlation with the peak position of the fragmentation functions. For doubly heavy tetraquarks ($\QXQq$), the parameter $\vqTTa_{\QXQq}$ is fixed at $4~\text{GeV}^2$, chosen according to the relation:
\begin{equation}
 \sqrt{\vqTTa_{\QXQq}} \simeq \frac{\sqrt{\vqTTa_{\TQQ}}}{2} \;.
\end{equation}
Here, $\TQQ$ represents fully heavy tetraquarks, such as $|c\bar{c}c\bar{c}\rangle$ or $|b\bar{b}b\bar{b}\rangle$, which are composed exclusively of heavy quarks and antiquarks. Their fragmentation dynamics are dominated by the interactions among the four heavy quarks in the absence of light-quark effects. In contrast, doubly heavy tetraquarks ($\QXQq$) involve both a heavy-quark pair ($Q\bar{Q}$) and a light-quark pair ($q\bar{q}$) in their lowest Fock state. The presence of light quarks in $\QXQq$ states reduces the binding effects compared to $\TQQ$ states, which justifies the scaling relation between $\vqTTa_{\QXQq}$ and $\vqTTa_{\TQQ}$. This choice ensures consistency in modeling the fragmentation functions and their peak positions for these two classes of tetraquarks.

\subsubsection{Initial Energy-Scale Inputs}
\label{sssec:FFs_XQq_Q}

In this section, we analyze the $z$-dependence of the initial-scale inputs for the $[Q \to \QXQq]$ fragmentation functions within the {\tt TQHL1.1} framework. To estimate the uncertainties around the lowest energy value, we follow a methodology similar to that used in our previous work on $\TQc$ states~\cite{Celiberto:2024mab}. In that study, the initial-scale FFs for $[g \to \TQc]$ were benchmarked by employing a simplified DGLAP evolution, considering only the gluon-to-gluon splitting kernel, $P_{gg}$.  For the $[Q \to \QXQq]$ FFs, we adopt a comparable approach, using a simplified DGLAP evolution where only the quark-to-quark splitting kernel, $P_{qq}$, is active. The factorization scale, $\mu_{F,0}$, is chosen as $3 m_Q + 2 m_q$, with variations tested in the range from $\mu_{F,0}/2$ to $2\mu_{F,0}$ to account for theoretical uncertainties.

\begin{figure*}[t]
\begin{subfigure}{0.35\textwidth}
    \centering
    \includegraphics[scale=0.42,clip]{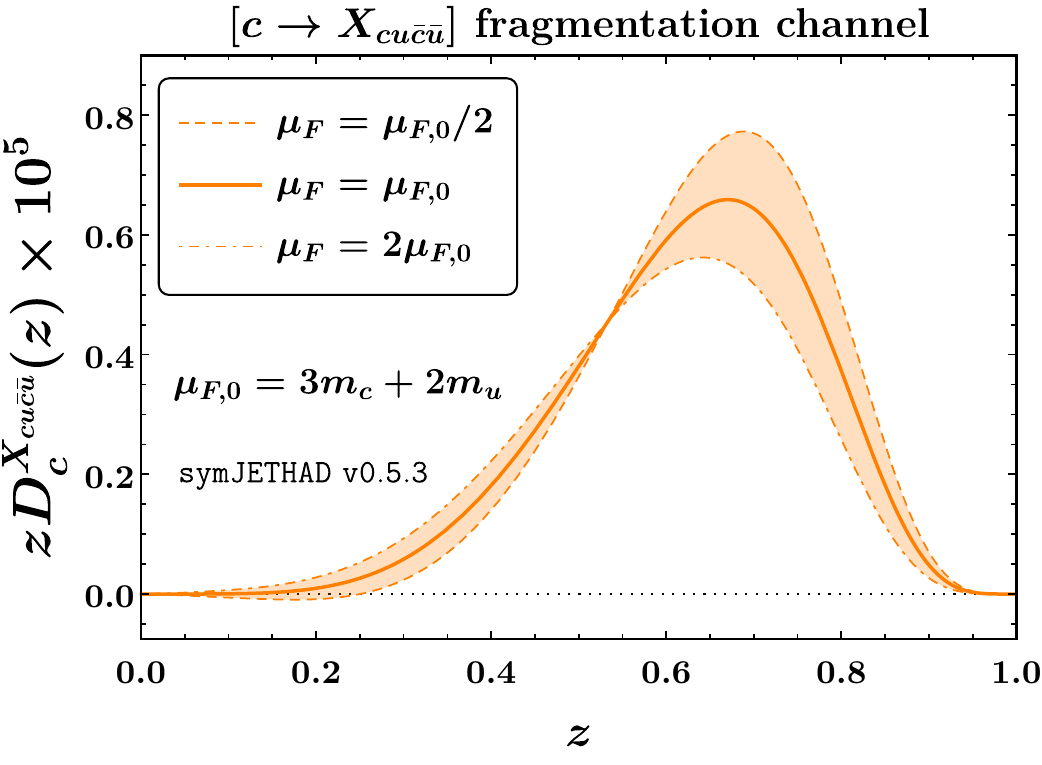}
\end{subfigure}
\hspace{0.5cm}
\begin{subfigure}{0.49\textwidth}
   \centering
    \includegraphics[scale=0.42,clip]{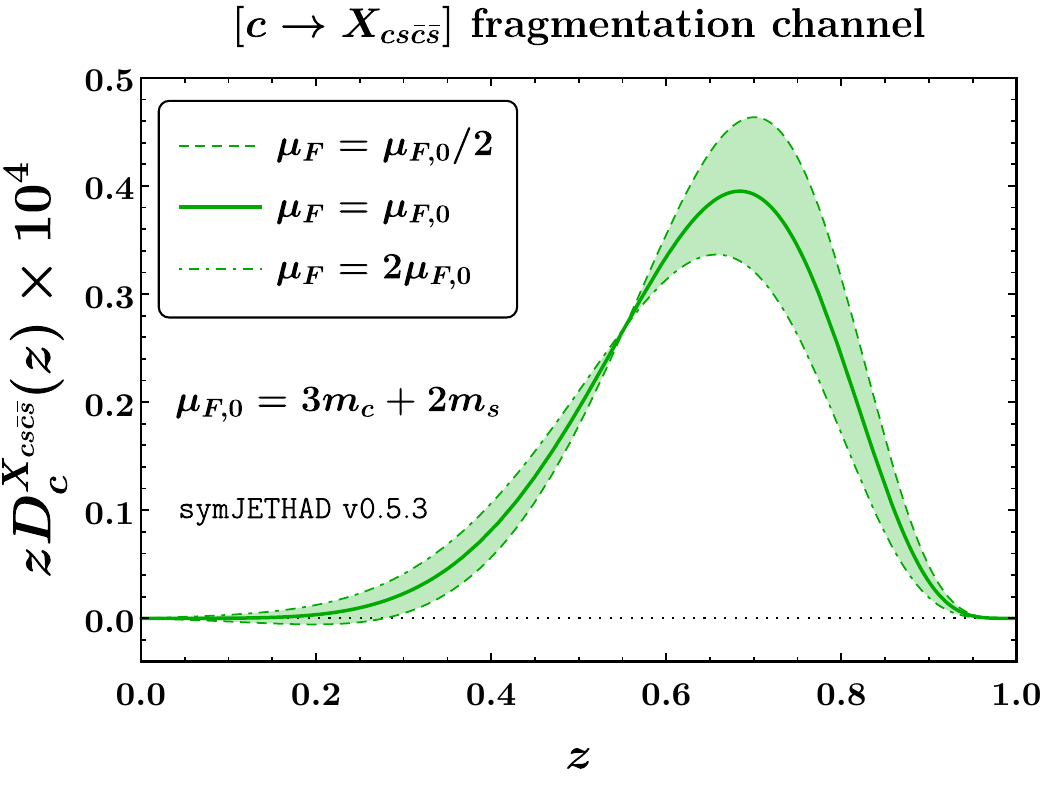}
\end{subfigure}
\vspace{0.25cm}
\begin{subfigure}{0.35\textwidth}
    \centering
    \includegraphics[scale=0.42,clip]{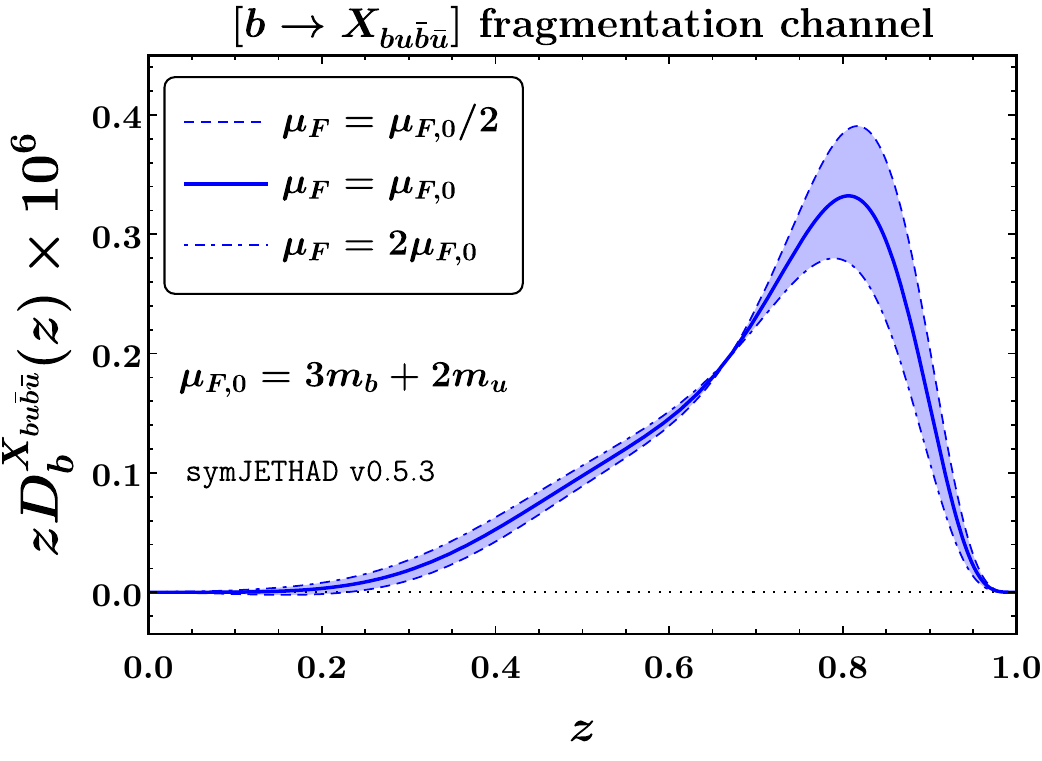}
\end{subfigure}
\hspace{0.5cm}
\begin{subfigure}{0.7\textwidth}
    \centering
    \includegraphics[scale=0.43,clip]{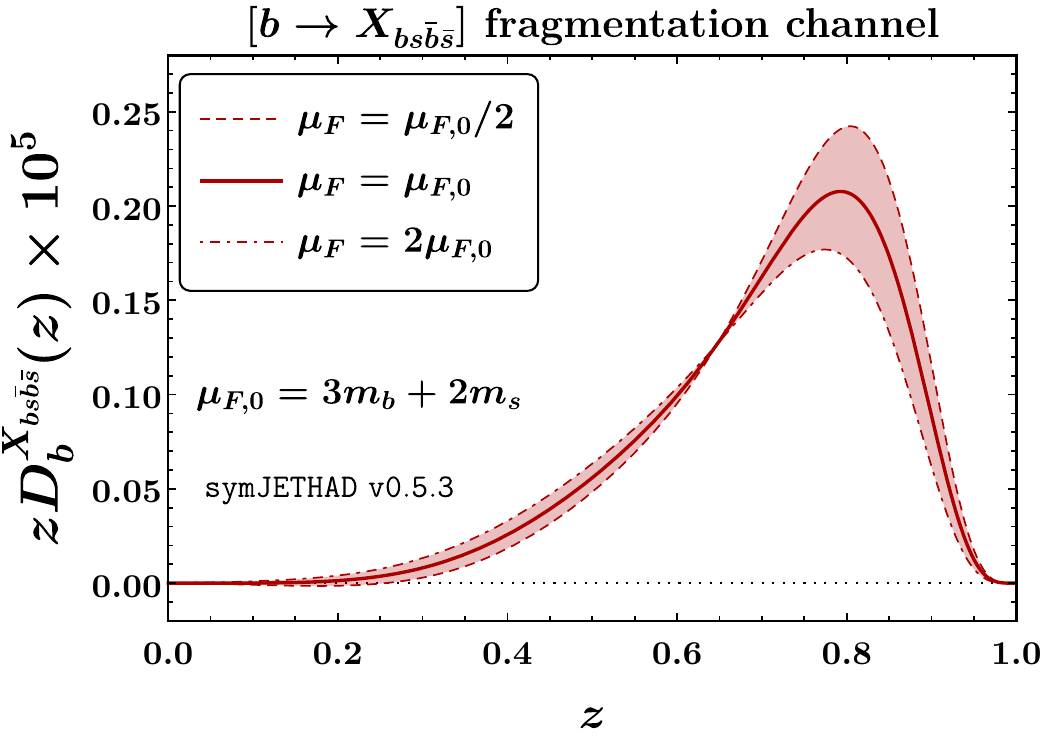}
\end{subfigure}
\caption{Collinear fragmentation of a constituent heavy quark into doubly charmed (top) and doubly bottomed (bottom) tetraquarks. The left and right panels correspond to the initial-scale inputs for the $[Q \to \QXQu]$ and $[Q \to \QXQs]$ channels, respectively. For illustrative purposes, an extended DGLAP evolution is performed over the range $(3m_Q+2m_q)/2$ to $2(3m_Q+2m_q)$.}
\label{fig:XQq_FF_initial-scale_Q}
\end{figure*}

The $z$-dependence of these FFs, shown in Fig.~\ref{fig:XQq_FF_initial-scale_Q}, reveals a distinct peak in the range $0.65 < z < 0.85$, while the FFs approach zero as $z \to 0$ or $z \to 1$. This behavior aligns with theoretical expectations for doubly heavy tetraquarks. 

Additionally, it is observed that the FFs for doubly bottomed tetraquarks are approximately five times larger than those for doubly charmed tetraquarks, with peaks occurring at higher $z$ values. This difference is linked to the ratio ${\cal R}_{q/Q} = m_q / m_Q$, which influences the polynomial coefficients $\rho_l^{(Q,q)}(z)$ and modulates the FFs based on the mass of the fragmenting heavy quark. These results provide a solid basis for modeling the fragmentation dynamics of doubly heavy tetraquarks.

\subsubsection{The {\tt TQHL1.1} functions}
\label{sssec:FFs_TQHL11}
The final step in constructing the {\tt TQHL1.1} fragmentation functions (FFs) for doubly heavy tetraquarks is to perform a consistent DGLAP evolution starting from the initial-scale inputs $[Q \to \QXQq]$ defined in Section~\ref{sssec:FFs_XQq_Q}. The minimum invariant mass for the fragmentation process $[Q \to \QXQq]$ is set as $\mu_{F,0} = 3 m_Q + 2 m_q$, which acts as the threshold for $Q$-quark fragmentation. This value is also adopted for the $\bar{Q}$-antiquark channel.

Unlike the standard two-step DGLAP evolution, which begins with an expanded decoupled evolution ({\tt EDevo}), we directly proceed with the all-order evolution ({\tt AOevo}) since our model includes only the $[Q \to \QXQq]$ channel at the initial scale. Starting from the \textit{evolution-ready} scale $Q_0 \equiv \mu_{F,0}$, the {\tt TQHL1.1} FFs are constructed via numerical DGLAP evolution and made available in the LHAPDF format.
To perform this step, we make use of {\tt APFEL++}~\cite{Bertone:2013vaa,Carrazza:2014gfa,Bertone:2017gds}. In future, we plan to link our code also to {\tt EKO}~\cite{Candido:2022tld,Hekhorn:2023gul}.

Our methodology does not include initial-scale contributions from light partons or nonconstituent heavy quarks, as these are expected to emerge only at scales $\mu_F > Q_0$. Previous studies~\cite{Nejad:2021mmp} suggest that such contributions are negligible at the initial scale. While gluon channel inputs can influence the evolution of other parton channels, these effects are less critical for exploratory studies focused on the dominant $[Q \to \QXQq]$ channel. However, further research is needed to model additional parton fragmentation inputs.

\begin{figure*}[!t]
\hspace*{-0.4cm} 
\begin{subfigure}{0.38\textwidth}
    \includegraphics[scale=0.38,clip]{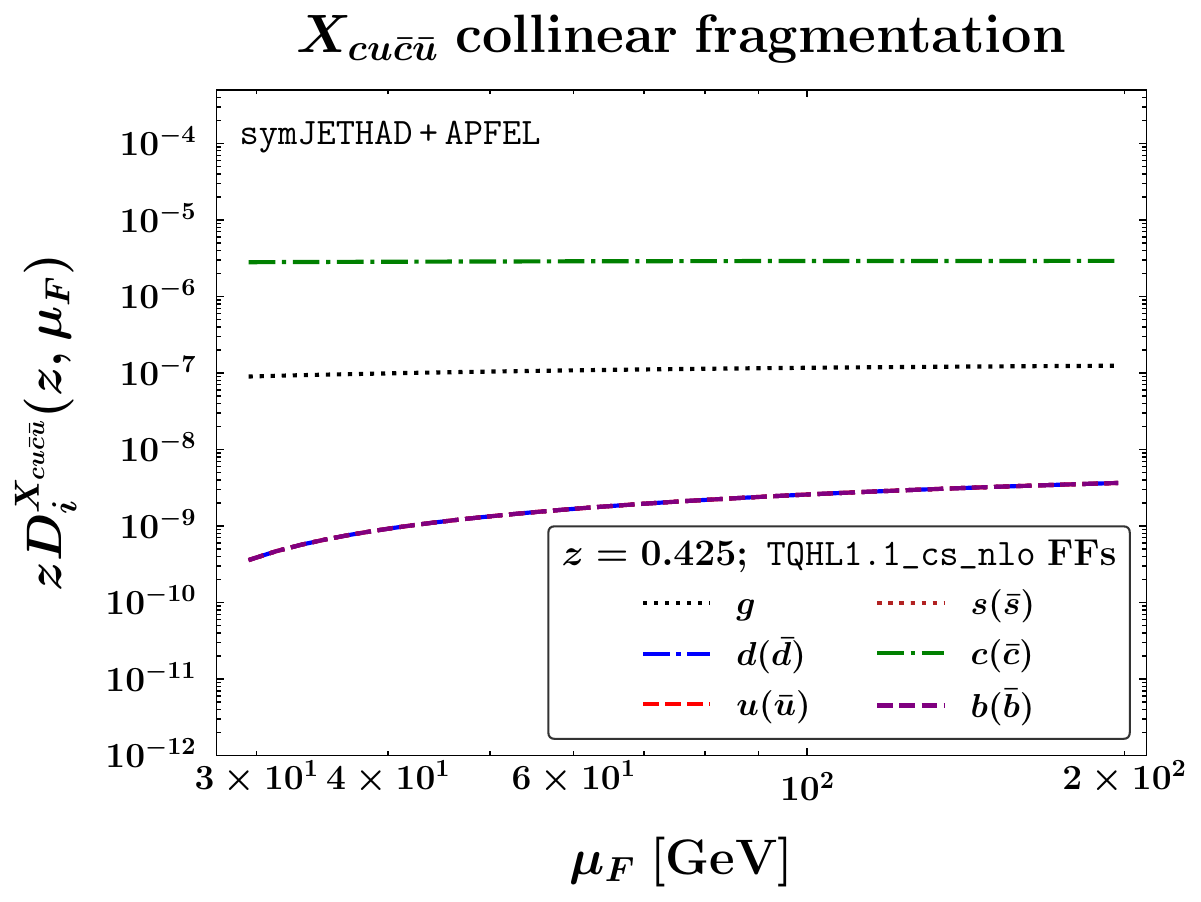}
\end{subfigure}
\hspace{1.8cm} 
\begin{subfigure}{0.38\textwidth}
    \includegraphics[scale=0.38,clip]{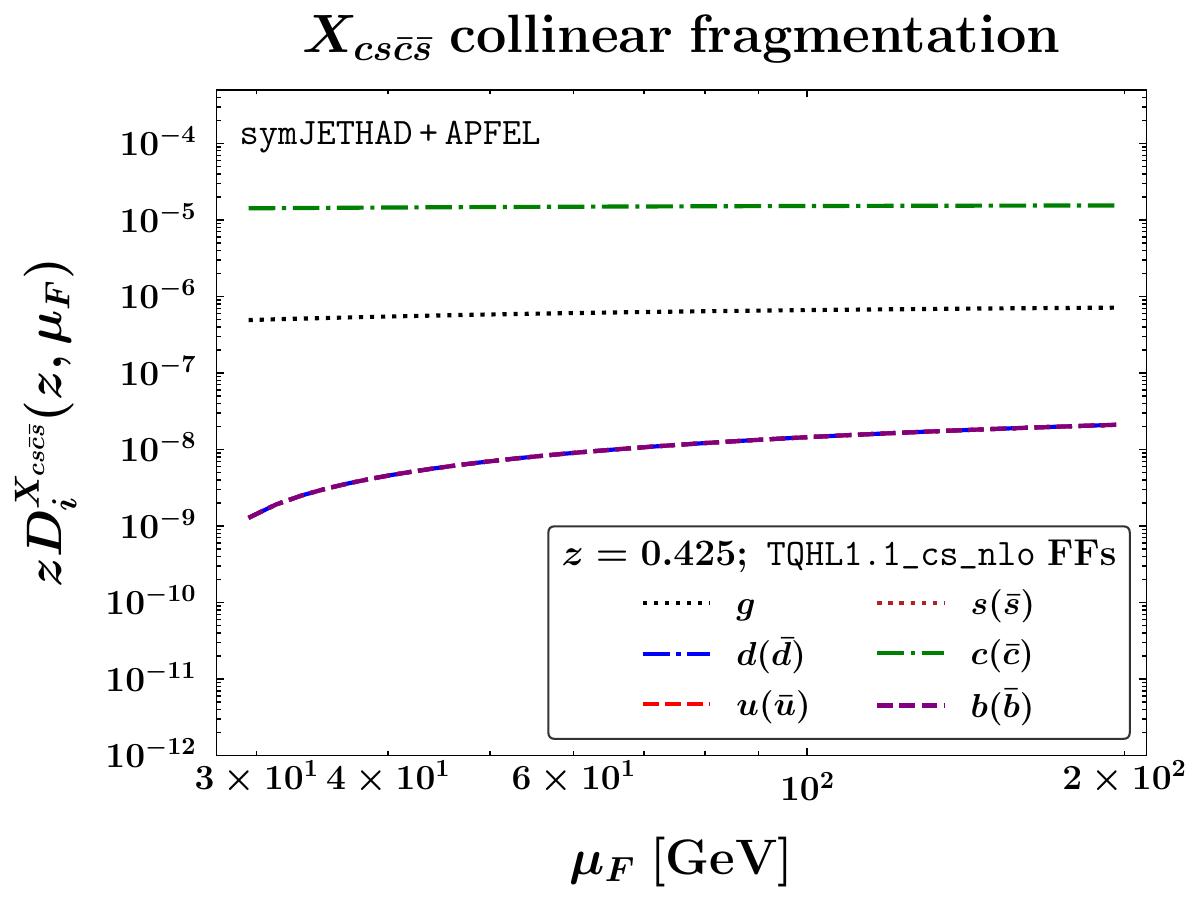}
\end{subfigure}
\caption{Dependence of {\tt TQHL1.1} collinear FFs on the factorization scale, illustrating the formation of $\QXcu$ (left) and $\QXcs$ (right) at $z = 0.425 \simeq \langle z \rangle$.}
\label{fig:NLO_FFs_Xcq}
\end{figure*}

\begin{figure*}[!t]
\hspace*{-0.4cm} 
\begin{subfigure}{0.38\textwidth}
    \includegraphics[scale=0.38,clip]{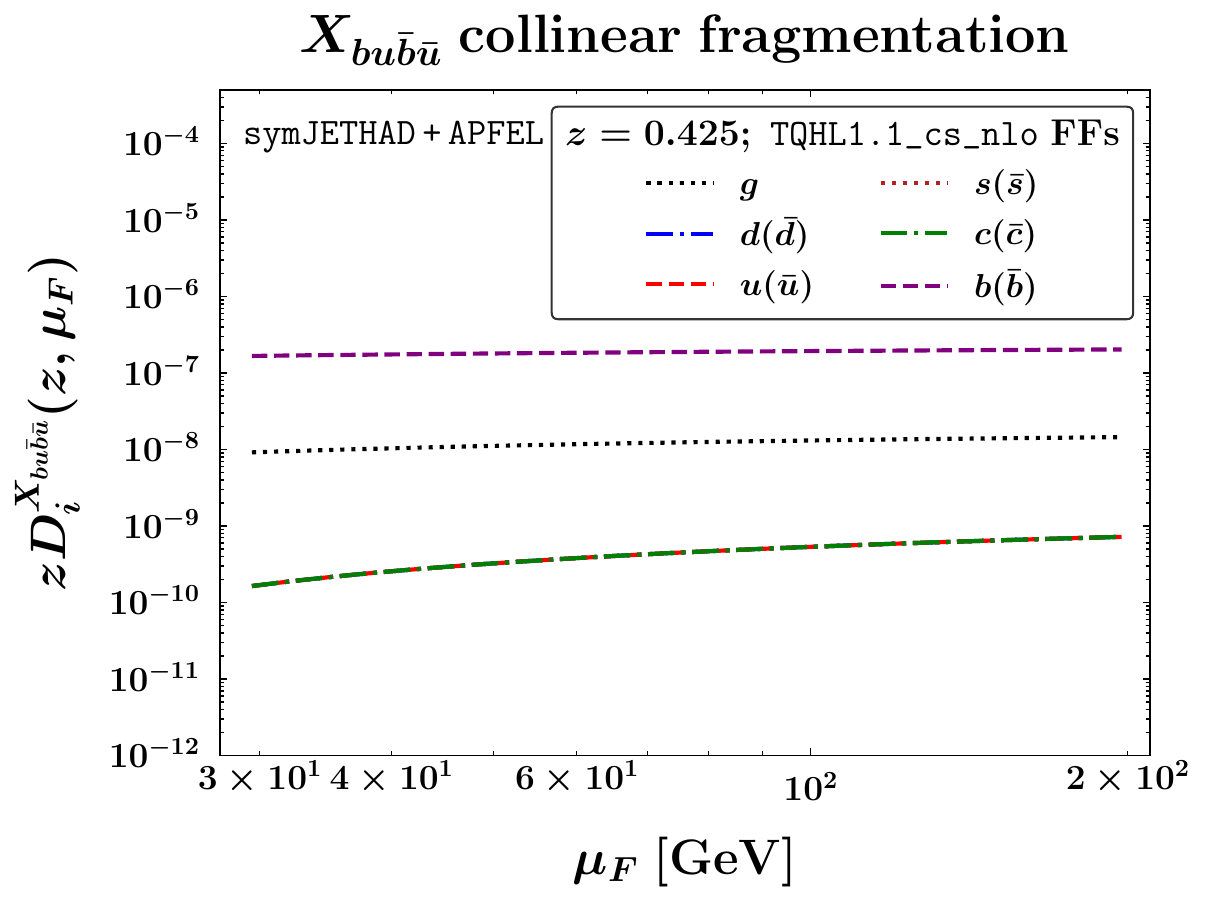}
\end{subfigure}
\hspace{1.8cm} 
\begin{subfigure}{0.38\textwidth}
    \includegraphics[scale=0.38,clip]{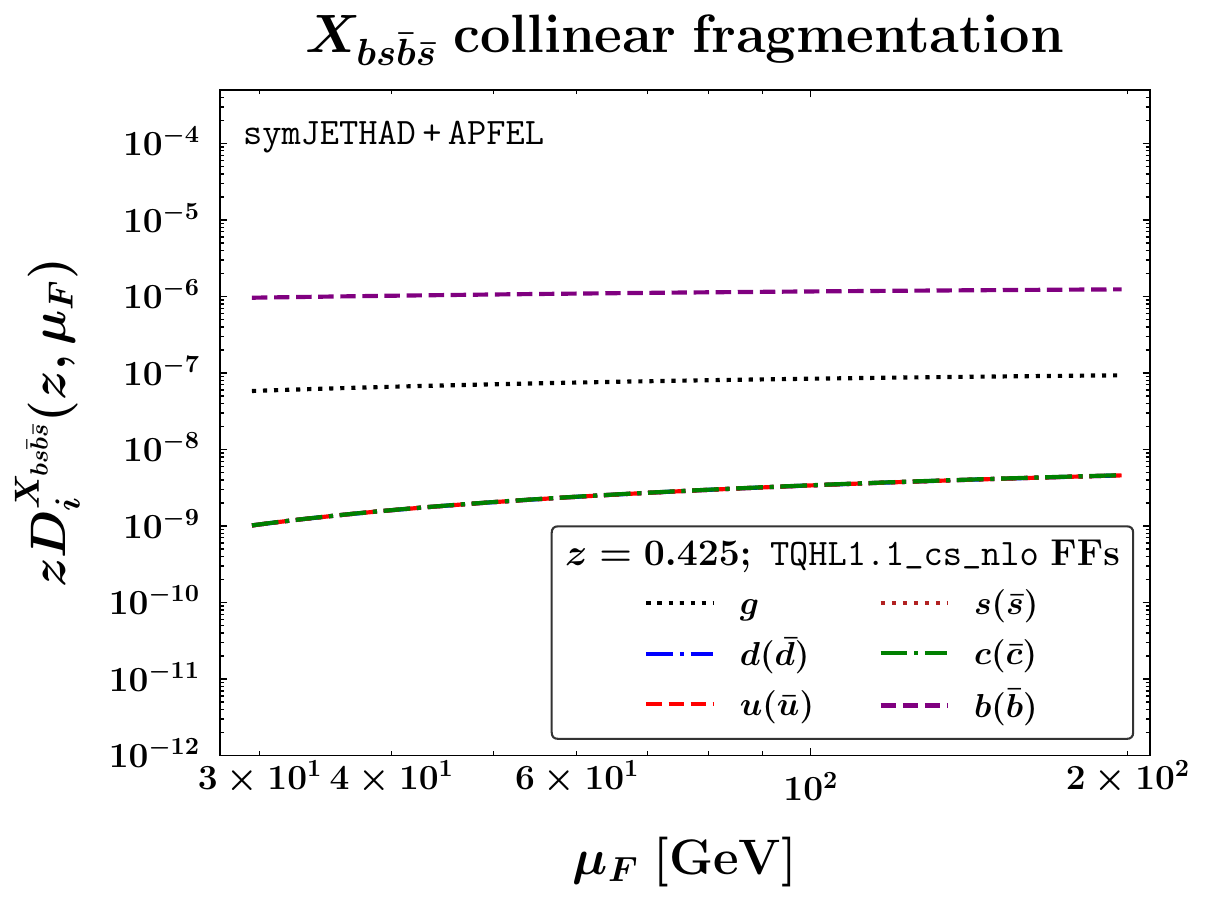}
\end{subfigure}
\caption{The factorization-scale dependence of the {\tt TQHL1.1} collinear fragmentation functions is shown for the formation of $\QXbu$ (left) and $\QXbs$ (right) at $z = 0.425 \simeq \langle z \rangle$.}
\label{fig:NLO_FFs_Xbq}
\end{figure*}

Figs.~\ref{fig:NLO_FFs_Xcq} and~\ref{fig:NLO_FFs_Xbq} display the $\mu_F$-dependence of the {\tt TQHL1.1} FFs for doubly charmed and doubly bottomed tetraquarks, respectively, at a representative momentum fraction $z = 0.425 \simeq \langle z \rangle$.
It roughly corresponds to the average value at which FFs are typically probed in semi-hard final states (see, \emph{e.g.}, Refs.~\cite{Celiberto:2016hae,Celiberto:2017ptm,Celiberto:2020wpk,Celiberto:2021dzy,Celiberto:2021fdp,Celiberto:2022dyf,Celiberto:2022keu,Celiberto:2022kxx,Celiberto:2024omj}).

The analysis reveals that the $[Q \to \QXQq]$ fragmentation channel dominates over all other parton channels, remaining consistently higher than the gluon channel across the energy range examined. The gradual increase of $[g \to \QXQq]$ FFs with $\mu_F$, as observed for other heavy-flavor systems, stabilizes high-energy resummed distributions, which will be explored in our phenomenological study (see Section~\ref{sec:results}).


\subsection{Fully heavy tetraquarks}
\label{ssec:FFs_T4Q}
NRQCD factorization provides a reliable framework for understanding double $\Jpsi$ resonances, interpreting them as fully charmed tetraquark states ($\TQc$). The production of $\TQc$ involves the emission of two charm and two anticharm quarks within a region of approximately $1/m_c$. This process is effectively described as a two-step convolution, separating the short-distance dynamics from the long-distance hadronization. A key calculation for the NRQCD-based input of the $[g \to \TQc]$ color-singlet $S$-wave fragmentation channel was presented in Ref.~\cite{Feng:2020riv}. For doubly heavy tetraquarks ($\QXQq$), the Suzuki model calculation was adapted in Ref.~\cite{Celiberto:2024mab} to describe the $[c \to \TQc]$ fragmentation input. In this work, we enhance the modeling of heavy-quark fragmentation using a more recent NRQCD-based computation from Ref.~\cite{Bai:2024ezn}.\\
\begin{figure*}[!t]
\centering
\includegraphics[width=0.475\textwidth]{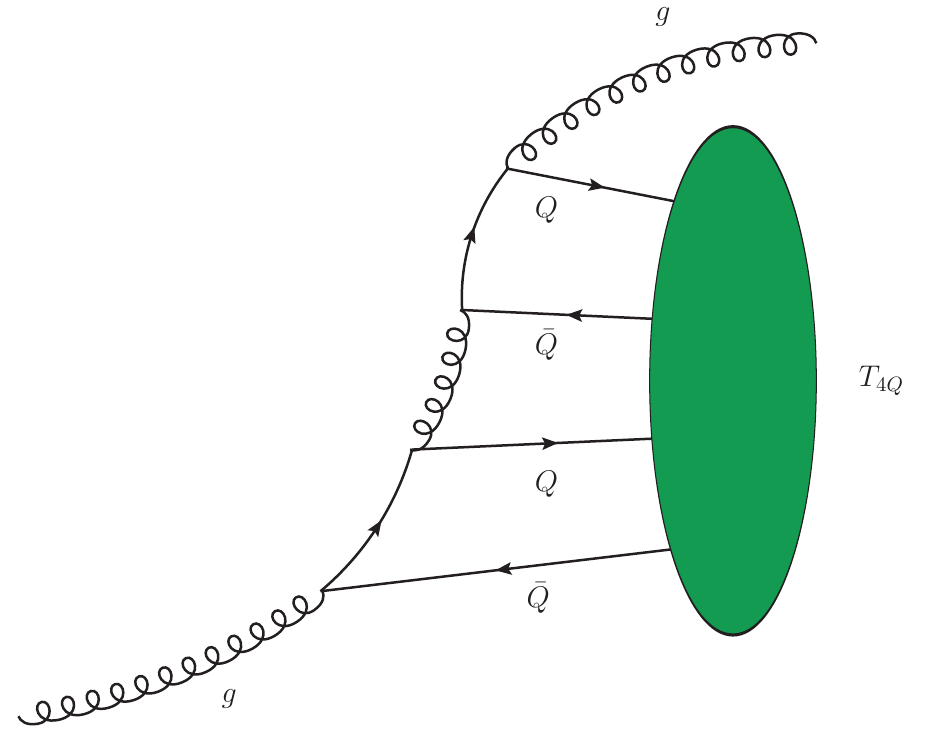}
\hspace{0.40cm}
\includegraphics[width=0.475\textwidth]{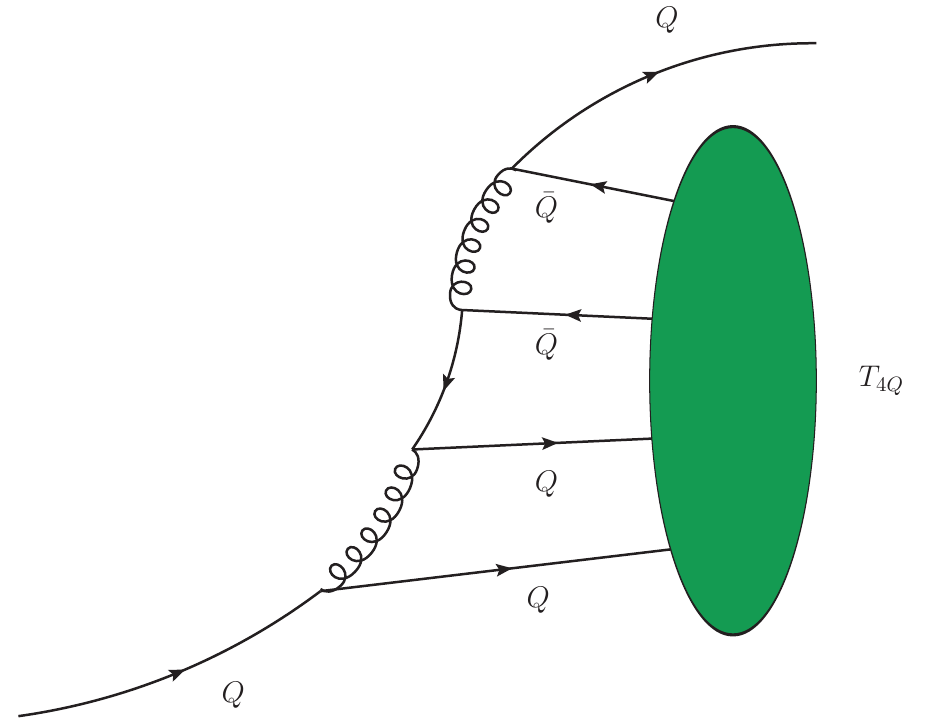}
\caption{LO representative diagrams for the collinear fragmentation process are depicted, illustrating the transformation of a gluon (left) or a constituent heavy quark (right) into a color-singlet $S$-wave fully heavy tetraquark. The green regions in the diagrams symbolize the non-perturbative hadronization contributions associated with the corresponding FFs.}
\label{fig:TQQ_FF_diagrams}
\end{figure*}
For fully heavy tetraquarks $\TQQ(J^{PC})$ with quantum numbers $J^{PC} = 0^{++}$ or $2^{++}$, the initial-scale input for the collinear fragmentation of a parton $i$ into the observed hadron is given by:
\begin{equation}
 \label{TQQ_FF_initial-scale}
 D^{\TQQ(J^{PC})}_i(z,\mu_{F,0}) \, = \,
 \frac{1}{m_Q^9}
 \sum_{[n]} 
 {\cal D}^{(J^{PC})}_i(z,[n])
 \,\langle {\cal O}^{\TQQ(J^{PC})}([n]) \rangle
 \;,
\end{equation}
where $m_Q$ is the quark mass ($m_c = 1.5$~GeV or $m_b = 4.9$~GeV). The term ${\cal D}^{(J^{PC})}_i(z,[n])$ represents the short distance coefficients describing the perturbative fragmentation process $[i \to (Q\bar{Q}Q\bar{Q})]$, while $\langle {\cal O}^{\TQQ(J^{PC})}([n]) \rangle$ corresponds to the LDMEs that describe the non-perturbative hadronization. The composite quantum number $[n]$ spans configurations $[3,3]$, $[6,6]$, $[3,6]$, and $[6,3]$.

To describe the color-singlet tetraquark state, we use the color diquark--anti\-diquark basis, which allows the state to be expressed as either a ($\bar{3} \otimes 3$) or a ($6 \otimes \bar{6}$) configuration. For $S$-wave states:
\begin{itemize}
    \item The ($\bar{3} \otimes 3$) configuration can have spins 0, 1, or 2.
    \item The ($6 \otimes \bar{6}$) configuration is limited to spin 0, in compliance with Fermi-Dirac statistics.
\end{itemize}
Symmetry relations for the coefficients and LDMEs include:
\begin{equation}
\begin{split}
 \label{TQQ_FF_initial-scale_symmetry}
 &{\cal D}^{(J^{PC})}_i(z,[3,6]) \, = \, {\cal D}^{(J^{PC})}_i(z,[6,3]) \,,
 \\
 &\langle {\cal O}^{\TQc(J^{PC})}([3,6]) \rangle \, = \, \langle {\cal O}^{\TQQ(J^{PC})}([6,3]) \rangle^*
 \;.
\end{split}
\end{equation}

\subsubsection{Short-distance coefficients}
\label{sssec:FFs_T4Q_SDCs}
The short distance coefficients describing the perturbative contribution to the fragmentation of a gluon into a $\TQQZpp$ state at the initial energy scale are provided in detail in Refs.~\cite{Feng:2020riv} and~\cite{Bai:2024ezn}:
\begin{equation}
\begin{split}
 \label{Dg_FF_SDC_0pp_33}
 {\cal D}&^{(0^{++})}_g(z,[3,3]) \,=\, 
 \frac{\pi^2 \as^4(4m_Q)}{497664 \, d^{\cal D}_g(z)} 
  \left[186624-430272 z+511072 z^2-425814 z^3\right. \\
 & +\, 217337 z^4-61915 z^5+7466 z^6 
 +\, 42(1-z)(2-z)(3-z) \\
 & \times\, (-144+634 z - 385 z^2+70 z^3 
  \ln (1-z)+36(2-z)(3-z) \\
 & \times\, (144-634 z+749 z^2-364 z^3 + 74 z^4) 
 \ln \left(1-\frac{z}{2}\right)+12(2-z)(3-z) \\
 & \times\, \left(72-362 z+361 z^2-136 z^3+23 z^4\right) \left. \ln \left(1-\frac{z}{3}\right)\right]
 \;,
\end{split}
\end{equation}
\begin{equation}
\begin{split}
 \label{Dg_FF_SDC_0pp_66}
 {\cal D}&^{(0^{++})}_g(z,[6,6]) \,=\,  
 \frac{\pi^2 \as^4(4m_Q)}{331776 \, d^{\cal D}_g(z)} 
 \left[186624-430272 z+617824 z^2-634902 z^3\right. \\
 & +\, 374489 z^4-115387 z^5+14378 z^6 -\, 6(1-z)(2-z)(3-z) \\
 & \times\, (-144-2166 z + 1015 z^2+70 z^3) 
 \ln (1-z)-156(2-z)(3-z) \\
 & \times\, (144-1242 z+1693 z^2-876 z^3 + 170 z^4) 
 \ln \left(1-\frac{z}{2}\right)+300(2-z)(3-z) \\
 & \times\, \left(72-714 z+953 z^2-472 z^3+87 z^4\right) 
 \left.\ln \left(1-\frac{z}{3}\right)\right]
 \;,
\end{split}
\end{equation}
\begin{equation}
\begin{split}
 \label{Dg_FF_SDC_0pp_36}
 {\cal D}&^{(0^{++})}_g(z,[3,6]) \,=\,  
 \frac{\pi^2 \as^4(4m_Q)}{165888 \, d^{\cal D}_g(z)} 
 \left[186624-430272 z+490720 z^2-394422 z^3\right. \\
 & +\, 199529 z^4-57547 z^5+7082 z^6 
  +\, 6(1-z)(2-z)(3-z) \\
 & \times\, (-432+3302 z - 1855 z^2+210 z^3) 
 \ln (1-z)-12(2-z)(3-z) \\
 & \times\, (720-2258 z+2329 z^2-1052 z^3 + 226 z^4) 
 \ln \left(1-\frac{z}{2}\right)+12(2-z)(3-z) \\
 & \times\, \left(936-4882 z+4989 z^2-1936 z^3+331 z^4\right) 
  \left.\ln \left(1-\frac{z}{3}\right)\right]
  \;,
\end{split}
\end{equation}
with $d^{\cal D}_g(z) = z (2-z)^2 (3-z)$. For the gluon-to-$\TQQTpp$ initial-scale perturbative fragmentation, only the $[3,3]$ term contributes. This is because the NRQCD operator for the $6 \otimes \bar{6}$ state is incompatible with Fermi--Dirac statistics for a diquark-antidiquark system in the $2^{++}$ configuration. As a result, both the $[6,6]$ term and the $[3,6]$ interference term are null. We express this as~\cite{Feng:2020riv,Bai:2024ezn}:
\begin{equation}
\begin{split}
 \label{Dg_FF_SDC_2pp_33}
\hspace{-0.00cm}
 {\cal D}^{(2^{++})}_g(z,[3,3]) \,&=\, 
 \frac{\pi^2 \as^4(4m_Q)}{622080 \, z \, d^{\cal D}_g(z)} 
 \left[\left(46656-490536 z \right.\right. 
 +\, 1162552 z^2-1156308 z^3 \\
 & \left.+\, 595421 z^4-170578 z^5+21212 z^6\right) 2z 
  +\, 3(1-z)(2-z)(3-z)\\
  & \times\,(-20304-31788 z)(1296+1044 z + 73036 z^2 +36574 z^3+7975 z^4) \\
 & \left.\times\, \ln (1-z)+33(2-z)(3-z)(1296+25)\right] -9224 z^2+9598 z^3\\
 &-3943z^4 \left.\, +725 z^5\right) \left.\times\, \ln \left(1-\frac{z}{3}\right)\right]
  \;,
\end{split}
\end{equation}
\begin{align}
&{\cal D}^{(2^{++})}_g(z,[6,6]) = 0 \\
&{\cal D}^{(2^{++})}_g(z,[3,6]) = 0\,.
\end{align}
The SDCs of the perturbative contribution to the constituent heavy-quark fragmentation into $\TQQZpp$ at the initial energy scale are defined as follows~\cite{Feng:2020riv,Bai:2024ezn}:
\begin{equation}
\begin{split}
 \label{DQ_FF_SDC_0pp_33}
 {\cal D}^{(0^{++})}_Q(z,[3,3]) \,&=\, 
 \frac{\pi^2 \as^4(5m_Q)}{559872 \, d^{\cal D}_Q(z)}\left[ -264 (z-4) (11 z-12) (z^2-16 z+16) \right. \\
 & \times\, (13 z^4-57 z^3-656 z^2+1424z-512) (3 z-4)^5\\
 &\times\,\log (z^2-16 z+16)+6 (11 z-12)(z^2-16 z+16) \\
 & \times\, (1273 z^5-16764 z^4+11840 z^3 +247808z^2-472320 z+171008) \\
 & \times\, (3 z-4)^5 \log (4-3 z) - 3 (11 z-12)(z^2-16 z+16) \\
 & \times\, (129 z^5 - 7172 z^4+49504 z^3-108416z^2 + 73984 z-9216)\\
 & \times\, (3 z-4)^5 \log\left[\left(4-\frac{11z}{3}\right)(4-z)\right] + 16 (z-1)\\
 & \times\, (657763 z^{12}-10028192z^{11}+ 188677968 z^{10}-2600899712 z^9\\
 & +\, 18018056448 z^8-71685000192z^7 + 179414380544 z^6\\
 & -\,294834651136 z^5 + 321642168320z^4-229388845056 z^3\\
 & +\, 102018056192 z^2-25480396800z \left. + 2717908992)\right]\;,
\end{split}
\end{equation}
\begin{equation}
\begin{split}
 \label{DQ_FF_SDC_0pp_66}
 {\cal D}^{(0^{++})}_Q(z,[6,6]) \,&=\, \frac{\pi^2 \as^4(5m_Q)}{373248 \, d^{\cal D}_Q(z)}\left[ -120 (z-4) (11 z-12) (z^2-16z+16) \right. \\
 & \times\, (35 z^4-535 z^3+3472 z^2-4240z+512)(3 z-4)^5 \log (z^2-16 z+16)  \\
 & -\, 30 (11 z-12)(z^2-16 z+16)(3395 z^5-48020 z^4+126144 z^3 \\
 & -\, 75776^2-38656 z+62464)(3 z-4)^5 \log (4-3 z) + 75 (11 z-12)\\
 & \times\, (z^2-16 z+16)(735 z^5 - 10684 z^4+34208 z^3-44160z^2 \\
 & +\, 20224 z+9216) (3 z-4)^5\log\left[\left(4-\frac{11z}{3}\right)(4-z)\right]\\
 & +\, 16 (z-1) (7916587 z^{12}-263987840z^{11}+3125201872 z^{10}\\
 & -\, 16993694336 z^9+51814689024 z^8-99638283264^7 \\
 & +\, 133459423232 z^6-140136398848 z^5 + 127161204736z^4\\
 & -\, 96695746560 z^3+53372518400 z^2-17930649600z \left. +2717908992)\right]
 \;,
\end{split}
\end{equation}
\begin{equation}
\begin{split}
 \label{DQ_FF_SDC_0pp_36}
 \hspace{-0.00cm}
 {\cal D}^{(0^{++})}_Q(z,[3,6]) \,&=\,\frac{\pi^2 \as^4(5m_Q)}{186624 \sqrt{6} \, d^{\cal D}_Q(z)}\left[ 24 (z-4) (11 z-12) (z^2-16z+16) \right. \\
 & \times\, (225 z^4-3085 z^3+17456 z^2 - 19760z+1536)(3 z-4)^5\\
 & \times\, \log (z^2-16 z+16) - 6 (11 z-12)(z^2-16 z+16) \\
 & \times\, (555 z^5+52428 z^4-363328 z^3+616448z^2-270080 z+70656) \\
 & \times\, (3 z-4)^5 \log (4-3 z) + 75 (11 z-12)(z^2-16 z+16) \\
 & \times\, (1245 z^5 -84308 z^4 + 601696z^3-1333120z^2 + 914688 z-119808)\\
 & \times\, (3 z-4)^5\log\left[\left(4-\frac{11z}{3}\right)(4-z)\right]+16 (z-1)(1829959z^{12}\\
 & -\, 44960912 z^{11} + 285792656 z^{10}-1090093952z^9 + 5123084544 z^8\\
 & -\, 24390724608 z^7 +77450817536 z^6-153897779200z^5\\
 & +\, 194102034432 z^4-155643543552 z^3 +77091307520 z^2\\
 & -\, 21705523200z \left. +2717908992)\right]\;,
\end{split}
\end{equation}
with $d^{\cal D}_Q(z) = (4-3 z)^6(z-4)^2 z(11 z-12)(z^2-16 z+16)$; whereas the $[3,3]$ SDC,
\begin{equation}
\begin{split}
 \label{DQ_FF_SDC_2pp_33}
 {\cal D}^{(2^{++})}_Q(z,[3,3]) \,&=\, 
 \frac{\pi^2 \as^4(5m_Q)}{2799360 \, z \, d^{\cal D}_Q(z)}\left[ 672 (z-4) (11 z-12) (z^2-16 z+16) \right. \\
 & \times\, (47z^5+12186 z^4-44608 z^3 + 40000 z^2 -7936 z+4608) \\
 & \times\, (3 z-4)^5 \log (z^2-16 z+16)  +6 (11 z-12)(z^2-16 z+16) \\
 & \times\, (107645 z^6-1088988 z^5+7805536 z^4  -20734976 z^3 +8933504z^2 \\
 & -\, 6013952 z+1695744)(3 z-4)^5 \log (4-3 z) -33 (11 z-12)\\
 & \times\, (z^2-16 z+16)(3581 z^5 - 53216 z^4-326176 z^3+419456z^2 \\
 & -\, 6912 z+55296) (3 z-4)^6 \log\left[\left(4-\frac{11z}{3}\right)(4-z)\right]+ 16 (z-1)\\
 & \times\, (96449507 z^{12} - 158520388z^{11} -26228206896 z^{10}\\
 & +\, 281743037888 z^9 -\, 1355257362432 z^8 + 1355257362432 z^7 \\
 & -\, 6637452959744 z^6 + 7595797282816 z^5 -5643951472640 z^4 \\
 & +\, 2662988513280 z^3 -788934950912 z^2 + 161828831232 z \\
 & \left. -\, 24461180928)\right]\;,
\end{split}
\end{equation}
is the only initial-scale contribution to $\TQQTpp$ perturbative fragmentation due to Fermi–Dirac statistics~\cite{Feng:2020riv,Bai:2024ezn}.

\subsubsection{Long-Distance Matrix Elements}
\label{sssec:FFs_T4Q_LDMEs}
The long-distance matrix elements, $\langle {\cal O}^{\TQQ(J^{PC})}([n]) \rangle$, encapsulate the nonperturbative contributions to the initial-scale fragmentation functions. Since experimental data and lattice QCD studies on tetraquarks are limited, potential models are used to estimate these LDMEs. Three potential-based models have been proposed~\cite{Feng:2020riv}. Among these, the second model~\cite{Lu:2020cns}, which includes relativistic effects, is chosen for our {\tt TQ4Q1.1} FFs due to its stability and consistency. For fully charmed states, the LDMEs are given as~\cite{Feng:2020riv}:
\begin{equation}
\begin{split}
\label{LDMEs_T4c}
 {\cal O}^{\TQcZpp}([3,3]) &\,=\, 0.0347\mbox{ GeV}^9 \;, \\
 {\cal O}^{\TQcZpp}([6,6]) &\,=\, 0.0128\mbox{ GeV}^9 \;, \\
 {\cal O}^{\TQcZpp}([3,6]) &\,=\, 0.0211\mbox{ GeV}^9 \;, \\
 {\cal O}^{\TQcTpp}([3,3]) &\,=\, 0.072\mbox{ GeV}^9 \;, \\
 {\cal O}^{\TQcTpp}([3,6]) &\,=\, {\cal O}^{\TQcTpp}([6,6]) \,=\, 0 \;.
\end{split}
\end{equation}

For fully bottomed states, exact LDME values are unavailable. As an approximation, a $\TQb$ tetraquark is modeled as a compact diquark-antidiquark cluster dominated by color Coulomb forces. Following dimensional analysis, the ratio of LDMEs between $\TQb$ and $\TQc$ states is estimated as (Ref.~\cite{Feng:2023agq}):
\begin{equation}
\label{LDMEs_T4b}
 \frac{\langle {\cal O}^{\TQb(J^{PC})}([n]) \rangle}{\langle {\cal O}^{\TQc(J^{PC})}([n]) \rangle} \simeq \left( \frac{m_b \, \as^{(b)}}{m_c \, \as^{(c)}} \right)^{\!9} \simeq 400 \,,
\end{equation}
where $\as^{(Q)}$ represents the strong coupling, approximated as $\as(m_Q v_{\cal Q}) \sim v_{\cal Q}$, with $v_Q$ being the relative velocity between the heavy quarks.

\subsubsection{Initial Energy-Scale Inputs}
\label{sssec:FFs_T4Q_gQ}

This subsection presents the initial energy-scale inputs for gluon and quark fragmentation channels in the {\tt TQ4Q1.1} FFs for fully heavy tetraquarks ($\TQQ$). 

\begin{figure*}[!t]
\centering
\hspace{-1.4cm}
\begin{subfigure}{0.42\textwidth}
    \centering
    \includegraphics[scale=0.43,clip]{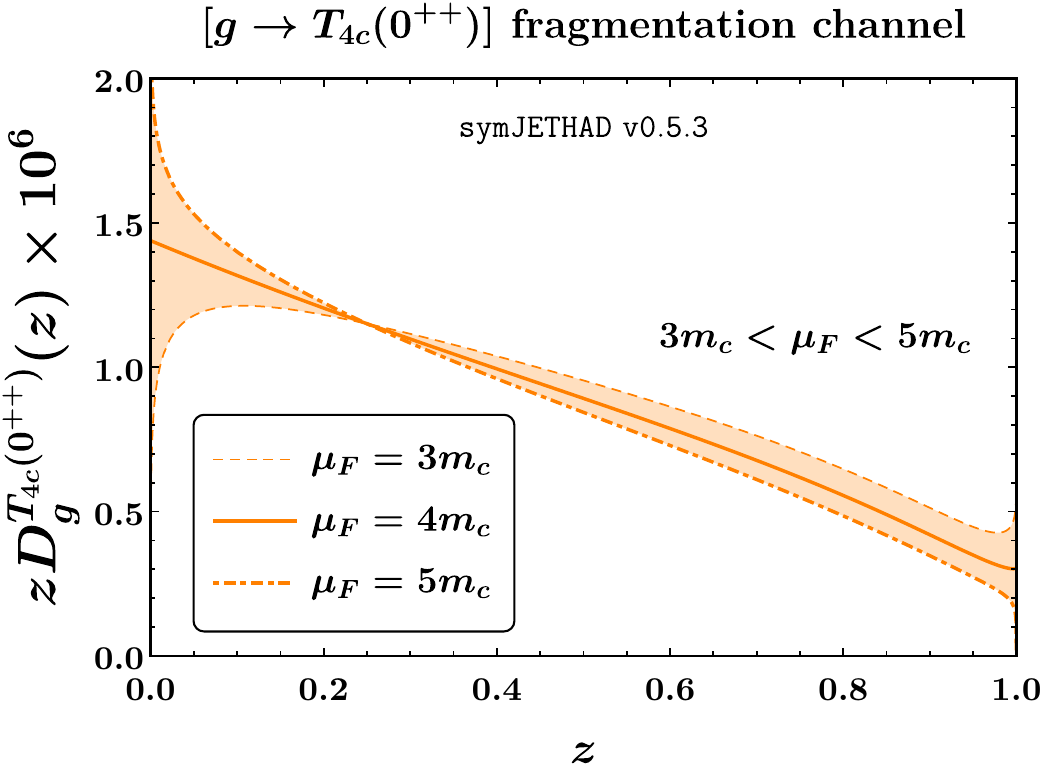}
\end{subfigure}
\hspace{1.2cm}
\begin{subfigure}{0.42\textwidth}
    \centering
    \includegraphics[scale=0.43,clip]{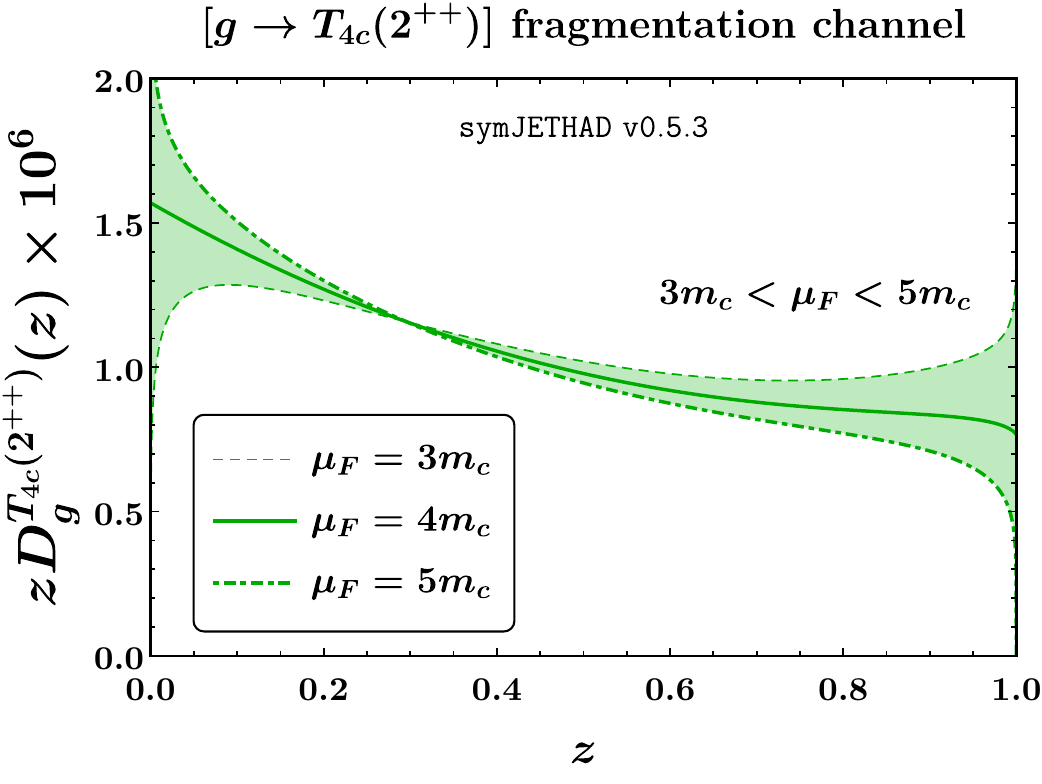}
\end{subfigure}
\vspace{0.5cm} 
\hspace{-1.4cm}
\begin{subfigure}{0.42\textwidth}
    \centering
    \includegraphics[scale=0.43,clip]{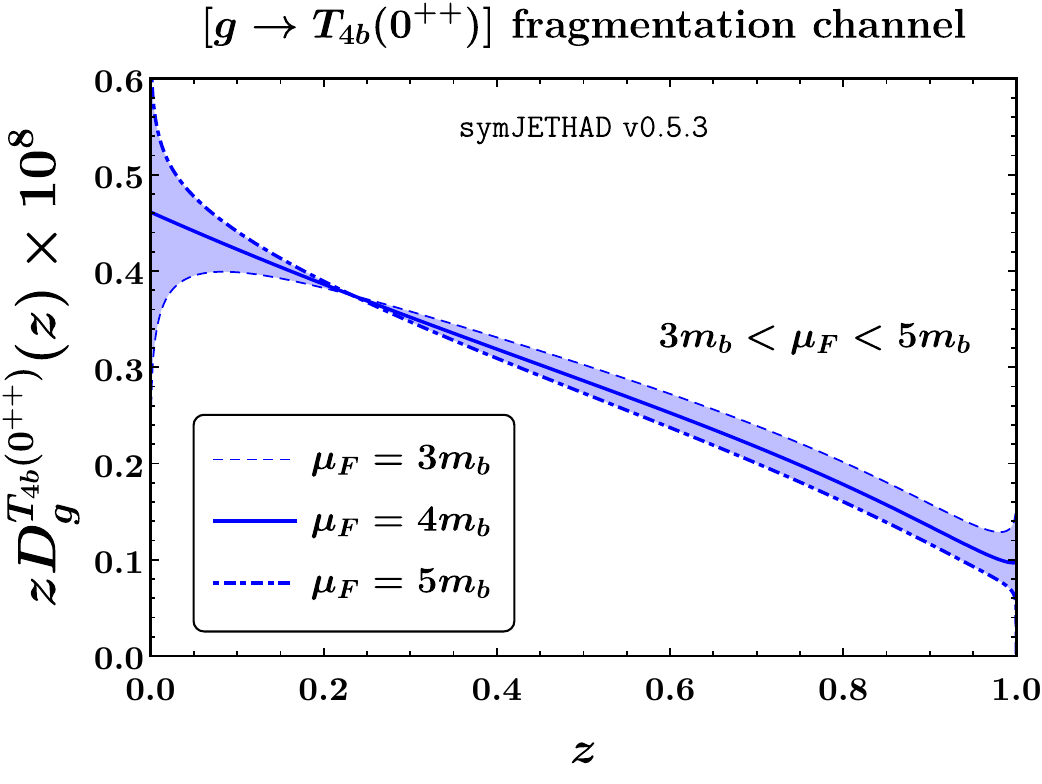}
\end{subfigure}
\hspace{1.2cm}
\begin{subfigure}{0.42\textwidth}
    \centering
    \includegraphics[scale=0.43,clip]{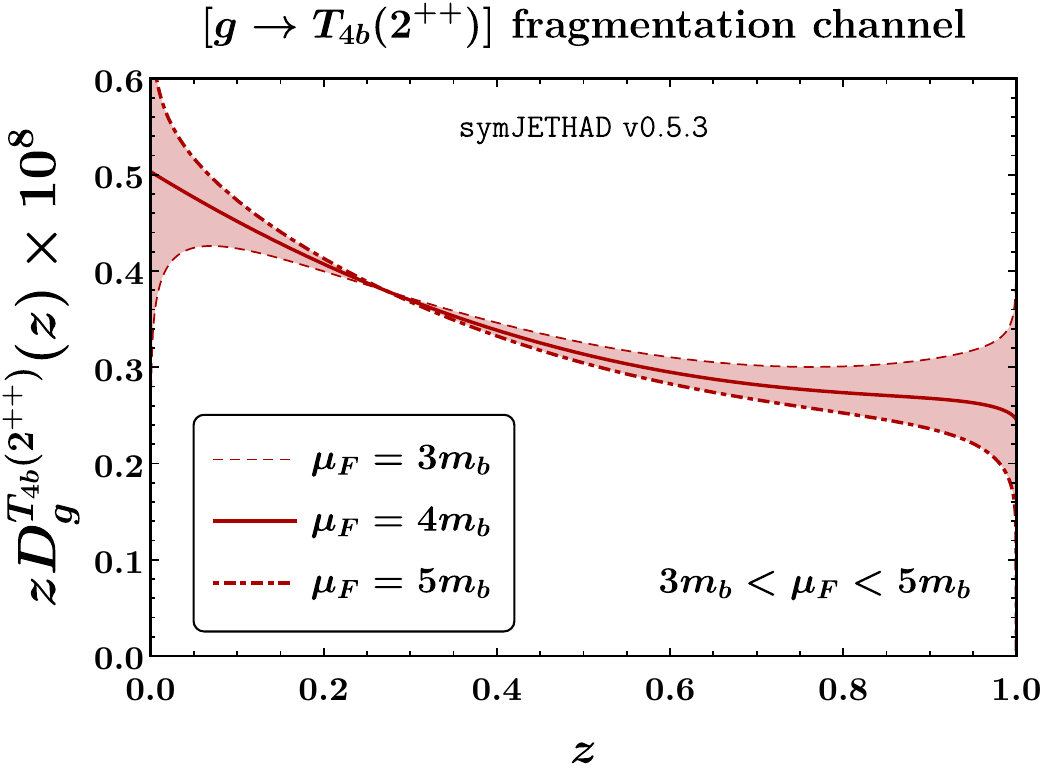}
\end{subfigure}
\caption{Collinear fragmentation of a gluon into fully charmed (upper) and fully bottomed (lower) tetraquarks. The left and right panels correspond to the initial-scale inputs for the $[g \to \TQQZpp]$ and $[g \to \TQQTpp]$ channels, respectively. For illustration, the DGLAP evolution is expanded over the range $3m_Q$ to $5m_Q$.}
\label{fig:TQQ_FF_initial-scale_gluon}
\end{figure*}

For the gluon-to-$\TQQ$ channel, the initial-scale inputs $[g \to \TQQ]$ are calculated using DGLAP evolution with the gluon-to-gluon splitting kernel, $P_{gg}$. The factorization scale is set to $\mu_{F,0} = 4 m_Q$, with a variation range of $\pm 2 m_Q$. Fig.~\ref{fig:TQQ_FF_initial-scale_gluon} shows the $z$-dependence of these inputs, where the $\TQb$ channels follow the same pattern as the $\TQc$ ones but differ in magnitude due to the \emph{Ansatz} in Eq.~\eqref{LDMEs_T4b}. Notably, the FFs do not vanish as $z \to 1$, which is unusual for collinear factorization and warrants further investigation.

\begin{figure*}[!t]
\centering
\hspace{-1.4cm}
\begin{subfigure}{0.42\textwidth}
    \centering
    \includegraphics[scale=0.43,clip]{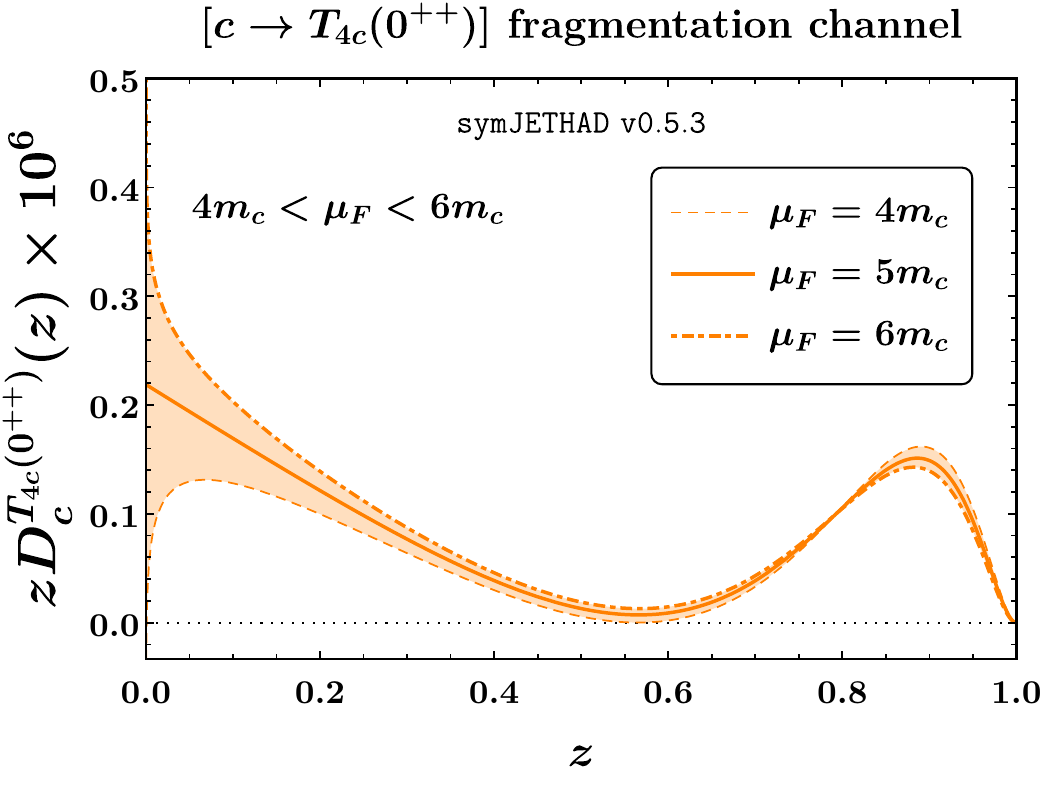}
\end{subfigure}
\hspace{1.2cm}
\begin{subfigure}{0.42\textwidth}
    \centering
    \includegraphics[scale=0.43,clip]{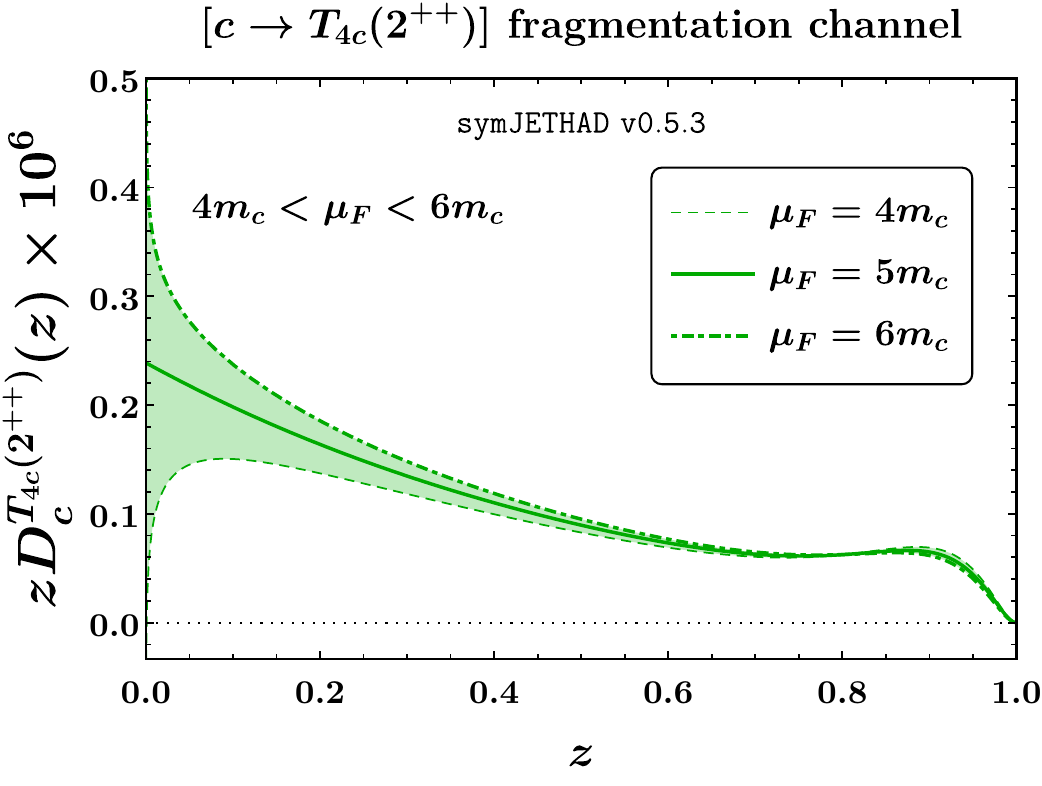}
\end{subfigure}
\vspace{0.5cm} 
\hspace{-1.4cm}
\begin{subfigure}{0.42\textwidth}
    \centering
    \includegraphics[scale=0.43,clip]{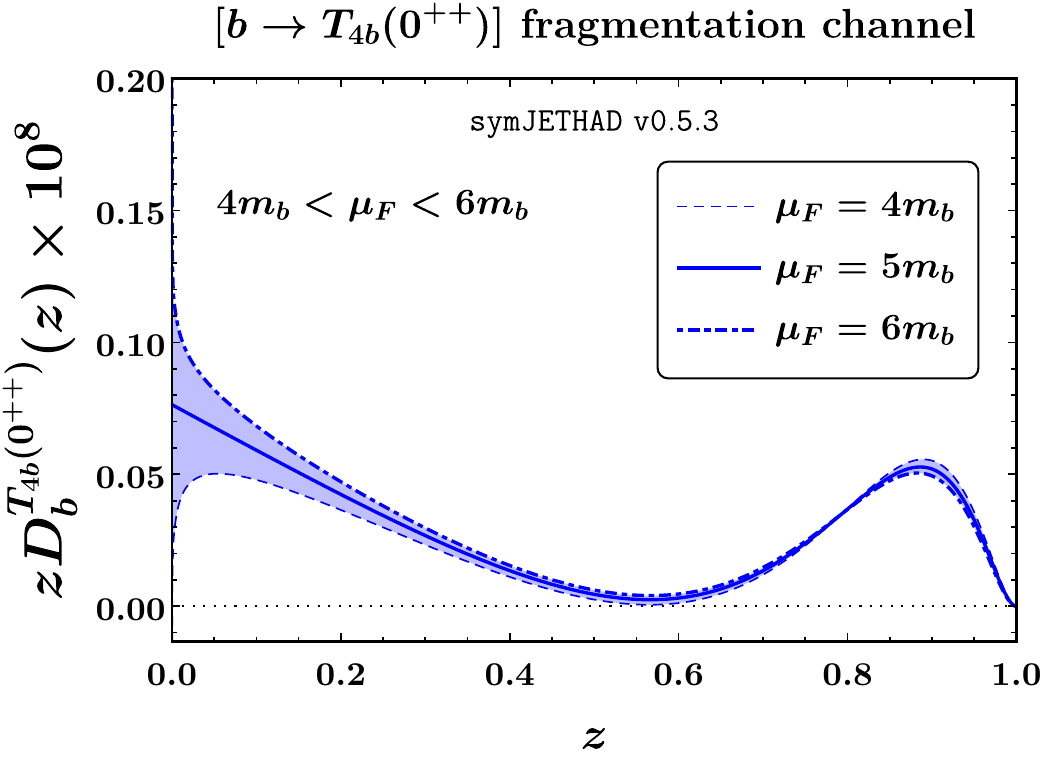}
\end{subfigure}
\hspace{1.2cm}
\begin{subfigure}{0.42\textwidth}
    \centering
    \includegraphics[scale=0.43,clip]{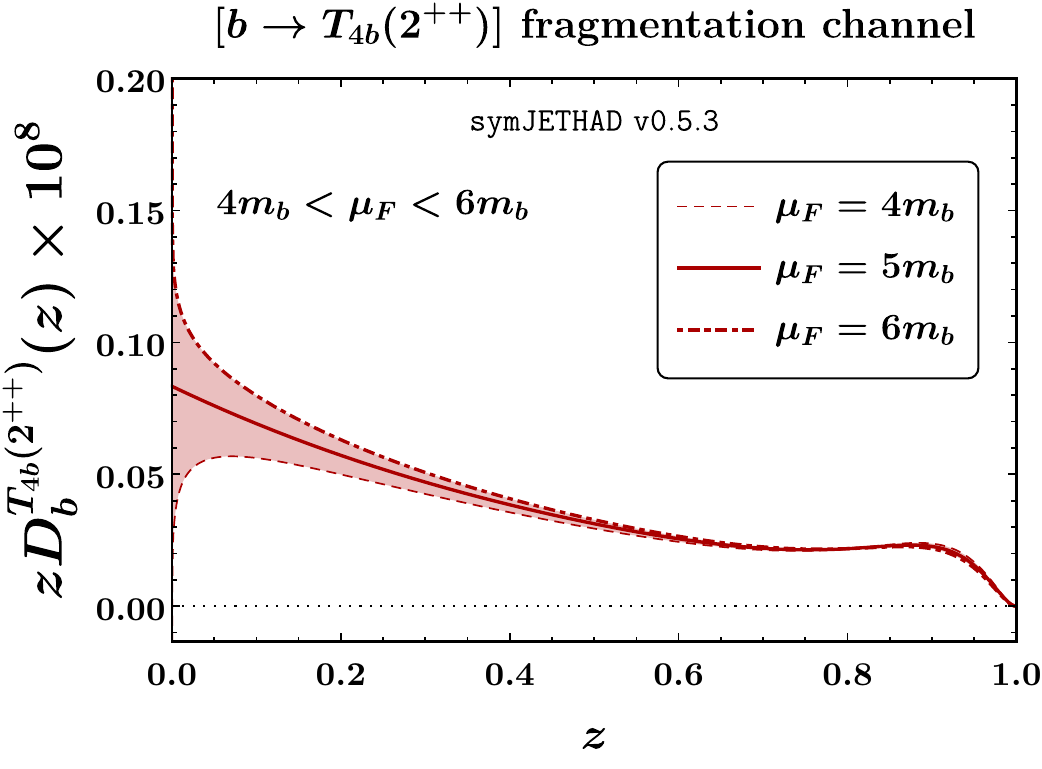}
\end{subfigure}
\caption{Collinear fragmentation of a constituent heavy quark into fully charmed (upper row) and fully bottomed (lower row) tetraquarks. The left and right panels correspond to the initial-scale inputs for the $[Q \to \TQQZpp]$ and $[Q \to \TQQTpp]$ channels, respectively. The DGLAP evolution is performed over the extended range $4m_Q$ to $6m_Q$.}
\label{fig:TQQ_FF_initial-scale_Q}
\end{figure*}

For the quark-to-$\TQQ$ channel, inputs $[Q \to \TQQ]$ are computed using a DGLAP evolution including both $P_{qq}$ and $P_{gg}$ kernels. The factorization scale is set to $\mu_{F,0} = 5 m_Q$, with a variation range of $\pm 2 m_Q$. Fig.~\ref{fig:TQQ_FF_initial-scale_Q} reveals distinct patterns: the $[Q \to \TQQZpp]$ FFs exhibit a peak at $z > 0.8$, while the $[Q \to \TQQTpp]$ FFs display a plateau-like shape.

\subsubsection{The {\tt TQ4Q1.1} functions}
\label{sssec:FFs_TQ4Q11}
The final step in constructing the {\tt TQ4Q1.1} fragmentation functions for $\TQQ$ involves a consistent DGLAP evolution of the initial-scale inputs introduced earlier. Unlike light-hadron fragmentation, both the heavy-quark and gluon channels exhibit evolution thresholds, arising from the minimal invariant masses for the splittings: $\mu_{F,0}(g \to \TQQ) = 4 m_Q$ for gluons and $\mu_{F,0}(Q \to \TQQ) = 5 m_Q$ for heavy quarks. These thresholds are determined by the $[g \to (Q\bar{Q}Q\bar{Q})]$ and $[Q,\bar{Q} \to (Q\bar{Q}Q\bar{Q}) + Q,\bar{Q}]$ perturbative splittings.

The DGLAP evolution is performed using the novel Heavy-Flavor Non-Relativ\-istic Evolution ({\HFNRevo}) scheme~\cite{Celiberto:2024mex,Celiberto:2024bxu,Celiberto:2024rxa} in two steps. First, starting with the gluon channel at $\mu_{F,0} = 4 m_Q$, only the gluon-to-gluon splitting kernel ($P_{gg}$) is active. This evolution generates gluon contributions up to the heavy-quark threshold $\mu_{F,0} = 5 m_Q$ and is analytically implemented via the {\symJethad} plugin. In the second step, the gluon FF evolved to $Q_0 = 5 m_Q$ is combined with the heavy-quark input, and a full numerical DGLAP evolution is performed to generate the NLO {\tt TQ4Q1.1} FFs, which are released in {\tt LHAPDF} format.

The absence of initial-scale inputs for nonconstituent light and heavy quarks could seem like a limitation. However, based on analogies with NRQCD studies of pseudoscalar and vector quarkonia, these channels are expected to be suppressed compared to gluon and charm contributions.
\begin{figure*}[!t]
   \hspace{-0.7cm}
   \includegraphics[scale=0.39,clip]{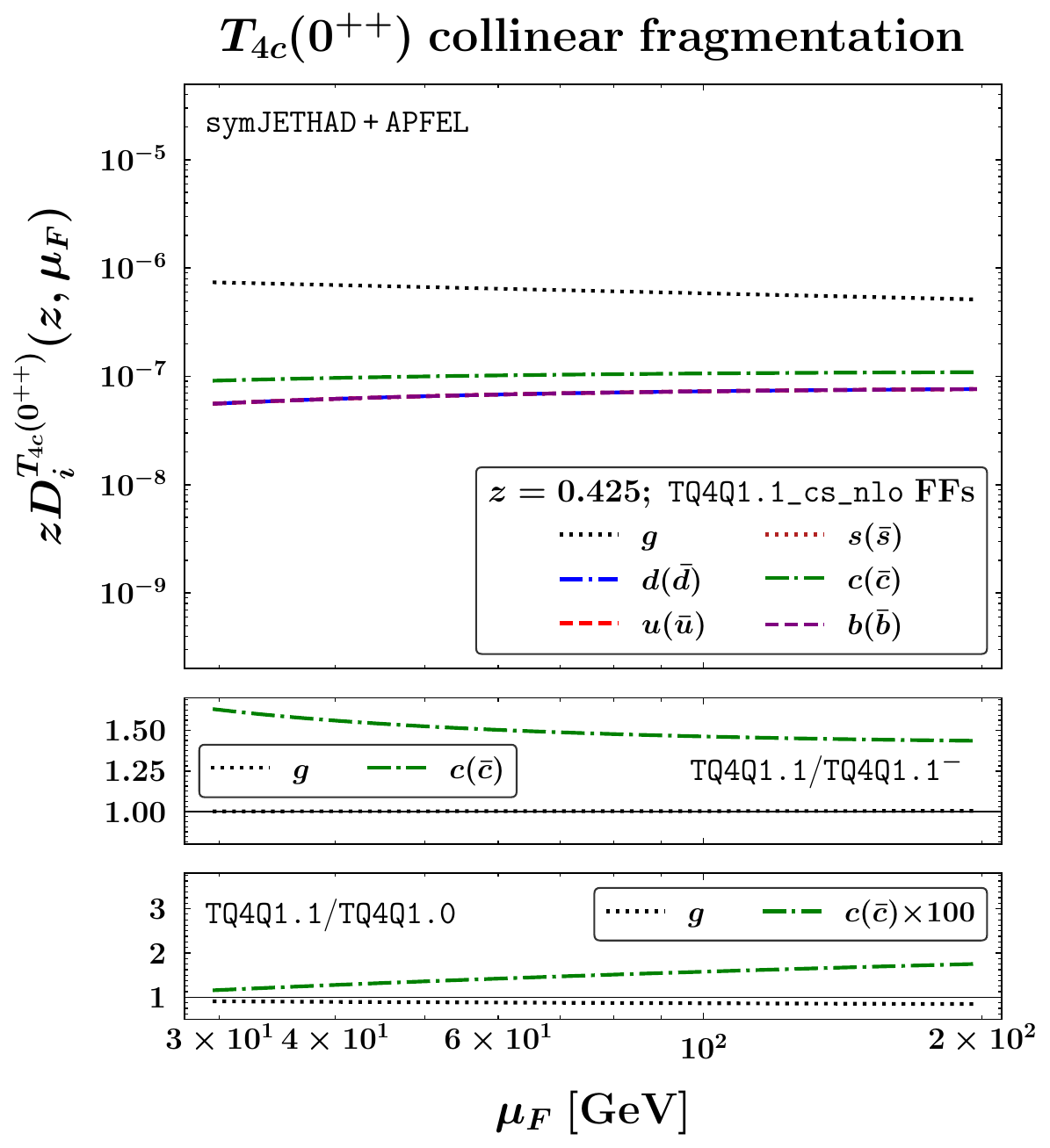}
   \hspace{-0.32cm}
   \includegraphics[scale=0.39,clip]{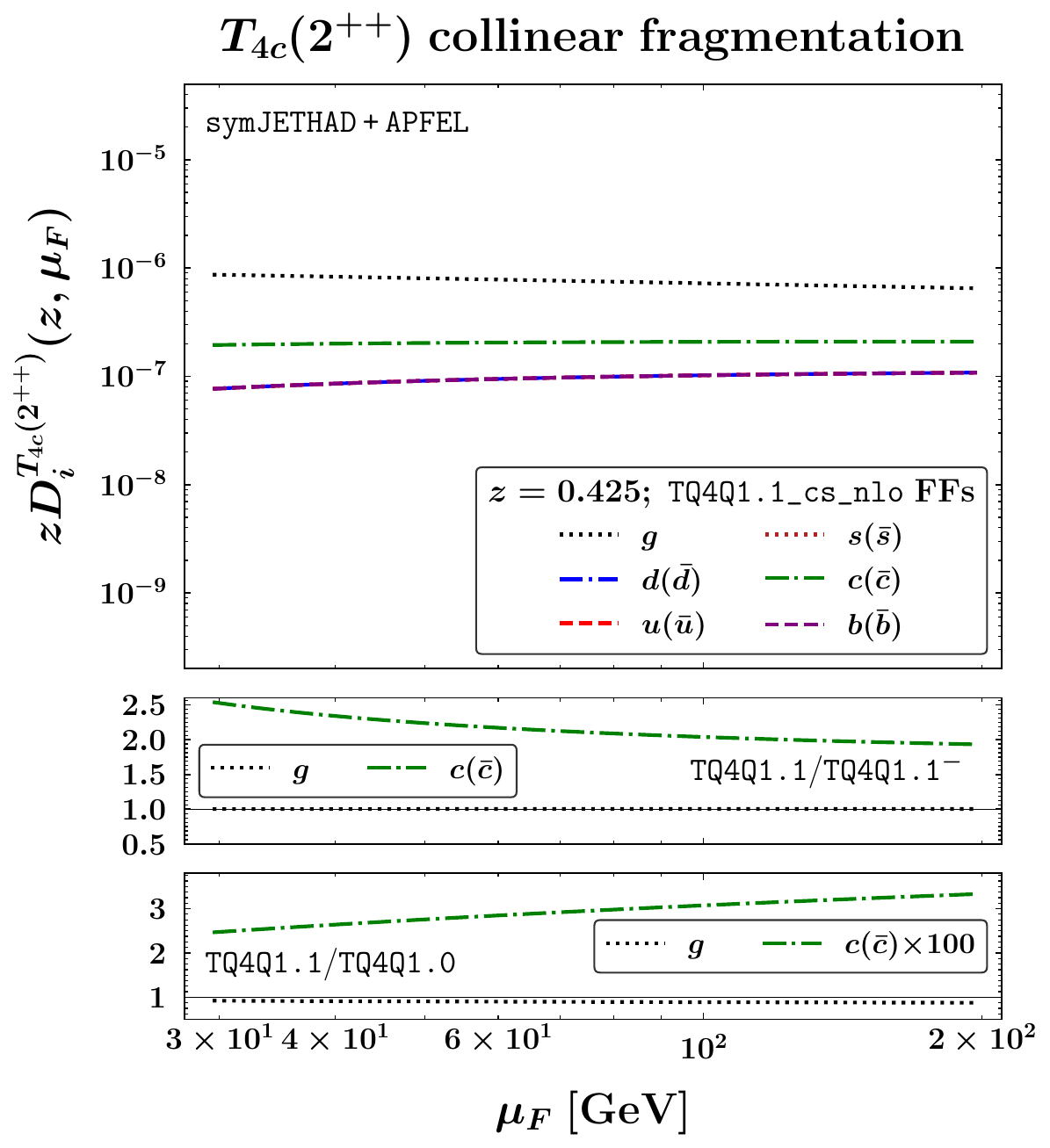}
\caption{Factorization-scale dependence of {\tt TQ4Q1.1} collinear FFs for $\TQcZpp$ (left) and $\TQcTpp$ (right) formation at $z = 0.425 \simeq \langle z \rangle$. The first ancillary panels display the ratio of {\tt TQ4Q1.1} to {\tt TQ4Q1.1}$^-$. The second ancillary panels compare {\tt TQ4Q1.1} with {\tt TQ4Q1.0}, where the charm ratio in the latter is scaled down by a factor of 100.}
\label{fig:NLO_FFs_Tc0_Tc2}
\end{figure*}
\begin{figure*}[!t]
   \hspace{-0.7cm}
   \includegraphics[scale=0.39,clip]{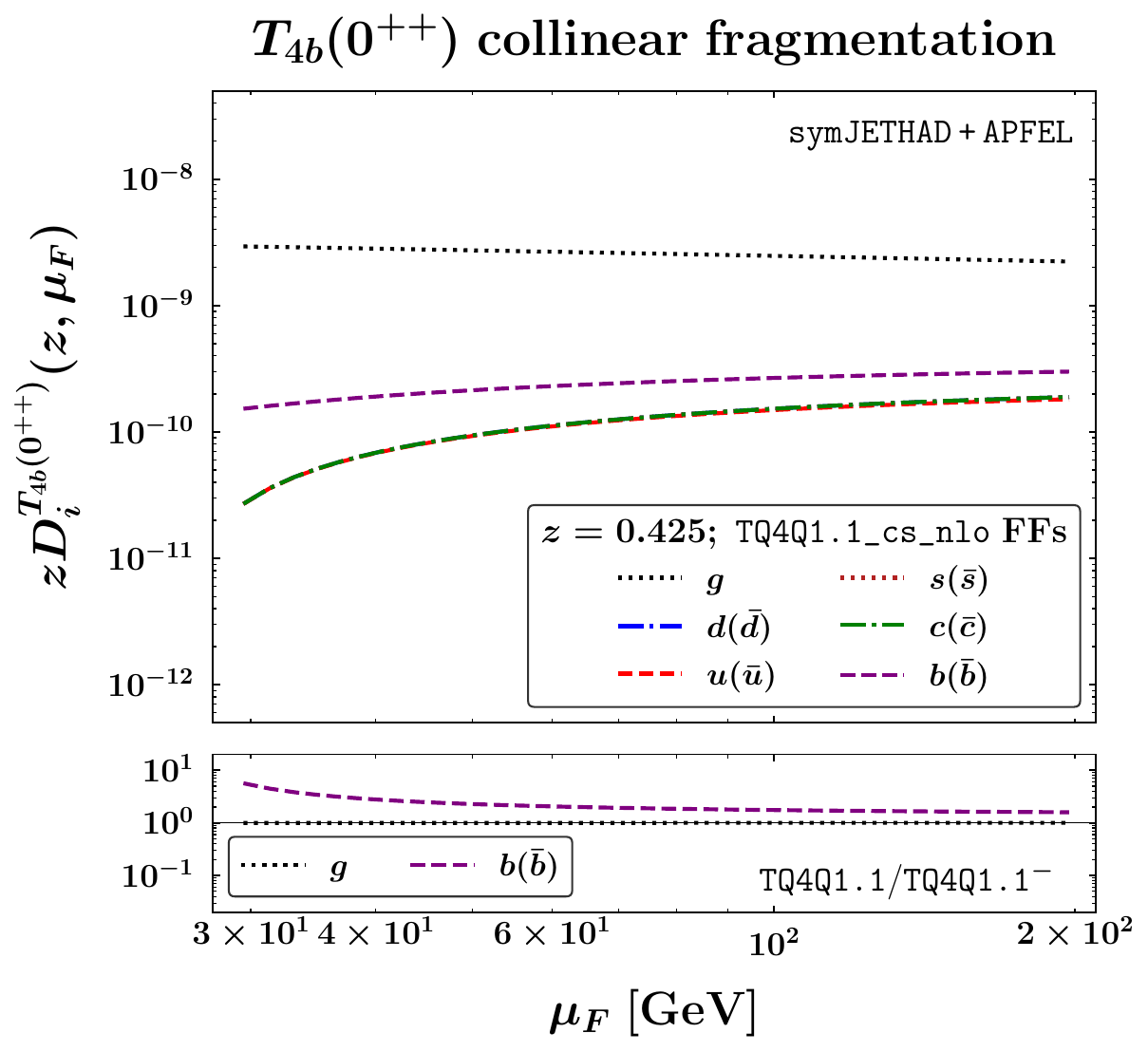}
   \hspace{-0.32cm}
   \includegraphics[scale=0.39,clip]{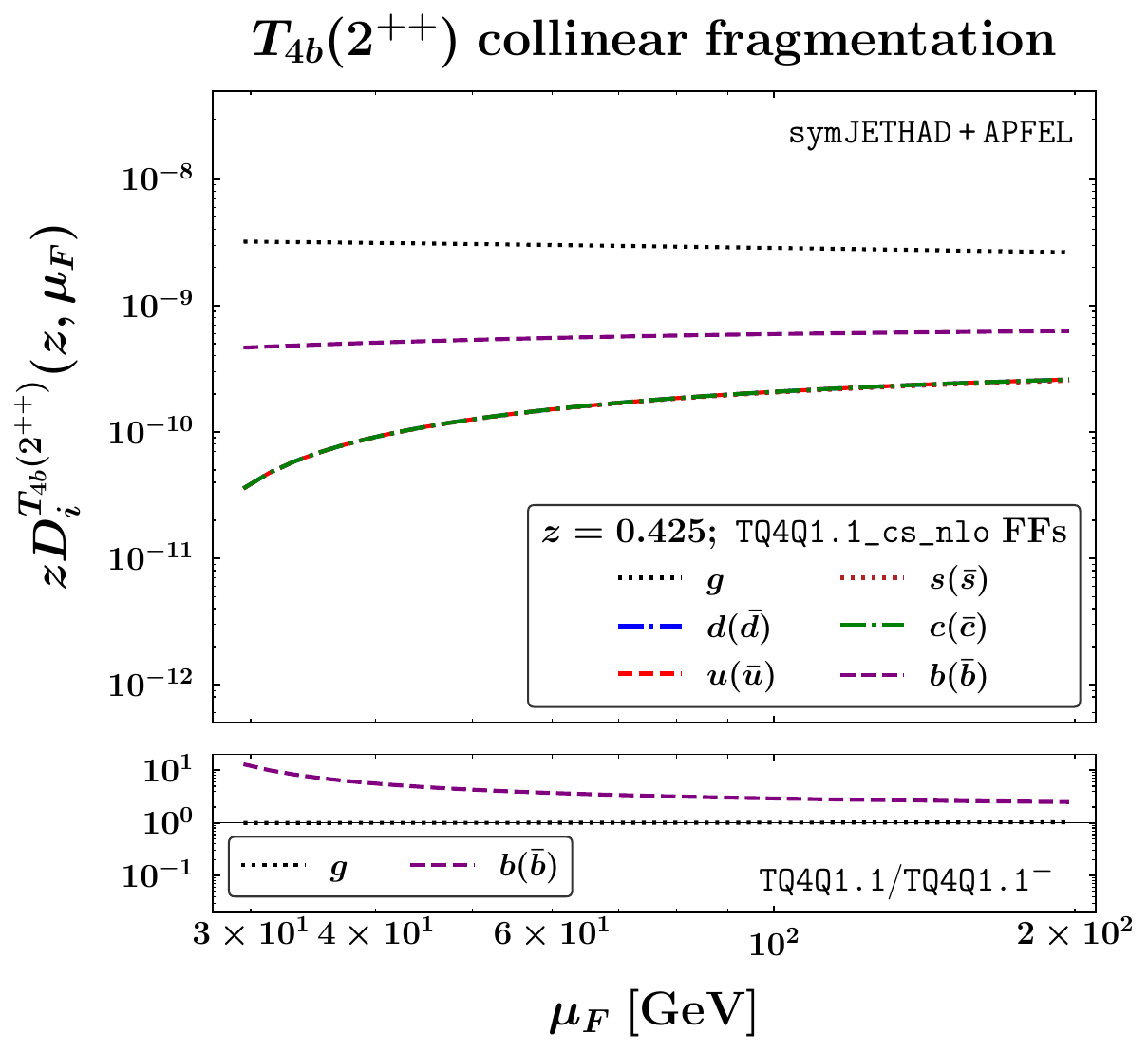}
\caption{Factorization-scale dependence of {\tt TQ4Q1.1} collinear FFs for $\TQbZpp$ (left) and $\TQbTpp$ (right) formation at $z = 0.425 \simeq \langle z \rangle$. The lower panels display the ratio of {\tt TQ4Q1.1} to {\tt TQ4Q1.1}$^-$ functions.}
\label{fig:NLO_FFs_Tb0_Tb2}
\end{figure*}

Figs.~\ref{fig:NLO_FFs_Tc0_Tc2} and~\ref{fig:NLO_FFs_Tb0_Tb2} illustrate the factorization-scale dependence of the {\tt TQ4Q1.1} FFs for $\TQcZpp$, $\TQcTpp$, $\TQbZpp$, and $\TQbTpp$. As before, the analysis is performed at $z = 0.425$, which represents an average value for semi-hard final states. The results show that the heavy-quark FF consistently dominates, being 5 to 20 times larger than other channels. Although the gluon FF contribution is smaller, it remains important due to the larger gluon PDF, which significantly enhances gluon-driven processes in hadron colliders.

For comparison, supplementary FF sets ({\tt TQ4Q1.1$^-$}), excluding the $[Q \to \TQQ]$ input, were constructed. The analysis indicates that the inclusion of the heavy-quark channel at the initial scale increases the FF magnitude by up to 10 times for $[Q \to \TQQ]$, while the gluon FF remains almost unchanged.

Finally, the {\tt TQ4Q1.1} gluon FFs display a gradual decline with increasing $\mu_F$, a behavior mirrored in other channels, such as gluon-to-$B_c^{(*)}$ and gluon-to-$\QXQq$ fragmentation. This soft $\mu_F$ dependence acts as a stabilizing factor for high-energy resummed distributions, making the {\tt TQ4Q1.1} FFs versatile tools for studies across collider environments.

\section{Tetraquark-plus-jet in hybrid factorization at NLLA/NLO$^+$}
\sectionmark{Tetraquark-plus-jet in HF at NLLA/NLO$^+$}
\label{ssec:NLL_cross_section}

In this section, we examine the semi-inclusive production of a bottomonium-like tetraquark ($\B$) and a light jet in high-energy proton-proton collisions, a process relevant for LHC and FCC phenomenology, and the formula for the cross section will be provided. The reaction of interest is:
\begin{equation}
\label{eq:process}
 {\rm p}(p_a) + {\rm p}(p_b) \,\to\, {\B}(\kappa_1, y_1) + {\cal X} + {\rm jet}(\kappa_2 , y_2) \;,
\end{equation}
where $\B$ is produced with momentum $\kappa_1$, rapidity $y_1$ and azimuthal angle $\varphi_1$, while a light jet, with momentum $\kappa_2$, rapidity $y_2$ and azimuthal angle $\varphi_2$, is detected. Both objects are characterized by high transverse momenta ($|\vec \kappa_{1,2}| \gg \Lambda_{\rm QCD}$) and are separated by a large rapidity gap, $\DY \equiv y_1 - y_2$. An undetected gluon system, ${\cal X}$, is inclusively produced. This setup provides a valuable testing ground for hybrid factorization, combining collinear and high-energy approaches.

\subsection{Resummed cross section}
The transverse momenta are decomposed in terms of Sudakov vectors:
\begin{equation}
\label{Sudakov}
 \kappa_{1,2} = x_{1,2} \, p_{a,b} - \frac{\kappa_{1,2 \perp}^2}{x_{1,2} s} \, p_{b,a} + \kappa_{{1,2 \perp}} \;,
\end{equation}
where $p_{a,b}$ represent the parent proton momenta, with $p_{a,b}^2 = 0$ and $(p_a \cdot p_b) = s/2$. We also have $\kappa_{1,2 \perp}^2 \equiv - \vec \kappa_{1,2}^{\,2}$. The longitudinal fractions $x_{1,2}$ and the rapidity separation $\DY$ are related by:
\begin{equation}
\label{Delta_Y}
 x_{1,2} = \frac{|\vec \kappa_{1,2}|}{\sqrt{s}} e^{\pm y_{1,2}}, \quad \text{d}y_{1,2} = \pm \frac{d x_{1,2}}{x_{1,2}} , \quad \DY = \ln\frac{x_1 x_2 s}{|\vec \kappa_1| |\vec \kappa_2|}.
\end{equation}
In a pure collinear framework, the LO differential cross section is expressed as:
\begin{eqnarray}
\label{sigma_collinear}
\frac{\drv\sigma^{\rm LO}_{\rm [coll.]}}{\drv x_1\drv x_2\drv ^2\vec \kappa_1\drv ^2\vec \kappa_2}
= \sum_{i,j=q,{\bar q},g}\int_0^1 \drv x_a \!\! \int_0^1 \drv x_b\ f_i\left(x_a\right) f_j\left(x_b\right) \nonumber \\
\quad\times \, \int_{x_1}^1 \frac{\drv \zeta}{\zeta} \, D^{\B}_i\left(\frac{x_1}{\zeta}\right) 
\frac{\drv {\hat\sigma}_{i,j}\left(\hat s\right)}
{\drv x_1\drv x_2\drv ^2\vec \kappa_1\drv ^2\vec \kappa_2}\;,
\end{eqnarray}
where $f_{i,j}$ are proton PDFs, $D^{\B}_i$ are $\B$ FFs, $\zeta$ is the outgoing parton’s momentum fraction, and $\drv \hat{\sigma}_{i,j}$ represents the partonic cross section. Here, $\hat{s} = x_a x_b s$ is the partonic center-of-mass energy squared. The indices \(i\) and \(j\) run over all partons, except for the \(t\) quark, which does not undergo hadronization.

Conversely, in hybrid factorization, the cross section combines collinear PDFs and FFs with high-energy dynamics through the BFKL formalism. The cross section is recast as:
\begin{equation}
\label{dsigma_Fourier}
\frac{\drv \sigma}{\drv y_1 \drv y_2 \drv |\vec \kappa_1| \drv |\vec \kappa_2| \drv \varphi_1 \drv \varphi_2} \!=\!
\left[ {\cal C}_0 + 2 \sum_{m=1}^\infty \cos (m \varphi)\,
 {\cal C}_m \right],
\end{equation}
where ${\cal C}_m$ are azimuthal-angle coefficients derived using NLLA-accurate BFKL resummation and $\varphi \equiv \varphi_1 - \varphi_2 - \pi$.
Utilizing the $\overline{\text{MS}}$ scheme for renormalization and applying the BFKL formalism, we obtain (refer to Ref.~\cite{Caporale:2012ih} for detailed computations)
\begin{equation}
\label{Cm_NLLp_MSb}
 \CmNLLp \!\!\!\!=\!\! \frac{x_1 x_2}{|\vec \kappa_1| |\vec \kappa_2|} 
 \int_{-\infty}^{+\infty} \!\!\! \drv \nu \, e^{{\DY} \bar \alpha_s(\mu_R)
 \chi^{\rm NLO}(m,\nu)} 
 \end{equation}
\[ 
 \times \, \alpha_s^2(\mu_R) \biggl\{ \E_\B^{\rm NLO}(m,\nu,|\vec \kappa_1|, x_1)[\E_J^{\rm NLO}(m,\nu,|\vec \kappa_2|,x_2)]^* 
\]
\[
 +
 \left.
 \alpha_s^2(\mu_R) \DY \frac{\beta_0}{4 \pi} \,
 \chi(m,\nu)
 \left[\ln\left(|\vec \kappa_1| |\vec \kappa_2|\right) + \frac{i}{2} \, \frac{\drv}{\drv \nu} \ln\frac{\E_\B}{\E_J^*}\right]
 \right\}
 .
\]
where $\bar \alpha_s(\mu_R) \equiv \alpha_s(\mu_R) N_c/\pi$ represents the QCD running coupling, $N_c = 3$ denotes the number of colors, and $\beta_0 = 11N_c/3 - 2 n_f/3$ is the leading coefficient of the QCD $\beta$-function.
We consider a two-loop running coupling approach with $\alpha_s\left(m_Z\right)=0.118$ and dynamically varying the number of active quark flavors, $n_f$. The `$+$` superscript in $\CmNLLp$ highlights contributions exceeding NLLA accuracy, originating from two key sources: exponentiated NLO corrections to the high-energy kernel and the interplay of NLO corrections to the impact factors.
The high-energy kernel in the exponent of Eq.~\eqref{Cm_NLLp_MSb} encompasses the resummation of logarithms of energy at NLLA accuracy:
\begin{eqnarray}
 \label{chi}
 \chi^{\rm NLO}(m,\nu) = \chi(m,\nu) + \bar\alpha_s \hat \chi(m,\nu) \;,
\end{eqnarray}
where $\chi(m,\nu)$ are the eigenvalues of the leading-order kernel:
\begin{eqnarray}
 \label{kernel_LO}
 \chi\left(m,\nu\right) = -2\gamma_{\rm E} - 2 \, {\rm Re} \left\{ \psi\left(\frac{1+m}{2} + i \nu \right) \right\} \, ,
\end{eqnarray}
with $\gamma_{\rm E}$ denoting the Euler-Mascheroni constant and $\psi(z) \equiv \Gamma^\prime
(z)/\Gamma(z)$ being the logarithmic derivative of the Gamma function. 
The $\hat\chi(m,\nu)$ term in Eq.~\eqref{chi} corresponds to the next-to-leading-order correction to the kernel:
\begin{eqnarray}
\label{chi_NLO}
\hat \chi\left(m,\nu\right) = \bar\chi(m,\nu)+\frac{\beta_0}{8 N_c}\chi(m,\nu)\left\{-\chi(m,\nu)+10/3+2\ln\left(\frac{\mu_R^2}{|\vec \kappa_1| |\vec \kappa_2|}\right)\right\},
\end{eqnarray}
with the characteristic $\bar\chi(m,\nu)$ function as evaluated in Ref.~\cite{Kotikov:2000pm}.
The two formulations
\begin{eqnarray}
\label{EFs}
\E_{\B,J}^{\rm NLO}(m,\nu,|\vec \kappa_{1,2}|,x_{1,2}) =
\E_{\B,J} +
\alpha_s(\mu_R) \, \hat \E_{\B,J}
\end{eqnarray}
represent, respectively, the next-to-leading order emission functions for the bottom-tetra-\-quark and the light-jet, derived in Mellin space after projecting onto the eigenfunctions of the leading-order kernel.
For the $\B$-particle emission function, we use the NLO calculation detailed in Ref.~\cite{Ivanov:2012iv}.
Although originally developed for investigating light-hadron production, this formulation adapts to our VFNS framework for heavy-flavored tetraquarks, assuming that transverse momenta remain well above the heavy-quark thresholds in the DGLAP evolution.
At leading order, the expression is:
\begin{equation}
\begin{split}
 \label{LOBEF}
 \E_\B(m,\nu,|\vec \kappa_1|,x_1) = \upsilon_c \, |\vec \kappa_1|^{2i\nu-1}\int_x^1 \frac{\drv \xi}{\xi} \; \hat{x}^{1-2i\nu} 
 \Big[\tau_c f_g(\xi)D_g^\B\left(\hat{x}\right)
 +\sum_{i=q,\bar q}f_i(\xi)D_i^\B\left(\hat{x}\right)\Big] \;,
\end{split}
\end{equation}
where $\hat{x} = x/\xi$, $\upsilon_c = 2 \sqrt{C_F/C_A}$, and $\tau_c = C_A/C_F$, with $C_F = (N_c^2-1)/(2N_c)$ and $C_A = N_c$ representing the Casimir factors associated with gluon emission by quarks and gluons, respectively.
The complete NLO formula for $\E_\B^{\rm NLO}$ is documented in Ref.~\cite{Ivanov:2012iv}.
For the light-jet, the leading-order emission function takes the form:
\begin{equation}
 \label{LOJEF}
 c_J(n,\nu,|\vec \kappa|,x) = \upsilon_c
 |\vec \kappa|^{2i\nu-1}\,\hspace{-0.05cm}\Big[\tau_c f_g(x)
 +\hspace{-0.15cm}\sum_{j=q,\bar q}\hspace{-0.10cm}f_j(x)\Big] \;.
\end{equation}
Its next-to-leading-order correction is determined by combining Eq.~(36) in Ref.~\cite{Caporale:2012ih}.

Equations~(\ref{Cm_NLLp_MSb}) and~(\ref{LOJEF}) succinctly demonstrate the implementation of our hybrid factorization scheme, merging collinear and high-energy dynamics. Within this framework, the cross section is expressed in terms of the BFKL formalism, where the gluon Green's function and emission functions are pivotal. The gluon Green's function handles the resummation of large logarithms in the high-energy limit, while the emission functions incorporate parton distribution functions (PDFs) and fragmentation functions (FFs), effectively bridging collinear factorization with high-energy behavior.

If all NLO terms in Eq.~\eqref{Cm_NLLp_MSb} are neglected, the pure leading-logarithmic (LL) limit of the azimuthal coefficients is recovered:
\begin{equation}
\begin{split}
 \label{Cm_LL_MSb}
 &\CmLL = \frac{x_1 x_2}{|\vec \kappa_1| |\vec \kappa_2|} 
 \int_{-\infty}^{+\infty} \drv \nu \, e^{{\DY} \bar \alpha_s(\mu_R)\chi(m,\nu)}\alpha_s^2(\mu_R) \, \E_\B(m,\nu,|\vec \kappa_1|, x_1)[\E_J(m,\nu,|\vec \kappa_2|,x_2)]^* \;.
\end{split}
\end{equation}

To rigorously compare predictions from high-energy resummation with those derived using a purely collinear, DGLAP-inspired framework, it is crucial to calculate observables using both approaches. However, the absence of numerical tools for fixed-order NLO distributions in semi-hard hadron-plus-jet production limits direct comparisons. To address this, we adopt an alternative strategy.

Our approach truncates the high-energy series at NLO accuracy to approximate the signal from a pure NLO calculation. By expanding azimuthal coefficients up to ${\cal O}(\alpha_s^3)$, we derive a high-energy fixed-order ($\HENLOp$) expression. This $\HENLOp$ approximation serves as a practical tool to compare BFKL resummation effects with fixed-order predictions. The $\HENLOp$ azimuthal coefficients in the $\MSb$ scheme are:
\begin{align}
\label{Cm_HENLOp_MSb}
 \CmHENLOp &= \frac{e^{\DY}}{s} 
 \int_{-\infty}^{+\infty} \drv \nu \, 
 \alpha_s^2(\mu_R)
 \left[ 1 + \bar \alpha_s(\mu_R) \DY \chi(m,\nu) \right]
 \\[0.75em] \nonumber
 \hspace{0.50cm}&\times \,
 \E_\B^{\rm NLO}(m,\nu,|\vec \kappa_1|, x_1)[\E_J^{\rm NLO}(m,\nu,|\vec \kappa_2|,x_2)]^* \;.
\end{align}
We set the renormalization ($\mu_R$) and factorization ($\mu_F$) scales to \emph{natural} energies based on the final-state kinematics, using $\mu_R = \mu_F \equiv \mu_N$, where:
\begin{eqnarray}
\label{mu_N}
 \mu_N = m_{\B \perp} + |\vec{\kappa}_2| \;,
\end{eqnarray}
with $m_{\B \perp} = \sqrt{m_\B^2 + |\vec{\kappa}_1|^2}$ representing the transverse mass of the bottomed tetraquark and $|\vec{\kappa}_2|$ denoting the transverse momentum of the light-jet. We also vary $\mu_R$ and $\mu_F$ within a range of $\mu_N/2$ to $2\mu_N$, controlled by the $C_\mu$ parameter. This variation evaluates the sensitivity of our results to energy-scale changes, providing robust estimates of theoretical uncertainties in our predictions.

\section{Phenomenology}
\label{sec:results}

The predictions presented in this section were obtained using the \textsc{Python} + \textsc{For\-tran}
{\Jethad} multimodular framework~\cite{Celiberto:2020wpk,Celiberto:2022rfj,Celiberto:2023fzz,Celiberto:2024mrq,Celiberto:2024swu}. 
Proton PDFs were modeled using the {\tt NNPDF4.0} NLO set~\cite{NNPDF:2021uiq,NNPDF:2021njg} provided by {\tt LHAPDF v6.5.4}~\cite{Buckley:2014ana}. 
To account for the impact of MHOUs, the factorization and renormalization scales, $\mu_F$ and $\mu_R$, were varied around the kinematics-driven \emph{natural} scale by a factor ranging from $1/2$ to $2$, controlled by the parameter $C_\mu$. 
The uncertainty bands in the plots represent the combined effects of MHOUs and numerical integration errors, which were consistently kept below 1\% through the optimized integration routines of {\Jethad}.

\subsection{Rapidity-interval rates}
\label{ssec:DY}

The rapidity-interval rate measures the cross section as a function of the rapidity difference, $\DY = y_1 - y_2$, between the tetraquark and a jet. 
It is defined by integrating the first azimuthal-angle coefficient, ${\cal C}_{0}^{\rm [resum]}$, over the transverse momenta and rapidities of the final-state particles:
\begin{equation}
\begin{split}
\label{DY_distribution}
 \hspace{-0.12cm}
 \frac{\drv \sigma(\DY, s)}{\drv \DY} &=
 \int_{y_1^{\rm min}}^{y_1^{\rm max}} \!\!\!\!\! \drv y_1
 \int_{y_2^{\rm min}}^{y_2^{\rm max}} \!\!\!\!\! \drv y_2
 \, \,
 \delta (\DY - (y_1 - y_2))
 \int_{|\vec \kappa_1|^{\rm min}}^{|\vec \kappa_1|^{\rm max}} 
 \!\!\!\!\! \drv |\vec \kappa_1|
 \int_{|\vec \kappa_2|^{\rm min}}^{|\vec \kappa_2|^{\rm max}} 
 \!\!\!\!\! \drv |\vec \kappa_2|
 \, \,
 {\cal C}_{0}^{\rm [resum]}
 \;.
\end{split}
\end{equation}
The `${\rm [resum]}$' superscript inclusively refers to $\NLLp$, $\LL$, or $\HENLOp$.

The transverse momenta of the hadron range between 30 and 120~GeV, while the jet spans 50 to 120~GeV, consistent with LHC analyses. 
To enhance high-energy signals and suppress unwanted effects, asymmetric transverse momentum windows and specific rapidity intervals are used: $|y_1| < 2.4$ for the hadron (barrel detection) and $|y_2| < 4.7$ for the jet (extended to endcap regions).

\begin{figure*}[!t]
    \hspace{-0.4cm}
   \includegraphics[scale=0.38,clip]{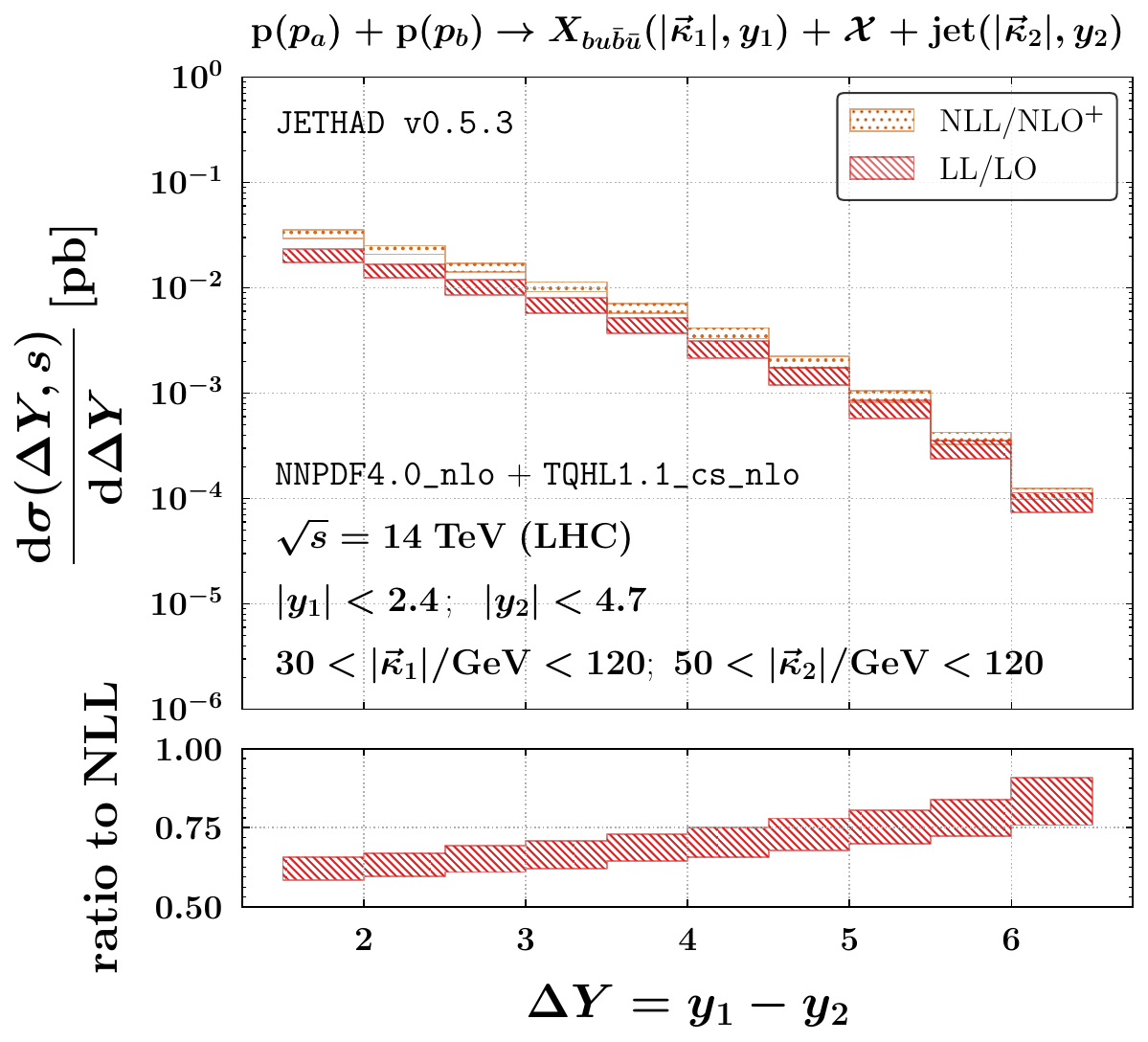}
   \includegraphics[scale=0.38,clip]{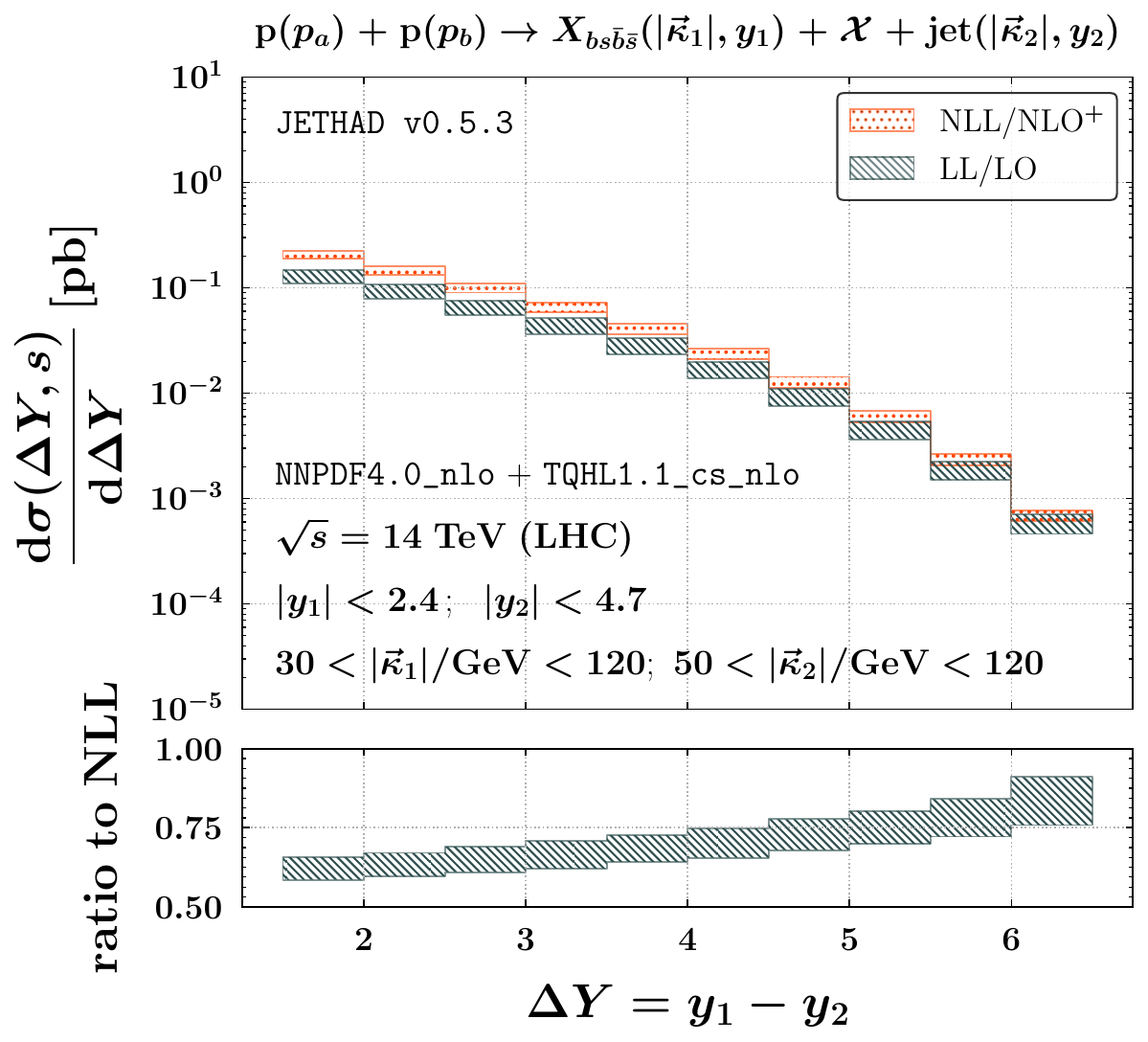}
   \vspace{0.50cm}
       \hspace{-0.4cm}
   \includegraphics[scale=0.38,clip]{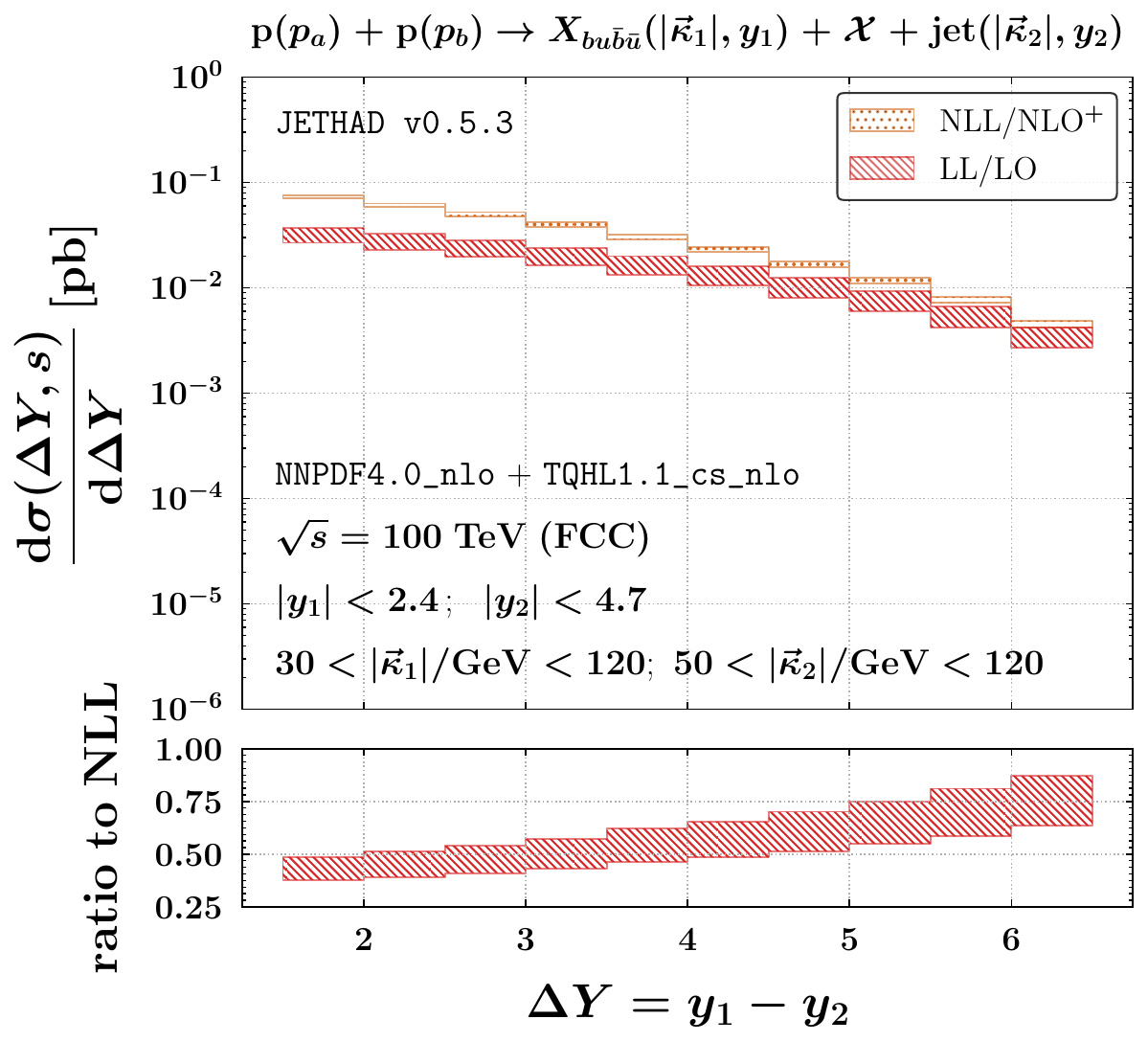}
   \includegraphics[scale=0.38,clip]{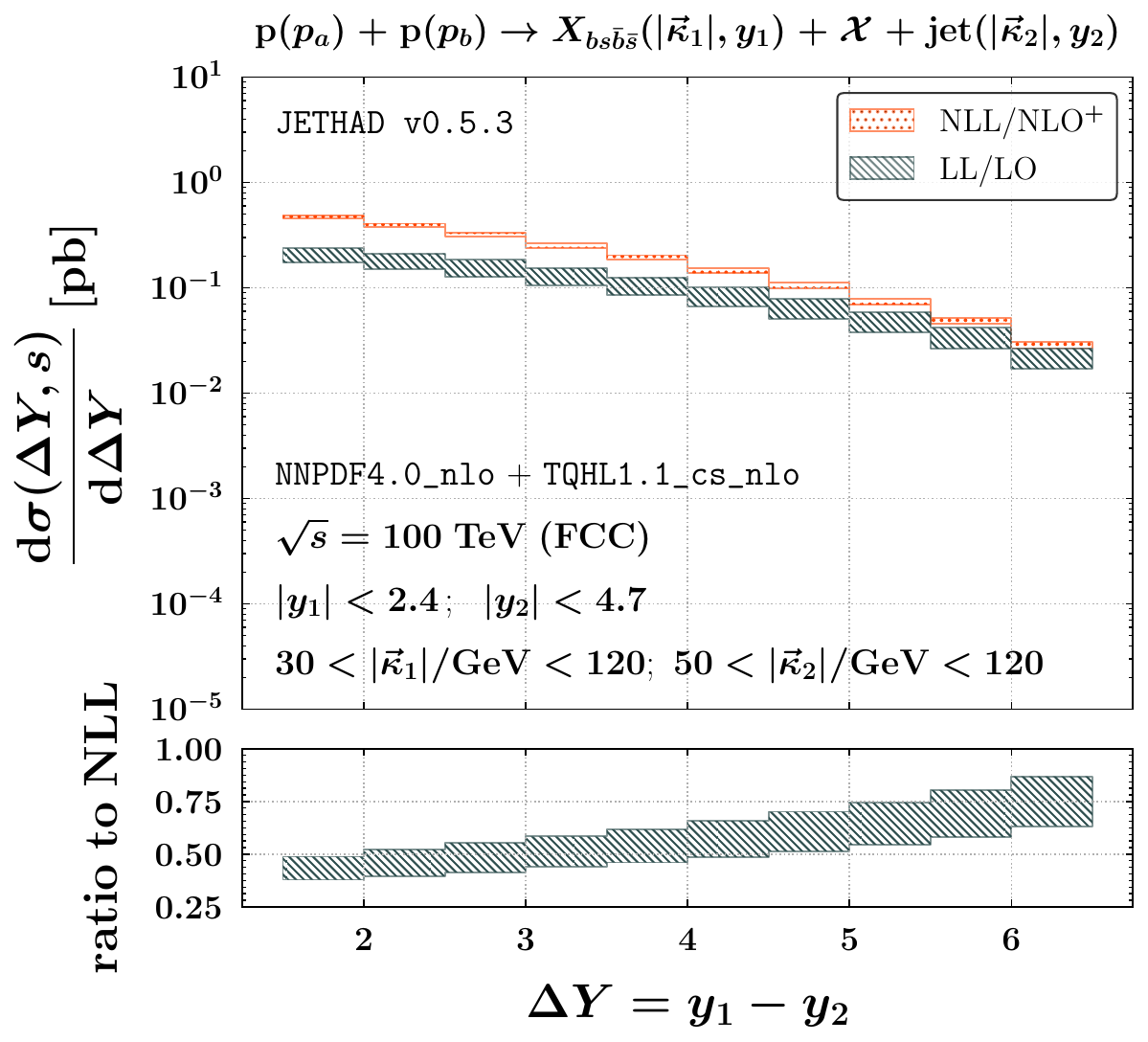}
\caption{Rapidity-interval distributions for $\QXbu$ (left) and $\QXbs$ (right) combined with jet hadroproduction are presented for $\sqrt{s} = 14$ TeV (LHC, top) and $100$ TeV (nominal FCC, bottom). The auxiliary panels beneath the main plots display the ratio of $\LL$ to $\NLLp$ predictions. Uncertainty bands represent the combined impact of MHOUs and the numerical integration over the multidimensional phase space.}
\label{fig:I_Xbq}
\end{figure*}

\begin{figure*}[!t]
   \hspace{-0.4cm}
   \includegraphics[scale=0.38,clip]{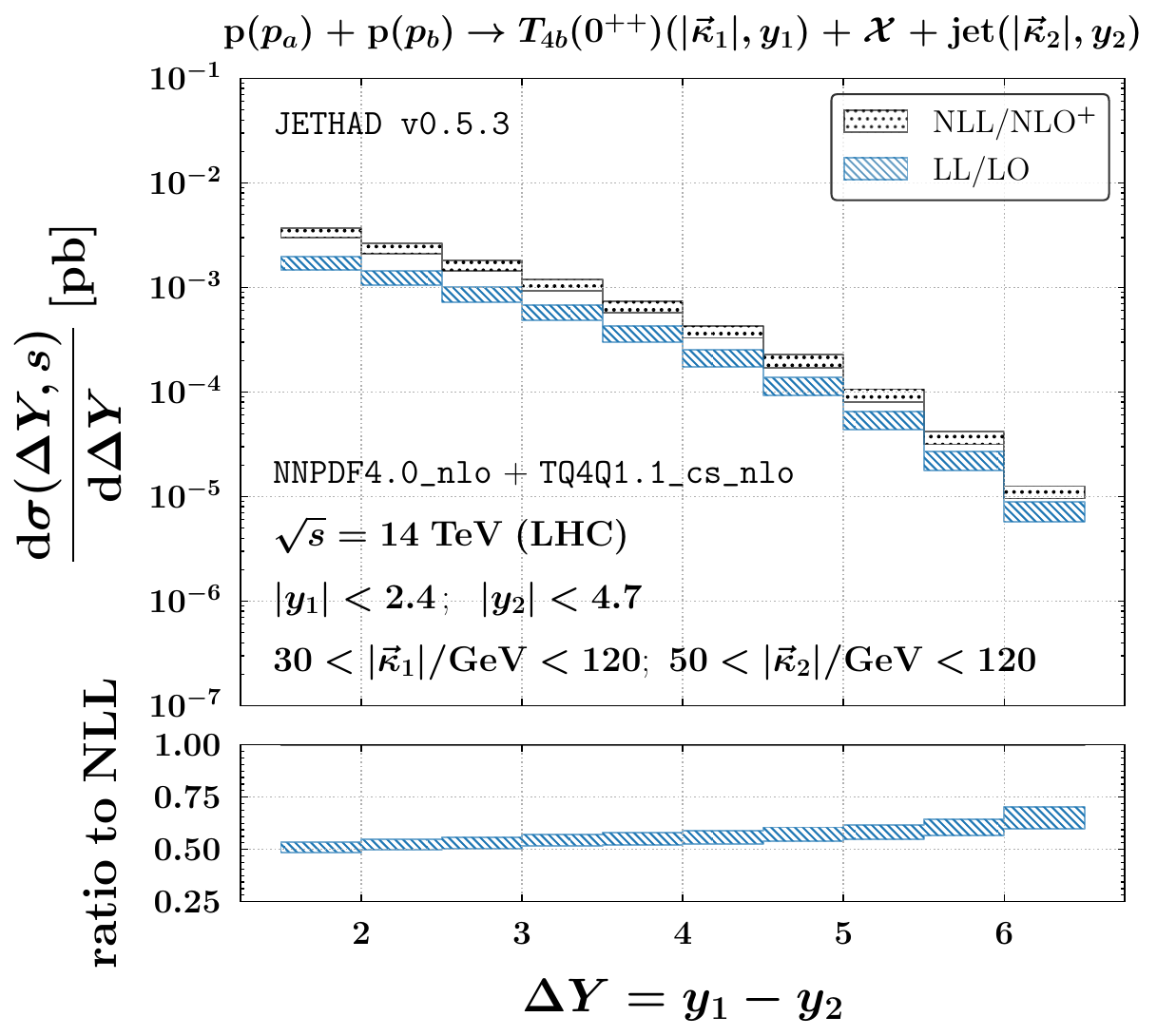}
   \includegraphics[scale=0.38,clip]{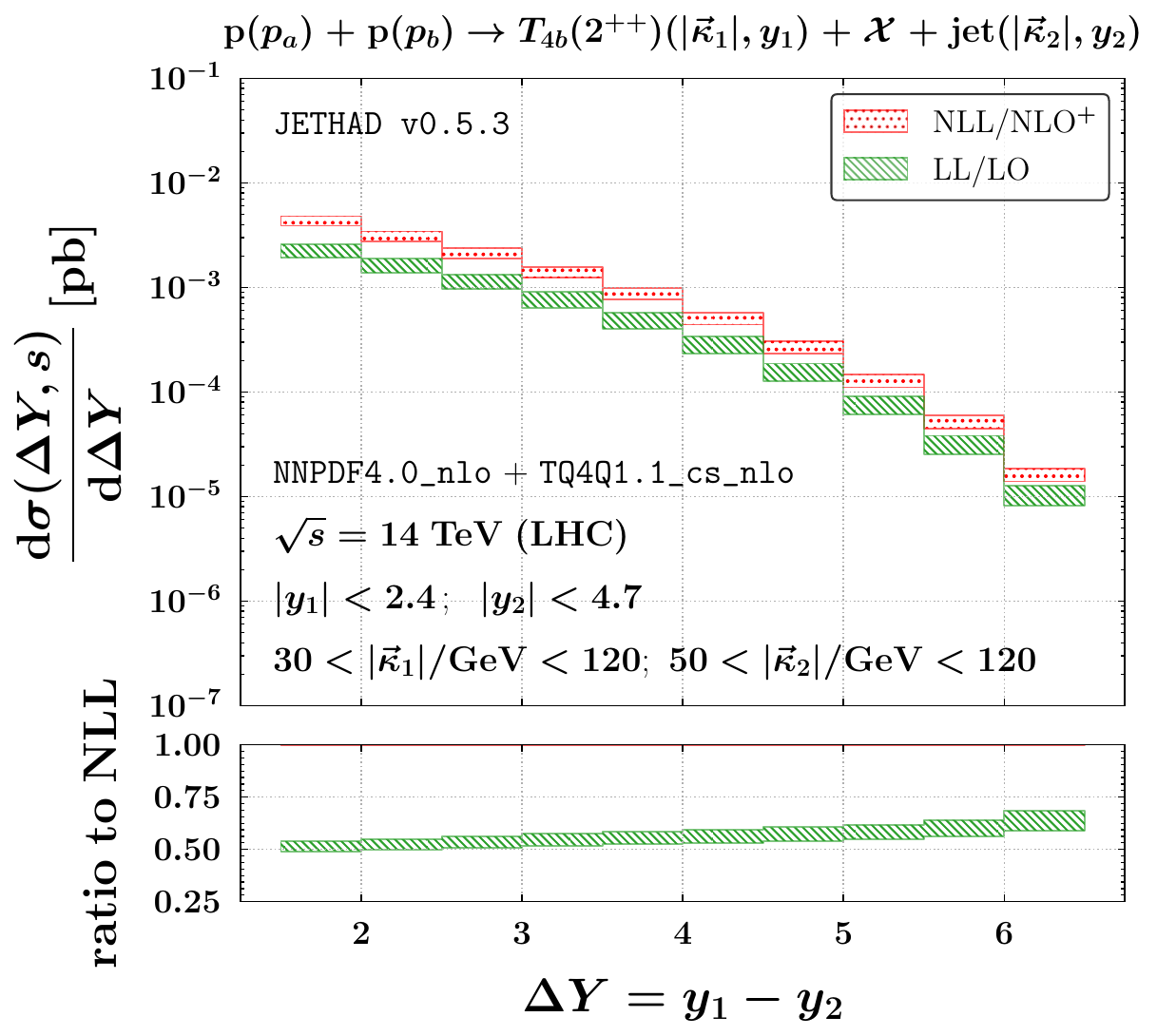}
   \vspace{0.50cm}
   \hspace{-0.4cm}
   \includegraphics[scale=0.38,clip]{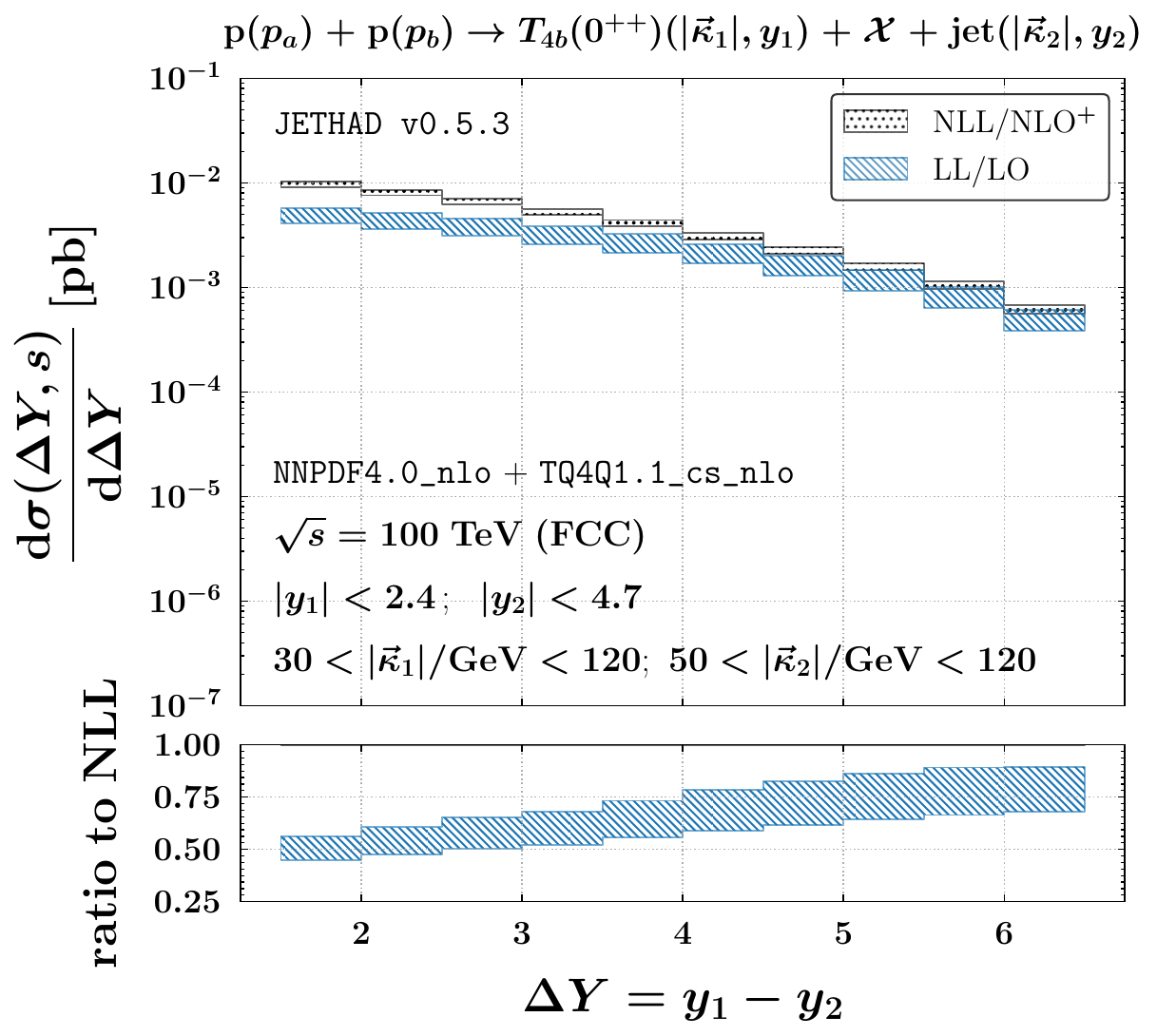}
   \includegraphics[scale=0.38,clip]{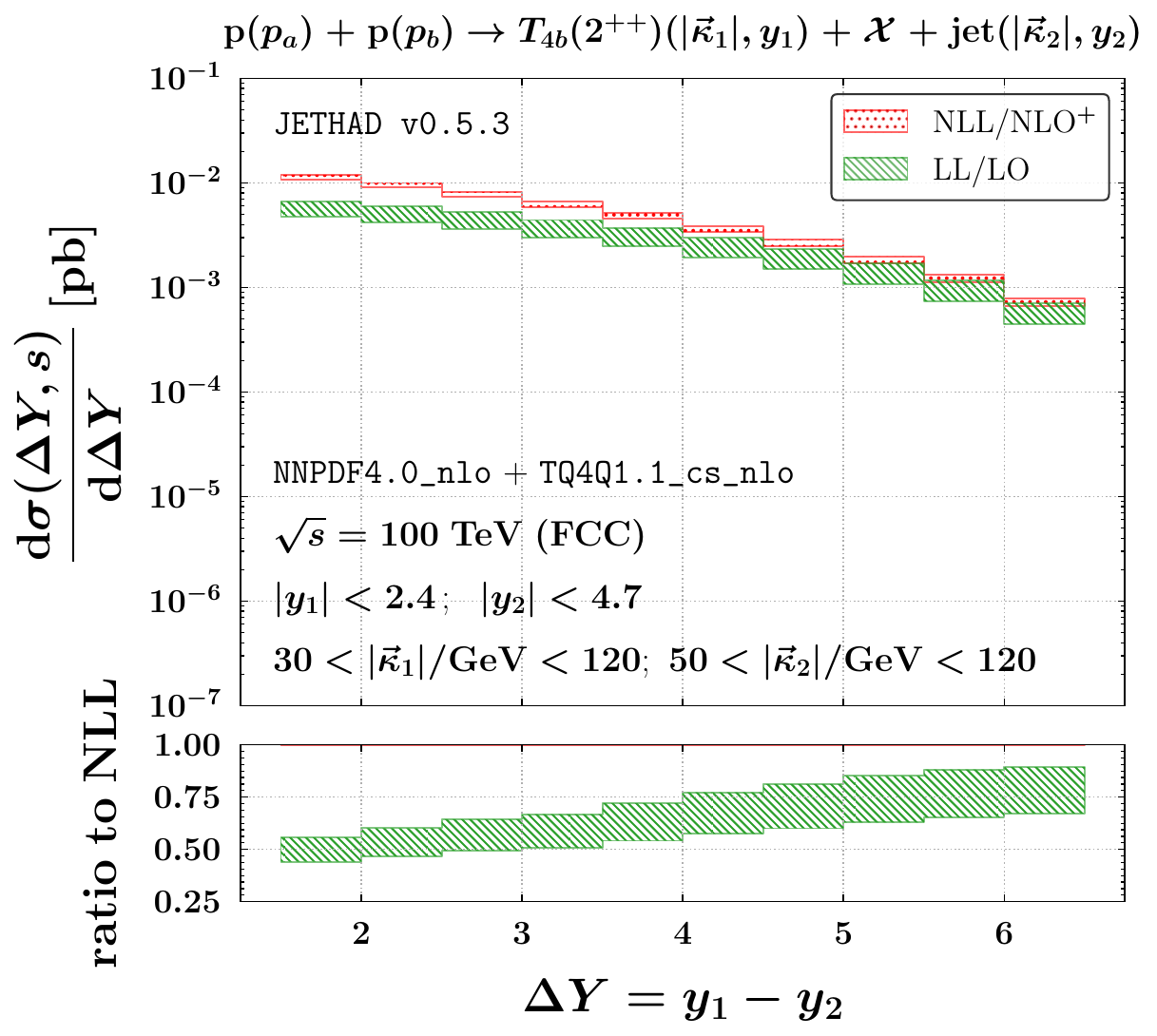}
\caption{Distributions of rapidity intervals for $\TQbZpp$ (left) and $\TQbTpp$ (right) combined with jet hadroproduction are shown for $\sqrt{s} = 14$ TeV (LHC, top) and $100$ TeV (nominal FCC, bottom). The supplementary panels beneath the main plots illustrate the ratio of $\LL$ to $\NLLp$ predictions. The uncertainty bands account for the combined influence of MHOUs and the numerical integration across the multidimensional phase space.}
\label{fig:I_T4b}
\end{figure*}

Figures~\ref{fig:I_Xbq} and~\ref{fig:I_T4b} display the rapidity-interval rates for doubly bottomed ($\QXbu$, $\QXbs$) and fully bottomed ($\TQbZpp$, $\TQbTpp$) tetraquarks, respectively, at $\sqrt{s} = 14$~TeV (LHC) and $100$~TeV (FCC). 
The ancillary panels show the ratio of leading-logarithm (LL) to next-to-leading-logarithm-plus ($\NLLp$) predictions. 
Rates generally decrease with $\DY$, due to a balance between the energy-dependent rise of the hard factor and the damping effect from collinear convolutions with PDFs and FFs.
\clearpage

\subsection{Transverse-momentum distributions}
\label{ssec:pT}
\begin{figure*}[!t]
   \hspace{-0.7cm}
   \includegraphics[scale=0.38,clip]{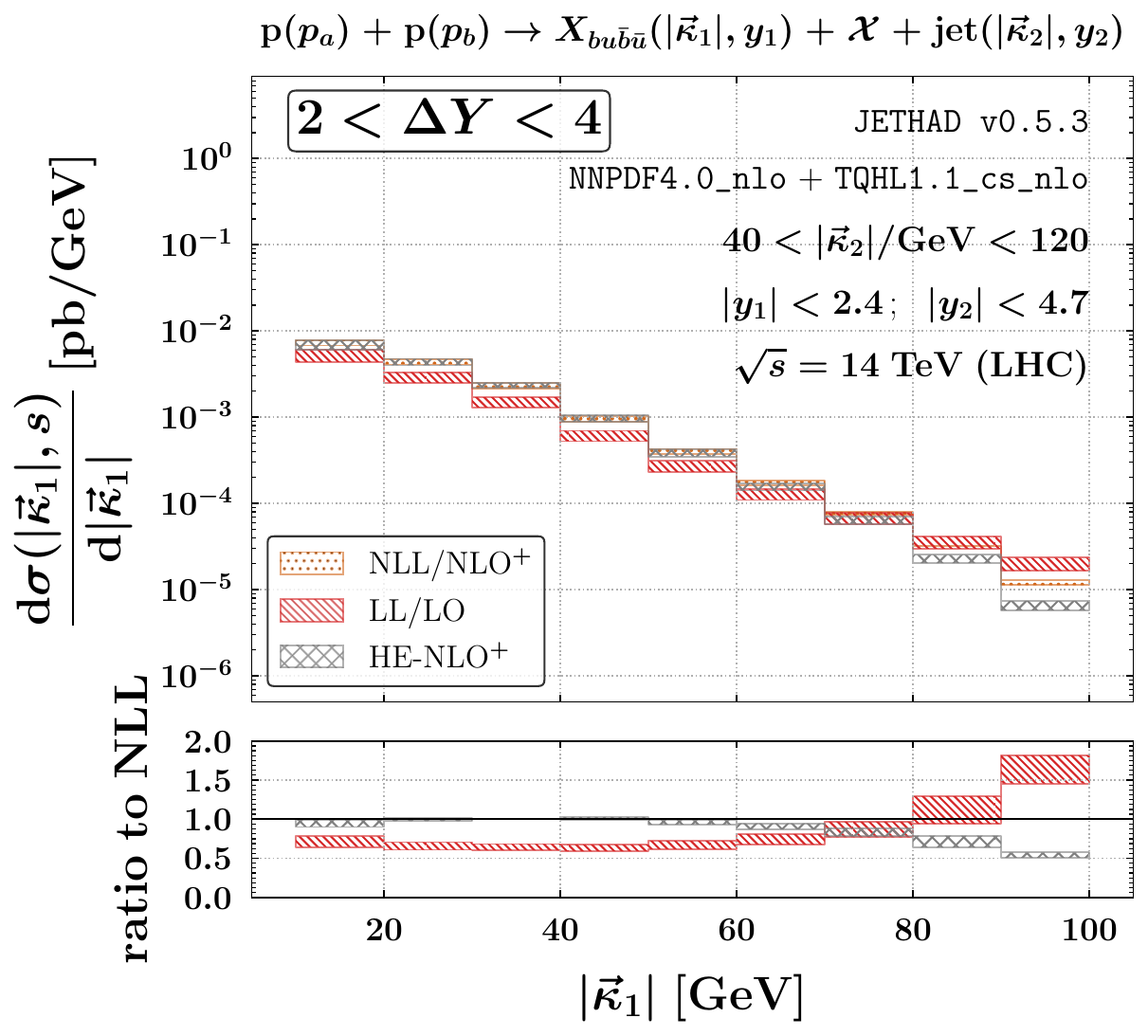}
   \includegraphics[scale=0.38,clip]{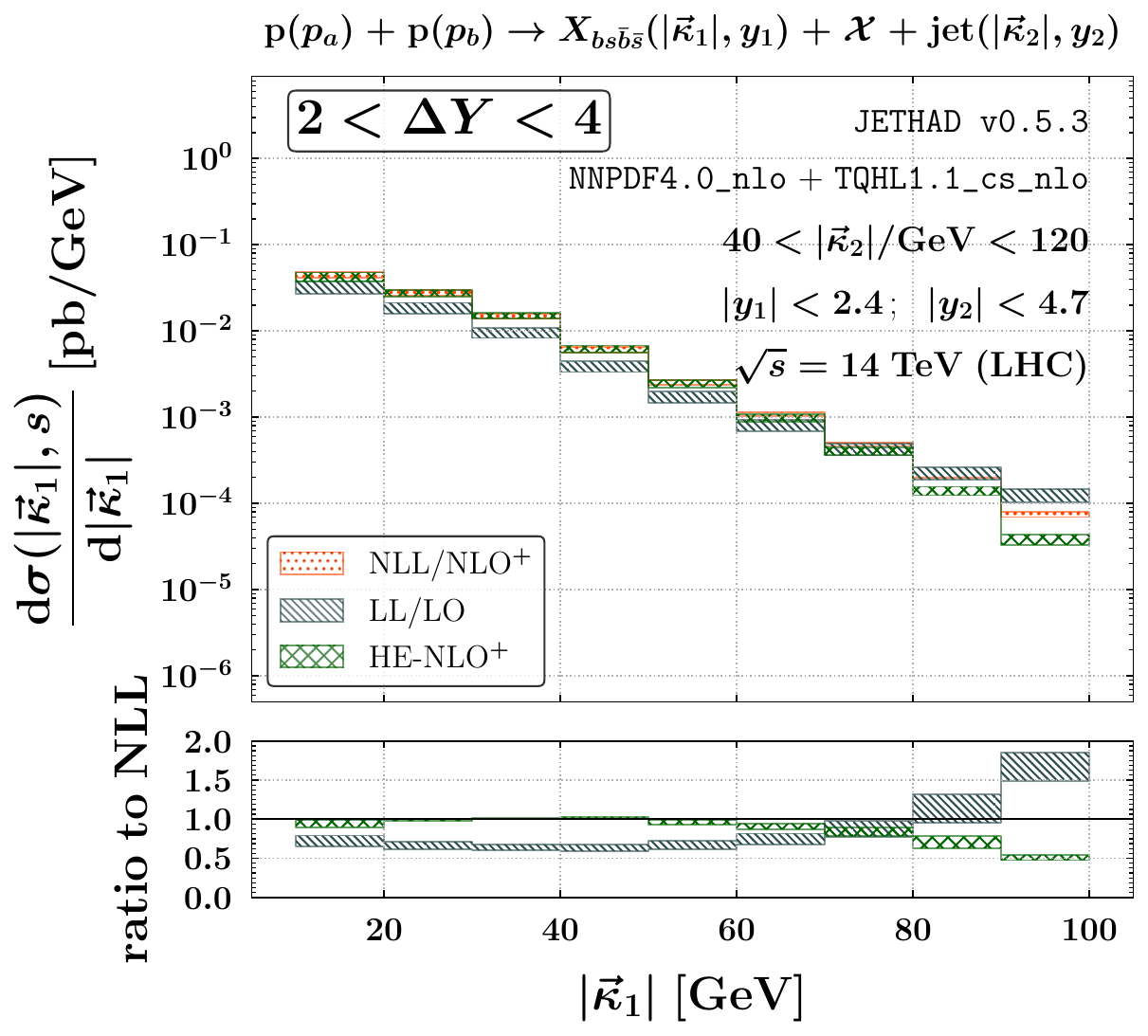}
   \vspace{0.50cm}
   \hspace{-0.7cm}
   \includegraphics[scale=0.38,clip]{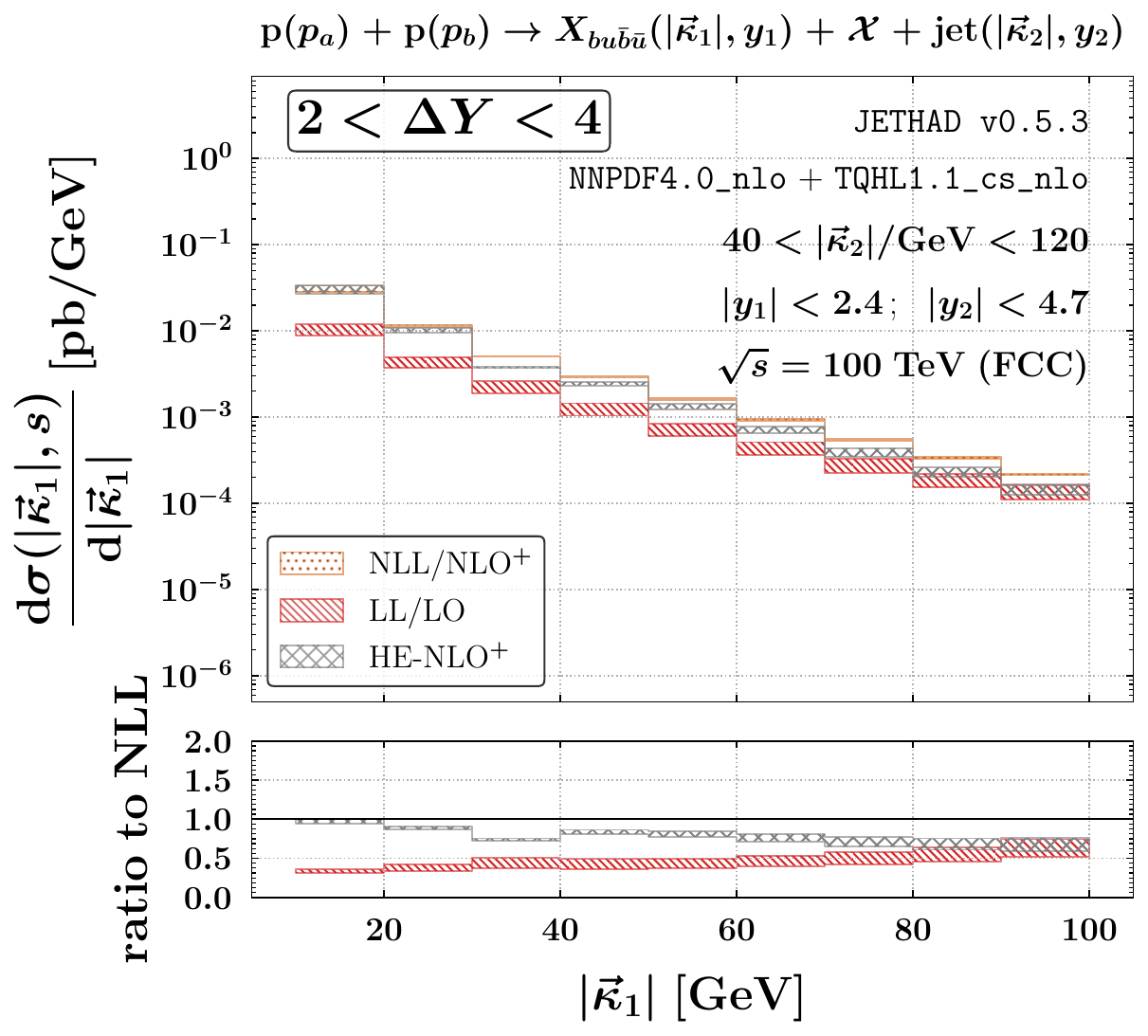}
   \includegraphics[scale=0.38,clip]{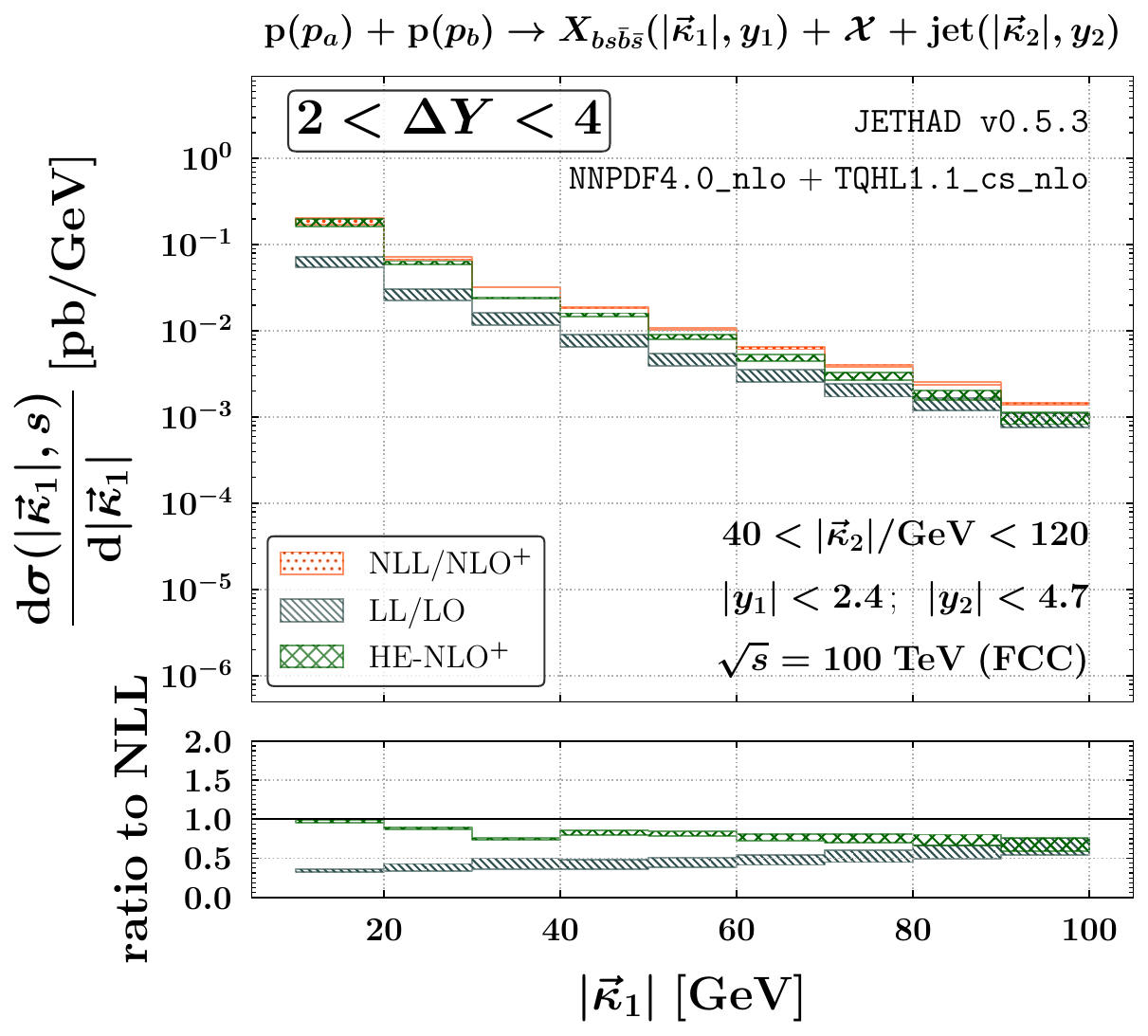}
\caption{Momentum distributions for $\QXbu$ (left panels) and $\QXbs$ (right panels) combined with jet production at $\sqrt{s} = 14$ TeV (upper plots, LHC) or $100$ TeV (lower plots, nominal FCC), within the range $2 < \DY < 4$. 
Secondary panels beneath the main graphs display the ratio of $\LL$ to $\NLLp$ predictions. The uncertainty bands account for the cumulative impact of MHOUs and multidimensional numerical integration over phase space.}
\label{fig:I-k1b-S_Xbq}
\end{figure*}

\begin{figure*}[!t]
   \hspace{-0.7cm}
   \includegraphics[scale=0.38,clip]{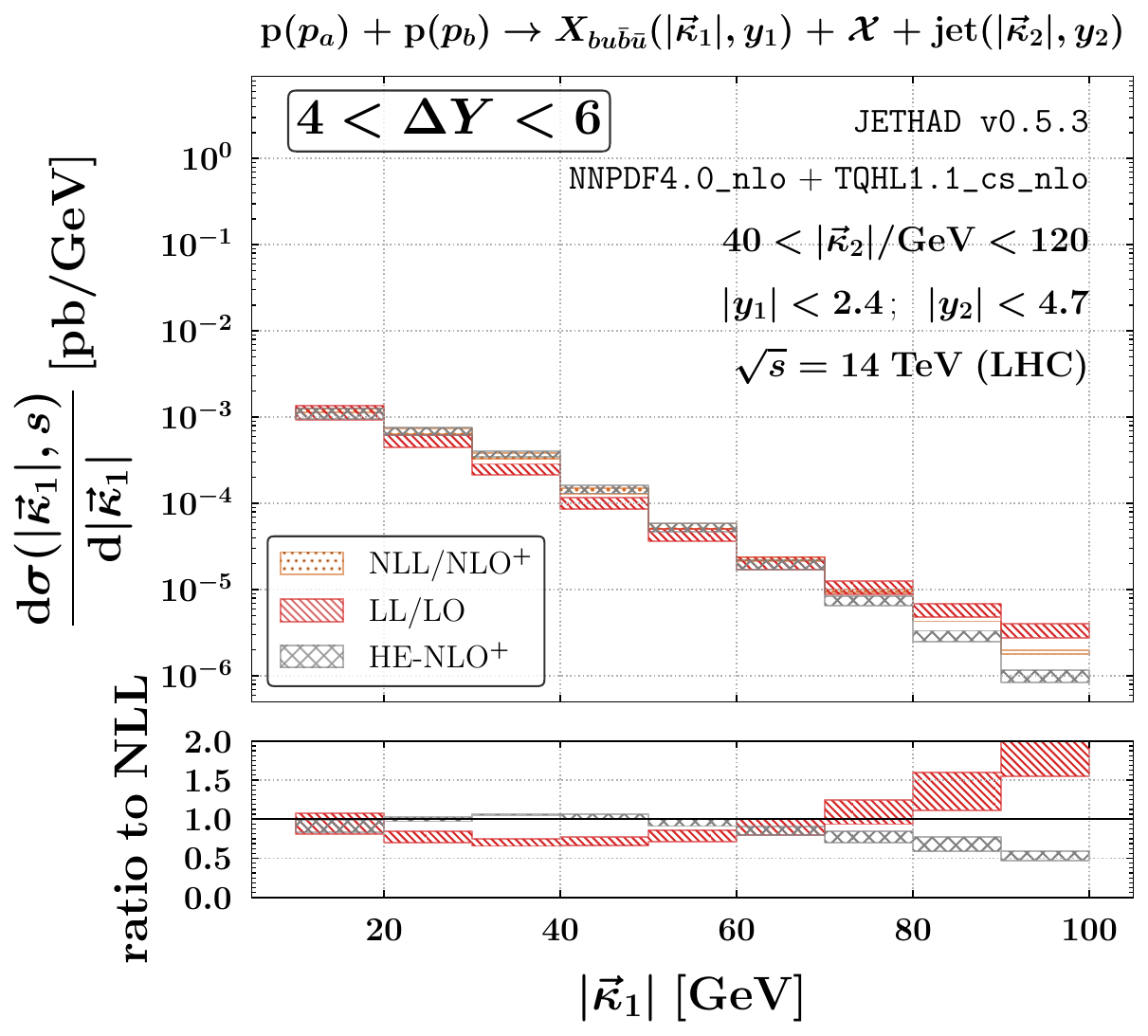}
   \includegraphics[scale=0.38,clip]{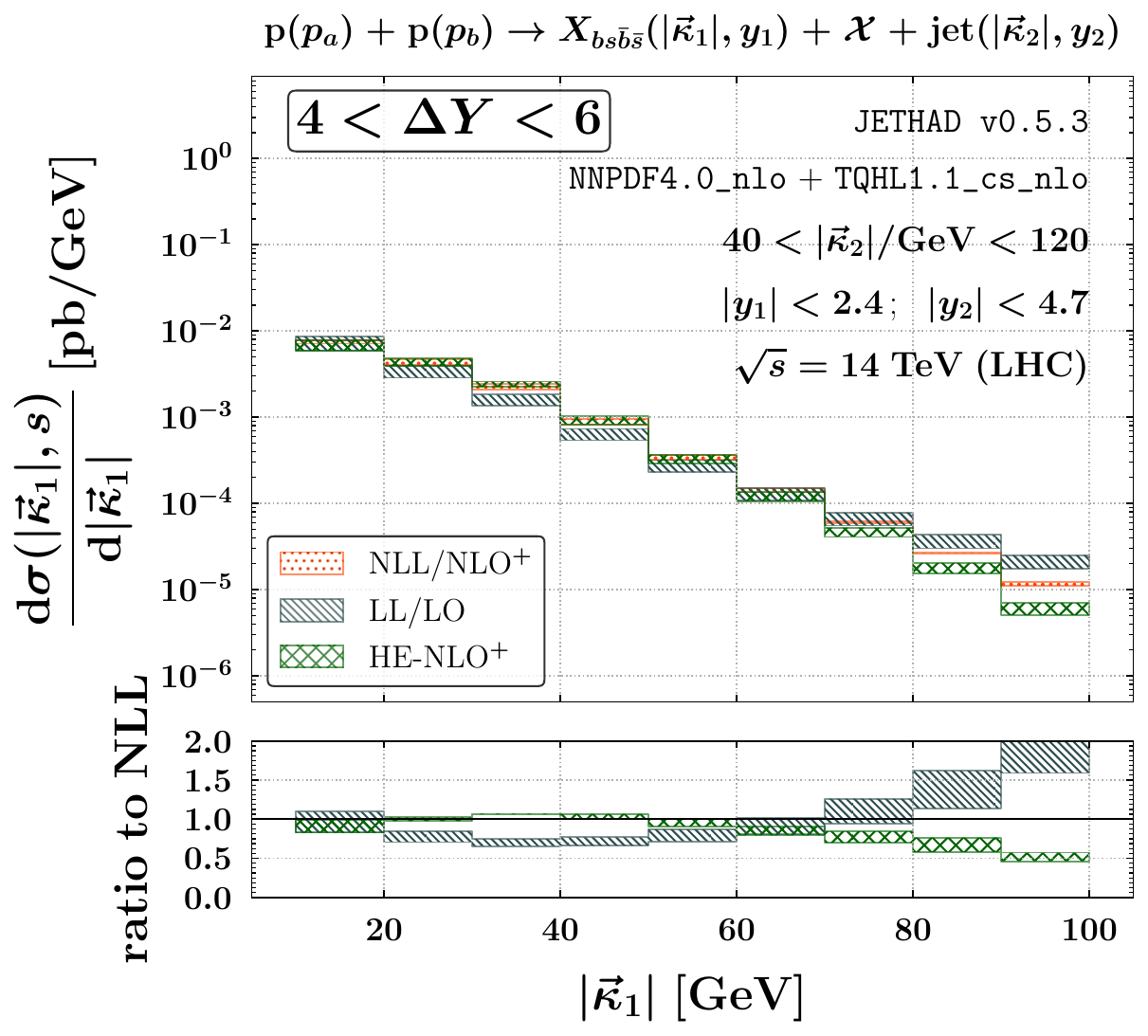}
   \vspace{0.50cm}
   \hspace{-0.7cm}
   \includegraphics[scale=0.38,clip]{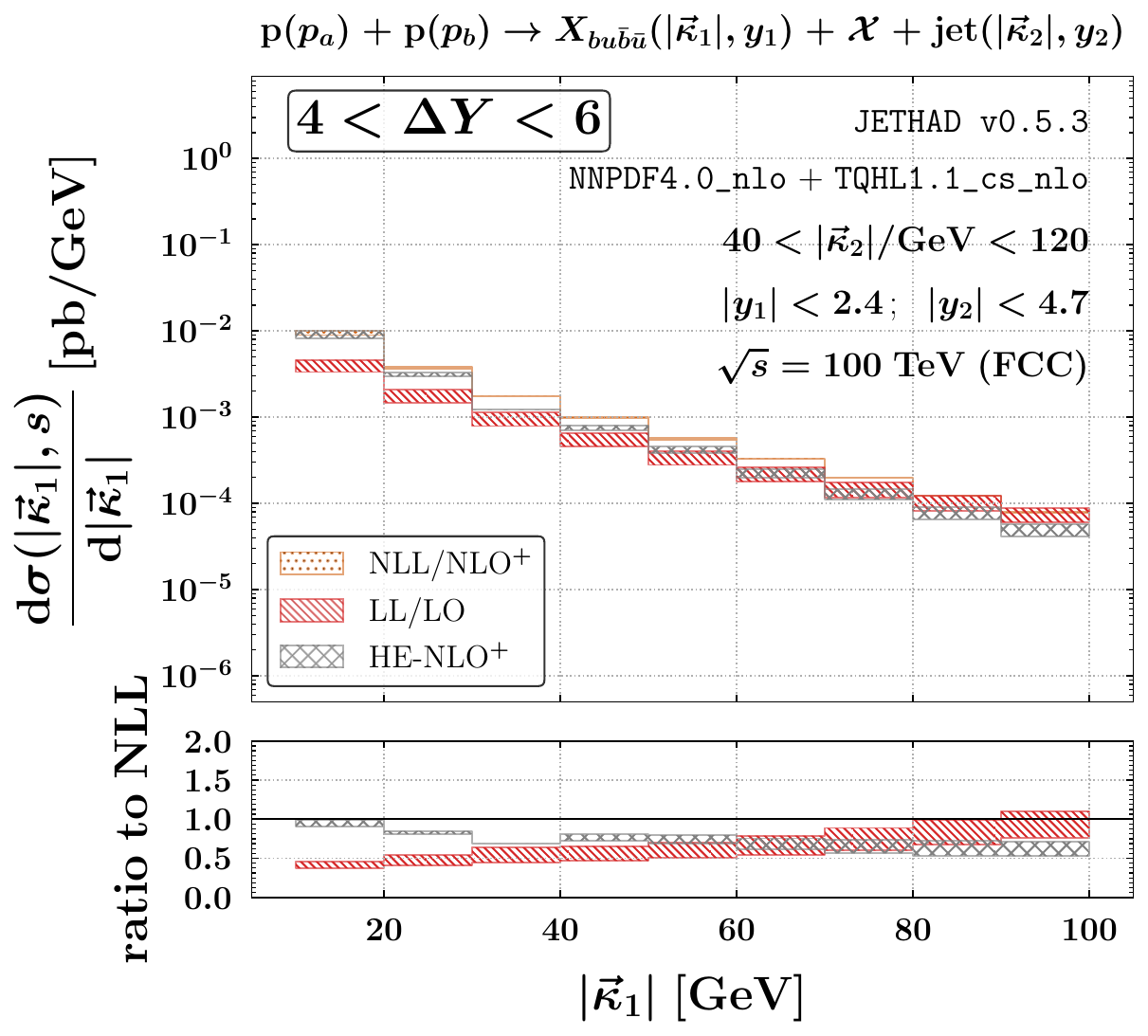}
   \includegraphics[scale=0.38,clip]{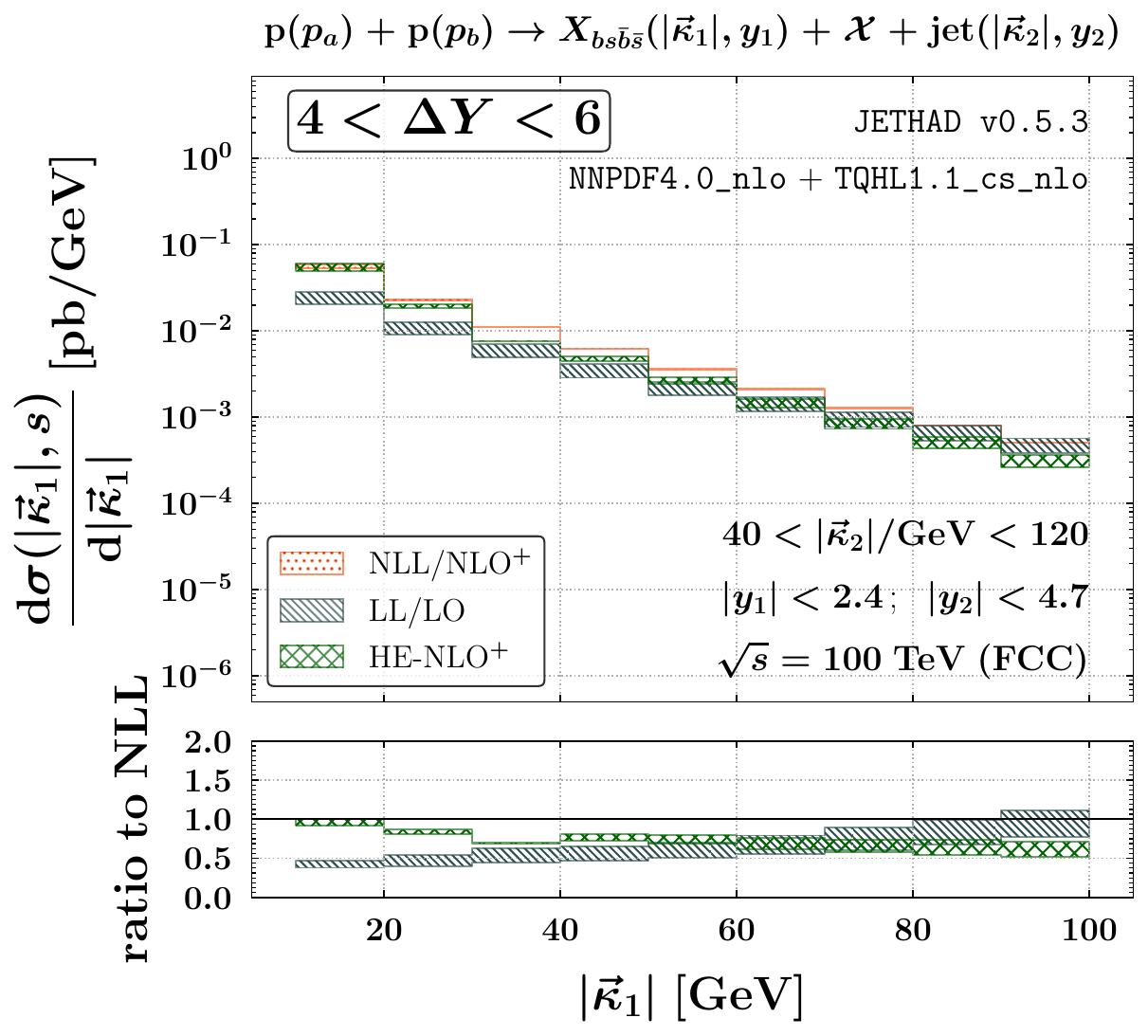}
\caption{Momentum distributions for $\QXbu$ (left panels) and $\QXbs$ (right panels) combined with jet production at $\sqrt{s} = 14$ TeV (upper plots, LHC) or $100$ TeV (lower plots, nominal FCC), within the range $4 < \DY < 6$. 
Supporting panels below the primary graphs present the ratio of $\LL$ to $\NLLp$ predictions. 
The uncertainty bands reflect the combined influence of MHOUs and the multidimensional numerical integration over the phase space.}
\label{fig:I-k1b-M_Xbq}
\end{figure*}

\begin{figure*}[!t]
   \hspace{-0.7cm}
   \includegraphics[scale=0.38,clip]{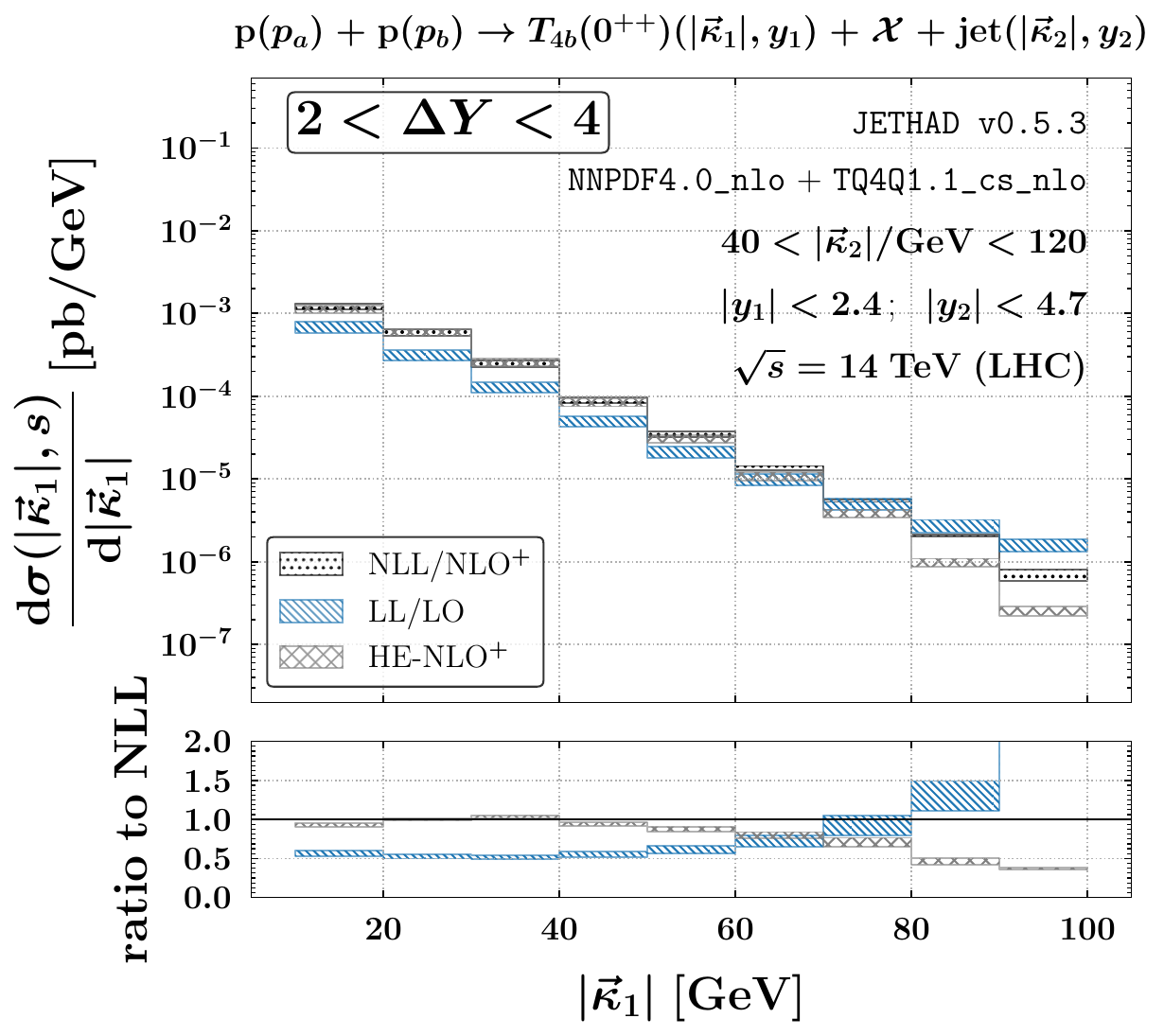}
   \includegraphics[scale=0.38,clip]{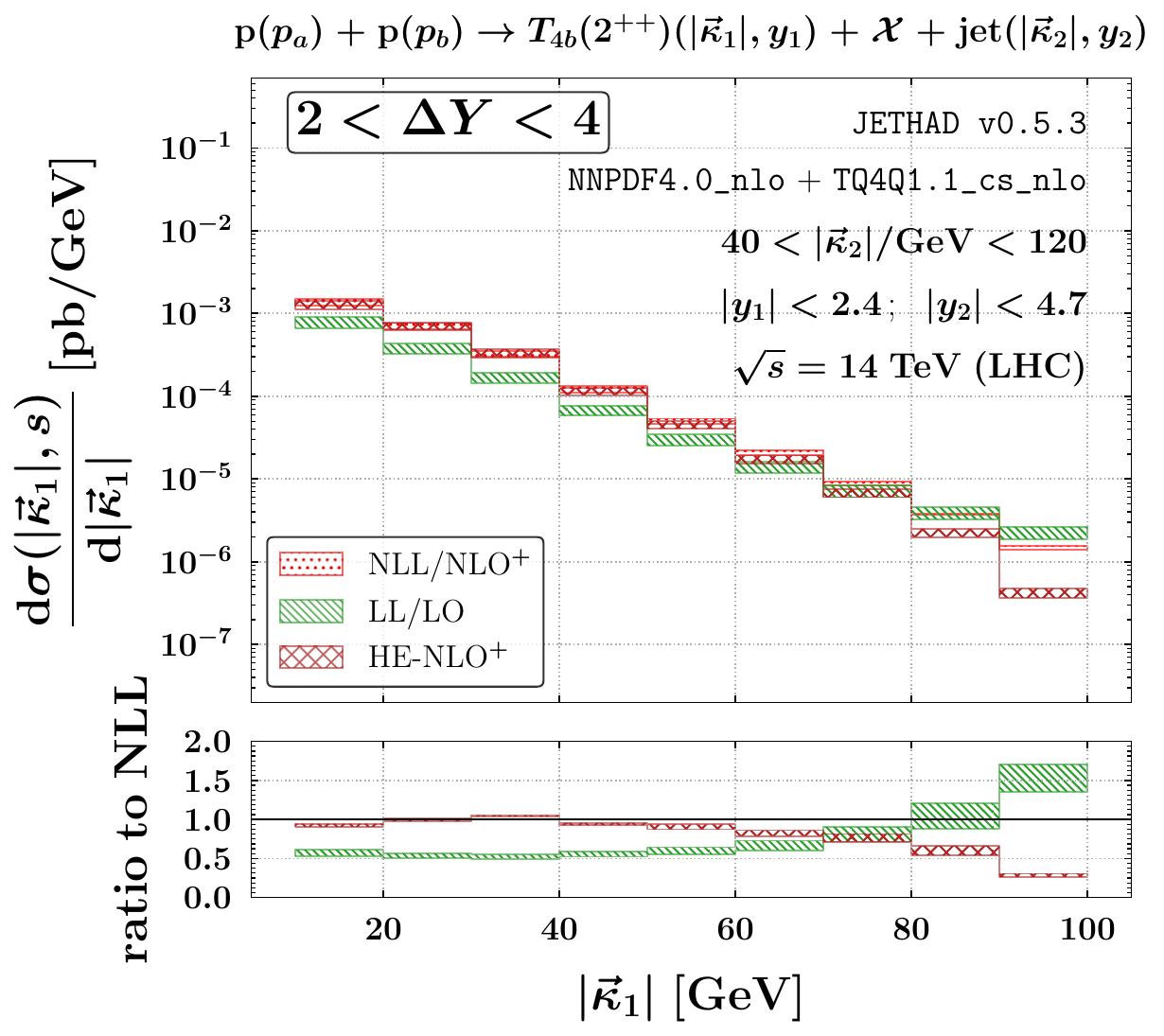}
   \vspace{0.50cm}
   \hspace{-0.7cm}
   \includegraphics[scale=0.38,clip]{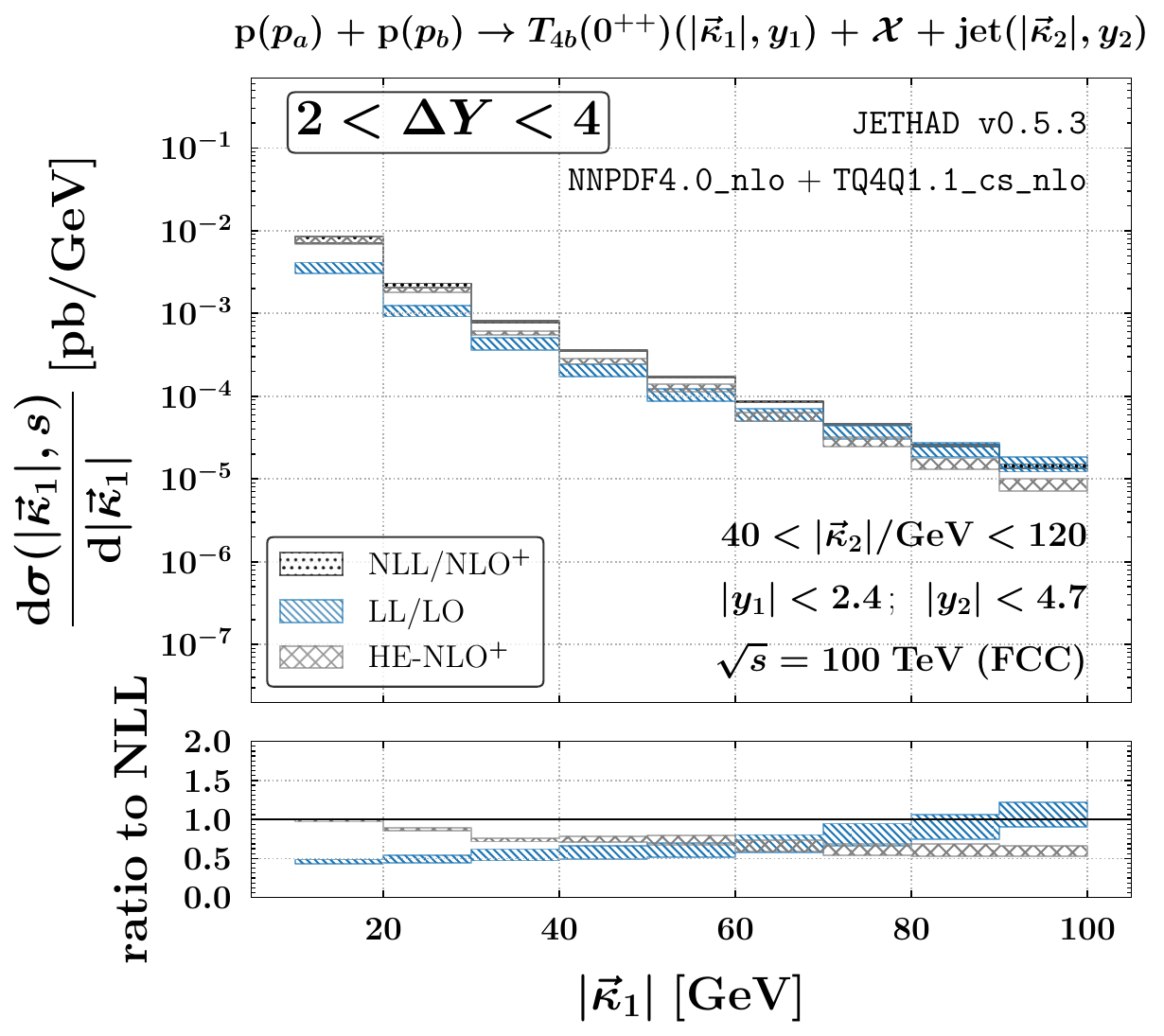}
   \includegraphics[scale=0.38,clip]{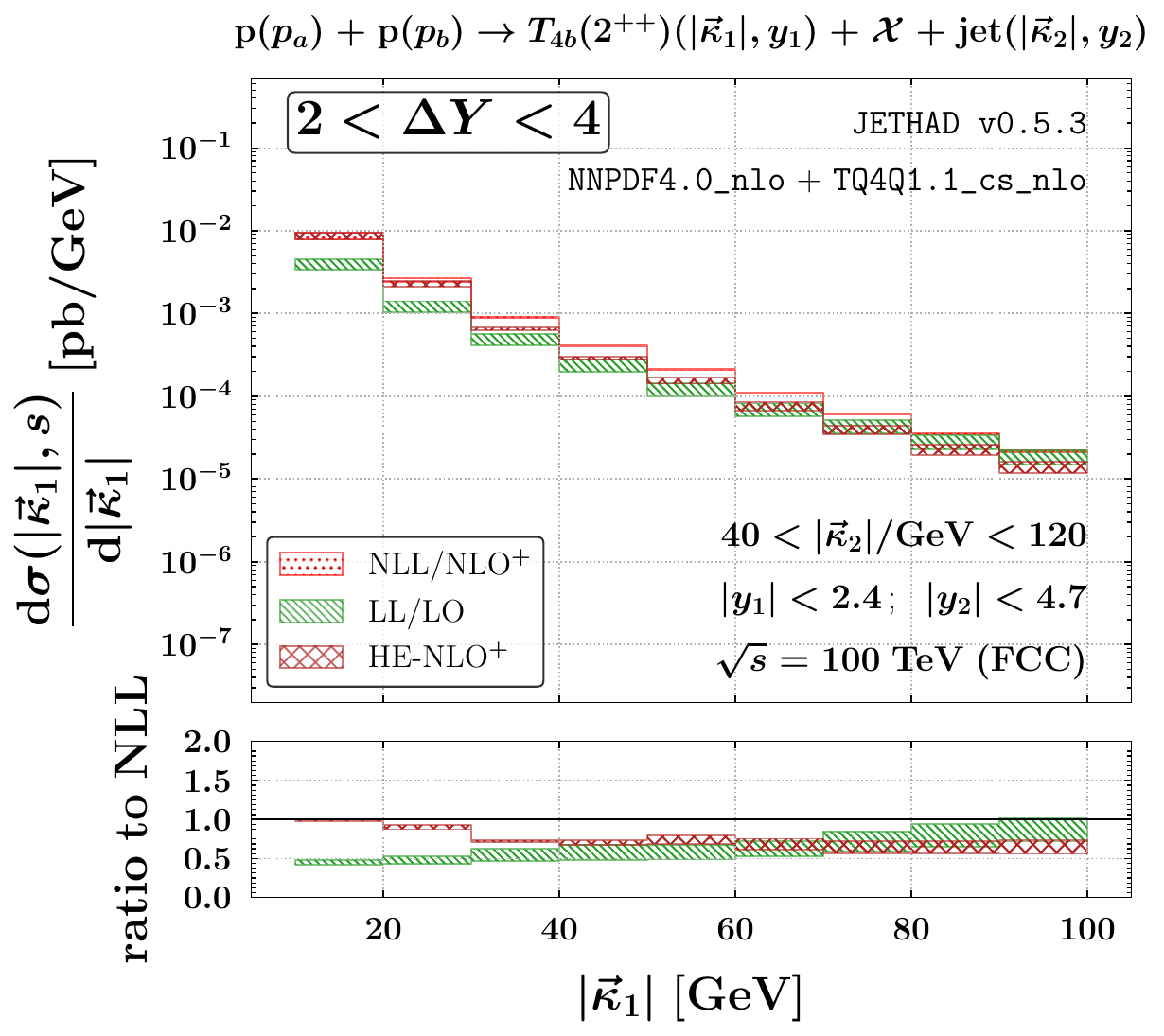}
\caption{Momentum distributions for $\TQbZpp$ (left panels) and $\TQbTpp$ (right panels) combined with jet production at $\sqrt{s} = 14$ TeV (upper plots, LHC) or $100$ TeV (lower plots, nominal FCC), within the range $2 < \DY < 4$. 
Auxiliary panels beneath the main graphs show the ratio of $\LL$ to $\NLLp$ predictions. 
The uncertainty bands encompass the overall impact of MHOUs and multidimensional numerical integration over phase space.}
\label{fig:I-k1b-S_T4b}
\end{figure*}

\begin{figure*}[!t]
   \hspace{-0.7cm}
   \includegraphics[scale=0.38,clip]{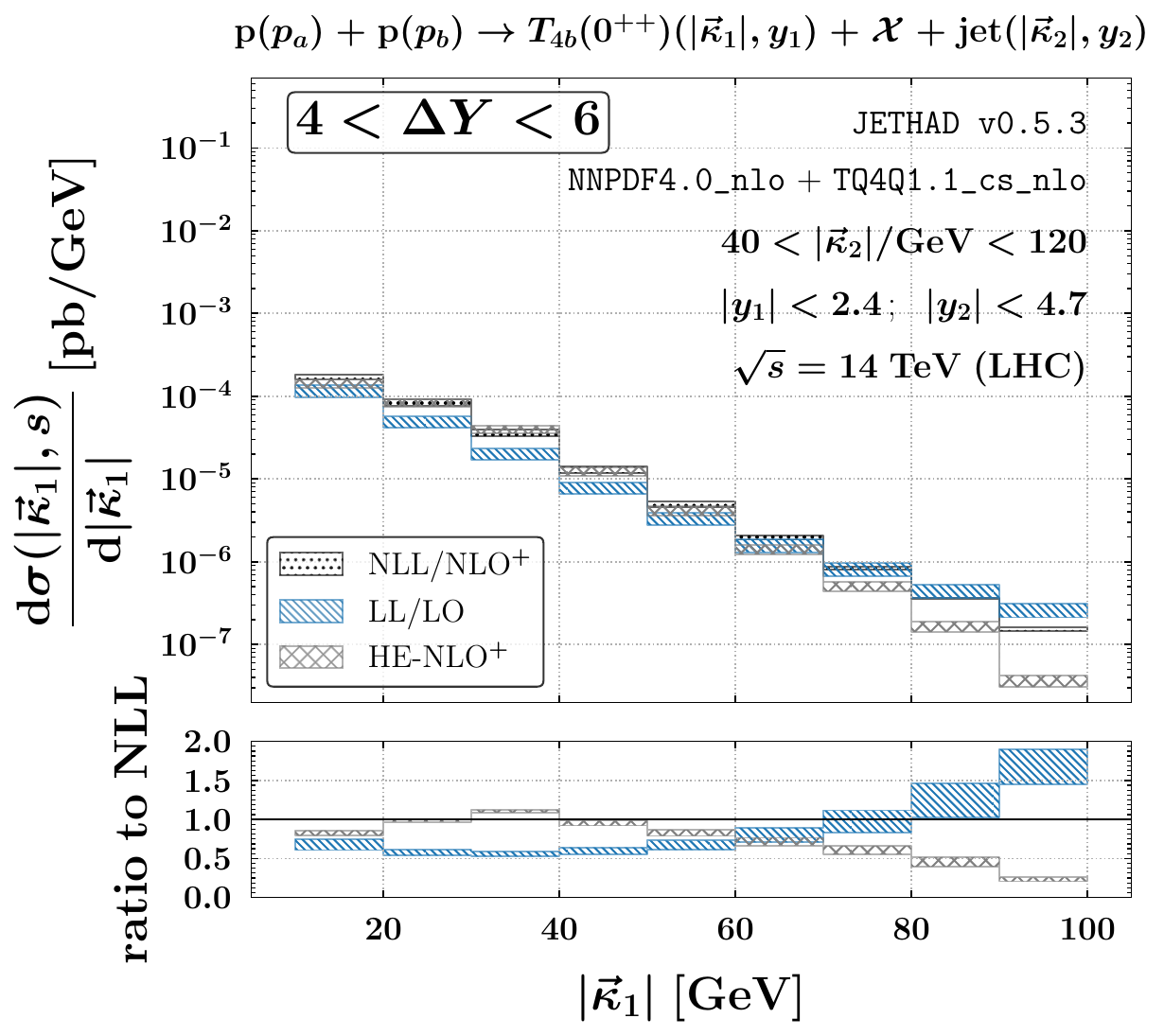}
   \includegraphics[scale=0.38,clip]{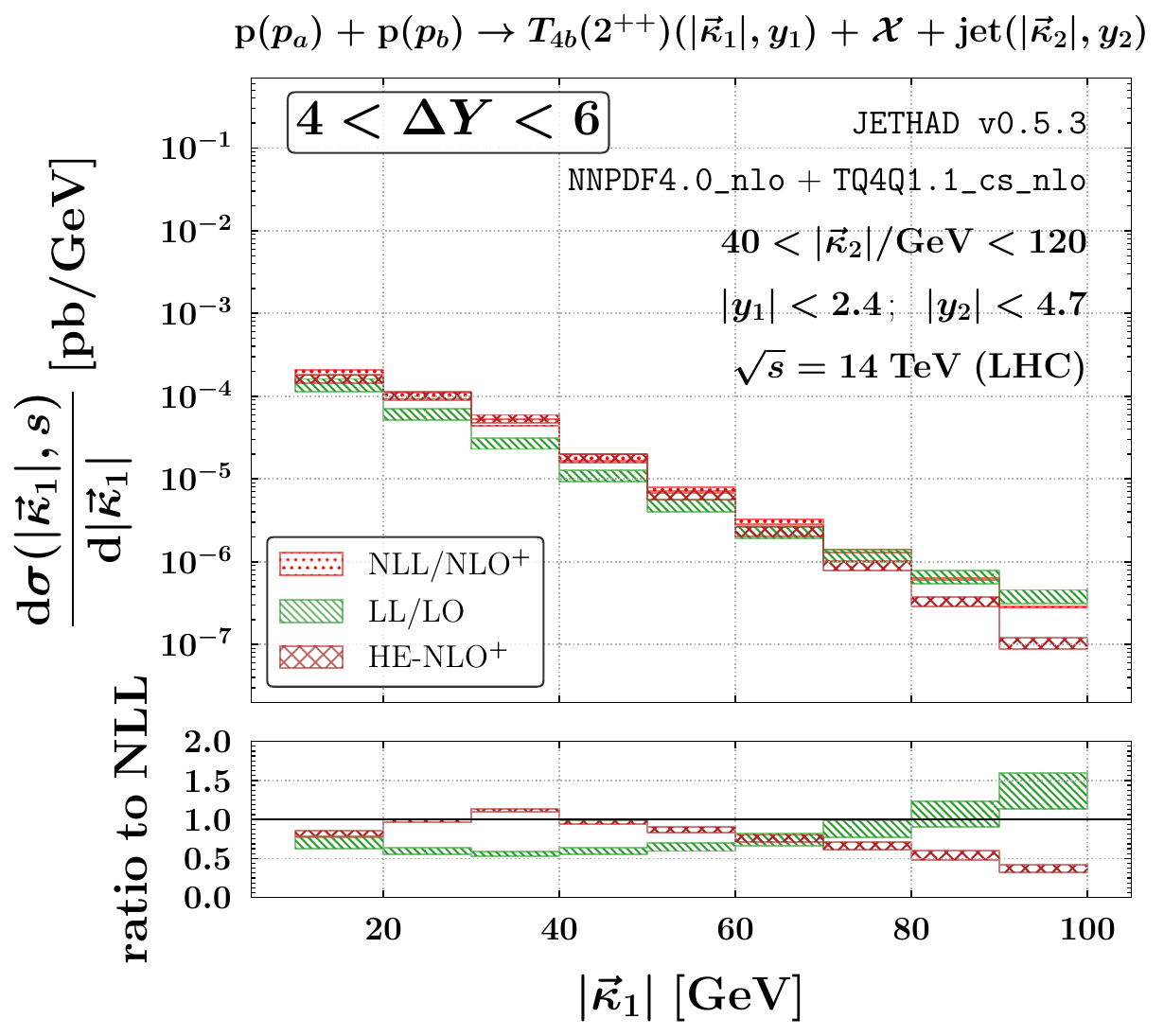}
   \vspace{0.50cm}
   \hspace{-0.7cm}
   \includegraphics[scale=0.38,clip]{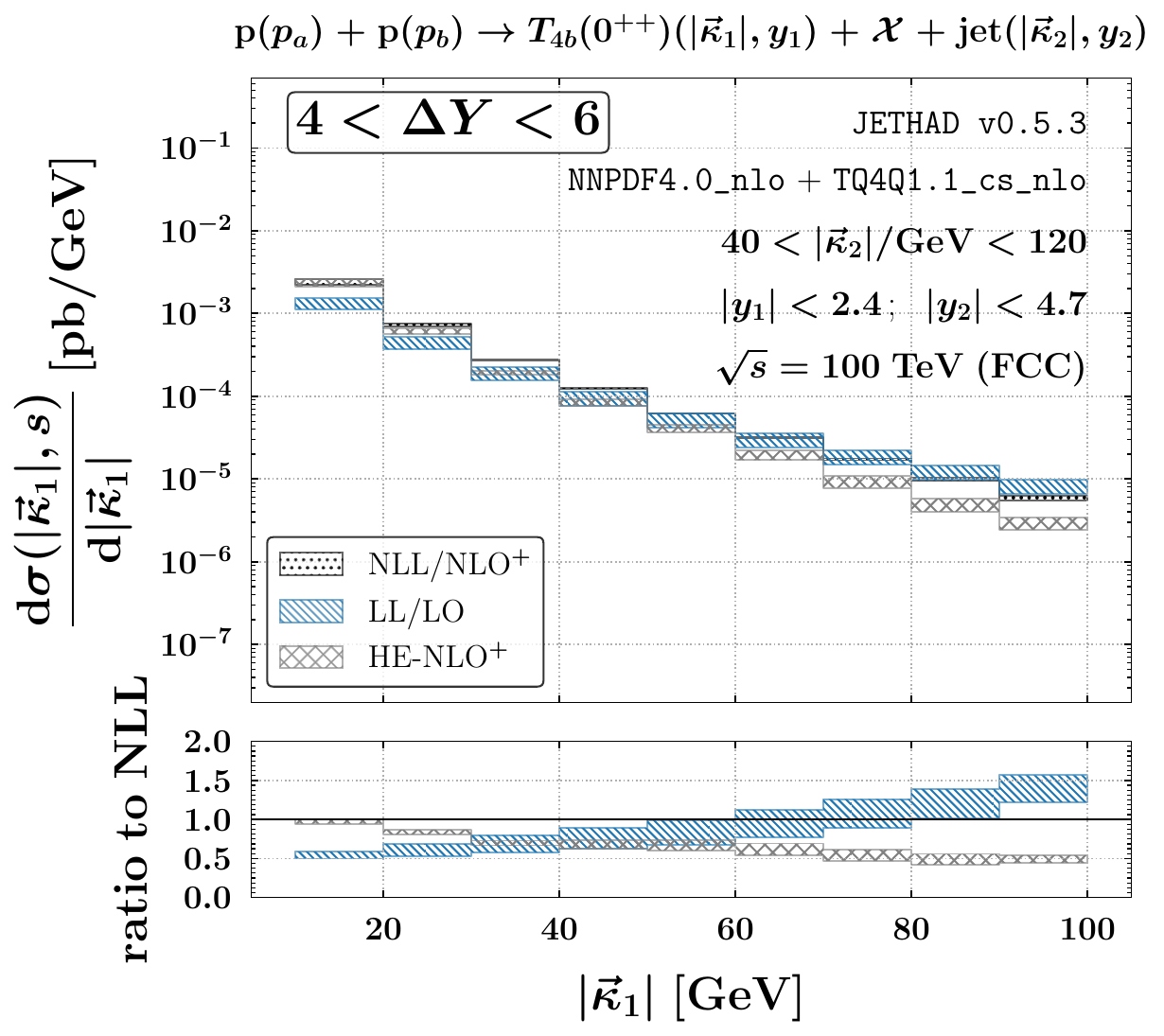}
   \includegraphics[scale=0.38,clip]{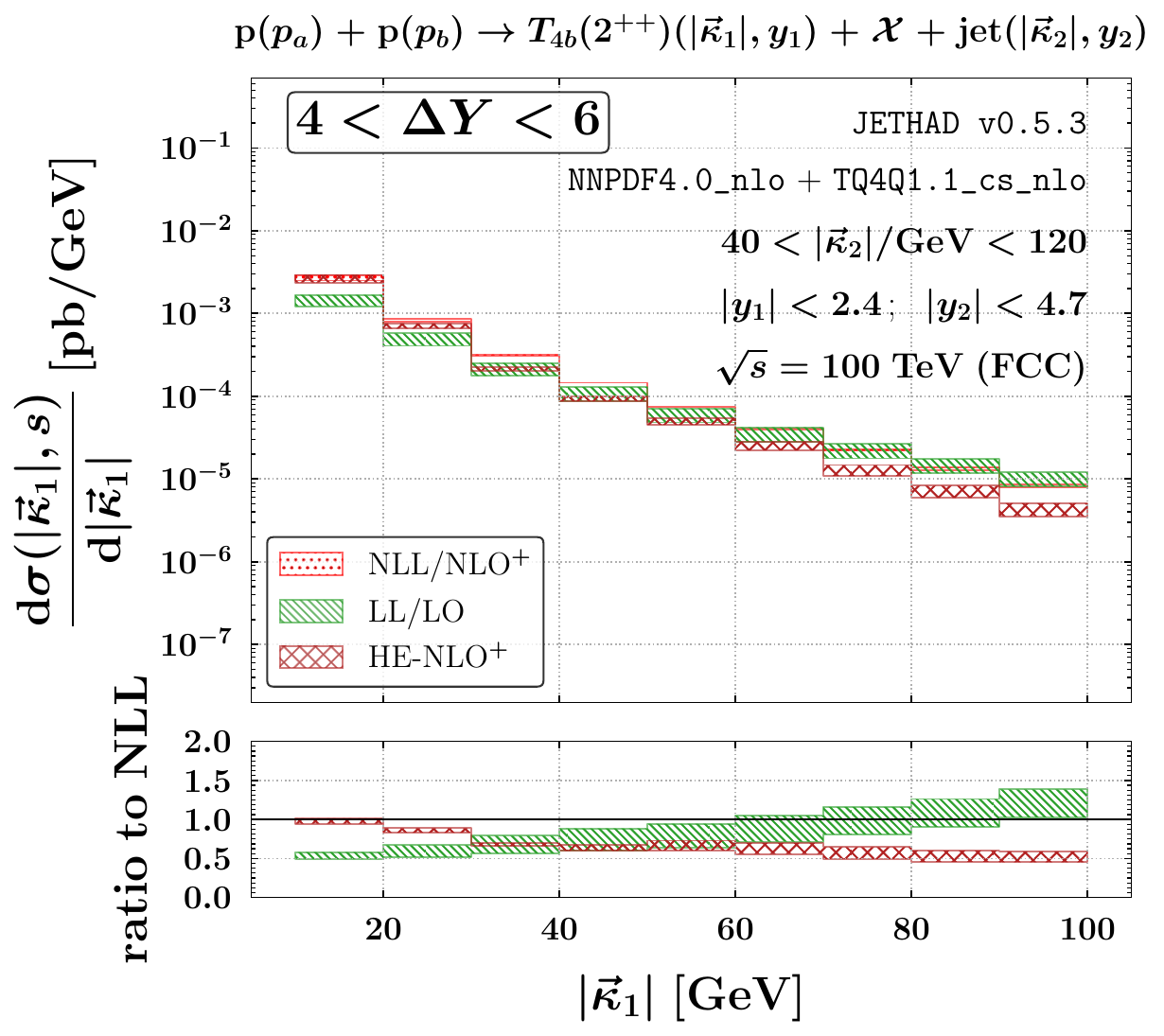}
\caption{Momentum distributions for $\TQbZpp$ (left panels) and $\TQbTpp$ (right panels) with jet production at $\sqrt{s} = 14$ TeV (upper plots, LHC) or $100$ TeV (lower plots, nominal FCC), within the range $4 < \DY < 6$. 
Supplementary panels below the primary plots display the ratio of $\LL$ to $\NLLp$ predictions. 
The uncertainty bands represent the combined effects of MHOUs and multidimensional phase-space numerical integration.}
\label{fig:I-k1b-M_T4b}
\end{figure*}

In this Section, we analyze the high-energy behavior of differential cross sections with respect to the transverse momentum of the tetraquark, $\vec \kappa_1$, and integrated over the jet transverse-momentum interval $40~\text{GeV} < |\vec{\kappa}_2| < 120~\text{GeV}$, as well as over two distinct $\Delta Y$ bins. Specifically, we define:
\begin{equation}
\begin{split}
\label{pT1_distribution}
 &\frac{\drv \sigma(|\vec \kappa_1|, s)}{\drv |\vec \kappa_1|} = 
 \int_{\DY^{\rm min}}^{\DY^{\rm max}} \drv \DY
 \int_{y_1^{\rm min}}^{y_1^{\rm max}} \drv y_1
 \int_{y_2^{\rm min}}^{y_2^{\rm max}}  \drv y_2
 \delta (\DY - (y_1 - y_2))
 \int_{|\vec \kappa_2|^{\rm min}}^{|\vec \kappa_2|^{\rm max}} 
 \!\!\drv |\vec \kappa_2|
 \, \,
 {\cal C}_{0}^{\rm [resum]}
 \;.
\end{split}
\end{equation}
Figures~\ref{fig:I-k1b-S_Xbq} and~\ref{fig:I-k1b-M_Xbq} present the outcomes for the $\QXbu$ (left plots) or $\QXbs$ (right plots) in addition to the light-jet $\vec \kappa_1$ distributions within the ranges $2 < \DY < 4$ and $4 < \DY < 6$, respectively.
Similarly, the graphs in Fig.~\ref{fig:I-k1b-S_T4b} and~\ref{fig:I-k1b-M_T4b} illustrate the predictions for the $\TQbZpp$ (left plots) or $\TQbTpp$ (right plots) combined with the light-jet $\vec \kappa_1$ distributions within $2 < \DY < 4$ and $4 < \DY < 6$, respectively.
The upper (lower) graphs in these figures correspond to results obtained at the 14~TeV~LHC (100~TeV~FCC).
We consistently use transverse-momentum bins with a size of 10~GeV.
Supplementary panels located directly beneath the main graphs emphasize the ratio between $\LL$ or $\HENLOp$ predictions and $\NLLp$ ones.\\

A common feature across all our distributions is the rapid decrease as $|\vec \kappa_1|$ increases. 
The findings demonstrate significant robustness to MHOUs, with uncertainty bands having a maximum width of 20\%.
We note that the $\HENLOp$ to $\NLLp$ ratios are generally less than one, decreasing further as $\vec \kappa_1$ becomes larger. 
This reduction is less pronounced at standard FCC energy levels compared to conventional LHC ones. 
Conversely, the $\LL$ to $\NLLp$ ratio displays an almost opposite behavior: it starts below one in the low-$|\vec \kappa_1|$ range but steadily increases as $|\vec \kappa_1|$ grows, eventually peaking at values between 1.5 and 2.

\section{Conclusion and outlook}
In this chapter, we investigated the semi-inclusive production of bottomonium-like tetraquarks, focusing on both doubly bottomed and fully bottomed states. Using a VFNS-fragmentation approach, we developed two new families of collinear fragmentation functions, {\tt TQHL1.1} and {\tt TQ4Q1.1}, improving upon earlier models. These functions were evolved using DGLAP equations to ensure consistency across energy scales.\\
These datasets extend and replace the corresponding {\tt 1.0} versions presented in recent studies~\cite{Celiberto:2023rzw,Celiberto:2024mab}. The {\tt TQHL1.1} series models the fragmentation of $\QXbq$ tetraquarks, developed using an enhanced version of the Suzuki model tailored for the heavy-quark channel. On the other hand, the {\tt TQ4Q1.1} series represents the fragmentation of $\TQb$ states, incorporating initial-scale inputs for both the gluon and heavy-quark channels. These inputs were derived within the framework of potential NRQCD. To ensure consistency at threshold, the inputs underwent DGLAP evolution, leveraging the fundamental capabilities of the innovative {\HFNRevo} methodology~\cite{Celiberto:2024mex,Celiberto:2024bxu,Celiberto:2024rxa}.\\
We applied a hybrid $\NLLp$ formalism to predict rapidity and transverse momentum distributions at energies ranging from the LHC to the FCC. Our approach demonstrated natural stability and convergence, highlighting the robustness of our methodology. Future work will integrate additional resummation techniques~\cite{Hatta:2020bgy,Hatta:2021jcd,Caucal:2022ulg,Taels:2022tza,Dasgupta:2014yra,Dasgupta:2016bnd,Banfi:2012jm,Banfi:2015pju,Liu:2017pbb,Luisoni:2015xha,Caletti:2021oor,Reichelt:2021svh} and extend the study to single inclusive detections at forward rapidity, probing the small-$x$ dynamics in the proton.
This research contributes to the understanding of exotic matter formation and paves the way for future explorations at upcoming colliders. Further improvements will include quantifying uncertainties, incorporating color-octet contributions, and investigating potential intrinsic bottom components\footnote{The recent discovery~\cite{Ball:2022qks} of intrinsic charm quarks~\cite{Brodsky:1980pb,Brodsky:2015fna} within the proton (see also Refs.~\cite{Jimenez-Delgado:2014zga,Ball:2016neh,Hou:2017khm,Guzzi:2022rca} for additional studies) and the latest findings on its valence distribution~\cite{NNPDF:2023tyk} open new avenues for investigating the possible presence of an intrinsic bottom component. This development expands our understanding of both conventional and exotic bottom physics.}, broadening our knowledge of both ordinary and exotic bottom physics.

\chapter{Diffractive di-hadron production in the saturation regime}
\chaptermark{Production in the saturation regime}

\section{The saturation regime and the CGC}
The BFKL equation is renowned for its ability to predict the rapid increase in the $\gamma^{*} p$ cross section as the center-of-mass energy $s$ grows. In the regime of high $s$, which corresponds to the small Bjorken-$x$ limit, the most effective framework for studying QCD processes is $k_t$-factorization. Considering the total cross section for Deep-Inelastic-Scattering (DIS), we have
\begin{align}
\sigma_{\gamma^{*} P}(x) = \frac{1}{(2 \pi)^{D-2}} \hspace{-0.1 cm}&\int \hspace{-0.1 cm} \frac{d^{D-2} k_{\perp}}{\vec{k}^{2}} \frac{\Phi_{\gamma^{*} \gamma^{*}} (\vec{k} \; )}{\vec{k}^{2}} \hspace{-0.1 cm} \int \hspace{-0.1 cm} \frac{d^{D-2} k_{\perp}'}{\vec{k}^{'2}} \frac{\Phi_{P P}(-\vec{k}')}{\vec{k}^{'2}} \hspace{-0.1 cm}\nonumber\\ \times\,&\int_{\delta-i\infty}^{\delta+i\infty} \hspace{-0.05 cm} \frac{d\omega}{2 \pi i} \left( \frac{1}{x} \right)^{\omega}  G_{\omega}(\vec{q}_A, \vec{q}_B) ,
\label{Int:Eq:TotCrossDIS}
\end{align}  
where $ x = {Q^2}/{s} $. It is essential to clarify the presence of two impact factors in this equation: one describes the $\gamma^{}$-$\gamma^{}$ transition, while the other corresponds to the proton-proton transition. The latter encapsulates non-perturbative contributions and cannot be treated like the parton impact factors discussed earlier. By modeling this impact factor based on physical arguments, we define its convolution with the Green function in the transverse momentum $\vec{k}'$ as $\mathcal{F}(x, \vec{k})$, leading to:
\begin{equation}
\label{Int:Eq:RiseOfCrossSection}
    \sigma_{\gamma^{*} P}(x) = \Phi_{\gamma^{*} \gamma^{*}} (\vec{k} \; ) \otimes_{ \vec{k} } \mathcal{F} (x, \vec{k}) \; ,
\end{equation}
where $\otimes_{ \vec{k} }$ means convolution in the transverse momenta $\vec{k}$. The term $\mathcal{F}(x, \vec{k})$, known as the \textit{unintegrated gluon density} (UGD), is related to the collinear gluon parton distribution function,  $f_g$, through
\begin{equation}
    f_g (x, Q^2) = \int \frac{d^2 k}{\pi \vec{k}^{\; 2}} \mathcal{F} (x, \vec{k}) \theta (Q^2 - \vec{k}^{\; 2}) \; .
\end{equation}
From Eq.~(\ref{Int:Eq:RiseOfCrossSection}), the cross section predicted by the BFKL equation exhibits a growth pattern as
\begin{equation}
    \sigma_{\gamma^{*} P}(x) \sim \left( \frac{1}{x} \right)^{\omega_0} = \left( \frac{s}{Q^2} \right)^{\omega_0} \; .
\end{equation}
This behavior contradicts the Froissart bound, which imposes~\cite{Froissart:1961ux}
\begin{equation*}
    \sigma_{tot} < c \ln^2 s,
\end{equation*}
with $c$ as a constant. Such a violation cannot be resolved through fixed-order radiative corrections, as it reflects the unbounded rise of the UGD at low $x$. This leads to a scenario where the target becomes increasingly dense with gluons until it reaches an infinitely dense state.

To address this issue, \textit{saturation} effects must be incorporated to curb the growth. The concept was initially proposed by Gribov, Levin, and Ryskin (GLR)~\cite{Gribov:1983ivg}, introducing a non-linear modification of the BFKL equation to account for gluon recombination effects. This approach utilizes the double-logarithmic approximation, which resums terms proportional to $\alpha_s \ln(\frac{1}{x}) \ln (\frac{Q^2}{\mu^2}) $.
A more contemporary perspective on saturation physics is represented by the Balitsky-JIMWLK (B-JIMWLK) equations, derived in the Shockwave framework by Balitsky~\cite{Balitsky:2001re,Balitsky:1995ub,Balitsky:1998kc,Balitsky:1998ya} and in the Color Glass Condensate (CGC) formalism by Jalilian Marian, Iancu, McLerran, Weigert, Leonidov and Kovner~\cite{Jalilian-Marian:1997qno,Jalilian-Marian:1997jhx,Jalilian-Marian:1997ubg,Jalilian-Marian:1998tzv,Kovner:2000pt,Weigert:2000gi,Iancu:2000hn,Iancu:2001ad,Ferreiro:2001qy}. In the large-$N$ limit, this hierarchy simplifies to the Balitsky-Kovchegov (BK) equation, as recovered by Kovchegov using the dipole formalism~\cite{Kovchegov:1999yj,Kovchegov:1999ua}. Notably, in the double logarithmic limit, the GLR equation can also be retrieved.

Saturation physics introduces a new scale, the \textit{saturation scale}, given by
\begin{equation}
\label{Int:Eq:SaturationScale}
    Q_s^2 (x) \equiv \left(\frac{A}{x}\right)^{\frac{1}{3}} \Lambda_{\rm{QCD}}^2 \; ,
\end{equation}
where $A$ is the mass number of the dense target. This scale highlights that saturation effects grow not only with decreasing $x$ but also with increasing $A$, making them more prominent in large nuclei. Consequently, while very small values $x$ are needed to observe saturation in a proton, these effects become significant at higher $x$ for large nuclei.\\

The semi-classical effective framework for QCD at small $x$ serves as a well-established methodology for investigating saturation phenomena at the Large Hadron Collider and the upcoming Electron-Ion Collider. Much like the BFKL framework, achieving precision in the CGC formalism necessitates NLO accuracy. Numerous NLO computations have been conducted for inclusive processes~\cite{Roy:2019hwr,vanHameren:2023oiq,Caucal:2023fsf,Chirilli:2011km,Beuf:2021srj,Taels:2022tza,Taels:2023czt}. Beyond inclusive channels, there has recently been growing interest in exclusive~\cite{Boussarie:2016bkq,Mantysaari:2022kdm,Siddikov:2024bre,Benic:2024pqe,Boussarie:2024bdo,Boussarie:2024pax,Boussarie:2025unw} and semi-inclusive~\cite{Marquet:2009ca,Hatta:2022lzj,Iancu:2021rup} diffractive processes. In the context of diffractive processes, the CGC provides a robust tool for examining the five-dimensional Wigner distribution~\cite{Marquet:2009ca,Hatta:2022lzj}. Prominent probes for studying this distribution include:
\begin{enumerate}
    \item Diffractive dijet production~\cite{Hatta:2016dxp,Boussarie:2014lxa,Boussarie:2016ogo,Boussarie:2019ero}, $\gamma^{*} P \rightarrow {\rm jet}_1 + {\rm jet}_2 + X + P'$;
    \item Semi-inclusive diffractive deep inelastic scattering (SIDDIS)~\cite{Marquet:2009ca,Hatta:2022lzj,Fucilla:2023mkl,Fucilla:2024yfl}, $\gamma^{*} P \rightarrow h + X + P'$;
    \item Diffractive di-hadron production~\cite{Fucilla:2022wcg,Fucilla:2022szm,Fucilla:2023dgx}, $\gamma^{*} P \rightarrow h_1 + h_2 + X + P'$.
\end{enumerate} 
In this chapter, we focus on diffractive di-hadron production, presenting an initial leading-order analysis. The development of the related plots is ongoing and will be presented in a future publication. Nonetheless, we provide the explicit structure of the cross section within these pages.

\section{Diffractive di-hadron production}
\begin{figure}[t]
   \centering
   \includegraphics[scale=0.5]{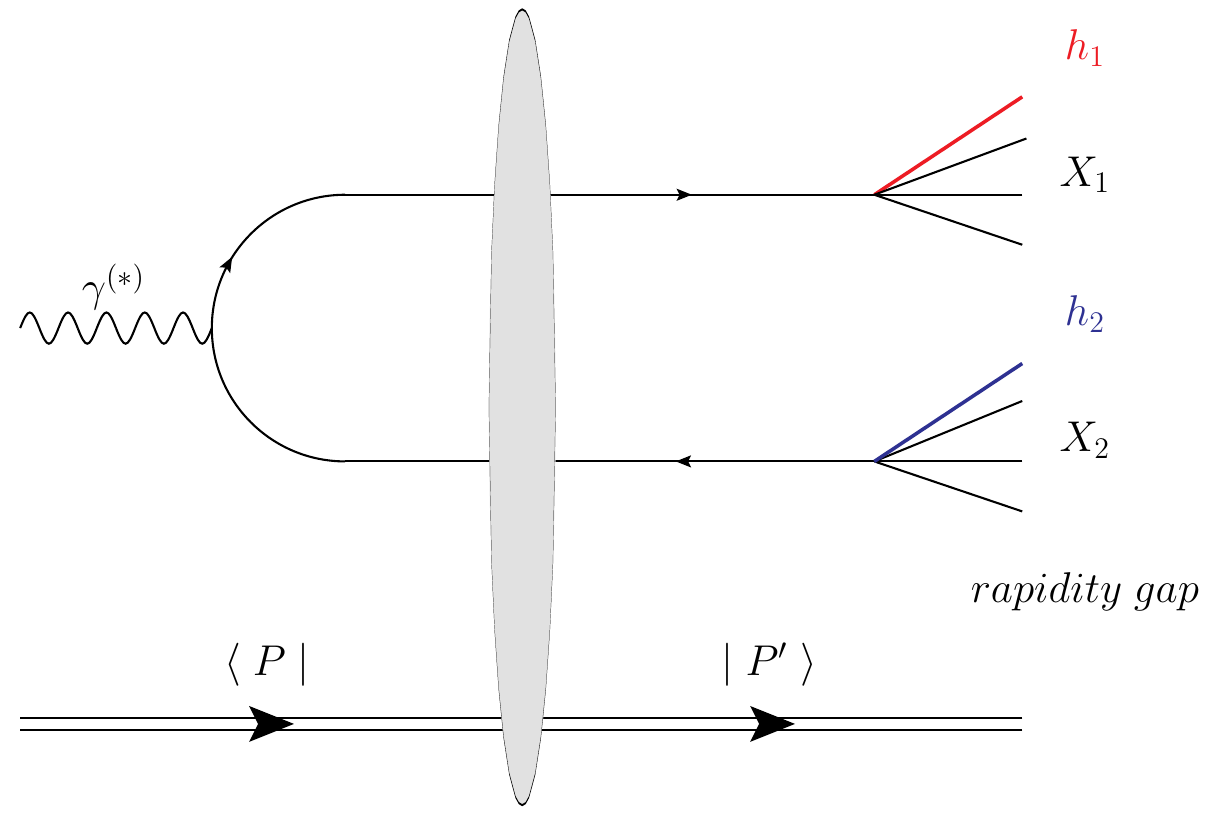}
   \caption{Amplitude of the process (\ref{process}) at LO. The grey blob symbolizes the QCD shockwave. The double line symbolizes the target, which remains intact in the figure, but could just as well break. The quark and antiquark fragment into the systems $(h_1 X_1)$ and $(h_2 X_2)$. The two tagged hadrons $h_1$ and $h_2$ are drawn in red and blue. Figure from~\cite{Fucilla:2022wcg}.}
   \label{fig:process}
\end{figure}

The primary aim of this section is to outline the framework necessary to compute the LO cross section for the semi-inclusive diffractive production of di-hadron~\cite{Fucilla:2022wcg},
\begin{equation}
\label{process}
    \gamma^{(*)}(p_\gamma) + P(p_0) \rightarrow h_1(p_{h_1}) + h_2(p_{h_2})  + X + P'(p'_0)\,,
\end{equation}
in the high-energy limit. Here, $P$ denotes a target, which can either be a nucleon or a nucleus; for simplicity, we will refer to it as a proton in the remainder of this discussion. The initial photon acts as the probing particle, also referred to as the projectile. This study considers both photoproduction scenarios, including ultraperipheral collisions, and electroproduction cases, such as those explored at EIC.
The schematic representation of this process, including the assumed separation of the diffractive system $(X h_1 h_2)$ from the outgoing nucleon or nucleus by a rapidity gap, is illustrated in Fig.~\ref{fig:process}. 

\subsubsection{Kinematics}
We adopt a light-cone basis defined by the vectors $n_1$ and $n_2$, where $n_1 \cdot n_2 = 1$, establishing the $+/-$ directions. In this framework, any four-vector can be expressed through the Sudakov decomposition as
\begin{equation}
p^\mu = p^+ n_1^\mu + p^- n_2^\mu + p_\perp^\mu,
\end{equation}
where $p^+$ and $p^-$ are the light-cone components, and $p_\perp^\mu$ represents the transverse part. The scalar product of two vectors then takes the form
\begin{equation}
    \begin{aligned}
    p \cdot q &= p^+ q^- + p^- q^+ + p_\perp \cdot q_\perp \\ 
    &= p^+ q^- + p^- q^+ -\vec{p} \cdot \vec{q}\,,
    \end{aligned} 
\end{equation}
where the transverse part is expressed as the negative scalar product of the transverse Eucledian vectors.

For simplicity, a reference frame is chosen in which the target moves with ultra-relativistic velocity. In this setup, the squared center-of-mass energy is given by $s = (p_\gamma + p_0)^2 \sim 2 p_\gamma^+ p_0^-$, satisfying the condition $s \gg \Lambda_{\text{QCD}}^2$. Moreover, $s$ dominates over all other relevant energy scales. In this frame, particles originating from the projectile predominantly move in the $n_1$ (or $+$) direction, while those associated with the target have a significant momentum component along the $n_2$ (or $-$) direction.

In this kinematic setup, the photon, with virtuality $Q$, is considered to move forward and, as a result, carries no transverse momentum. The photon momentum is expressed as
\begin{equation}
\vec{p}_{\gamma}=0,\quad p_{\gamma}^{\mu}=p_{\gamma}^{+}n_{1}^{\mu}+\frac{p_{\gamma}^{2}}{2p_{\gamma}^{+}}n_{2}^{\mu},\quad-p_{\gamma}^{2}\equiv Q^{2}\geq 0. \label{photonk}
\end{equation}
The transverse polarization of the photon is denoted as $\varepsilon_T$, while the longitudinal polarization vector is given by
\begin{equation}
\varepsilon_{L}^{\alpha}=\frac{1}{\sqrt{-p_{\gamma}^{2}}}\left(  p_{\gamma
}^{+}n_{1}^{\alpha}-\frac{p_{\gamma}^{2}}{2p_{\gamma}^{+}}n_{2}^{\alpha
}\right)  ,\quad\varepsilon_{L}^{+}=\frac{p_{\gamma}^{+}}{Q},\quad
\varepsilon_{L}^{-}=\frac{Q}{2p_{\gamma}^{+}}.
\end{equation}
The momentum of the produced hadrons $h_i$ is written as
\begin{equation}
\label{ph}
p^\mu_{h_i}=p^+_{h_i} n_1^\mu + \frac{m_{h_i}^2 + \vec{p}_{h_i}^{\,2}}{2 p^+_{h_i}} n_2^\mu + p^\mu_{h_i\perp}\quad (i=1,2) \,,
\end{equation}
where $m_{h_i}$ represents the mass of the hadron and $\vec{p}_{h_i}$ denotes its transverse momentum in Euclidean space.
For the fragmenting quark, which has a virtuality $p_q^2$, the momentum is expressed as
\begin{equation}\label{pq}
p^\mu_q=p^+_q n_1^\mu + \frac{p_q^2+\vec{p}_{q}^{\,2}}{2 p^+_{q}} n_2^\mu + p^\mu_{q\perp}\,,
\end{equation}
and a similar expression is used for an antiquark with virtuality $p_{\bar{q}}^2$:
\begin{equation}
\label{pqbar}
p^\mu_{\bar{q}}=p^+_{\bar{q}} n_1^\mu + \frac{p^2_{\bar{q}}+\vec{p}_{\bar{q}}^{\,2}}{2 p^+_{\bar{q}}} n_2^\mu + p^\mu_{\bar{q}\perp}\,.
\end{equation}
From this point onward, the notation $p_{ij} = p_i - p_j$ is adopted to represent differences between momenta.

\subsubsection{Collinear factorization}
The kinematical region under consideration~\cite{Fucilla:2022wcg} is defined by the condition $\vec{p}_{h_1}^{\,2} \sim \vec{p}_{h_2}^{\,2} \gg \Lambda_{\text{QCD}}^2$. In this context, the hadron momenta serve as the hard scale, enabling the application of perturbative QCD and collinear factorization. 
A further constraint, $\vec{p}^{\, 2} \gg \vec{p}_{h_{1,2}}^{\,2}$, is imposed, where $\vec{p}$ denotes the relative transverse momentum of the two hadrons. This ensures that the produced hadrons are sufficiently separated in angle (or, equivalently, have a large invariant mass), thereby eliminating the need to account for di-hadron unpolarized fragmentation functions. Each hadron is assumed to result from two distinct fragmentation cascades that are well separated.

After collinear factorization, the quark and antiquark in the hard subprocess are treated as on-shell particles. To facilitate further computations, the longitudinal momentum fractions $x_q$ and $x_{\bar{q}}$ are introduced, defined as
\begin{equation}
\label{xq-xqbar}
p_q^+ = x_q p^+_{\gamma} \quad \text{and} \quad p_{\bar{q}}^+ = x_{\bar{q}} p^+_{\gamma}\,.
\end{equation}
Similarly, the longitudinal momentum fraction of each produced hadron is expressed as
\begin{equation}
\label{xh}
p_{h_i}^+ = x_{h_i} p^+_{\gamma} \,.
\end{equation}

\subsubsection{Shockwave approach}
The shockwave formalism offers a powerful tool for studying gluonic saturation at high energies. To set the stage, the key technical elements necessary for understanding this framework are outlined below~\cite{Fucilla:2022wcg}.

In the context of the photon impact factor, the gluonic field $A$ is divided into external background fields $b$ and internal fields $\mathcal{A}$, depending on their $+$-momentum. Fields with $+$-momentum below the rapidity cut-off $e^\eta p_\gamma^+$ (where $\eta < 0$) are classified as external, while those above the cut-off are considered internal. The light-cone gauge $n_2 \cdot A = 0$ is employed throughout.

The external fields, after being subjected to a strong boost from the target rest frame to the current reference frame, assume the form
\begin{equation}
    b^\mu (x) = b^-(x_\perp) \delta (x^+) n_2^\mu \,.
\end{equation}

The resummation of all-order interactions with these external fields results in the formation of a high-energy Wilson line. This Wilson line, which represents the shockwave, is localized at $x^- = 0$ and is expressed as
\begin{equation}
    U_{\vec{z}} = \mathcal{P} \exp \left(i g \int d z^+ b^-(z)\right)\,,
\end{equation}
where $\mathcal{P}$ denotes the path-ordering operator commonly used in such calculations.

The small-$x$ factorization framework is applicable in this context, where the scattering amplitude is expressed as the convolution of the projectile impact factor with the non-perturbative matrix elements of Wilson line operators acting on the target states. 
One of these operators is the dipole operator, which, in the fundamental representation of $SU(N_c)$, is given by
\begin{equation}
\left[\operatorname{Tr} \left(U_1 U_2^\dag\right)-N_c\right]\left(\vec{p_1},\vec{p}_2\right) = \int d^d \vec{z}_{1} d^d \vec{z}_{2\perp} e^{- i \vec{p}_1 \cdot \vec{z}_1} e^{- i \vec{p}_2 \cdot \vec{z}_2} \left[\operatorname{Tr} \left(U_{\vec{z}_1} U_{\vec{z}_2}^\dag\right)-N_c\right]\,,
\end{equation}
where $\vec{z}_{1,2}$ represent the transverse positions of the quark and antiquark produced from the photon, and $\vec{p}_{1,2}$ are their respective transverse momenta acquired from the shockwave.

The matrix elements of the proton are parameterized using a general function $F$, as defined in Ref.~\cite{Boussarie:2016ogo},
\begin{eqnarray}
\left\langle P^{\prime}\left(p_{0}^{\prime}\right)\left|T\left(\operatorname{Tr}\left(U_{\frac{z_{\perp}}{2}} U_{-\frac{z_{\perp}}{2}}^{\dagger}\right)-N_{c}\right)\right| P\left(p_{0}\right)\right\rangle
& \equiv & 2 \pi \delta\left(p_{00^{\prime}}^{-}\right) F_{p_{0 \perp} p_{0 \perp}^{\prime}}\left(z_{\perp}\right) \nonumber \\
   & \equiv & 2 \pi \delta\left(p_{00^{\prime}}^{-}\right) F\left(z_{\perp}\right) ,
\label{Eq:Fzperp}
\end{eqnarray}
and the corresponding Fourier transform (FT) is written as
\begin{equation}
\label{eq: FT F}
\int d^{d} z_{\perp} e^{i\left(z_{\perp} \cdot p_{\perp}\right)} F\left(z_{\perp}\right) \equiv \mathbf{F}\left(p_{\perp}\right).
\end{equation}

\subsection{LO cross section}
In the framework of QCD collinear factorization, the total cross section at leading-twist and leading order is given by the expression (see Ref.~\cite{Collins:2011zzd}, chap.~12, and Ref.~\cite{Altarelli:1979kv}):
\begin{equation}
\frac{d \sigma_{0JI}^{h_1 h_2}}{d x_{h_1} d x_{h_2}} = \sum_{q} \int_{x_{h_1}}^1 \frac{d x_q}{x_q} \int_{x_{h_2}}^1 \frac{d x_{\bar{q}}}{x_{\bar{q}}} D_q^{h_1}\left(\frac{x_{h_1}}{x_q},\mu_F\right) D_{\bar{q}}^{h_2}\left(\frac{x_{h_2}}{x_{\bar{q}}}, \mu_F\right) \frac{d\hat{\sigma}_{JI}}{d x_q d x_{\bar{q}}} + (h_1 \leftrightarrow h_2) \; ,
\label{eq:coll_facto}
\end{equation}
where $q$ indicates the quark flavor types ($q=u,d,s,c,b$), and $J,I=L,T$ refer to the photon polarization. Specifically, $J$ labels the polarization in the complex conjugated amplitude, while $I$ refers to the polarization in the amplitude itself, as this expression involves the modulus square of the amplitude. The variables $x_q$ and $x_{\bar{q}}$ represent the longitudinal momentum fractions of the photon carried by the fragmenting quark and antiquark, respectively. Similarly, $x_{h_1}$ and $x_{h_2}$ denote the longitudinal momentum fractions of the photon carried by the produced hadrons. The parameter $\mu_F$ is the factorization scale, $D_{q(\bar{q})}^h$ represents the FF for a quark (antiquark) producing a hadron, and $d\hat{\sigma}$ corresponds to the partonic cross section, which describes the subprocess
\begin{equation}
\label{partonic_LO}
     \gamma^{(*)}(p_\gamma) + P(p_0) \rightarrow q (p_q) + \bar{q}(p_{\bar{q}}) + P'(p'_0)\,.
\end{equation}
Our approach employs a hybrid method combining high-energy and collinear factorization, where the fixed-order cross section in Eq.~(\ref{eq:coll_facto}) is replaced with a high-energy resummed one. Here, we focus on the photoproduction cross section, which can be experimentally investigated at the LHC in Ultra Peripheral Collisions. In this context, the photon is restricted to transverse polarization. Starting from Eq.~(2.32) of Ref.~\cite{Fucilla:2022wcg},
\begin{equation*}
\frac{d\sigma_{0TT}^{h_1 h_2}}{dx_{h_1} dx_{h_2} d^d \! p_{h_1\perp} d^d \! p_{h_2\perp}} 
= \frac{\alpha_{\mathrm{em}}}{(2\pi)^{4(d-1)} N_c} 
\sum_q \int^1_{x_{h_1}} \frac{dx_q}{x_q} \int^1_{x_{h_2}} \frac{dx_{\bar{q}}}{x_{\bar{q}}} 
\left( \frac{x_q}{x_{h_1}} \right)^d \left( \frac{x_{\bar{q}}}{x_{h_2}} \right)^d 
\end{equation*}
\begin{equation*}
\times \delta(1 - x_q - x_{\bar{q}}) Q_q^2 
D_q^{h_1} \left( \frac{x_{h_1}}{x_q} \right) 
D_{\bar{q}}^{h_2} \left( \frac{x_{h_2}}{x_{\bar{q}}} \right) 
\mathcal{F}_{TT} + (h_1 \leftrightarrow h_2),
\end{equation*}
we introduce the explicit GBW parametrization~\cite{Golec-Biernat:1998zce} for the target matrix element, \( F(z_{\perp}) \), as defined in Eq.~(\ref{Eq:Fzperp}):
\begin{align}
    F\left(z_{\perp}\right) = N_c \sigma_0 \left(1- e^{-\frac{\Vec{z}^2}{4 R_0^2 }}\right) \;,
\end{align}
with 
\begin{align}
    R_0 & = \frac{1}{Q_0} \left(\frac{x_P}{x_0}\right)^{\frac{\lambda}{2}} \,,\quad
    x_P = \frac{Q^2 + M^2 -t}{Q^2 + s } \;,\nonumber \\
    Q_0=1 \mathrm{GeV}, \quad \sigma_0& =23.03 \mathrm{mb}, \quad \lambda=0.288, \quad x_0=3.04 \times 10^{-4} \;.
\end{align}
In the back-to-back limit of the two hadrons, we obtain
\begin{equation}
    \begin{split}
& \frac{d \sigma_{0 TT}^{q\bar{q} \rightarrow h_1 h_2 } }{d y_1 d y_2  d |p_{h}|} \left(q, s, y_1, y_2, |p_{h}|, \mu^2 \right)  \\ 
&=  \frac{\alpha_{em}(\mu^2) e^{-y_1} e^{-y_2} s}{\pi }Q_q^2 N_c\sigma_0^2\int_{x_{h_1}}^{1-x_{h_2} } d x_q  D_q^{h_1} \left( \frac{x_{h_1}}{x_q}\right) D_{\bar{q}}^{h_2} \left( \frac{x_{h_2}}{1- x_q}\right) 
 \\ &\times \frac{x_q (1-x_q) 
 (2 x_q^2 -2 x_q +1)}{C^2}\int d^2 \vec z_{1}  \int d^2 \vec z_{2} \frac{\vec{z}_1\cdot\vec{z}_2}{z_1^2 z_2^2} (1-e^{-A z_{1}^2}) (1-e^{-A z_{2}^2})\\
 &\times J_0 \left[ |\vec z_1-\vec z_2| \right] + ( h_1 \leftrightarrow h_2)\, ,
\end{split}
\end{equation}
where
\begin{align}
    &C=\left(\frac{x_q}{2 x_{h_1}} + \frac{1-x_q}{2 x_{h_2}} \right)|p_{h}|\;,\\
    &A=\frac{Q_0^2 x_0^\lambda}{4 x_q^\lambda(1-x_q)^\lambda(e^{-y_1}+e^{-y_2})^{2\lambda}C^2} \; .
\end{align}
The integration over the variables \(z_{1}\) and \(z_{2}\) can be performed analytically, significantly reducing the complexity of the expression. As a result, the expression is streamlined into a more manageable form:
\begin{equation}
    \begin{split}
& \frac{d \sigma_{0 TT}^{q\bar{q} \rightarrow h_1 h_2 } }{d y_1 d y_2  d |p_{h}|} \left(q, s, y_1, y_2, |p_{h}|, \mu^2 \right)  \\
 &=  4Q_q^2 N_c\sigma_0^2\int_{x_{h_1}}^{1-x_{h_2} } d x_q \frac{\alpha_{em}(\mu^2) \pi e^{-y_1} e^{-y_2} s}{e^{{1}/{2A}} } D_q^{h_1} \left( \frac{x_{h_1}}{x_q}\right) D_{\bar{q}}^{h_2} \left( \frac{x_{h_2}}{1- x_q}\right) 
 \\&\times \frac{x_q (1-x_q) 
 (2 x_q^2 -2 x_q +1)}{C^2} + ( h_1 \leftrightarrow h_2)\,.
\end{split}
\end{equation}
This is the final expression that serves as the starting point for beginning the numerical analysis and generating the plots. This work is still in progress.

\section{Conclusions and outlook}
To conclude this chapter, we outline an ongoing effort to extend the results presented here, originally derived in the collinear factorization framework, by considering the full transverse momentum dependence of the produced hadrons. The goal is to express the dihadron production cross-section in a form that accounts for gluon saturation effects in the target, using the dipole picture and a model for the dipole amplitude \( F(z_\perp) \) such as the GBW parameterization.

The analytical derivation reformulates the cross-section in terms of hadron rapidities and transverse momenta, and introduces angular correlations via Bessel functions. The photon is treated as real, in the context of ultra-peripheral collisions, and a sum over transverse polarizations is performed to simplify the tensor structure. The final expression involves multidimensional integrals that encode both the kinematic and dynamic structure of the process.

To complement the analytical development, a \textit{Mathematica} code has been implemented to perform the integrations and to produce numerical predictions in selected kinematic regimes. The script incorporates the relevant parton distribution and fragmentation functions, as well as the GBW dipole model. This computational framework will support a future phenomenological analysis, allowing for a comparison between saturation and non-saturation scenarios and guiding the interpretation of possible experimental observations.

\newpage 

\thispagestyle{empty} 
\mbox{} 

\newpage 
\chapter*{Conclusions}
\addcontentsline{toc}{chapter}{Conclusions} 
\markboth{CONCLUSIONS}{CONCLUSIONS} 

In the present thesis, the theoretical and computational tools developed across the chapters have provided valuable insights into high-energy QCD processes. From the refinement of the BFKL formalism to the study of Higgs impact factors, exotic matter production, and diffractive processes in the saturation regime, the results presented here aim to contribute to ongoing efforts in both theoretical and experimental research.\\

The first chapter reviews the BFKL formalism, starting with Regge theory and the concept of Reggeization to describe high-energy scattering. It covers the gluon Regge trajectory, factorization of amplitudes, and the BFKL equation in the leading (LLA) and next-to-leading (NLLA) logarithmic approximations, including quasi-multi-Regge kinematics (QMRK).\\

In the second chapter, the real corrections to the Higgs impact factor at next-to-leading order, considering a physical top-quark mass, have been presented. The computations were conducted using the framework of high-energy factorization and involved the inclusion of both quark- and gluon-initiated contributions. These results are expected to improve the theoretical description of forward Higgs boson production and provide crucial insights into Higgs phenomenology.

The findings highlighted the importance of incorporating finite-top-mass effects in order to achieve higher precision. The leading-order impact factor was revisited, and the next-to-leading-order real corrections to the impact factor were computed. The computations demonstrated consistency with previous works conducted in the infinite-top-mass approximation, while also emphasizing significant modifications introduced by the finite-top-mass dependence.

The phase-space integrals were regularized to handle divergences associated with collinear, soft, and rapidity regions. The infrared divergences present in the real corrections are expected to cancel with those in the virtual corrections. Similarly, the rapidity divergences, originating from the separation of multi-Regge and quasi-multi-Regge kinematics, were addressed using appropriate subtraction terms.

The asymptotic behavior of the real contributions was analyzed in the collinear, soft, and rapidity limits. In particular, it was observed that the soft divergences were confined to specific diagrams, and their contribution was factorized to facilitate cancellation against corresponding virtual terms. Additionally, the rapidity divergence was managed through the inclusion of the BFKL counter-term, ensuring the finiteness of the final result.

Gauge invariance was verified explicitly by demonstrating the cancellation of contributions in the limit where the Reggeon transverse momentum approached zero. This crucial check confirmed the consistency of the computation and the robustness of the theoretical framework employed.\\

In the third chapter, the semi-inclusive production of exotic matter in the semi-hard regime was investigated, focusing on the production of doubly bottomed and fully bottomed tetraquarks. A hybrid factorization framework was employed, combining collinear and BFKL dynamics to describe the relevant processes. Variable-Flavor Number Schemes (VFNS) were introduced as a tool to handle the transition between different kinematic regimes, allowing heavy quarks to be treated in a consistent manner across varying energy scales.

Fragmentation functions specifically tailored for tetraquark states were constructed, and two updated parametrizations, \texttt{TQHL1.1} and \texttt{TQ4Q1.1}, were developed. These functions were shown to enable precise predictions of tetraquark production in collider experiments at 14~TeV and 100~TeV energies.

The production mechanisms of tetraquarks in association with jets were studied in the context of hybrid factorization at NLLA/NLO$^+$ accuracy. Azimuthal coefficients were derived using BFKL resummation, and the role of gluon and heavy-quark fragmentation channels was highlighted. Observables such as rapidity-interval and transverse-momentum distributions were analyzed to explore their sensitivity to the underlying dynamics.

The findings of this chapter were intended to pave the way for future experimental and theoretical studies on exotic matter. The constructed fragmentation functions and hybrid factorization approach were demonstrated to provide a robust framework for investigating the dynamics of exotic hadrons at high-energy colliders.\\

In the last chapter, the diffractive di-hadron production process in the saturation regime was analyzed within the framework of QCD at high energies. The semi-classical effective theory, based on the Color Glass Condensate (CGC) formalism, was utilized to describe the saturation effects and their influence on the cross section. The analysis began with an introduction to the saturation regime, emphasizing the role of the saturation scale, $Q_s$, as a key parameter in describing the density of gluons in the target. The diffractive di-hadron production process was then formulated within the framework of collinear factorization, with the kinematical setup carefully outlined. To facilitate the computation of cross sections, the light-cone basis and Sudakov decomposition were applied, allowing for a systematic treatment of momentum components. The shockwave formalism was employed to resum all-order interactions with external gluonic fields, leading to the factorization of the scattering amplitude into projectile and target components using the small-$x$ framework. Furthermore, the LO cross section for diffractive di-hadron production was derived, incorporating fragmentation functions and partonic cross sections. The dependency on the transverse momenta and rapidity gaps was highlighted, and the resulting expression serves as the starting point for numerical studies.

These results provide the theoretical foundation for understanding diffractive processes in the saturation regime, with applications to experimental measurements at current and future facilities, such as the EIC. Future work will focus on the numerical evaluation of the cross section and the development of phenomenological predictions, including the inclusion of NLO corrections for improved precision.

\newpage 

\thispagestyle{empty} 
\mbox{} 

\newpage 

\listoffigures  
\addcontentsline{toc}{chapter}{List of Figures} 

\printbibliography
\addcontentsline{toc}{chapter}{Bibliography} 

\end{document}